\documentclass[11pt]{article}

\usepackage[T1]{fontenc}
\usepackage{lmodern}
\usepackage[letterpaper,margin=.85in]{geometry}
\usepackage{amsmath,amssymb,mathtools}
\usepackage{dsfont}
\numberwithin{equation}{section}
\usepackage{booktabs,array}
\usepackage{enumitem}
\usepackage{microtype}
\usepackage{graphicx,adjustbox,etoolbox}
\usepackage{float}
\usepackage{placeins}
\usepackage[usenames,dvipsnames,svgnames]{xcolor}
\usepackage{xurl}
\usepackage[sort,compress]{cite}
\usepackage{hyperref}
\usepackage{tikz}
\usetikzlibrary{arrows.meta,positioning,fit,calc,backgrounds,shapes.geometric}

\hypersetup{
  colorlinks=true,
  linkcolor=MidnightBlue,
  citecolor=MidnightBlue,
  urlcolor=RoyalBlue
}
\setlist{itemsep=.2em,topsep=.35em,leftmargin=2em}

\makeatletter

\newcommand{\sdpaper@authorlist}{}
\newcommand{\sdpaper@affiliationlist}{}
\newcommand{\sdpaper@emaillist}{}
\newif\ifsdpaper@firstemail
\sdpaper@firstemailtrue
\renewcommand{\author}[1]{\g@addto@macro\sdpaper@authorlist{#1\space}}
\newcommand{\affiliation}[1]{\g@addto@macro\sdpaper@affiliationlist{#1\par}}
\newcommand{\emailAdd}[1]{%
  \ifsdpaper@firstemail
    \gdef\sdpaper@emaillist{\sdpaper@email{#1}}%
    \global\sdpaper@firstemailfalse
  \else
    \g@addto@macro\sdpaper@emaillist{,\space\sdpaper@email{#1}}%
  \fi}
\newcommand{\sdpaper@email}[1]{%
  \href{mailto:#1}{\textcolor{ForestGreen!75!black}{\texttt{#1}}}}
\renewcommand{\maketitle}{%
  \begingroup
  \thispagestyle{empty}%
  \setlength{\parindent}{0pt}%
  \setlength{\parskip}{0pt}%
  \noindent\hspace*{.04\linewidth}%
  \begin{minipage}{.92\linewidth}
    \raggedright
    {\sffamily\bfseries\boldmath\fontsize{20}{24}\selectfont \@title\par}
    \vspace{1.25cm}
    \hrule height 1.15pt
    \vspace{1.05cm}
    {\sffamily\bfseries\large \sdpaper@authorlist\par}
    \vspace{.45em}
    {\itshape \sdpaper@affiliationlist}
    \vspace{.35em}
    {\itshape E-mail:}\enspace \sdpaper@emaillist\par
  \end{minipage}
  \par
  \endgroup}
\makeatother

\renewenvironment{abstract}{%
  \par\vspace{1.65em}
  \noindent\hspace*{.04\linewidth}%
  \begin{minipage}{.92\linewidth}
  \setlength{\parindent}{0pt}%
  \setlength{\parskip}{0pt}%
  \noindent\textsc{Abstract:}\quad\ignorespaces
}{%
  \par\end{minipage}\par
}

\newcommand{\sixj}{6j}

\newcommand{\NP}{N_{\mathrm P}}
\newcommand{\Nsix}{N_{6j}}

\newsavebox{\crawPillowTenElevenViewsBox}
\newsavebox{\crawOpenSixjOneNineViewsBox}
\newsavebox{\crawOpenSixjKFourOutputsBox}
\newsavebox{\crawPillowTwelveNineteenViewsBox}
\newsavebox{\crawPillowTwelveNineteenOutputsBox}
\newsavebox{\crawFMoveThreeClassMapBox}
\newsavebox{\crawStepNinePillowRawSurvivorsBox}

\newif\ifSDCornerLabels
\newif\ifcrawShowOriginalNumbers
\SDCornerLabelsfalse
\crawShowOriginalNumberstrue

\pgfkeys{
  /sd figure labels/.is family,
  /sd figure labels,
  position/.is choice,
  position/original/.code={\SDCornerLabelsfalse},
  position/upper-left/.code={\SDCornerLabelstrue},
  source numbers/.is if=crawShowOriginalNumbers
}
\newcommand{\SDFigureLabels}[1]{%
  \pgfkeys{/sd figure labels/.cd,#1}}

\newcommand{\crawOriginalNumber}[1]{%
  \ifcrawShowOriginalNumbers
    \ifblank{#1}{}{\node[craw number] at (-5.18,1.87) {#1};}%
  \fi}

\newcommand{\SDSurfaceTitle}[1]{%
  \ifSDCornerLabels
    \node[craw note,font=\scriptsize\bfseries,text=black!78,
      anchor=west,inner sep=0pt] at (-5.18,1.87) {#1};%
  \else
    \node[craw note,font=\scriptsize\bfseries,text=black!78]
      at (0,-1.94) {#1};%
  \fi}

\newcommand{\SDThetaTitle}[1]{%
  \ifSDCornerLabels
    \node[sdemb theta title,anchor=south west] at (0,2.90) {#1};%
  \else
    \node[sdemb theta title,anchor=south] at (2,2.90) {#1};%
  \fi}

\newcommand{\SDKFourTitle}[1]{%
  \ifSDCornerLabels
    \node[sdemb kfour title,anchor=west,inner sep=0pt] at (-2.55,1.98) {#1};%
  \else
    \node[sdemb kfour title] at (0,1.98) {#1};%
  \fi}

\newcommand{\SDMiniTitle}[1]{%
  \ifSDCornerLabels
    \node[font=\scriptsize\bfseries,anchor=west,inner sep=0pt] at (-2.55,1.99) {#1};%
  \else
    \node[font=\scriptsize\bfseries] at (0,1.99) {#1};%
  \fi}

\definecolor{CZeroLeft}{RGB}{28,28,28}
\definecolor{CZeroRight}{RGB}{50,165,35}
\definecolor{CZeroSurfaceEdge}{RGB}{130,130,130}
\definecolor{CZeroSurfaceFill}{RGB}{244,239,224}

\tikzset{
  c2sheet/.style={draw=CZeroSurfaceEdge,fill=CZeroSurfaceFill,
    line width=1.05pt,line cap=round,line join=round},
  c2rim/.style={draw=CZeroSurfaceEdge,fill=white,line width=.95pt},
  c2torusrim/.style={draw=CZeroSurfaceEdge,fill=white,
    line width=1.02pt,line cap=round,line join=round},
  c2seamfront/.style={draw=CZeroSurfaceEdge,line width=1.02pt,
    line cap=round,line join=round},
  c2seamback/.style={draw=CZeroSurfaceEdge,line width=1.02pt,
    dash pattern=on .25pt off 2.05pt,line cap=round,line join=round},
  c2left/.style={draw=CZeroLeft,line width=1.12pt,
    line cap=round,line join=round},
  c2right/.style={draw=CZeroRight,line width=1.12pt,
    line cap=round,line join=round},
  c2leftback/.style={draw=CZeroLeft,line width=1.22pt,
    dash pattern=on 3.6pt off 2.2pt,line cap=round,line join=round},
  c2rightback/.style={draw=CZeroRight,line width=1.22pt,
    dash pattern=on 3.6pt off 2.2pt,line cap=round,line join=round},
  c2label/.style={font=\scriptsize,inner sep=.8pt},
  c2tiny/.style={font=\tiny,inner sep=.6pt},
  c2cardtitle/.style={font=\footnotesize\bfseries,align=center},
  c2reject/.style={draw=BrickRed,line width=1.25pt,line cap=round},
  c2pair/.style={<->,>=Latex,draw=ForestGreen,line width=.8pt},
  c2pairlabel/.style={font=\scriptsize,fill=white,inner sep=1pt,
    text=ForestGreen!65!black},
  c2highlight/.style={draw=BurntOrange,line width=.8pt,
    rounded corners=2pt},
  c2annotation/.style={draw=Purple,line width=.72pt,
    rounded corners=2pt}
}

\newcommand{\CZeroClosedStandard}{%
  \draw[c2leftback]
    (CZeroTopRear)--(-.38,1.42)
    ..controls(-2.20,1.50)and(-2.20,-1.50)..(-.38,-1.42)--(CZeroBottomRear);
  \draw[c2rightback]
    (CZeroBottomRear)--(.38,-1.42)
    ..controls(2.20,-1.50)and(2.20,1.50)..(.38,1.42)--(CZeroTopRear);
  \draw[c2left]
    (.159,1.20)--(-.38,1.20)
    ..controls(-1.62,1.20)and(-1.70,.45)..(-.38,.20)--(.153,.20);
  \draw[c2right]
    (.153,.20)--(.38,.20)
    ..controls(1.70,.45)and(1.62,1.20)..(.38,1.20)--(.159,1.20);
  \draw[c2left]
    (.153,-.20)--(-.38,-.20)
    ..controls(-1.70,-.45)and(-1.62,-1.20)..(-.38,-1.20)--(.159,-1.20);
  \draw[c2right]
    (.159,-1.20)--(.38,-1.20)
    ..controls(1.62,-1.20)and(1.70,-.45)..(.38,-.20)--(.153,-.20);
}

\newcommand{\CZeroClosedVariant}[1]{%
  \ifcase#1\relax
    \CZeroClosedStandard
  \or 
    \draw[c2leftback] (CZeroTopRear)--(-.38,1.42)
      ..controls(-2.20,1.50)and(-2.20,-1.50)..(-.38,-1.42)--(CZeroBottomRear);
    \draw[c2rightback] (CZeroBottomRear)--(.38,-1.42)
      ..controls(2.20,-1.50)and(2.20,1.50)..(.38,1.42)--(CZeroTopRear);
    \draw[c2left] (.17,1.30)--(-.38,1.30)
      ..controls(-1.58,1.25)and(-1.66,.48)..(-.38,.20)--(-.05,.20)
      ..controls(.52,.20)and(.55,.02)..(.16,.02);
    \draw[c2right] (.16,.02)--(.38,.02)
      ..controls(1.62,.40)and(1.60,1.25)..(.38,1.30)--(.17,1.30);
    \draw[c2left] (.11,-.20)--(-.38,-.20)
      ..controls(-1.70,-.45)and(-1.62,-1.26)..(-.38,-1.10)--(.10,-1.10);
    \draw[c2right] (.10,-1.10)--(.38,-1.10)
      ..controls(1.62,-1.26)and(1.70,-.45)..(.38,-.20)--(.11,-.20);
  \or 
    \draw[c2leftback] (CZeroTopRear)--(-.38,1.42)
      ..controls(-2.20,1.50)and(-2.20,-1.50)..(-.38,-1.42)--(CZeroBottomRear);
    \draw[c2rightback] (CZeroBottomRear)--(.38,-1.42)
      ..controls(2.20,-1.50)and(2.20,1.50)..(.38,1.42)--(CZeroTopRear);
    \draw[c2left] (.17,1.30)--(-.38,1.30)
      ..controls(-1.62,1.26)and(-1.70,.45)..(-.38,.20)--(.11,.20);
    \draw[c2right] (.11,.20)--(.38,.20)
      ..controls(1.70,.45)and(1.62,1.26)..(.38,1.30)--(.17,1.30);
    \draw[c2left] (.11,-.20)--(-.38,-.20)
      ..controls(-1.65,-.42)and(-1.58,-1.22)..(-.38,-1.10)--(-.04,-1.10)
      ..controls(.52,-1.06)and(.55,-.78)..(.15,-.78);
    \draw[c2right] (.15,-.78)--(.38,-.78)
      ..controls(1.62,-.90)and(1.55,-1.20)..(.38,-1.10)--(.10,-1.10);
  \else
    \CZeroClosedStandard
  \fi
}

\newcommand{\OpenSDHalfSurface}{%
  \path[opensd sheet]
    (-3.35,.48)..controls(-2.63,.56)and(-2.48,1.63)..(-1.42,1.63)
    --(0,1.63)
    arc[start angle=90,end angle=270,x radius=.16,y radius=.38]
    ..controls(-.73,.78)and(-.73,.55)..(0,.46)
    arc[start angle=90,end angle=270,x radius=.16,y radius=.46]
    ..controls(-.73,-.55)and(-.73,-.78)..(0,-.87)
    arc[start angle=90,end angle=270,x radius=.16,y radius=.38]
    --(-1.42,-1.63)
    ..controls(-2.48,-1.63)and(-2.63,-.56)..(-3.35,-.48)--cycle;
  \draw[opensd hole] (-3.35,0) ellipse[x radius=.15,y radius=.48];
  \foreach \yc/\ry in {1.25/.38,0/.46,-1.25/.38}{
    \path[fill=CZeroSurfaceFill,draw=none]
      (0,\yc) ellipse[x radius=.16,y radius=\ry];
    \draw[opensd seam back] (0,{\yc+\ry})
      arc[start angle=90,end angle=270,x radius=.16,y radius=\ry];
    \draw[opensd seam front] (0,{\yc-\ry})
      arc[start angle=-90,end angle=90,x radius=.16,y radius=\ry];
  }
}

\newcommand{\OpenSDExternalRoot}{%
  \foreach \yy in {.28,0,-.28}{
    \draw[opensd wire] (-4.05,\yy)--(-3.20,\yy);
    \fill[RoyalBlue] (-3.20,\yy) circle[radius=.035];
  }
}

\newcommand{\OpenSDCompletionBlock}{%
  \filldraw[opensd blob] (.34,-1.70) rectangle (1.05,1.70);
  \foreach \yy in {.28,0,-.28}
    \draw[opensd wire] (1.05,\yy)--(1.75,\yy);
}

\newcommand{\OpenSDProjectedCompletionBlock}[1]{%
  \path[opensd blob]
    (.34,-1.66)
    ..controls(.17,-1.29)and(.23,-.88)..(.34,-.58)
    ..controls(.45,-.27)and(.27,.06)..(.40,.36)
    ..controls(.52,.67)and(.18,1.08)..(.39,1.46)
    ..controls(.58,1.78)and(1.02,1.73)..(1.20,1.48)
    ..controls(1.45,1.15)and(1.30,.82)..(1.43,.50)
    ..controls(1.56,.18)and(1.55,-.17)..(1.42,-.50)
    ..controls(1.29,-.83)and(1.45,-1.17)..(1.20,-1.45)
    ..controls(.98,-1.70)and(.53,-1.82)..(.34,-1.66)--cycle;
  \node[font=\small] at (.86,0) {#1};
  \foreach \yy in {.28,0,-.28}
    \draw[opensd wire] (1.44,\yy)--(2.08,\yy);
}

\newcommand{\OpenSDPillowHalf}{%
  \OpenSDHalfSurface
  \OpenSDExternalRoot
  \draw[opensd front] (-3.20,.28)
    ..controls(-2.48,.34)and(-2.25,1.45)..(-.34,1.48)--(.13,1.48);
  \draw[opensd front] (-3.20,0)
    ..controls(-2.42,.08)and(-2.16,1.24)..(-.34,1.30)--(.16,1.30);
  \draw[opensd front] (-3.20,-.28)
    ..controls(-2.48,-.34)and(-2.25,-1.45)..(-.34,-1.48)--(.13,-1.48);
  \draw[opensd front] (.13,1.02)--(-.34,1.08)
    ..controls(-1.42,1.08)and(-1.48,.36)..(-.34,.25)--(.135,.22);
  \draw[opensd front] (.16,0)--(-.34,0)
    ..controls(-1.53,-.12)and(-1.62,-1.15)..(-.34,-1.30)--(.16,-1.30);
  \draw[opensd front] (.135,-.22)--(-.34,-.25)
    ..controls(-1.38,-.39)and(-1.34,-1.06)..(-.34,-1.08)--(.13,-1.02);
  \foreach \xx/\yy in
    {.13/1.48,.16/1.30,.13/1.02,.135/.22,.16/0,.135/-.22,
     .13/-1.02,.16/-1.30,.13/-1.48}{
    \draw[opensd wire] (\xx,\yy)--(.34,\yy);
    \fill[RoyalBlue] (\xx,\yy) circle[radius=.035];
  }
  \OpenSDCompletionBlock
}

\newcommand{\OpenSDSixJHalf}{%
  \OpenSDHalfSurface
  \OpenSDExternalRoot
  \draw[opensd rear] (-.16,1.30)--(-.34,1.30)
    ..controls(-1.92,1.39)and(-1.92,-1.39)..(-.34,-1.30)--(-.16,-1.30);
  \draw[opensd front] (-3.20,.28)
    ..controls(-2.48,.34)and(-2.25,1.45)..(-.34,1.48)--(.13,1.48);
  \draw[opensd front] (-3.20,0)--(.16,0);
  \draw[opensd front] (-3.20,-.28)
    ..controls(-2.48,-.34)and(-2.25,-1.45)..(-.34,-1.48)--(.13,-1.48);
  \draw[opensd front] (.13,1.02)--(-.34,1.08)
    ..controls(-1.42,1.08)and(-1.48,.36)..(-.34,.25)--(.135,.22);
  \draw[opensd front] (.135,-.22)--(-.34,-.25)
    ..controls(-1.48,-.36)and(-1.42,-1.08)..(-.34,-1.08)--(.13,-1.02);
  \foreach \xx/\yy in
    {.13/1.48,-.16/1.30,.13/1.02,.135/.22,.16/0,.135/-.22,
     .13/-1.02,-.16/-1.30,.13/-1.48}{
    \draw[opensd wire] (\xx,\yy)--(.34,\yy);
    \fill[RoyalBlue] (\xx,\yy) circle[radius=.035];
  }
  \OpenSDCompletionBlock
}

\newcommand{\OpenBWeightedSDFigure}{%
\begin{tikzpicture}[x=1cm,y=1cm,baseline=(current bounding box.center)]
  \tikzset{
    opensd sheet/.style={draw=CZeroSurfaceEdge,fill=CZeroSurfaceFill,
      line width=.72pt,line cap=round,line join=round},
    opensd hole/.style={draw=CZeroSurfaceEdge,fill=white,line width=.72pt},
    opensd seam front/.style={draw=CZeroSurfaceEdge,line width=.68pt},
    opensd seam back/.style={draw=CZeroSurfaceEdge,line width=.68pt,
      dash pattern=on .18pt off 1.55pt,line cap=round},
    opensd front/.style={draw=black!88,line width=.92pt,
      line cap=round,line join=round,
      preaction={draw=CZeroSurfaceFill,line width=1.85pt}},
    opensd rear/.style={draw=black!88,line width=.92pt,
      dash pattern=on 3pt off 1.9pt,line cap=round,line join=round,
      preaction={draw=CZeroSurfaceFill,line width=1.85pt,solid}},
    opensd wire/.style={draw=black!88,line width=.84pt,line cap=round},
    opensd blob/.style={draw=black!46,fill=black!40,line width=.55pt}
  }

  \begin{scope}[shift={(-6.05,0)}]
    \foreach \yy in {.17,0,-.17}
      \draw[opensd wire] (-.82,\yy)--(.82,\yy);
    \filldraw[opensd blob] (-.29,-.52) rectangle (.29,.52);
  \end{scope}
  \node[font=\scriptsize] at (-6.05,.78)
    {$\langle B^{(abc)}_{IJM}B^{(cba)}_{JIM}\rangle$};

  \node[font=\large] at (-4.88,0) {$=$};
  \node[font=\normalsize] at (-4.42,0) {$-\hbar$};

  \begin{scope}[shift={(-3.35,0)}]
    \foreach \yy in {.17,0,-.17}
      \draw[opensd wire] (-.62,\yy)--(.62,\yy);
  \end{scope}
  \node[font=\scriptsize] at (-3.35,-.67) {$C_0(IJM)$};

  \node[font=\large] at (-2.50,0) {$+$};
  \begin{scope}[shift={(-.55,0)}]
    \begin{scope}[x=.88cm,y=.78cm,scale=.48]
      \OpenSDPillowHalf
    \end{scope}
  \end{scope}
  \node[font=\scriptsize\bfseries] at (-1.05,-1.03) {Pillow};

  \node[font=\large] at (.47,0) {$-$};
  \begin{scope}[shift={(2.45,0)}]
    \begin{scope}[x=.88cm,y=.78cm,scale=.48]
      \OpenSDSixJHalf
    \end{scope}
  \end{scope}
  \node[font=\scriptsize\bfseries] at (1.95,-1.03) {open-$6j$};
\end{tikzpicture}%
}

\newcommand{\OpenSDTargetCylinder}{%
  \path[opensd cylinder]
    (-1.25,.60)--(1.25,.60)
    arc[start angle=90,end angle=-90,x radius=.20,y radius=.60]
    --(-1.25,-.60)
    arc[start angle=270,end angle=90,x radius=.20,y radius=.60]--cycle;
  \draw[opensd rim back] (1.25,.60)
    arc[start angle=90,end angle=270,x radius=.20,y radius=.60];
  \draw[opensd wire] (-1.081,.32)--(1.419,.32);
  \draw[opensd wire] (-1.050,0)--(1.450,0);
  \draw[opensd wire] (-1.081,-.32)--(1.419,-.32);
  \filldraw[opensd rim,fill=white]
    (-1.25,0) ellipse[x radius=.20,y radius=.60];
  \draw[opensd rim] (1.25,-.60)
    arc[start angle=-90,end angle=90,x radius=.20,y radius=.60];
}

\newcommand{\OpenBTopologyProjectionFigure}{%
\begin{tikzpicture}[x=1cm,y=1cm,baseline=(current bounding box.center)]
  \tikzset{
    opensd sheet/.style={draw=CZeroSurfaceEdge,fill=CZeroSurfaceFill,
      line width=.72pt,line cap=round,line join=round},
    opensd hole/.style={draw=CZeroSurfaceEdge,fill=white,line width=.72pt},
    opensd seam front/.style={draw=CZeroSurfaceEdge,line width=.68pt},
    opensd seam back/.style={draw=CZeroSurfaceEdge,line width=.68pt,
      dash pattern=on .18pt off 1.55pt,line cap=round},
    opensd front/.style={draw=black!88,line width=.92pt,
      line cap=round,line join=round,
      preaction={draw=CZeroSurfaceFill,line width=1.85pt}},
    opensd rear/.style={draw=black!88,line width=.92pt,
      dash pattern=on 3pt off 1.9pt,line cap=round,line join=round,
      preaction={draw=CZeroSurfaceFill,line width=1.85pt,solid}},
    opensd wire/.style={draw=black!88,line width=.84pt,line cap=round},
    opensd blob/.style={draw=black!46,fill=black!40,line width=.55pt},
    opensd cylinder/.style={draw=CZeroSurfaceEdge,fill=CZeroSurfaceFill,
      line width=.72pt,line join=round},
    opensd rim/.style={draw=CZeroSurfaceEdge,fill=none,line width=.72pt},
    opensd rim back/.style={draw=CZeroSurfaceEdge,fill=none,line width=.72pt,
      dash pattern=on 1.8pt off 1.35pt,line cap=round}
  }

  \begin{scope}[shift={(-5.05,.28)}]
    \begin{scope}[scale=.82]
      \OpenSDTargetCylinder
    \end{scope}
  \end{scope}
  \node[font=\scriptsize,align=center] at (-5.05,1.14)
    {one target\\cylinder $C_0$};

  \node[font=\large] at (-3.35,.28) {$=$};

  \node[font=\normalsize] at (-2.72,.28) {$\displaystyle\sum_{p=1}^{\NP}$};
  \begin{scope}[shift={(-.72,.28)}]
    \begin{scope}[x=.88cm,y=.78cm,scale=.48]
      \renewcommand{\OpenSDCompletionBlock}{%
        \OpenSDProjectedCompletionBlock{$p$}}
      \OpenSDPillowHalf
    \end{scope}
  \end{scope}
  \node[font=\scriptsize,align=center] at (-.98,1.30)
    {Pillow completion\\giving $C_0$};

  \node[font=\large] at (.53,.28) {$-$};

  \node[font=\normalsize] at (1.22,.28)
    {$\displaystyle\sum_{q=1}^{\Nsix}$};
  \begin{scope}[shift={(3.25,.28)}]
    \begin{scope}[x=.88cm,y=.78cm,scale=.48]
      \renewcommand{\OpenSDCompletionBlock}{%
        \OpenSDProjectedCompletionBlock{$q$}}
      \OpenSDSixJHalf
    \end{scope}
  \end{scope}
  \node[font=\scriptsize,align=center] at (2.98,1.30)
    {open-$6j$ completion\\giving $C_0$};

  \node[font=\small,anchor=base] at (-5.05,-1.10)
    {$\displaystyle
      \underbrace{1}_{\substack{\text{one target}\\[-1pt]\text{cylinder}}}$};
  \node[font=\small,anchor=base] at (-3.35,-1.10) {$=$};
  \node[font=\small,anchor=base] at (-.98,-1.10)
    {$\displaystyle
      \underbrace{\NP}_{\substack{\text{Pillow copies}\\[-1pt]\text{of that cylinder}}}$};
  \node[font=\small,anchor=base] at (.53,-1.10) {$-$};
  \node[font=\small,anchor=base] at (2.98,-1.10)
    {$\displaystyle
      \underbrace{\Nsix}_{\substack{\text{open-}6j\text{ copies}\\[-1pt]\text{of that cylinder}}}$};
\end{tikzpicture}%
}

\providecolor{czheLeft}{RGB}{28,28,28}
\providecolor{czheRight}{RGB}{50,165,35}
\providecolor{czheEdge}{RGB}{130,130,130}
\providecolor{czheFill}{RGB}{244,239,224}

\tikzset{
  czhe sheet/.style={draw=czheEdge,fill=czheFill,line width=1.05pt,
    line cap=round,line join=round},
  czhe cuff/.style={draw=czheEdge,fill=white,line width=.95pt},
  czhe torus rim/.style={draw=czheEdge,fill=white,line width=1.02pt,
    line cap=round,line join=round},
  czhe seam fill/.style={fill=czheFill,draw=none},
  czhe seam rear/.style={draw=czheEdge,line width=1.02pt,
    dash pattern=on .12pt off 2.10pt,line cap=round,line join=round},
  czhe seam front/.style={draw=czheEdge,line width=1.02pt,
    line cap=round,line join=round},
  czhe left/.style={draw=czheLeft,line width=1.02pt,
    line cap=round,line join=round},
  czhe right/.style={draw=czheRight,line width=1.02pt,
    line cap=round,line join=round},
  czhe left rear/.style={draw=czheLeft,line width=1.08pt,
    dash pattern=on 3.6pt off 2.25pt,line cap=round,line join=round},
  czhe right rear/.style={draw=czheRight,line width=1.08pt,
    dash pattern=on 3.6pt off 2.25pt,line cap=round,line join=round},
  czhe rear handle/.style={draw=czheEdge,line width=1.05pt,
    dash pattern=on 2.2pt off 1.6pt,line cap=round,line join=round},
  czhe rear handle fill/.style={fill=czheEdge!14,draw=none},
  czhe title/.style={font=\scriptsize\bfseries,align=center}
}

\newcommand{\czheSurface}{%
  \path[czhe sheet]
    (-5.20,.52)
    ..controls(-4.20,.62)and(-3.92,1.74)..(-2.45,1.74)
    --(2.45,1.74)
    ..controls(3.92,1.74)and(4.20,.62)..(5.20,.52)
    --(5.20,-.52)
    ..controls(4.20,-.62)and(3.92,-1.74)..(2.45,-1.74)
    --(-2.45,-1.74)
    ..controls(-3.92,-1.74)and(-4.20,-.62)..(-5.20,-.52)
    --cycle;
  \draw[czhe cuff] (-5.20,0) ellipse[x radius=.17,y radius=.52];
  \draw[czhe cuff] ( 5.20,0) ellipse[x radius=.17,y radius=.52];

  \path[czhe seam fill] (0, 1.33) ellipse[x radius=.17,y radius=.41];
  \path[czhe seam fill] (0, 0)    ellipse[x radius=.17,y radius=.48];
  \path[czhe seam fill] (0,-1.33) ellipse[x radius=.17,y radius=.41];
  \draw[czhe seam rear] (0,1.74)
    arc[start angle=90,end angle=270,x radius=.17,y radius=.41];
  \draw[czhe seam front] (0,.92)
    arc[start angle=-90,end angle=90,x radius=.17,y radius=.41];
  \draw[czhe seam rear] (0,.48)
    arc[start angle=90,end angle=270,x radius=.17,y radius=.48];
  \draw[czhe seam front] (0,-.48)
    arc[start angle=-90,end angle=90,x radius=.17,y radius=.48];
  \draw[czhe seam rear] (0,-.92)
    arc[start angle=90,end angle=270,x radius=.17,y radius=.41];
  \draw[czhe seam front] (0,-1.74)
    arc[start angle=-90,end angle=90,x radius=.17,y radius=.41];

  \path (0,1.33) ++({.17*cos(38)},{.41*sin(38)})
    coordinate (czheTopOpen);
  \path (0,1.33) ++({.17*cos(8)},{.41*sin(8)})
    coordinate (czheTopGlobalFront);
  \path (0,1.33) ++({-.17*cos(8)},{.41*sin(8)})
    coordinate (czheTopGlobalRear);
  \path (0,1.33) ++({.17*cos(-38)},{.41*sin(-38)})
    coordinate (czheTopLocal);
  \path (0,0) ++({.17*cos(32)},{.48*sin(32)})
    coordinate (czheMiddleUpper);
  \coordinate (czheMiddleOpen) at (.17,0);
  \path (0,0) ++({.17*cos(-32)},{.48*sin(-32)})
    coordinate (czheMiddleLowerFront);
  \path (0,-1.33) ++({.17*cos(38)},{.41*sin(38)})
    coordinate (czheBottomLocalFront);
  \path (0,-1.33) ++({.17*cos(-8)},{.41*sin(-8)})
    coordinate (czheBottomGlobalFront);
  \path (0,-1.33) ++({-.17*cos(-8)},{.41*sin(-8)})
    coordinate (czheBottomGlobalRear);
  \path (0,-1.33) ++({.17*cos(-38)},{.41*sin(-38)})
    coordinate (czheBottomOpen);

  \draw[czhe torus rim] (0,.70)
    ellipse[x radius=.72,y radius=.22];
  \draw[czhe torus rim] (0,-.70)
    ellipse[x radius=.72,y radius=.22];
}

\newcommand{\czheEyeFour}{%
  \draw[czhe torus rim,rotate around={-18:(1.50,.50)}]
    (1.50,.50) ellipse[x radius=.22,y radius=.10];}
\newcommand{\czheEyeFive}{%
  \draw[czhe torus rim,rotate around={18:(1.55,-.48)}]
    (1.55,-.48) ellipse[x radius=.22,y radius=.10];}

\newcommand{\czheRearHandleUpper}{%
  \path[czhe rear handle fill]
    (1.25,.80)..controls(2.00,.84)and(2.08,.16)..(1.25,.20)
    --(1.48,.36)..controls(1.72,.36)and(1.72,.64)..(1.48,.64)--cycle;
  \draw[czhe rear handle]
    (1.25,.80)..controls(2.00,.84)and(2.08,.16)..(1.25,.20);
  \draw[czhe rear handle]
    (1.48,.64)..controls(1.72,.64)and(1.72,.36)..(1.48,.36);}
\newcommand{\czheRearHandleLower}{%
  \path[czhe rear handle fill]
    (1.66,-.18)..controls(2.42,-.14)and(2.48,-.86)..(1.66,-.82)
    --(1.90,-.64)..controls(2.10,-.64)and(2.10,-.36)..(1.90,-.36)--cycle;
  \draw[czhe rear handle]
    (1.66,-.18)..controls(2.42,-.14)and(2.48,-.86)..(1.66,-.82);
  \draw[czhe rear handle]
    (1.90,-.36)..controls(2.10,-.36)and(2.10,-.64)..(1.90,-.64);}

\newcommand{\czheRearThroughUpperBridge}{%
  \draw[czhe right rear,draw=RoyalBlue]
    (czheBottomGlobalRear)--(.38,-1.39)
    ..controls(2.02,-1.43)and(2.00,-.38)..(.94,.03)
    ..controls(1.05,.12)and(1.20,.22)..(1.365,.28)
    ..controls(1.90,.26)and(1.86,.74)..(1.365,.72)
    ..controls(1.10,.86)and(.80,1.25)..(.38,1.39)
    --(czheTopGlobalRear);}

\newcommand{\czheRearThroughLowerBridge}{%
  \draw[czhe right rear,draw=RoyalBlue]
    (czheBottomGlobalRear)--(.38,-1.39)
    ..controls(1.30,-1.55)and(2.40,-1.42)..(2.45,-1.00)
    ..controls(2.35,-.82)and(2.05,-.76)..(1.78,-.73)
    ..controls(2.29,-.75)and(2.26,-.25)..(1.78,-.27)
    ..controls(2.15,-.14)and(2.48,.05)..(2.50,.34)
    ..controls(2.53,.66)and(2.48,.86)..(2.30,1.05)
    ..controls(2.02,1.40)and(1.12,1.45)..(.38,1.39)
    --(czheTopGlobalRear);}

\newcommand{\czheBlackRear}{%
  \draw[czhe left rear]
    (czheTopGlobalRear)--(-.38,1.39)
    ..controls(-2.20,1.50)and(-2.20,-1.50)..(-.38,-1.39)
    --(czheBottomGlobalRear);}

\newcommand{\czheBlackFront}{%
  \draw[czhe left]
    (-5.03,.29)..controls(-3.85,.30)and(-3.10,1.60)..(-.38,1.60)
    --(czheTopOpen);
  \draw[czhe left] (-5.03,0)--(czheMiddleOpen);
  \draw[czhe left]
    (-5.03,-.29)..controls(-3.85,-.30)and(-3.10,-1.60)..(-.38,-1.60)
    --(czheBottomOpen);
  \draw[czhe left]
    (czheTopLocal)--(-.38,1.17)
    ..controls(-1.62,1.18)and(-1.70,.45)..(-.38,.28)
    --(czheMiddleUpper);
  \draw[czhe left]
    (czheMiddleLowerFront)--(-.38,-.28)
    ..controls(-1.70,-.45)and(-1.62,-1.18)..(-.38,-1.17)
    --(czheBottomLocalFront);}

\newcommand{\czheOpenGreen}{%
  \draw[czhe right,draw=ForestGreen]
    (czheTopOpen)--(.38,1.60)
    ..controls(3.10,1.60)and(3.85,.30)..(5.03,.29);
  \draw[czhe right,draw=BrickRed] (czheMiddleOpen)--(5.03,0);
  \draw[czhe right,draw=Orange]
    (czheBottomOpen)--(.38,-1.60)
    ..controls(3.10,-1.60)and(3.85,-.30)..(5.03,-.29);}

\newcommand{\czheUpperBase}{%
  \draw[czhe right,draw=Purple]
    (czheMiddleUpper)--(.38,.28)
    ..controls(1.70,.45)and(1.62,1.18)..(.38,1.17)
    --(czheTopLocal);}
\newcommand{\czheLowerBase}{%
  \draw[czhe right,draw=Magenta]
    (czheBottomLocalFront)--(.38,-1.17)
    ..controls(1.62,-1.18)and(1.70,-.45)..(.38,-.28)
    --(czheMiddleLowerFront);}

\newcommand{\czheRearFour}{%
  \draw[czhe right rear,draw=RoyalBlue]
    (czheBottomGlobalRear)--(.38,-1.39)
    ..controls(2.06,-1.44)and(2.05,-.40)..(.94,.03)
    ..controls(.78,.17)and(.79,.42)..(.90,.58)
    ..controls(1.09,.87)and(1.34,1.18)..(.38,1.39)
    --(czheTopGlobalRear);}
\newcommand{\czheRearFive}{%
  \draw[czhe right rear,draw=RoyalBlue]
    (czheBottomGlobalRear)--(.38,-1.39)
    ..controls(1.34,-1.18)and(1.09,-.87)..(.90,-.58)
    ..controls(.79,-.42)and(.78,-.17)..(.94,-.03)
    ..controls(2.05,.40)and(2.06,1.44)..(.38,1.39)
    --(czheTopGlobalRear);}

\newcommand{\czheUpperFour}{%
  \draw[czhe right,draw=Purple]
    (czheMiddleUpper)--(.38,.22)
    ..controls(.87,.07)and(1.51,.05)..(1.93,.15)
    ..controls(2.20,.23)and(2.22,.60)..(1.98,.82)
    ..controls(1.61,1.18)and(.82,1.21)..(.38,1.17)
    --(czheTopLocal);}
\newcommand{\czheLowerFive}{%
  \draw[czhe right,draw=Magenta]
    (czheBottomLocalFront)--(.38,-1.17)
    ..controls(.82,-1.21)and(1.61,-1.18)..(1.98,-.82)
    ..controls(2.22,-.60)and(2.20,-.23)..(1.93,-.15)
    ..controls(1.51,-.05)and(.87,-.07)..(.38,-.22)
    --(czheMiddleLowerFront);}

\newcommand{\czheLowerAroundRearBridge}{%
  \draw[czhe right,draw=Magenta]
    (czheBottomLocalFront)--(.38,-1.17)
    ..controls(.85,-1.25)and(1.85,-1.23)..(2.45,-.92)
    ..controls(2.82,-.70)and(2.82,-.30)..(2.45,-.08)
    ..controls(1.85,.03)and(.85,-.02)..(.38,-.22)
    --(czheMiddleLowerFront);}

\newcommand{\czheEndpointCell}[2]{%
  \begin{minipage}[t]{.485\linewidth}\centering
  \begin{tikzpicture}[x=.555cm,y=.555cm,
      baseline=(current bounding box.center)]
    \path[use as bounding box] (-5.72,-2.14) rectangle (5.72,2.17);
    \czheSurface
    \ifnum#1=4\relax\czheEyeFour\else\czheEyeFive\fi
    \czheBlackRear
    \ifnum#1=4\relax\czheRearFour\else\czheRearFive\fi
    \czheBlackFront
    \czheOpenGreen
    \ifnum#1=4\relax
      \czheUpperFour\czheLowerBase
    \else
      \czheUpperBase\czheLowerFive
    \fi
    \node[czhe title,anchor=south west] at (-5.18,1.82) {#2};
  \end{tikzpicture}
  \end{minipage}%
}

\newcommand{\CZeroHandleEndpointFour}{%
  \czheEndpointCell{4}{(a) 6j-7-3: view I}}
\newcommand{\CZeroHandleEndpointFive}{%
  \czheEndpointCell{5}{(e) 6j-7-3: view II}}

\newcommand{\czheMigrationMiddleCell}[2]{%
  \begin{minipage}[t]{.485\linewidth}\centering
  \begin{tikzpicture}[x=.555cm,y=.555cm,
      baseline=(current bounding box.center)]
    \path[use as bounding box] (-5.72,-2.14) rectangle (5.72,2.17);
    \czheSurface
    \ifnum#1=2\relax\czheRearHandleUpper\else\czheRearHandleLower\fi
    \czheBlackRear
    \ifnum#1=2\relax
      \czheRearThroughUpperBridge
    \else
      \czheRearThroughLowerBridge
    \fi
    \czheBlackFront
    \czheOpenGreen
    \ifnum#1=2\relax
      \czheUpperFour\czheLowerBase
    \else\ifnum#1=3\relax
      \czheUpperBase\czheLowerBase
    \else
      \czheUpperBase\czheLowerAroundRearBridge
    \fi\fi
    \node[czhe title,anchor=south west] at (-5.18,1.82) {#2};
  \end{tikzpicture}
  \end{minipage}%
}

\newcommand{\CZeroHandleStageTwo}{%
  \czheMigrationMiddleCell{2}{(b) lift the handle behind the surface}}
\newcommand{\CZeroHandleStageThree}{%
  \czheMigrationMiddleCell{3}{(c) carry the rear handle below}}
\newcommand{\CZeroHandleStageFour}{%
  \czheMigrationMiddleCell{4}{(d) move the lower loop around it}}

\providecolor{czrmWarm}{RGB}{244,239,224}
\providecolor{czrmGray}{RGB}{132,132,132}
\providecolor{czrmGreen}{RGB}{50,165,35}

\tikzset{
  czrm surface/.style={draw=czrmGray,fill=czrmWarm,line width=.9pt,
    line cap=round,line join=round},
  czrm hole/.style={draw=czrmGray,fill=white,line width=.82pt,
    line cap=round,line join=round},
  czrm cuff/.style={draw=czrmGray,fill=white,line width=.82pt},
  czrm torus rim/.style={draw=czrmGray,fill=white,line width=.88pt,
    line cap=round,line join=round},
  czrm seam rear/.style={draw=czrmGray,line width=.82pt,
    dash pattern=on 1.2pt off 1.6pt,line cap=round},
  czrm seam front/.style={draw=czrmGray,line width=.82pt,
    line cap=round},
  czrm left/.style={draw=black!88,line width=1.02pt,
    line cap=round,line join=round},
  czrm right/.style={draw=czrmGreen,line width=1.02pt,
    line cap=round,line join=round},
  czrm left back/.style={czrm left,dash pattern=on 3.5pt off 2.2pt},
  czrm right back/.style={czrm right,dash pattern=on 3.5pt off 2.2pt},
  czrm title/.style={font=\scriptsize\bfseries,align=center},
  czrm note/.style={font=\tiny,align=center,text=black!68},
  czrm arrow/.style={-{Latex[length=2mm,width=1.2mm]},
    draw=black!55,line width=.65pt}
}

\newcommand{\czrmSurface}{%
  \path[czrm surface]
    (-3.18,.46)
    ..controls(-2.55,.55)and(-2.42,1.48)..(-1.48,1.55)
    --(1.48,1.55)
    ..controls(2.42,1.48)and(2.55,.55)..(3.18,.46)
    --(3.18,-.46)
    ..controls(2.55,-.55)and(2.42,-1.48)..(1.48,-1.55)
    --(-1.48,-1.55)
    ..controls(-2.42,-1.48)and(-2.55,-.55)..(-3.18,-.46)--cycle;
}

\newcommand{\czrmCentralCuff}[2]{%
  \fill[fill=czrmWarm] (0,#1) ellipse[x radius=.15,y radius=#2];
  \draw[czrm seam rear]
    (0,{#1+#2}) arc[start angle=90,end angle=270,
      x radius=.15,y radius=#2];
  \draw[czrm seam front]
    (0,{#1-#2}) arc[start angle=-90,end angle=90,
      x radius=.15,y radius=#2];
}

\newcommand{\czrmCuffs}{%
  \draw[czrm cuff] (-3.18,0) ellipse[x radius=.13,y radius=.46];
  \draw[czrm cuff] ( 3.18,0) ellipse[x radius=.13,y radius=.46];
  \czrmCentralCuff{ 1.215}{.335}
  \czrmCentralCuff{ 0}{.22}
  \czrmCentralCuff{-1.215}{.335}
}

\newcommand{\czrmTorusRims}{%
  \draw[czrm torus rim]
    (0,.22) ..controls(-.72,.22)and(-.72,.88)..(0,.88)
    ..controls(.72,.88)and(.72,.22)..(0,.22);
  \draw[czrm torus rim]
    (0,-.22) ..controls(-.72,-.22)and(-.72,-.88)..(0,-.88)
    ..controls(.72,-.88)and(.72,-.22)..(0,-.22);
}

\newcommand{\czrmFrontThirdRim}[4]{%
  \begin{scope}[shift={(#1,#2)},rotate=#3]
    \draw[czrm torus rim] (0,0)
      ellipse[x radius=.18,y radius=#4];
  \end{scope}
}

\newcommand{\czrmRearThirdRim}[4]{%
  \begin{scope}[shift={(#1,#2)},rotate=#3]
    \draw[czrm seam rear,fill=white] (0,0)
      ellipse[x radius=.18,y radius=#4];
  \end{scope}
}

\newcommand{\czrmRetainedThirdRim}{%
  \czrmFrontThirdRim{1.12}{-.50}{18}{.08}
}

\newcommand{\czrmMigrationTorusRims}[1]{%
  \ifcase#1\relax
  \or
    \czrmTorusRims
    \czrmFrontThirdRim{1.12}{.43}{-18}{.08}
  \or
    \draw[czrm torus rim]
      (0,.22)..controls(-.72,.22)and(-.70,.92)..(.06,.90)
      ..controls(.80,.87)and(.72,.22)..(0,.22);
    \draw[czrm torus rim]
      (0,-.22)..controls(-.72,-.22)and(-.72,-.88)..(0,-.88)
      ..controls(.72,-.88)and(.72,-.22)..(0,-.22);
    \czrmFrontThirdRim{1.34}{.36}{-24}{.075}
  \or
    \draw[czrm torus rim]
      (0,.22)..controls(-.68,.22)and(-.65,.90)..(.08,.87)
      ..controls(.76,.84)and(.70,.22)..(0,.22);
    \draw[czrm torus rim]
      (0,-.22)..controls(-.70,-.22)and(-.68,-.91)..(.08,-.89)
      ..controls(.76,-.87)and(.70,-.22)..(0,-.22);
    \czrmRearThirdRim{1.26}{-.48}{-8}{.075}
  \or
    \draw[czrm torus rim]
      (0,.22)..controls(-.70,.22)and(-.70,.88)..(0,.88)
      ..controls(.70,.88)and(.70,.22)..(0,.22);
    \draw[czrm torus rim]
      (0,-.22)..controls(-.72,-.22)and(-.70,-.92)..(.08,-.90)
      ..controls(.80,-.88)and(.72,-.22)..(0,-.22);
    \czrmFrontThirdRim{1.22}{-.48}{22}{.072}
  \or
    \draw[czrm torus rim]
      (0,.22)..controls(-.72,.22)and(-.72,.88)..(0,.88)
      ..controls(.72,.88)and(.72,.22)..(0,.22);
    \draw[czrm torus rim]
      (0,-.22)..controls(-.72,-.22)and(-.72,-.88)..(0,-.88)
      ..controls(.72,-.88)and(.72,-.22)..(0,-.22);
    \czrmFrontThirdRim{1.10}{-.38}{18}{.075}
  \fi
}

\newcommand{\czrmEndLabels}{}

\newcommand{\czrmOpenWilson}[1]{%
  \ifcase#1\relax
  \or 
    \draw[czrm left] (-3.05,.25)..controls(-2.32,.28)and(-1.88,1.05)..(.15,1.215);
    \draw[czrm right] (.15,1.215)..controls(1.88,1.05)and(2.32,.28)..(3.05,.25);
    \draw[czrm left] (-3.05,0)--(.15,0);
    \draw[czrm right] (.15,0)--(3.05,0);
    \draw[czrm left] (-3.05,-.25)..controls(-2.32,-.28)and(-1.88,-1.05)..(.15,-1.215);
    \draw[czrm right] (.15,-1.215)..controls(1.88,-1.05)and(2.32,-.28)..(3.05,-.25);
  \or 
    \draw[czrm left] (-3.05,.25)..controls(-2.32,.28)and(-1.82,1.08)..(.15,1.215);
    \draw[czrm right] (.15,1.215)..controls(1.32,1.04)and(1.88,.48)..(3.05,.25);
    \draw[czrm left] (-3.05,0)..controls(-1.65,.02)and(-.85,.18)..(.15,0);
    \draw[czrm right] (.15,0)..controls(1.05,-.08)and(1.75,.08)..(3.05,0);
    \draw[czrm left] (-3.05,-.25)..controls(-2.28,-.30)and(-1.84,-1.03)..(.15,-1.215);
    \draw[czrm right] (.15,-1.215)..controls(1.90,-1.05)and(2.30,-.28)..(3.05,-.25);
  \or 
    \draw[czrm left] (-3.05,.25)..controls(-2.30,.28)and(-1.90,1.05)..(.15,1.215);
    \draw[czrm right] (.15,1.215)..controls(1.90,1.05)and(2.30,.28)..(3.05,.25);
    \draw[czrm left] (-3.05,0)..controls(-1.70,-.08)and(-1.05,.08)..(.15,0);
    \draw[czrm right] (.15,0)..controls(.85,-.18)and(1.65,-.02)..(3.05,0);
    \draw[czrm left] (-3.05,-.25)..controls(-1.88,-.48)and(-1.32,-1.04)..(.15,-1.215);
    \draw[czrm right] (.15,-1.215)..controls(1.82,-1.08)and(2.32,-.28)..(3.05,-.25);
  \or 
    \draw[czrm left] (-3.05,.25)..controls(-2.28,.30)and(-1.72,1.10)..(.15,1.215);
    \draw[czrm right] (.15,1.215)..controls(1.58,1.02)and(2.18,.48)..(3.05,.25);
    \draw[czrm left] (-3.05,0)--(.15,0);
    \draw[czrm right] (.15,0)--(3.05,0);
    \draw[czrm left] (-3.05,-.25)..controls(-2.18,-.48)and(-1.58,-1.02)..(.15,-1.215);
    \draw[czrm right] (.15,-1.215)..controls(1.72,-1.10)and(2.28,-.30)..(3.05,-.25);
  \else 
    \draw[czrm left] (-3.05,.25)..controls(-2.42,.42)and(-1.60,1.02)..(.15,1.215);
    \draw[czrm right] (.15,1.215)..controls(1.72,1.02)and(2.42,.38)..(3.05,.25);
    \draw[czrm left] (-3.05,0)..controls(-1.52,.05)and(-.72,-.12)..(.15,0);
    \draw[czrm right] (.15,0)..controls(.72,.12)and(1.52,-.05)..(3.05,0);
    \draw[czrm left] (-3.05,-.25)..controls(-2.42,-.38)and(-1.72,-1.02)..(.15,-1.215);
    \draw[czrm right] (.15,-1.215)..controls(1.60,-1.02)and(2.42,-.42)..(3.05,-.25);
  \fi
}

\newcommand{\czrmClosedWilson}[1]{%
  \draw[czrm left back] (-.15,1.215)--(-.28,1.05)
    ..controls(-1.48,1.28)and(-1.50,-1.28)..(-.28,-1.05)--(-.15,-1.215);
  \ifnum#1>4\relax
    \draw[czrm right back] (-.15,-1.215)--(.28,-1.05)
      ..controls(2.02,-1.28)and(2.00,1.28)..(.28,1.05)--(-.15,1.215);
  \else
    \draw[czrm right back] (-.15,-1.215)--(.28,-1.05)
      ..controls(1.50,-1.28)and(1.48,1.28)..(.28,1.05)--(-.15,1.215);
  \fi
  \ifnum#1=5\relax
    \draw[czrm left] (.096,.958)--(-.27,.98)
      ..controls(-1.20,.95)and(-1.28,.42)..(-.27,.205)--(.096,.169);
    \draw[czrm right] (.096,.169)--(.27,.205)
      ..controls(1.28,.42)and(1.20,.95)..(.27,.98)--(.096,.958);
    \draw[czrm left] (.096,-.169)--(-.27,-.205)
      ..controls(-1.28,-.42)and(-1.20,-.95)..(-.27,-.98)--(.096,-.958);
    \draw[czrm right] (.096,-.958)--(.27,-.98)
      ..controls(.82,-.94)and(.88,-.40)..(.27,-.205)--(.096,-.169);
  \else\ifnum#1=6\relax
    \draw[czrm left] (.096,.958)--(-.27,.98)
      ..controls(-1.20,.95)and(-1.28,.42)..(-.27,.205)--(.096,.169);
    \draw[czrm right] (.096,.169)--(.27,.205)
      ..controls(.88,.40)and(.82,.94)..(.27,.98)--(.096,.958);
    \draw[czrm left] (.096,-.169)--(-.27,-.205)
      ..controls(-1.28,-.42)and(-1.20,-.95)..(-.27,-.98)--(.096,-.958);
    \draw[czrm right] (.096,-.958)--(.27,-.98)
      ..controls(1.20,-.95)and(1.28,-.42)..(.27,-.205)--(.096,-.169);
  \else\ifnum#1=4\relax
    \draw[czrm left back] (-.096,.958)
      ..controls(-.78,1.00)and(-.96,.68)..(-.82,.38)
      ..controls(-.62,.08)and(.62,.08)..(.82,.38)
      ..controls(.96,.68)and(.78,1.00)..(.096,.958);
    \draw[czrm right] (.096,.958)
      ..controls(.38,1.17)and(.78,1.50)..(1.05,1.55);
    \draw[czrm right back] (1.05,1.55)
      ..controls(.72,1.36)and(.30,1.06)..(-.096,.958);
    \draw[czrm left] (.096,-.169)--(-.27,-.205)
      ..controls(-1.28,-.42)and(-1.20,-.95)..(-.27,-.98)--(.096,-.958);
    \draw[czrm right] (.096,-.958)--(.27,-.98)
      ..controls(1.20,-.95)and(1.28,-.42)..(.27,-.205)--(.096,-.169);
  \else\ifnum#1=2\relax
    \draw[czrm left] (.096,.958)--(-.27,.98)
      ..controls(-1.10,.93)and(-1.25,.45)..(-.27,.205)--(.096,.169);
    \draw[czrm right] (.096,.169)--(.27,.205)
      ..controls(1.40,.42)and(1.27,.95)..(.27,.98)--(.096,.958);
    \draw[czrm left] (.096,-.169)--(-.27,-.205)
      ..controls(-1.35,-.42)and(-1.20,-.93)..(-.27,-.98)--(.096,-.958);
    \draw[czrm right] (.096,-.958)--(.27,-.98)
      ..controls(1.12,-.93)and(1.23,-.45)..(.27,-.205)--(.096,-.169);
  \else\ifnum#1=3\relax
    \draw[czrm left] (.096,.958)--(-.27,.98)
      ..controls(-1.34,.95)and(-1.16,.42)..(-.27,.205)--(.096,.169);
    \draw[czrm right] (.096,.169)--(.27,.205)
      ..controls(1.08,.42)and(1.18,.93)..(.27,.98)--(.096,.958);
    \draw[czrm left] (.096,-.169)--(-.27,-.205)
      ..controls(-1.08,-.42)and(-1.18,-.93)..(-.27,-.98)--(.096,-.958);
    \draw[czrm right] (.096,-.958)--(.27,-.98)
      ..controls(1.34,-.95)and(1.16,-.42)..(.27,-.205)--(.096,-.169);
  \else
    \draw[czrm left] (.096,.958)--(-.27,.98)
      ..controls(-1.20,.95)and(-1.28,.42)..(-.27,.205)--(.096,.169);
    \draw[czrm right] (.096,.169)--(.27,.205)
      ..controls(1.28,.42)and(1.20,.95)..(.27,.98)--(.096,.958);
    \draw[czrm left] (.096,-.169)--(-.27,-.205)
      ..controls(-1.28,-.42)and(-1.20,-.95)..(-.27,-.98)--(.096,-.958);
    \draw[czrm right] (.096,-.958)--(.27,-.98)
      ..controls(1.20,-.95)and(1.28,-.42)..(.27,-.205)--(.096,-.169);
  \fi\fi\fi\fi\fi
}

\newcommand{\czrmSourceCard}[7]{%
  \begin{minipage}[t]{.485\linewidth}\centering
  \begin{tikzpicture}[x=.88cm,y=.72cm]
    \czrmSurface
    \czrmTorusRims
    \ifnum#7=3\relax\czrmRetainedThirdRim\fi
    \czrmCuffs
    \czrmClosedWilson{#5}
    \czrmOpenWilson{#4}
    \czrmEndLabels
    \node[czrm title,anchor=south west] at (-3.18,1.66)
      {#1 panel #2\quad $#3$};
  \end{tikzpicture}
  \par\vspace{-2pt}{\scriptsize #6\par}
  \end{minipage}%
}

\newcommand{\czrmTwoCards}[2]{%
  \noindent#1\hfill#2\par\medskip
}

\newcommand{\czrmMigrationSurface}[1]{%
  \ifnum#1=5\relax
    \path[czrm surface]
      (-3.18,.46)
      ..controls(-2.55,.55)and(-2.42,1.48)..(-1.48,1.55)
      --(1.48,1.55)
      ..controls(2.42,1.48)and(2.55,.55)..(3.18,.46)
      --(3.18,-.46)
      ..controls(2.55,-.55)and(2.42,-1.48)..(1.48,-1.55)
      --(-1.48,-1.55)
      ..controls(-2.42,-1.48)and(-2.55,-.55)..(-3.18,-.46)--cycle;
  \else\ifnum#1=2\relax
    \path[czrm surface]
      (-3.18,.46)..controls(-2.48,.58)and(-2.20,1.52)..(-1.10,1.55)
      ..controls(-.18,1.92)and(.80,1.72)..(1.42,1.50)
      ..controls(2.42,1.42)and(2.56,.55)..(3.18,.46)--(3.18,-.46)
      ..controls(2.55,-.55)and(2.42,-1.48)..(1.48,-1.55)--(-1.48,-1.55)
      ..controls(-2.42,-1.48)and(-2.55,-.55)..(-3.18,-.46)--cycle;
  \else\ifnum#1=3\relax
    \path[czrm surface]
      (-3.18,.46)..controls(-2.55,.55)and(-2.42,1.48)..(-1.48,1.55)
      ..controls(-.28,1.62)and(.45,1.48)..(1.18,1.32)
      ..controls(2.26,1.15)and(2.52,.52)..(3.18,.46)--(3.18,-.46)
      ..controls(2.54,-.58)and(2.28,-1.18)..(1.18,-1.35)
      ..controls(.42,-1.68)and(-.42,-1.76)..(-1.48,-1.55)
      ..controls(-2.42,-1.48)and(-2.55,-.55)..(-3.18,-.46)--cycle;
  \else\ifnum#1=4\relax
    \path[czrm surface]
      (-3.18,.46)..controls(-2.55,.55)and(-2.42,1.48)..(-1.48,1.55)--(1.48,1.55)
      ..controls(2.42,1.48)and(2.55,.55)..(3.18,.46)--(3.18,-.46)
      ..controls(2.45,-.58)and(2.18,-1.52)..(1.08,-1.55)
      ..controls(.18,-1.92)and(-.80,-1.72)..(-1.42,-1.50)
      ..controls(-2.42,-1.42)and(-2.56,-.55)..(-3.18,-.46)--cycle;
  \else
    \czrmSurface
  \fi\fi\fi\fi
  \ifnum#1=2\relax
  \else\ifnum#1=3\relax
  \else\ifnum#1=4\relax
  \else\ifnum#1=5\relax
  \else\ifnum#1=1\relax
  \fi\fi\fi\fi\fi
}

\newcommand{\czrmMigrationNetwork}[1]{%
  \ifcase#1\relax
  \or \czrmClosedWilson{6}\czrmOpenWilson{1}
  \or \czrmClosedWilson{6}\czrmOpenWilson{2}
  \or \czrmClosedWilson{5}\czrmOpenWilson{5}
  \or \czrmClosedWilson{5}\czrmOpenWilson{3}
  \or \czrmClosedWilson{5}\czrmOpenWilson{1}
  \fi
}

\tikzset{
  pics/sdwarmtorus/.style={code={
    \path[draw=none,fill=CZeroSurfaceFill,even odd rule]
      (0,0) ellipse[x radius=1.45,y radius=.78]
      (-.56,.08)
      .. controls (-.28,.32) and (.28,.32) .. (.56,.08)
      .. controls (.27,-.13) and (-.27,-.13) .. (-.56,.08)
      --cycle;
    \draw[CZeroSurfaceEdge,line width=.8pt,line cap=round]
      (0,0) ellipse[x radius=1.45,y radius=.78];
    \draw[CZeroSurfaceEdge,line width=.78pt,line cap=round]
      (-.56,.08)
      .. controls (-.28,.32) and (.28,.32) .. (.56,.08);
    \draw[CZeroSurfaceEdge,line width=.78pt,line cap=round]
      (-.70,.18)
      .. controls (-.65,.15) and (-.61,.11) .. (-.56,.08)
      .. controls (-.27,-.13) and (.27,-.13) .. (.56,.08)
      .. controls (.61,.11) and (.65,.15) .. (.70,.18);
  }},
  pics/sdwarmgenustwo/.style={code={
    \path[draw=none,fill=CZeroSurfaceFill,even odd rule]
      (-1.45,0)
      .. controls (-1.40,.58) and (-1.06,.78) .. (-.68,.78)
      .. controls (-.36,.78) and (-.30,.63) .. (0,.63)
      .. controls (.30,.63) and (.36,.78) .. (.68,.78)
      .. controls (1.06,.78) and (1.40,.58) .. (1.45,0)
      .. controls (1.40,-.58) and (1.06,-.78) .. (.68,-.78)
      .. controls (.36,-.78) and (.30,-.63) .. (0,-.63)
      .. controls (-.30,-.63) and (-.36,-.78) .. (-.68,-.78)
      .. controls (-1.06,-.78) and (-1.40,-.58) .. (-1.45,0)--cycle
      (-.80,.08)
      .. controls (-.65,.26) and (-.31,.26) .. (-.16,.08)
      .. controls (-.31,-.10) and (-.65,-.10) .. (-.80,.08)--cycle
      (.16,.08)
      .. controls (.31,.26) and (.65,.26) .. (.80,.08)
      .. controls (.65,-.10) and (.31,-.10) .. (.16,.08)--cycle;
    \draw[CZeroSurfaceEdge,line width=.8pt,line cap=round]
      (-1.45,0)
      .. controls (-1.40,.58) and (-1.06,.78) .. (-.68,.78)
      .. controls (-.36,.78) and (-.30,.63) .. (0,.63)
      .. controls (.30,.63) and (.36,.78) .. (.68,.78)
      .. controls (1.06,.78) and (1.40,.58) .. (1.45,0)
      .. controls (1.40,-.58) and (1.06,-.78) .. (.68,-.78)
      .. controls (.36,-.78) and (.30,-.63) .. (0,-.63)
      .. controls (-.30,-.63) and (-.36,-.78) .. (-.68,-.78)
      .. controls (-1.06,-.78) and (-1.40,-.58) .. (-1.45,0)--cycle;
    \draw[CZeroSurfaceEdge,line width=.76pt,line cap=round]
      (-.80,.08)
      .. controls (-.65,.26) and (-.31,.26) .. (-.16,.08);
    \draw[CZeroSurfaceEdge,line width=.76pt,line cap=round]
      (-.90,.17)
      .. controls (-.86,.14) and (-.83,.10) .. (-.80,.08)
      .. controls (-.65,-.10) and (-.31,-.10) .. (-.16,.08)
      .. controls (-.13,.10) and (-.10,.14) .. (-.06,.17);
    \draw[CZeroSurfaceEdge,line width=.76pt,line cap=round]
      (.16,.08)
      .. controls (.31,.26) and (.65,.26) .. (.80,.08);
    \draw[CZeroSurfaceEdge,line width=.76pt,line cap=round]
      (.06,.17)
      .. controls (.10,.14) and (.13,.10) .. (.16,.08)
      .. controls (.31,-.10) and (.65,-.10) .. (.80,.08)
      .. controls (.83,.10) and (.86,.14) .. (.90,.17);
  }},
  pics/sdgreentorus/.style={code={
    \path[draw=none,fill=ForestGreen!10,even odd rule]
      (0,0) ellipse[x radius=1.45,y radius=.78]
      (-.56,.08)
      .. controls (-.28,.32) and (.28,.32) .. (.56,.08)
      .. controls (.27,-.13) and (-.27,-.13) .. (-.56,.08)
      --cycle;
    \draw[CZeroSurfaceEdge,line width=.8pt,line cap=round]
      (0,0) ellipse[x radius=1.45,y radius=.78];
    \draw[CZeroSurfaceEdge,line width=.78pt,line cap=round]
      (-.56,.08)
      .. controls (-.28,.32) and (.28,.32) .. (.56,.08);
    \draw[CZeroSurfaceEdge,line width=.78pt,line cap=round]
      (-.70,.18)
      .. controls (-.65,.15) and (-.61,.11) .. (-.56,.08)
      .. controls (-.27,-.13) and (.27,-.13) .. (.56,.08)
      .. controls (.61,.11) and (.65,.15) .. (.70,.18);
  }},
  pics/sdshortcylinder/.style={code={
    \path[draw=ForestGreen!80!black,fill=ForestGreen!12,
      line width=.7pt,line join=round]
      (-.46,0)--(-.46,-.07)
      arc[start angle=180,end angle=360,x radius=.46,y radius=.11]
      --(.46,0)--cycle;
    \path[draw=ForestGreen!80!black,fill=ForestGreen!28,
      line width=.72pt]
      (0,0) ellipse[x radius=.46,y radius=.11];
    \draw[ForestGreen!80!black,line width=.55pt]
      (-.46,-.07)
      arc[start angle=180,end angle=360,x radius=.46,y radius=.11];
  }},
  pics/sdball/.style={code={
    \path[draw=CZeroSurfaceEdge,fill=CZeroSurfaceFill,line width=.8pt]
      (0,0) circle[radius=.72];
    \draw[CZeroSurfaceEdge!60,line width=.45pt]
      (-.67,.05) .. controls (-.28,-.19) and (.28,-.19) .. (.67,.05);
  }},
  pics/sdthreesphere/.style={code={
    \path[draw=none,fill=CZeroSurfaceFill]
      (0,.76) arc[start angle=90,end angle=270,radius=.76]--cycle;
    \path[draw=none,fill=ForestGreen!10]
      (0,-.76) arc[start angle=-90,end angle=90,radius=.76]--cycle;
    \draw[CZeroSurfaceEdge,line width=.8pt] (0,0) circle[radius=.76];
    \draw[CZeroSurfaceEdge,dash pattern=on 2.2pt off 1.5pt,line width=.55pt]
      (0,-.76)--(0,.76);
  }}
}

\newcommand{\SDOpenHandleFillingFigure}{%
\begin{tikzpicture}[x=1cm,y=1cm,baseline=(current bounding box.center)]
  \tikzset{
    hcarrow/.style={-{Latex[length=2mm,width=1.2mm]},
      draw=black!65,line width=.7pt},
    hcpanel/.style={font=\normalfont\fontsize{5}{6}\selectfont,align=center},
    hclabel/.style={font=\tiny,align=center}
  }

  \node[hcpanel] at (-3.55,1.55) {(a) pieces kept separate};
  \pic[scale=.82] at (-4.55,.32) {sdwarmtorus};
  \node[hclabel] at (-4.55,-.53) {solid torus \(V\)};
  \node[font=\normalsize] at (-3.10,.32) {\(+\)};
  \pic[scale=1.00] at (-2.18,.40) {sdshortcylinder};
  \node[hclabel] at (-2.18,-.53) {thickened disk\\\(D^2\times I\)};

  \draw[hcarrow] (-1.43,.32)--(-.45,.32)
    node[midway,above,hclabel]
      {fit it into\\the hole};

  \node[hcpanel] at (.88,1.64) {(b) fit the disk\\into the hole};
  \pic[scale=.82] at (.88,.32) {sdwarmtorus};
  \begin{scope}[shift={(.88,.32)}]
    \begin{scope}[scale=.82]
      \path[draw=none,fill=ForestGreen!45,fill opacity=.30]
        (-.515,.08)
        .. controls (-.245,-.09) and (.245,-.09) .. (.515,.08)
        -- (.515,-.06)
        .. controls (.245,-.23) and (-.245,-.23) .. (-.515,-.06)
        --cycle;
      \draw[ForestGreen!70!black,draw opacity=.45,line width=.65pt,
        line cap=round]
        (-.515,-.06)
        .. controls (-.245,-.23) and (.245,-.23) .. (.515,-.06);
      \draw[ForestGreen!70!black,draw opacity=.35,line width=.55pt]
        (-.515,.08)--(-.515,-.06)
        (.515,.08)--(.515,-.06);
      \path[draw=none,fill=ForestGreen!28]
        (-.56,.08)
        .. controls (-.28,.32) and (.28,.32) .. (.56,.08)
        .. controls (.27,-.13) and (-.27,-.13) .. (-.56,.08)
        --cycle;
      \draw[CZeroSurfaceEdge,line width=.78pt,line cap=round]
        (-.56,.08)
        .. controls (-.28,.32) and (.28,.32) .. (.56,.08);
      \draw[CZeroSurfaceEdge,line width=.78pt,line cap=round]
        (-.70,.18)
        .. controls (-.65,.15) and (-.61,.11) .. (-.56,.08)
        .. controls (-.27,-.13) and (.27,-.13) .. (.56,.08)
        .. controls (.61,.11) and (.65,.15) .. (.70,.18);
      \draw[ForestGreen!80!black,line width=.72pt,line cap=round]
        (-.515,.08)
        .. controls (-.255,.285) and (.255,.285) .. (.515,.08)
        .. controls (.245,-.09) and (-.245,-.09) .. (-.515,.08)
        --cycle;
    \end{scope}
  \end{scope}
  \node[hclabel] at (.88,-.53)
    {thin side sewn to the inner annulus};

  \draw[hcarrow] (2.16,.32)--(3.03,.32)
    node[midway,above,hclabel]{smooth\\the seam};

  \node[hcpanel] at (4.05,1.55) {(c) result};
  \pic[scale=1.05] at (4.05,.32) {sdball};
  \node[font=\scriptsize] at (4.05,.35) {\(B^3\)};
  \node[hclabel] at (4.05,-.53) {solid ball};
\end{tikzpicture}%
}

\newcommand{\SDOpenAttachingAnnulusFigure}{%
\begin{tikzpicture}[x=1cm,y=1cm,baseline=(current bounding box.center)]
  \tikzset{
    aaarrow/.style={-{Latex[length=2mm,width=1.2mm]},
      draw=black!65,line width=.7pt},
    aapanel/.style={font=\normalfont\fontsize{5}{6}\selectfont,align=center},
    aalabel/.style={font=\tiny,align=center},
    redloop/.style={draw=BrickRed!85!black,line width=.82pt,
      line cap=round},
    rededge/.style={draw=BrickRed!85!black,line width=.58pt},
    greenedge/.style={draw=ForestGreen!80!black,line width=.62pt}
  }

  \node[aapanel] at (-3.55,1.55) {(a) pieces kept separate};
  \pic[scale=.82] at (-4.55,.32) {sdwarmtorus};
  \begin{scope}[shift={(-4.55,.32)},scale=.82]
    \draw[redloop] (0,.05) ellipse[x radius=1.00,y radius=.46];
  \end{scope}
  \node[aalabel,text=BrickRed!85!black] at (-3.52,.78) {\(\lambda\)};
  \node[aalabel] at (-4.55,-.53)
    {red longitude\\\(\lambda\subset\partial V\)};
  \node[font=\normalsize] at (-3.10,.32) {\(+\)};
  \pic[scale=1.00] at (-2.18,.40) {sdshortcylinder};
  \node[aalabel] at (-2.18,-.53)
    {green thickened disk\\\(D^2\times I\)};

  \draw[aaarrow] (-1.43,.32)--(-.45,.32)
    node[midway,above,aalabel]{thicken\\and sew};

  \node[aapanel] at (.88,1.55) {(b) identify the annuli};
  \pic[scale=.82] at (.88,.32) {sdwarmtorus};
  \begin{scope}[shift={(.88,.32)},scale=.82]
    \def\aaCenterY{.05}
    \def\aaAnnulusOuterX{1.00}
    \def\aaAnnulusOuterY{.46}
    \def\aaAnnulusInnerX{.91}
    \def\aaAnnulusInnerY{.38}
    \path[draw=none,fill=ForestGreen!55,fill opacity=.30]
      (-\aaAnnulusOuterX,\aaCenterY)
      .. controls (-.80,.72) and (-.34,.94) .. (0,.97)
      .. controls (.34,.94) and (.80,.72) ..
        (\aaAnnulusOuterX,\aaCenterY)
      arc[start angle=0,end angle=-180,
        x radius=\aaAnnulusOuterX,y radius=\aaAnnulusOuterY]
      --cycle;
    \path[draw=none,fill=ForestGreen!55,fill opacity=.30]
      (-\aaAnnulusInnerX,\aaCenterY)
      .. controls (-.72,.57) and (-.30,.75) .. (0,.78)
      .. controls (.30,.75) and (.72,.57) ..
        (\aaAnnulusInnerX,\aaCenterY)
      arc[start angle=0,end angle=-180,
        x radius=\aaAnnulusInnerX,y radius=\aaAnnulusInnerY]
      --cycle;
    \path[even odd rule,draw=none,fill=BrickRed!85,fill opacity=.68]
      (0,\aaCenterY)
        ellipse[x radius=\aaAnnulusOuterX,y radius=\aaAnnulusOuterY]
      (0,\aaCenterY)
        ellipse[x radius=\aaAnnulusInnerX,y radius=\aaAnnulusInnerY];
    \draw[ForestGreen!80!black,line width=.72pt,line join=round]
      (-\aaAnnulusOuterX,\aaCenterY)
      .. controls (-.80,.72) and (-.34,.94) .. (0,.97)
      .. controls (.34,.94) and (.80,.72) ..
        (\aaAnnulusOuterX,\aaCenterY);
    \draw[ForestGreen!75!black,draw opacity=.95,line width=.68pt]
      (-\aaAnnulusInnerX,\aaCenterY)
      .. controls (-.72,.57) and (-.30,.75) .. (0,.78)
      .. controls (.30,.75) and (.72,.57) ..
        (\aaAnnulusInnerX,\aaCenterY);
    \draw[ForestGreen!80!black,line width=.58pt]
      (0,\aaCenterY)
        ellipse[x radius=\aaAnnulusOuterX,y radius=\aaAnnulusOuterY];
    \draw[ForestGreen!80!black,line width=.58pt]
      (0,\aaCenterY)
        ellipse[x radius=\aaAnnulusInnerX,y radius=\aaAnnulusInnerY];
  \end{scope}
  \node[aalabel] at (.88,-.53)
    {shared attaching annulus};

  \draw[aaarrow] (2.16,.32)--(3.03,.32)
    node[midway,above,aalabel]{smooth};

  \node[aapanel] at (4.05,1.55) {(c) result};
  \pic[scale=1.05] at (4.05,.32) {sdball};
  \node[font=\scriptsize] at (4.05,.35) {\(B^3\)};
  \node[aalabel] at (4.05,-.53) {solid ball};
\end{tikzpicture}%
}

\newcommand{\SDGenusTwoHandleFillingFigure}{%
\begin{tikzpicture}[x=1cm,y=1cm,baseline=(current bounding box.center)]
  \tikzset{
    gtarrow/.style={-{Latex[length=2mm,width=1.2mm]},
      draw=black!65,line width=.7pt},
    gtpanel/.style={font=\normalfont\fontsize{5}{6}\selectfont,align=center},
    gtlabel/.style={font=\tiny,align=center},
    gtredloop/.style={draw=BrickRed!85!black,line width=.82pt,
      line cap=round}
  }

  \node[gtpanel] at (-3.55,1.55) {(a) pieces kept separate};
  \pic[scale=.82] at (-4.55,.32) {sdwarmgenustwo};
  \begin{scope}[shift={(-4.55,.32)},scale=.82]
    \draw[gtredloop] (0,.05) ellipse[x radius=1.12,y radius=.50];
  \end{scope}
  \node[gtlabel,text=BrickRed!85!black] at (-3.52,.81) {red curve};
  \node[gtlabel] at (-4.55,-.53) {genus-two\\handlebody};
  \node[font=\normalsize] at (-3.10,.32) {\(+\)};
  \pic[scale=1.00] at (-2.18,.40) {sdshortcylinder};
  \node[gtlabel] at (-2.18,-.53)
    {green thickened disk\\\(D^2\times I\)};

  \draw[gtarrow] (-1.43,.32)--(-.45,.32)
    node[midway,above,gtlabel]{thicken\\and sew};

  \node[gtpanel] at (.88,1.55) {(b) identify the annuli};
  \pic[scale=.82] at (.88,.32) {sdwarmgenustwo};
  \begin{scope}[shift={(.88,.32)},scale=.82]
    \def\gtCenterY{.05}
    \def\gtOuterX{1.12}
    \def\gtOuterY{.50}
    \def\gtInnerX{1.00}
    \def\gtInnerY{.41}
    \path[draw=none,fill=ForestGreen!55,fill opacity=.30]
      (-\gtOuterX,\gtCenterY)
      .. controls (-.90,.72) and (-.38,.94) .. (0,.97)
      .. controls (.38,.94) and (.90,.72) ..
        (\gtOuterX,\gtCenterY)
      arc[start angle=0,end angle=-180,
        x radius=\gtOuterX,y radius=\gtOuterY]--cycle;
    \path[draw=none,fill=ForestGreen!55,fill opacity=.30]
      (-\gtInnerX,\gtCenterY)
      .. controls (-.79,.58) and (-.34,.76) .. (0,.79)
      .. controls (.34,.76) and (.79,.58) ..
        (\gtInnerX,\gtCenterY)
      arc[start angle=0,end angle=-180,
        x radius=\gtInnerX,y radius=\gtInnerY]--cycle;
    \path[even odd rule,draw=none,fill=BrickRed!85,fill opacity=.68]
      (0,\gtCenterY) ellipse[x radius=\gtOuterX,y radius=\gtOuterY]
      (0,\gtCenterY) ellipse[x radius=\gtInnerX,y radius=\gtInnerY];
    \draw[ForestGreen!80!black,line width=.72pt,line join=round]
      (-\gtOuterX,\gtCenterY)
      .. controls (-.90,.72) and (-.38,.94) .. (0,.97)
      .. controls (.38,.94) and (.90,.72) ..
        (\gtOuterX,\gtCenterY);
    \draw[ForestGreen!75!black,line width=.68pt]
      (-\gtInnerX,\gtCenterY)
      .. controls (-.79,.58) and (-.34,.76) .. (0,.79)
      .. controls (.34,.76) and (.79,.58) ..
        (\gtInnerX,\gtCenterY);
    \draw[ForestGreen!80!black,line width=.58pt]
      (0,\gtCenterY) ellipse[x radius=\gtOuterX,y radius=\gtOuterY];
    \draw[ForestGreen!80!black,line width=.58pt]
      (0,\gtCenterY) ellipse[x radius=\gtInnerX,y radius=\gtInnerY];
  \end{scope}
  \node[gtlabel] at (.88,-.53) {one handle canceled};

  \draw[gtarrow] (2.16,.32)--(3.00,.32)
    node[midway,above,gtlabel]{smooth};

  \node[gtpanel] at (4.25,1.55) {(c) result};
  \pic[scale=.82] at (4.25,.32) {sdwarmtorus};
  \node[gtlabel] at (4.25,-.53)
    {solid torus \(D^2\times S^1\)};
\end{tikzpicture}%
}

\tikzset{
  sdport/.style={circle,draw=black!70,line width=.45pt,
    minimum size=5.2mm,inner sep=0pt,font=\scriptsize},
  sds/.style={sdport,fill=cyan!18},
  sdt/.style={sdport,fill=magenta!18},
  sda/.style={sdport,fill=green!18},
  sdb/.style={sdport,fill=yellow!32},
  sdc/.style={sdport,fill=red!16},
  sdfixed/.style={draw=MidnightBlue,line width=.9pt},
  sdmatch/.style={draw=black!72,line width=.9pt},
  sdcycle/.style={draw=BrickRed,line width=1.35pt},
  sdcontext/.style={draw=black!20,line width=.8pt},
  sdpathone/.style={draw=ForestGreen,line width=1.2pt},
  sdpathtwo/.style={draw=RoyalBlue,line width=1.2pt},
  sdpaththree/.style={draw=Purple,line width=1.2pt},
  sdcompact/.style={minimum size=4.3mm,font=\scriptsize},
  sdselected/.style={line width=.8pt},
  sdpaneltitle/.style={font=\footnotesize\bfseries,align=center},
  sdarrow/.style={-{Latex[length=2mm,width=1.2mm]},
    draw=black!65,line width=.65pt},
  sdstepbox/.style={draw=MidnightBlue,fill=MidnightBlue!5,
    rounded corners=2pt,line width=.5pt,align=center,
    text width=2.5cm,minimum height=8mm,inner sep=3pt,font=\small},
  sdtest/.style={diamond,aspect=2.1,draw=BurntOrange,
    fill=BurntOrange!8,line width=.5pt,align=center,
    inner sep=1.5pt,font=\scriptsize},
  sdaccept/.style={sdstepbox,draw=ForestGreen,fill=ForestGreen!8},
  sdreject/.style={sdstepbox,draw=BrickRed,fill=BrickRed!6},
  sdbranch/.style={font=\scriptsize,fill=white,inner sep=1pt}
}

\newcommand{\sdcompactports}[1]{%
  \node[sds,sdcompact] (#1s1) at (-3.10,.42) {$s_1$};
  \node[sds,sdcompact] (#1s2) at (-2.90,0) {$s_2$};
  \node[sds,sdcompact] (#1s3) at (-3.10,-.42) {$s_3$};
  \node[sdt,sdcompact] (#1t1) at (3.10,.42) {$t_1$};
  \node[sdt,sdcompact] (#1t2) at (2.90,0) {$t_2$};
  \node[sdt,sdcompact] (#1t3) at (3.10,-.42) {$t_3$};
  \node[sda,sdcompact] (#1a1) at (-.30,1.45) {$a_1$};
  \node[sda,sdcompact] (#1a2) at (-.66,.92) {$a_2$};
  \node[sda,sdcompact] (#1a3) at (.06,.92) {$a_3$};
  \node[sdb,sdcompact] (#1b1) at (.35,.15) {$b_1$};
  \node[sdb,sdcompact] (#1b2) at (-.01,-.38) {$b_2$};
  \node[sdb,sdcompact] (#1b3) at (.71,-.38) {$b_3$};
  \node[sdc,sdcompact] (#1c1) at (-.30,-.92) {$c_1$};
  \node[sdc,sdcompact] (#1c2) at (-.66,-1.45) {$c_2$};
  \node[sdc,sdcompact] (#1c3) at (.06,-1.45) {$c_3$};
}
\newcommand{\sdedgelist}[3]{%
  \foreach \sdfrom/\sdto in {#3}{%
    \draw[#2] (#1\sdfrom)--(#1\sdto);%
  }%
}
\newcommand{\sdparallel}[4]{%
  \draw[#2,bend left=18] (#1#3) to (#1#4);%
  \draw[#2,bend right=18] (#1#3) to (#1#4);%
}
\newcommand{\sdoutlineports}[2]{%
  \foreach \sdv in {#2}{%
    \draw[black!75,line width=.7pt] (#1\sdv) circle[radius=2.15mm];%
  }%
}
\title{Checking the Schwinger--Dyson equation\\
in the purely open matrix--tensor model}

\author{Daniel L. Jafferis,$^{a}$}

\author{Liza Rozenberg,$^{a}$}

\author{Diandian Wang,$^{a,b}$}

\author{Cynthia Yan$^{a}$}

\affiliation{$^{a}$Jefferson Physical Laboratory,
Harvard University, Cambridge, MA 02138, USA}

\affiliation{$^{b}$Department of Physics,
Syracuse University, Syracuse, NY 13244, USA}

\emailAdd{jafferis@g.harvard.edu}
\emailAdd{erozenberg@g.harvard.edu}
\emailAdd{dwang90@syr.edu}
\emailAdd{xyan2@g.harvard.edu}

\begin{document}

%

\definecolor{crawLeft}{RGB}{28,28,28}
\definecolor{crawRight}{RGB}{50,165,35}
\definecolor{crawEdge}{RGB}{130,130,130}
\definecolor{crawFill}{RGB}{244,239,224}

\tikzset{
  craw sheet/.style={draw=crawEdge,fill=crawFill,line width=1.05pt,
    line cap=round,line join=round},
  craw hole/.style={draw=crawEdge,fill=white,line width=.95pt,
    line cap=round,line join=round},
  craw cuff/.style={draw=crawEdge,fill=white,line width=.95pt},
  craw torus rim/.style={draw=crawEdge,fill=white,line width=1.02pt,
    line cap=round,line join=round},
  craw seam fill/.style={fill=crawFill,draw=none},
  craw seam rear/.style={draw=crawEdge,line width=1.02pt,
    dash pattern=on .12pt off 2.10pt,line cap=round,line join=round},
  craw seam front/.style={draw=crawEdge,line width=1.02pt,
    line cap=round,line join=round},
  craw left/.style={draw=crawLeft,line width=1.02pt,
    line cap=round,line join=round},
  craw right/.style={draw=crawRight,line width=1.02pt,
    line cap=round,line join=round},
  craw left rear/.style={draw=crawLeft,line width=1.08pt,
    dash pattern=on 3.6pt off 2.25pt,line cap=round,line join=round},
  craw right rear/.style={draw=crawRight,line width=1.08pt,
    dash pattern=on 3.6pt off 2.25pt,line cap=round,line join=round},
  craw number/.style={circle,draw=RoyalBlue,fill=white,line width=.65pt,
    minimum size=5.4mm,inner sep=0pt,font=\scriptsize\bfseries,
    text=RoyalBlue},
  craw note/.style={font=\tiny,align=center,text=black!68},
  craw relation/.style={font=\scriptsize\bfseries,text=ForestGreen!75!black},
  craw reject/.style={draw=black!42,line width=1.65pt,line cap=round}
}

\newcommand{\crawOuterStandard}{%
  \path[craw sheet]
    (-5.20,.52)
    ..controls(-4.20,.62)and(-3.92,1.74)..(-2.45,1.74)
    --(2.45,1.74)
    ..controls(3.92,1.74)and(4.20,.62)..(5.20,.52)
    --(5.20,-.52)
    ..controls(4.20,-.62)and(3.92,-1.74)..(2.45,-1.74)
    --(-2.45,-1.74)
    ..controls(-3.92,-1.74)and(-4.20,-.62)..(-5.20,-.52)
    --cycle;
}

\newcommand{\crawExternalCuffs}{%
  \draw[craw cuff] (-5.20,0) ellipse[x radius=.17,y radius=.52];
  \draw[craw cuff] ( 5.20,0) ellipse[x radius=.17,y radius=.52];
}

\newcommand{\crawThreeSeams}{%
  \path[craw seam fill] (0, 1.33) ellipse[x radius=.17,y radius=.41];
  \path[craw seam fill] (0, 0)    ellipse[x radius=.17,y radius=.48];
  \path[craw seam fill] (0,-1.33) ellipse[x radius=.17,y radius=.41];
  \draw[craw seam rear]  (0, 1.74)
    arc[start angle=90,end angle=270,x radius=.17,y radius=.41];
  \draw[craw seam front] (0, .92)
    arc[start angle=-90,end angle=90,x radius=.17,y radius=.41];
  \draw[craw seam rear]  (0, .48)
    arc[start angle=90,end angle=270,x radius=.17,y radius=.48];
  \draw[craw seam front] (0,-.48)
    arc[start angle=-90,end angle=90,x radius=.17,y radius=.48];
  \draw[craw seam rear]  (0,-.92)
    arc[start angle=90,end angle=270,x radius=.17,y radius=.41];
  \draw[craw seam front] (0,-1.74)
    arc[start angle=-90,end angle=90,x radius=.17,y radius=.41];
  \path (0,1.33) ++({.17*cos(38)},{.41*sin(38)})
    coordinate (crawTopOpen);
  \path (0,1.33) ++({.17*cos(8)},{.41*sin(8)})
    coordinate (crawTopGlobalFront);
  \path (0,1.33) ++({-.17*cos(8)},{.41*sin(8)})
    coordinate (crawTopGlobalRear);
  \path (0,1.33) ++({.17*cos(-38)},{.41*sin(-38)})
    coordinate (crawTopLocal);
  \path (0,0) ++({.17*cos(32)},{.48*sin(32)})
    coordinate (crawMiddleUpper);
  \coordinate (crawMiddleOpen) at (.17,0);
  \coordinate (crawMiddleOpenRear) at (-.17,0);
  \path (0,0) ++({.17*cos(-32)},{.48*sin(-32)})
    coordinate (crawMiddleLowerFront);
  \path (0,0) ++({-.17*cos(-32)},{.48*sin(-32)})
    coordinate (crawMiddleLowerRear);
  \path (0,-1.33) ++({.17*cos(38)},{.41*sin(38)})
    coordinate (crawBottomLocalFront);
  \path (0,-1.33) ++({-.17*cos(38)},{.41*sin(38)})
    coordinate (crawBottomLocalRear);
  \path (0,-1.33) ++({.17*cos(-8)},{.41*sin(-8)})
    coordinate (crawBottomGlobalFront);
  \path (0,-1.33) ++({-.17*cos(-8)},{.41*sin(-8)})
    coordinate (crawBottomGlobalRear);
  \path (0,-1.33) ++({.17*cos(-38)},{.41*sin(-38)})
    coordinate (crawBottomOpen);
}

\newcommand{\crawTorusRimsStandard}{%
  \draw[craw torus rim] (0,.70)  ellipse[x radius=.72,y radius=.22];
  \draw[craw torus rim] (0,-.70) ellipse[x radius=.72,y radius=.22];
}

\newcommand{\crawThirdHoleStandard}{%
  \draw[craw torus rim,rotate around={-18:(1.25,.53)}]
    (1.25,.53) ellipse[x radius=.22,y radius=.10];}
\newcommand{\crawThirdHoleUpperInner}{%
  \draw[craw torus rim,rotate around={-20:(1.35,.35)}]
    (1.35,.35) ellipse[x radius=.22,y radius=.10];}
\newcommand{\crawThirdHoleLower}{%
  \draw[craw torus rim,rotate around={18:(1.37,-.40)}]
    (1.37,-.40) ellipse[x radius=.22,y radius=.10];}
\newcommand{\crawThirdHoleLowerOutside}{%
  \draw[craw torus rim,rotate around={18:(1.82,-.42)}]
    (1.82,-.42) ellipse[x radius=.22,y radius=.10];}
\newcommand{\crawThirdHoleNeck}{%
  \draw[craw torus rim,rotate around={22:(1.37,-.40)}]
    (1.37,-.40) ellipse[x radius=.22,y radius=.09];}
\newcommand{\crawThirdHoleUpper}{%
  \draw[craw torus rim,rotate around={-12:(1.25,.53)}]
    (1.25,.53) ellipse[x radius=.22,y radius=.09];}
\newcommand{\crawThirdHoleCentral}{%
  \draw[craw torus rim,rotate around={18:(1.06,.18)}]
    (1.06,.18) ellipse[x radius=.22,y radius=.09];}
\newcommand{\crawThirdHoleThreeUpper}{%
  \draw[craw torus rim,rotate around={-18:(1.45,.45)}]
    (1.45,.45) ellipse[x radius=.22,y radius=.10];}
\newcommand{\crawThirdHoleThreeLower}{%
  \draw[craw torus rim,rotate around={18:(1.55,-.45)}]
    (1.55,-.45) ellipse[x radius=.22,y radius=.10];}
\newcommand{\crawThirdHoleFourUpper}{%
  \draw[craw torus rim,rotate around={-18:(1.50,.50)}]
    (1.50,.50) ellipse[x radius=.22,y radius=.10];}
\newcommand{\crawThirdHoleFiveLower}{%
  \draw[craw torus rim,rotate around={18:(1.55,-.48)}]
    (1.55,-.48) ellipse[x radius=.22,y radius=.10];}
\newcommand{\crawThirdHoleSixUpper}{%
  \draw[craw torus rim,rotate around={55:(2.19,.38)}]
    (2.19,.38) ellipse[x radius=.19,y radius=.060];}
\newcommand{\crawThirdHoleSevenUpper}{%
  \draw[craw torus rim,rotate around={-16:(1.92,.38)}]
    (1.92,.38) ellipse[x radius=.22,y radius=.10];}
\newcommand{\crawThirdHoleEightLower}{%
  \draw[craw torus rim,rotate around={18:(1.82,-.28)}]
    (1.82,-.28) ellipse[x radius=.22,y radius=.10];}
\newcommand{\crawThirdHoleNineLower}{%
  \draw[craw torus rim,rotate around={18:(1.75,-.47)}]
    (1.75,-.47) ellipse[x radius=.22,y radius=.10];}
\newcommand{\crawThirdHoleHidden}{}

\newcommand{\crawSurfaceStandard}{%
  \crawOuterStandard
  \crawExternalCuffs
  \crawThreeSeams
  \crawTorusRimsStandard
}

\newcommand{\crawSurfaceUpperMigrate}{%
  \crawOuterStandard
  \crawExternalCuffs
  \crawThreeSeams
  \draw[craw torus rim] (1.35,.63) ellipse[x radius=.72,y radius=.19,
    rotate=-13];
  \draw[craw torus rim] (0,-.70) ellipse[x radius=.72,y radius=.22];
}

\newcommand{\crawSurfaceLowerMigrate}{%
  \crawOuterStandard
  \crawExternalCuffs
  \crawThreeSeams
  \draw[craw torus rim] (0,.70) ellipse[x radius=.72,y radius=.22];
  \draw[craw torus rim] (1.35,-.63) ellipse[x radius=.72,y radius=.19,
    rotate=13];
}

\newcommand{\crawSurfaceBridge}{%
  \crawOuterStandard
  \crawExternalCuffs
  \crawThreeSeams
  \draw[craw torus rim] (-.35,.68) ellipse[x radius=.66,y radius=.20,
    rotate=-8];
  \draw[craw torus rim] (-.35,-.68) ellipse[x radius=.66,y radius=.20,
    rotate=8];
}

\newcommand{\crawOpenBase}{%
  \draw[craw left]
    (-5.03,.29)..controls(-3.85,.30)and(-3.10,1.60)..(-.38,1.60)--(crawTopOpen);
  \draw[craw right]
    (crawTopOpen)--(.38,1.60)..controls(3.10,1.60)and(3.85,.30)..(5.03,.29);
  \draw[craw left] (-5.03,0)--(crawMiddleOpen);
  \draw[craw right] (crawMiddleOpen)--(5.03,0);
  \draw[craw left]
    (-5.03,-.29)..controls(-3.85,-.30)and(-3.10,-1.60)..(-.38,-1.60)--(crawBottomOpen);
  \draw[craw right]
    (crawBottomOpen)--(.38,-1.60)..controls(3.10,-1.60)and(3.85,-.30)..(5.03,-.29);
}

\newcommand{\crawOpenBCOnly}{%
  \draw[craw left] (-5.03,0)--(crawMiddleOpen);
  \draw[craw right] (crawMiddleOpen)--(5.03,0);
  \draw[craw left]
    (-5.03,-.29)..controls(-3.85,-.30)and(-3.10,-1.60)..(-.38,-1.60)--(crawBottomOpen);
  \draw[craw right]
    (crawBottomOpen)--(.38,-1.60)..controls(3.10,-1.60)and(3.85,-.30)..(5.03,-.29);
}
\newcommand{\crawOpenAOnly}{%
  \draw[craw left]
    (-5.03,.29)..controls(-3.85,.30)and(-3.10,1.60)..(-.38,1.60)--(crawTopOpen);
  \draw[craw right]
    (crawTopOpen)--(.38,1.60)..controls(3.10,1.60)and(3.85,.30)..(5.03,.29);
}

\newcommand{\crawCandidateThreeAUnderEye}{%
  \draw[craw left]
    (-5.03,.29)..controls(-3.85,.30)and(-3.10,1.60)..(-.38,1.60)--(crawTopOpen);
  \draw[craw right]
    (crawTopOpen)--(.38,1.60)
    ..controls(.66,1.47)and(.78,1.10)..(.80,.78)
    ..controls(.81,.53)and(.82,.27)..(1.04,.22)
    ..controls(1.25,.17)and(1.55,.17)..(1.78,.23)
    ..controls(2.58,.20)and(3.72,.24)..(5.03,.29);
}

\newcommand{\crawCandidateThreeCAboveLowerEye}{%
  \draw[craw left]
    (-5.03,-.29)..controls(-3.85,-.30)and(-3.10,-1.60)..(-.38,-1.60)--(crawBottomOpen);
  \draw[craw right]
    (crawBottomOpen)--(.38,-1.60)
    ..controls(1.00,-1.58)and(1.78,-1.39)..(2.10,-1.00)
    ..controls(2.32,-.72)and(2.29,-.48)..(2.12,-.34)
    ..controls(2.01,-.25)and(1.90,-.21)..(1.78,-.22)
    ..controls(2.73,-.17)and(4.03,-.24)..(5.03,-.29);
}
\newcommand{\crawOpenBOnly}{%
  \draw[craw left] (-5.03,0)--(crawMiddleOpen);
  \draw[craw right] (crawMiddleOpen)--(5.03,0);
}
\newcommand{\crawOpenCOnly}{%
  \draw[craw left]
    (-5.03,-.29)..controls(-3.85,-.30)and(-3.10,-1.60)..(-.38,-1.60)--(crawBottomOpen);
  \draw[craw right]
    (crawBottomOpen)--(.38,-1.60)..controls(3.10,-1.60)and(3.85,-.30)..(5.03,-.29);
}
\newcommand{\crawCandidateElevenA}{%
  \coordinate (crawElevenTopTurn) at (1.55,1.74);
  \coordinate (crawElevenShoulderTurn) at (4.792,.627);
  \draw[craw left]
    (-5.03,.29)..controls(-3.85,.30)and(-3.10,1.60)..(-.38,1.60)--(crawTopOpen);
  \draw[craw right]
    (crawTopOpen)..controls(.58,1.60)and(1.12,1.70)..(crawElevenTopTurn);
  \draw[craw right rear]
    (crawElevenTopTurn)..controls(1.05,1.50)and(.50,1.42)..(crawTopGlobalRear);
  \draw[craw left rear]
    (crawTopGlobalRear)--(-.38,1.39)
    ..controls(-2.20,1.50)and(-2.20,-1.50)..(-.38,-1.39)--(crawBottomGlobalRear);
  \draw[craw right rear]
    (crawBottomGlobalRear)..controls(1.55,-1.45)and(3.65,-.20)..(crawElevenShoulderTurn);
  \draw[craw right]
    (crawElevenShoulderTurn)..controls(4.90,.57)and(4.92,.33)..(5.03,.29);
}

\newcommand{\crawCandidateTwelveB}{%
  \draw[craw left] (-5.03,0)--(crawMiddleOpen);
  \draw[craw right]
    (crawMiddleOpen)..controls(.48,-.02)and(.50,-.20)..(crawMiddleLowerFront);
  \draw[craw left]
    (crawMiddleLowerFront)--(-.38,-.28)
    ..controls(-1.70,-.45)and(-1.62,-1.18)..(-.38,-1.17)--(crawBottomLocalFront);
  \draw[craw right]
    (crawBottomLocalFront)..controls(1.30,-1.14)and(3.30,-.44)..(5.03,0);
}

\newcommand{\crawCandidateThirteenB}{%
  \draw[craw left] (-5.03,0)--(crawMiddleOpen);
  \draw[craw right]
    (crawMiddleOpen)..controls(.48,.02)and(.50,.20)..(crawMiddleUpper);
  \draw[craw left]
    (crawMiddleUpper)--(-.38,.28)
    ..controls(-1.70,.45)and(-1.62,1.18)..(-.38,1.17)--(crawTopLocal);
  \draw[craw right]
    (crawTopLocal)..controls(1.30,1.14)and(3.30,.44)..(5.03,0);
}

\newcommand{\crawCandidateFourteenC}{%
  \draw[craw left]
    (-5.03,-.29)..controls(-3.85,-.30)and(-3.10,-1.60)..(-.38,-1.60)--(crawBottomOpen);
  \draw[craw right]
    (crawBottomOpen)..controls(.48,-1.58)and(.46,-1.24)..(crawBottomLocalFront);
  \draw[craw left]
    (crawBottomLocalFront)--(-.38,-1.17)
    ..controls(-1.62,-1.18)and(-1.70,-.45)..(-.38,-.28)--(crawMiddleLowerFront);
  \draw[craw right]
    (crawMiddleLowerFront)..controls(1.30,-.34)and(3.32,-.30)..(5.03,-.29);
}

\newcommand{\crawCandidateFifteenA}{%
  \draw[craw left]
    (-5.03,.29)..controls(-3.85,.30)and(-3.10,1.60)..(-.38,1.60)--(crawTopOpen);
  \draw[craw right]
    (crawTopOpen)..controls(.48,1.58)and(.46,1.24)..(crawTopLocal);
  \draw[craw left]
    (crawTopLocal)--(-.38,1.17)
    ..controls(-1.62,1.18)and(-1.70,.45)..(-.38,.28)--(crawMiddleUpper);
  \draw[craw right]
    (crawMiddleUpper)..controls(1.30,.34)and(3.32,.30)..(5.03,.29);
}

\newcommand{\crawCandidateTenC}{%
  \coordinate (crawTenBottomTurn) at (1.55,-1.74);
  \coordinate (crawTenShoulderTurn) at (4.792,-.627);
  \draw[craw left]
    (-5.03,-.29)..controls(-3.85,-.30)and(-3.10,-1.60)..(-.38,-1.60)
    --(crawBottomOpen);
  \draw[craw right]
    (crawBottomOpen)..controls(.58,-1.60)and(1.12,-1.70)
    ..(crawTenBottomTurn);
  \draw[craw right rear]
    (crawTenBottomTurn)..controls(1.05,-1.50)and(.50,-1.42)
    ..(crawBottomGlobalRear);
  \draw[craw left rear]
    (crawBottomGlobalRear)--(-.38,-1.39)
    ..controls(-2.20,-1.50)and(-2.20,1.50)..(-.38,1.39)
    --(crawTopGlobalRear);
  \draw[craw right rear]
    (crawTopGlobalRear)..controls(1.55,1.45)and(3.65,.20)
    ..(crawTenShoulderTurn);
  \draw[craw right]
    (crawTenShoulderTurn)..controls(4.90,-.57)and(4.92,-.33)
    ..(5.03,-.29);
}

\newcommand{\crawUBase}{%
  \draw[craw left]
    (crawTopLocal)--(-.38,1.17)
    ..controls(-1.62,1.18)and(-1.70,.45)..(-.38,.28)--(crawMiddleUpper);
  \draw[craw right]
    (crawMiddleUpper)--(.38,.28)
    ..controls(1.70,.45)and(1.62,1.18)..(.38,1.17)--(crawTopLocal);
}

\newcommand{\crawDBase}{%
  \draw[craw left]
    (crawMiddleLowerFront)--(-.38,-.28)
    ..controls(-1.70,-.45)and(-1.62,-1.18)..(-.38,-1.17)--(crawBottomLocalFront);
  \draw[craw right]
    (crawBottomLocalFront)--(.38,-1.17)
    ..controls(1.62,-1.18)and(1.70,-.45)..(.38,-.28)--(crawMiddleLowerFront);
}

\newcommand{\crawGlobalRear}{%
  \draw[craw left rear]
    (crawTopGlobalRear)--(-.38,1.39)
    ..controls(-2.20,1.50)and(-2.20,-1.50)..(-.38,-1.39)--(crawBottomGlobalRear);
  \draw[craw right rear]
    (crawBottomGlobalRear)--(.38,-1.39)
    ..controls(2.20,-1.50)and(2.20,1.50)..(.38,1.39)--(crawTopGlobalRear);
}

\newcommand{\crawGlobalFront}{%
  \draw[craw left]
    (crawTopGlobalFront)--(-.38,1.39)
    ..controls(-2.20,1.50)and(-2.20,-1.50)..(-.38,-1.39)--(crawBottomGlobalFront);
  \draw[craw right]
    (crawBottomGlobalFront)--(.38,-1.39)
    ..controls(2.20,-1.50)and(2.20,1.50)..(.38,1.39)--(crawTopGlobalFront);
}

\newcommand{\crawGlobalRearRightKink}{%
  \draw[craw left rear]
    (crawTopGlobalRear)--(-.38,1.39)
    ..controls(-2.20,1.50)and(-2.20,-1.50)..(-.38,-1.39)--(crawBottomGlobalRear);
  \draw[craw right rear]
    (crawBottomGlobalRear)--(.38,-1.39)
    ..controls(2.10,-1.45)and(2.35,-.86)..(1.42,-.60)
    ..controls(.92,-.43)and(1.02,-.14)..(1.72,-.05)
    ..controls(2.42,.05)and(2.16,1.42)..(.38,1.39)--(crawTopGlobalRear);
}

\newcommand{\crawGlobalRearLeftKink}{%
  \draw[craw left rear]
    (crawTopGlobalRear)--(-.38,1.39)
    ..controls(-2.16,1.42)and(-2.42,.05)..(-1.72,-.05)
    ..controls(-1.02,-.14)and(-.92,-.43)..(-1.42,-.60)
    ..controls(-2.35,-.86)and(-2.10,-1.45)..(-.38,-1.39)--(crawBottomGlobalRear);
  \draw[craw right rear]
    (crawBottomGlobalRear)--(.38,-1.39)
    ..controls(2.20,-1.50)and(2.20,1.50)..(.38,1.39)--(crawTopGlobalRear);
}

\newcommand{\crawGlobalRearRightNeck}{%
  \draw[craw left rear]
    (crawTopGlobalRear)--(-.38,1.39)
    ..controls(-2.20,1.50)and(-2.20,-1.50)..(-.38,-1.39)--(crawBottomGlobalRear);
  \draw[craw right rear]
    (crawBottomGlobalRear)--(.38,-1.39)
    ..controls(2.10,-1.46)and(2.55,-.88)..(3.55,-.49)
    ..controls(2.78,-.38)and(2.55,-.12)..(3.42,.10)
    ..controls(2.78,.34)and(2.32,1.45)..(.38,1.39)--(crawTopGlobalRear);
}

\newcommand{\crawGlobalRearLeftOfUpperEye}{%
  \draw[craw left rear]
    (crawTopGlobalRear)--(-.38,1.39)
    ..controls(-2.20,1.50)and(-2.20,-1.50)..(-.38,-1.39)--(crawBottomGlobalRear);
  \draw[craw right rear]
    (crawBottomGlobalRear)--(.38,-1.39)
    ..controls(2.10,-1.42)and(2.02,-.42)..(.91,-.03)
    ..controls(.78,.09)and(.79,.33)..(.87,.49)
    ..controls(1.18,.82)and(1.92,1.16)..(.38,1.39)--(crawTopGlobalRear);
}

\newcommand{\crawGlobalRearLeftOfUpperEyeNeck}{%
  \draw[craw left rear]
    (crawTopGlobalRear)--(-.38,1.39)
    ..controls(-2.20,1.50)and(-2.20,-1.50)..(-.38,-1.39)--(crawBottomGlobalRear);
  \draw[craw right rear]
    (crawBottomGlobalRear)--(.38,-1.39)
    ..controls(2.18,-1.43)and(2.55,-.76)..(3.35,-.48)
    ..controls(2.43,-.34)and(1.48,-.18)..(.92,.02)
    ..controls(.78,.15)and(.80,.34)..(.88,.49)
    ..controls(1.50,.84)and(2.82,.98)..(3.38,.38)
    ..controls(2.50,.28)and(1.96,1.31)..(.38,1.39)--(crawTopGlobalRear);
}

\newcommand{\crawGlobalRearLeftOfLowerEye}{%
  \draw[craw left rear]
    (crawTopGlobalRear)--(-.38,1.39)
    ..controls(-2.20,1.50)and(-2.20,-1.50)..(-.38,-1.39)--(crawBottomGlobalRear);
  \draw[craw right rear]
    (crawBottomGlobalRear)--(.38,-1.39)
    ..controls(1.92,-1.16)and(1.08,-.82)..(.72,-.50)
    ..controls(.66,-.33)and(.69,-.09)..(.82,.03)
    ..controls(2.02,.42)and(2.10,1.42)..(.38,1.39)--(crawTopGlobalRear);
}

\newcommand{\crawUTongueRight}{%
  \draw[craw left]
    (crawTopLocal)--(-.38,1.30)
    ..controls(-1.62,1.26)and(-1.70,.45)..(-.38,.20)--(crawMiddleUpper);
  \draw[craw right]
    (crawMiddleUpper)..controls(.70,.17)and(1.22,.13)..(1.28,.35)
    ..controls(1.34,.55)and(.90,.52)..(.62,.43)
    ..controls(1.62,.58)and(1.66,1.25)..(.38,1.30)--(crawTopLocal);
}

\newcommand{\crawUOutsideEyeTongue}{%
  \draw[craw left]
    (crawTopLocal)--(-.38,1.17)
    ..controls(-1.62,1.18)and(-1.70,.45)..(-.38,.28)--(crawMiddleUpper);
  \draw[craw right]
    (crawMiddleUpper)--(.38,.27)
    ..controls(.65,.24)and(.83,.18)..(.96,.20)
    ..controls(1.04,.23)and(1.02,.39)..(.91,.43)
    ..controls(.74,.50)and(.72,.83)..(.90,1.02)
    ..controls(.83,1.12)and(.58,1.16)..(.38,1.17)--(crawTopLocal);
}

\newcommand{\crawUOutsideEyeBase}{%
  \draw[craw left]
    (crawTopLocal)--(-.38,1.17)
    ..controls(-1.62,1.18)and(-1.70,.45)..(-.38,.28)--(crawMiddleUpper);
  \draw[craw right]
    (crawMiddleUpper)--(.38,.27)
    ..controls(.57,.28)and(.76,.37)..(.80,.55)
    ..controls(.86,.82)and(.70,1.10)..(.38,1.17)--(crawTopLocal);
}

\newcommand{\crawDOutsideEyeTongue}{%
  \draw[craw left]
    (crawMiddleLowerFront)--(-.38,-.28)
    ..controls(-1.70,-.45)and(-1.62,-1.18)..(-.38,-1.17)--(crawBottomLocalFront);
  \draw[craw right]
    (crawBottomLocalFront)--(.38,-1.17)
    ..controls(.61,-1.12)and(.88,-.88)..(1.02,-.66)
    ..controls(1.16,-.58)and(1.54,-.58)..(1.72,-.64)
    ..controls(1.54,-.35)and(1.15,-.17)..(.92,-.18)
    ..controls(.72,-.19)and(.55,-.24)..(.38,-.27)--(crawMiddleLowerFront);
}

\newcommand{\crawDOutsideEyeBase}{%
  \draw[craw left]
    (crawMiddleLowerFront)--(-.38,-.28)
    ..controls(-1.70,-.45)and(-1.62,-1.18)..(-.38,-1.17)--(crawBottomLocalFront);
  \draw[craw right]
    (crawBottomLocalFront)--(.38,-1.17)
    ..controls(.76,-1.10)and(1.08,-.82)..(1.04,-.55)
    ..controls(1.01,-.40)and(.72,-.29)..(.38,-.27)--(crawMiddleLowerFront);
}

\newcommand{\crawULobeAroundEye}{%
  \draw[craw left]
    (crawTopLocal)--(-.38,1.17)
    ..controls(-1.62,1.18)and(-1.70,.45)..(-.38,.28)--(crawMiddleUpper);
  \draw[craw right]
    (crawMiddleUpper)--(.38,.22)
    ..controls(.86,.08)and(1.42,.06)..(1.76,.14)
    ..controls(2.10,.22)and(2.12,.53)..(1.92,.76)
    ..controls(1.55,1.17)and(.82,1.20)..(.38,1.17)--(crawTopLocal);
}

\newcommand{\crawDLobeAroundEye}{%
  \draw[craw left]
    (crawMiddleLowerFront)--(-.38,-.28)
    ..controls(-1.70,-.45)and(-1.62,-1.18)..(-.38,-1.17)--(crawBottomLocalFront);
  \draw[craw right]
    (crawBottomLocalFront)--(.38,-1.17)
    ..controls(.82,-1.20)and(1.55,-1.17)..(1.92,-.76)
    ..controls(2.12,-.53)and(2.10,-.22)..(1.76,-.14)
    ..controls(1.42,-.06)and(.86,-.08)..(.38,-.22)--(crawMiddleLowerFront);
}

\newcommand{\crawUTongueLeft}{%
  \draw[craw left]
    (crawTopLocal)--(-.38,1.30)
    ..controls(-1.66,1.25)and(-1.62,.58)..(-.62,.43)
    ..controls(-.90,.52)and(-1.34,.55)..(-1.28,.35)
    ..controls(-1.22,.13)and(-.70,.17)..(crawMiddleUpper);
  \draw[craw right]
    (crawMiddleUpper)--(.38,.20)
    ..controls(1.70,.45)and(1.62,1.26)..(.38,1.30)--(crawTopLocal);
}

\newcommand{\crawUTopEar}{%
  \draw[craw left]
    (crawTopLocal)--(-.38,1.30)
    ..controls(-1.62,1.26)and(-1.70,.45)..(-.38,.20)--(crawMiddleUpper);
  \draw[craw right]
    (crawMiddleUpper)--(.38,.20)
    ..controls(1.58,.43)and(1.52,1.18)..(.92,1.30)
    ..controls(1.62,1.36)and(1.44,1.64)..(.78,1.58)
    ..controls(.48,1.54)and(.40,1.40)..(.38,1.30)--(crawTopLocal);
}

\newcommand{\crawDBottomEar}{%
  \draw[craw left]
    (crawMiddleLowerFront)--(-.38,-.20)
    ..controls(-1.70,-.45)and(-1.62,-1.26)..(-.38,-1.10)--(crawBottomLocalFront);
  \draw[craw right]
    (crawBottomLocalFront)--(.38,-1.10)
    ..controls(.40,-1.40)and(.48,-1.54)..(.78,-1.58)
    ..controls(1.44,-1.64)and(1.62,-1.36)..(.92,-1.30)
    ..controls(1.52,-1.18)and(1.58,-.43)..(.38,-.20)--(crawMiddleLowerFront);
}

\newcommand{\crawDRear}{%
  \draw[craw left rear]
    (crawMiddleLowerRear)--(-.38,-.20)
    ..controls(-1.70,-.45)and(-1.62,-1.26)..(-.38,-1.10)--(crawBottomLocalRear);
  \draw[craw right rear]
    (crawBottomLocalRear)--(.38,-1.10)
    ..controls(1.62,-1.26)and(1.70,-.45)..(.38,-.20)--(crawMiddleLowerRear);
}

\newcommand{\crawDRearOutsideEye}{%
  \draw[craw left rear]
    (crawMiddleLowerRear)--(-.38,-.20)
    ..controls(-1.70,-.45)and(-1.62,-1.26)..(-.38,-1.10)--(crawBottomLocalRear);
  \draw[craw right rear]
    (crawBottomLocalRear)--(.38,-1.10)
    ..controls(.76,-1.10)and(1.08,-.82)..(1.04,-.55)
    ..controls(1.01,-.40)and(.72,-.29)..(.38,-.20)--(crawMiddleLowerRear);
}

\newcommand{\crawURearBridgeRight}{%
  \draw[craw left]
    (crawTopLocal)--(-.38,1.30)
    ..controls(-1.62,1.26)and(-1.70,.45)..(-.38,.20)--(crawMiddleUpper);
  \draw[craw right] (crawMiddleUpper)..controls(.55,.22)and(.82,.34)..(.88,.54);
  \draw[craw right rear]
    (.88,.54)..controls(.98,.73)and(.92,.94)..(.82,1.06);
  \draw[craw right]
    (.82,1.06)..controls(.70,1.22)and(.52,1.30)..(crawTopLocal);
}

\newcommand{\crawDRearBridgeRight}{%
  \draw[craw left]
    (crawMiddleLowerFront)--(-.38,-.20)
    ..controls(-1.70,-.45)and(-1.62,-1.26)..(-.38,-1.10)--(crawBottomLocalFront);
  \draw[craw right]
    (crawBottomLocalFront)..controls(.52,-1.30)and(.70,-1.22)..(.82,-1.06);
  \draw[craw right rear]
    (.82,-1.06)..controls(.92,-.94)and(.98,-.73)..(.88,-.54);
  \draw[craw right] (.88,-.54)..controls(.82,-.34)and(.55,-.22)..(crawMiddleLowerFront);
}

\newcommand{\crawUMigratedRight}{%
  \draw[craw left]
    (crawTopLocal)--(-.38,1.30)
    ..controls(-1.50,1.24)and(-1.62,.48)..(-.38,.28)--(crawMiddleUpper);
  \draw[craw right]
    (crawMiddleUpper)..controls(.82,.12)and(1.58,.14)..(2.42,.18)
    ..controls(3.08,.20)and(3.48,.16)..(3.62,.28)
    ..controls(3.80,.41)and(3.74,.61)..(3.48,.67)
    ..controls(2.78,.72)and(2.26,.78)..(1.64,.91)
    ..controls(1.08,1.04)and(.70,1.23)..(crawTopLocal);
}

\newcommand{\crawDMigratedRight}{%
  \draw[craw left]
    (crawMiddleLowerFront)--(-.38,-.20)
    ..controls(-1.62,-.48)and(-1.50,-1.24)..(-.38,-1.17)--(crawBottomLocalFront);
  \draw[craw right]
    (crawBottomLocalFront)..controls(.70,-1.23)and(1.08,-1.04)..(1.64,-.91)
    ..controls(2.26,-.78)and(2.78,-.72)..(3.48,-.67)
    ..controls(3.74,-.61)and(3.80,-.41)..(3.62,-.28)
    ..controls(3.48,-.16)and(3.08,-.20)..(2.42,-.18)
    ..controls(1.58,-.14)and(.82,-.12)..(crawMiddleLowerFront);
}

\newcommand{\crawUBridgeLong}{%
  \draw[craw left]
    (crawTopLocal)--(-.38,1.30)
    ..controls(-1.45,1.23)and(-1.55,.55)..(-.38,.28)--(crawMiddleUpper);
  \draw[craw right]
    (crawMiddleUpper)..controls(1.15,.22)and(2.34,.20)..(3.47,.22)
    ..controls(3.70,.24)and(3.76,.48)..(3.53,.61)
    ..controls(2.35,.68)and(1.20,.88)..(crawTopLocal);
}

\newcommand{\crawDBridgeLong}{%
  \draw[craw left]
    (crawMiddleLowerFront)--(-.38,-.28)
    ..controls(-1.55,-.55)and(-1.45,-1.23)..(-.38,-1.17)--(crawBottomLocalFront);
  \draw[craw right]
    (crawBottomLocalFront)..controls(1.20,-.88)and(2.35,-.68)..(3.53,-.61)
    ..controls(3.76,-.48)and(3.70,-.24)..(3.47,-.22)
    ..controls(2.34,-.20)and(1.15,-.22)..(crawMiddleLowerFront);
}

%
\newcommand{\crawSixSevenBlackRear}{%
  \draw[craw left rear]
    (crawTopGlobalRear)--(-.38,1.39)
    ..controls(-2.20,1.50)and(-2.20,-1.50)..(-.38,-1.39)
    --(crawBottomGlobalRear);}

\newcommand{\crawSixSevenBlackFront}{%
  \draw[craw left]
    (-5.03,.29)..controls(-3.85,.30)and(-3.10,1.60)..(-.38,1.60)
    --(crawTopOpen);
  \draw[craw left] (-5.03,0)--(crawMiddleOpen);
  \draw[craw left]
    (-5.03,-.29)..controls(-3.85,-.30)and(-3.10,-1.60)..(-.38,-1.60)
    --(crawBottomOpen);
  \draw[craw left]
    (crawTopLocal)--(-.38,1.17)
    ..controls(-1.62,1.18)and(-1.70,.45)..(-.38,.28)
    --(crawMiddleUpper);
  \draw[craw left]
    (crawMiddleLowerFront)--(-.38,-.28)
    ..controls(-1.70,-.45)and(-1.62,-1.18)..(-.38,-1.17)
    --(crawBottomLocalFront);}

\newcommand{\crawSixSevenGreenABase}{%
  \draw[craw right]
    (crawTopOpen)--(.38,1.60)
    ..controls(3.10,1.60)and(3.85,.30)..(5.03,.29);}
\newcommand{\crawSixSevenGreenBBase}{%
  \draw[craw right] (crawMiddleOpen)--(5.03,0);}
\newcommand{\crawSixSevenGreenCBase}{%
  \draw[craw right]
    (crawBottomOpen)--(.38,-1.60)
    ..controls(3.10,-1.60)and(3.85,-.30)..(5.03,-.29);}
\newcommand{\crawSixSevenGreenUBase}{%
  \draw[craw right]
    (crawMiddleUpper)--(.38,.28)
    ..controls(1.70,.45)and(1.62,1.18)..(.38,1.17)
    --(crawTopLocal);}
\newcommand{\crawSixSevenGreenDBase}{%
  \draw[craw right]
    (crawBottomLocalFront)--(.38,-1.17)
    ..controls(1.62,-1.18)and(1.70,-.45)..(.38,-.28)
    --(crawMiddleLowerFront);}

\newcommand{\crawSixSevenGreenUCompact}{%
  \draw[craw right]
    (crawMiddleUpper)--(.38,.28)
    ..controls(.58,.28)and(.79,.39)..(.82,.59)
    ..controls(.85,.84)and(.67,1.10)..(.38,1.17)
    --(crawTopLocal);}
\newcommand{\crawSixSevenGreenDCompact}{%
  \draw[craw right]
    (crawBottomLocalFront)--(.38,-1.17)
    ..controls(.67,-1.10)and(.85,-.84)..(.82,-.59)
    ..controls(.79,-.39)and(.58,-.28)..(.38,-.28)
    --(crawMiddleLowerFront);}

\newcommand{\crawSixSevenGreenGOuter}{%
  \draw[craw right rear]
    (crawBottomGlobalRear)--(.38,-1.39)
    ..controls(2.28,-1.52)and(2.30,1.52)..(.38,1.39)
    --(crawTopGlobalRear);}

\newcommand{\crawSixSevenGreenAThree}{%
  \draw[craw right]
    (crawTopOpen)--(.38,1.60)
    ..controls(.78,1.55)and(1.08,1.12)..(1.10,.72)
    ..controls(1.11,.50)and(1.04,.34)..(1.15,.30)
    ..controls(1.32,.25)and(1.63,.25)..(1.82,.30)
    ..controls(2.70,.27)and(3.86,.27)..(5.03,.29);}

\newcommand{\crawSixSevenGreenCThree}{%
  \draw[craw right]
    (crawBottomOpen)--(.38,-1.60)
    ..controls(.78,-1.55)and(1.08,-1.12)..(1.10,-.72)
    ..controls(1.11,-.50)and(1.04,-.34)..(1.15,-.30)
    ..controls(1.36,-.25)and(1.70,-.25)..(1.92,-.30)
    ..controls(2.76,-.27)and(3.88,-.27)..(5.03,-.29);}

\newcommand{\crawSixSevenGreenGUpperLeft}{%
  \draw[craw right rear]
    (crawBottomGlobalRear)--(.38,-1.39)
    ..controls(2.06,-1.44)and(2.05,-.40)..(.94,.03)
    ..controls(.78,.17)and(.79,.42)..(.90,.58)
    ..controls(1.09,.87)and(1.34,1.18)..(.38,1.39)
    --(crawTopGlobalRear);}
\newcommand{\crawSixSevenGreenGLowerLeft}{%
  \draw[craw right rear]
    (crawBottomGlobalRear)--(.38,-1.39)
    ..controls(1.34,-1.18)and(1.09,-.87)..(.90,-.58)
    ..controls(.79,-.42)and(.78,-.17)..(.94,-.03)
    ..controls(2.05,.40)and(2.06,1.44)..(.38,1.39)
    --(crawTopGlobalRear);}

\newcommand{\crawSixSevenGreenULobeFour}{%
  \draw[craw right]
    (crawMiddleUpper)--(.38,.22)
    ..controls(.87,.07)and(1.51,.05)..(1.93,.15)
    ..controls(2.20,.23)and(2.22,.60)..(1.98,.82)
    ..controls(1.61,1.18)and(.82,1.21)..(.38,1.17)
    --(crawTopLocal);}
\newcommand{\crawSixSevenGreenDLobeFive}{%
  \draw[craw right]
    (crawBottomLocalFront)--(.38,-1.17)
    ..controls(.82,-1.21)and(1.61,-1.18)..(1.98,-.82)
    ..controls(2.22,-.60)and(2.20,-.23)..(1.93,-.15)
    ..controls(1.51,-.05)and(.87,-.07)..(.38,-.22)
    --(crawMiddleLowerFront);}

\newcommand{\crawSixSevenGreenULobeSix}{%
  \draw[craw right]
    (crawMiddleUpper)--(.38,.22)
    ..controls(.90,.02)and(1.90,-.02)..(2.42,.08)
    ..controls(2.78,.15)and(2.82,.67)..(2.52,.92)
    ..controls(1.95,1.25)and(.90,1.22)..(.38,1.17)
    --(crawTopLocal);}
\newcommand{\crawSixSevenGreenULobeSeven}{%
  \draw[craw right]
    (crawMiddleUpper)--(.38,.22)
    ..controls(.86,.07)and(1.76,.03)..(2.29,.13)
    ..controls(2.61,.20)and(2.64,.53)..(2.41,.72)
    ..controls(1.88,1.12)and(.87,1.20)..(.38,1.17)
    --(crawTopLocal);}

\newcommand{\crawSixSevenGreenGSixLeft}{%
  \draw[craw right rear]
    (crawBottomGlobalRear)--(.38,-1.39)
    ..controls(1.48,-1.42)and(1.51,-.34)..(1.16,.02)
    ..controls(1.05,.16)and(1.07,.39)..(1.17,.57)
    ..controls(1.32,.82)and(1.30,1.18)..(.38,1.39)
    --(crawTopGlobalRear);}

\newcommand{\crawSixSevenGreenGSixNeck}{%
  \draw[craw right rear]
    (crawBottomGlobalRear)--(.38,-1.39)
    ..controls(1.55,-1.47)and(3.44,-.92)..(3.64,-.50)
    ..controls(3.90,-.05)and(3.62,.78)..(3.16,1.04)
    ..controls(2.58,1.31)and(1.25,1.40)..(.38,1.39)
    --(crawTopGlobalRear);}


\newcommand{\crawSixSevenGreenASixLeftRear}{%
  \draw[craw right rear]
    (2.28,1.74)
    ..controls(2.12,1.24)and(1.68,.82)..(1.65,.46)
    ..controls(1.63,.22)and(1.78,.15)..(2.05,.16)
    ..controls(2.70,.20)and(3.40,.88)..(3.715,1.372);}
\newcommand{\crawSixSevenGreenASixLeftFront}{%
  \draw[craw right]
    (crawTopOpen)--(.38,1.60)
    ..controls(1.02,1.68)and(1.82,1.74)..(2.28,1.74);
  \draw[craw right]
    (3.715,1.372)..controls(4.05,.89)and(4.56,.36)..(5.03,.29);}

\newcommand{\crawSixSevenGreenBSixRightRear}{%
  \draw[craw right rear]
    (2.781,-1.720)
    ..controls(2.50,-.78)and(2.20,.28)..(1.96,.78)
    ..controls(2.06,.90)and(2.27,.86)..(2.32,.66)
    ..controls(2.80,.24)and(3.35,-.70)..(3.503,-1.495);}
\newcommand{\crawSixSevenGreenBSixRightFront}{%
  \draw[craw right]
    (crawMiddleOpen)--(.48,-.02)
    ..controls(1.27,-.05)and(2.30,-.63)..(2.781,-1.720);
  \draw[craw right]
    (3.503,-1.495)..controls(3.87,-.83)and(4.46,-.10)..(5.03,0);}

\newcommand{\crawSixSevenGreenCSixRightRear}{%
  \draw[craw right rear]
    (2.10,-1.74)
    ..controls(1.85,-.78)and(1.82,.42)..(1.68,1.15)
    ..controls(1.78,1.27)and(2.03,1.25)..(2.08,1.10)
    ..controls(2.62,.75)and(3.57,-.25)..(3.814,-1.305);}
\newcommand{\crawSixSevenGreenCSixRightFront}{%
  \draw[craw right]
    (crawBottomOpen)--(.42,-1.60)
    ..controls(.96,-1.67)and(1.67,-1.74)..(2.10,-1.74);
  \draw[craw right]
    (3.814,-1.305)..controls(4.12,-.78)and(4.61,-.36)..(5.03,-.29);}

\newcommand{\crawSixSevenGreenGSeven}{%
  \draw[craw right rear]
    (crawBottomGlobalRear)--(.38,-1.39)
    ..controls(1.48,-1.43)and(1.44,-.45)..(1.02,-.05)
    ..controls(.88,.10)and(.88,.41)..(1.00,.59)
    ..controls(1.18,.86)and(1.28,1.18)..(.38,1.39)
    --(crawTopGlobalRear);}

\newcommand{\crawSixSevenGreenCSevenRear}{%
  \draw[craw right rear]
    (2.10,-1.74)
    ..controls(1.68,-.92)and(1.48,.25)..(1.58,.82)
    ..controls(1.68,1.12)and(2.20,1.12)..(2.46,.80)
    ..controls(2.92,.35)and(3.55,-.45)..(3.814,-1.305);}
\newcommand{\crawSixSevenGreenCSevenFront}{%
  \draw[craw right]
    (crawBottomOpen)--(.42,-1.60)
    ..controls(.96,-1.67)and(1.67,-1.74)..(2.10,-1.74);
  \draw[craw right]
    (3.814,-1.305)..controls(4.12,-.78)and(4.61,-.36)..(5.03,-.29);}

\newcommand{\crawSixSevenGreenAEightRear}{%
  \draw[craw right rear]
    (2.45,1.74)
    ..controls(2.42,1.20)and(2.30,.58)..(1.78,.10)
    ..controls(1.45,-.13)and(1.43,-.46)..(1.70,-.62)
    ..controls(2.02,-.80)and(2.49,-.61)..(2.79,-.35)
    ..controls(3.13,-.04)and(3.31,.48)..(3.715,1.372);}
\newcommand{\crawSixSevenGreenAEightFront}{%
  \draw[craw right]
    (crawTopOpen)--(.38,1.60)
    ..controls(1.05,1.67)and(1.88,1.74)..(2.45,1.74);
  \draw[craw right]
    (3.715,1.372)..controls(4.02,.88)and(4.57,.36)..(5.03,.29);}

\newcommand{\crawSixSevenGreenCEight}{%
  \draw[craw right]
    (crawBottomOpen)--(.38,-1.60)
    ..controls(1.42,-1.63)and(2.66,-1.31)..(2.78,-.78)
    ..controls(2.84,-.48)and(2.64,-.24)..(2.38,-.16)
    ..controls(3.10,-.16)and(4.08,-.24)..(5.03,-.29);}
\newcommand{\crawSixSevenGreenDLobeEight}{%
  \draw[craw right]
    (crawBottomLocalFront)--(.38,-1.17)
    ..controls(.88,-1.23)and(1.92,-1.17)..(2.40,-.73)
    ..controls(2.68,-.48)and(2.61,-.12)..(2.28,-.04)
    ..controls(1.68,.08)and(.89,-.07)..(.38,-.22)
    --(crawMiddleLowerFront);}
\newcommand{\crawSixSevenGreenGEight}{%
  \draw[craw right rear]
    (crawBottomGlobalRear)--(.38,-1.39)
    ..controls(1.49,-1.15)and(1.44,-.71)..(1.34,-.48)
    ..controls(1.23,-.31)and(1.26,-.10)..(1.43,.02)
    ..controls(2.04,.40)and(2.05,1.42)..(.38,1.39)
    --(crawTopGlobalRear);}

\newcommand{\crawSixSevenGreenCNineRear}{%
  \draw[craw right rear]
    (2.10,-1.74)
    ..controls(1.72,-1.08)and(1.32,-.78)..(1.30,-.43)
    ..controls(1.29,-.15)and(1.58,-.05)..(1.92,-.08)
    ..controls(2.62,-.12)and(3.50,-.58)..(3.814,-1.305);}
\newcommand{\crawSixSevenGreenCNineFront}{%
  \draw[craw right]
    (crawBottomOpen)--(.42,-1.60)
    ..controls(.96,-1.67)and(1.67,-1.74)..(2.10,-1.74);
  \draw[craw right]
    (3.814,-1.305)..controls(4.12,-.78)and(4.61,-.36)..(5.03,-.29);}
\newcommand{\crawSixSevenGreenDLobeNine}{%
  \draw[craw right]
    (crawBottomLocalFront)--(.38,-1.17)
    ..controls(.84,-1.24)and(1.72,-1.22)..(2.18,-.88)
    ..controls(2.48,-.66)and(2.46,-.31)..(2.14,-.16)
    ..controls(1.62,.08)and(.86,-.06)..(.38,-.22)
    --(crawMiddleLowerFront);}
\newcommand{\crawSixSevenGreenGNine}{%
  \draw[craw right rear]
    (crawBottomGlobalRear)--(.38,-1.39)
    ..controls(1.23,-1.39)and(1.19,-.88)..(.98,-.61)
    ..controls(.84,-.46)and(.83,-.25)..(.91,-.09)
    ..controls(.98,.06)and(1.05,.23)..(1.06,.41)
    ..controls(1.10,.74)and(1.24,1.18)..(.38,1.39)
    --(crawTopGlobalRear);}

\newcommand{\crawRouteSixSevenThreeLeft}{%
  \crawSixSevenBlackRear \crawSixSevenGreenGOuter
  \crawSixSevenBlackFront \crawSixSevenGreenAThree
  \crawSixSevenGreenBBase \crawSixSevenGreenCBase
  \crawSixSevenGreenUCompact \crawSixSevenGreenDBase}
\newcommand{\crawRouteSixSevenThreeRight}{%
  \crawSixSevenBlackRear \crawSixSevenGreenGOuter
  \crawSixSevenBlackFront \crawSixSevenGreenABase
  \crawSixSevenGreenBBase \crawSixSevenGreenCThree
  \crawSixSevenGreenUBase \crawSixSevenGreenDCompact}
\newcommand{\crawRouteSixSevenFour}{%
  \crawSixSevenBlackRear \crawSixSevenGreenGUpperLeft
  \crawSixSevenBlackFront \crawSixSevenGreenABase
  \crawSixSevenGreenBBase \crawSixSevenGreenCBase
  \crawSixSevenGreenULobeFour \crawSixSevenGreenDBase}
\newcommand{\crawRouteSixSevenFive}{%
  \crawSixSevenBlackRear \crawSixSevenGreenGLowerLeft
  \crawSixSevenBlackFront \crawSixSevenGreenABase
  \crawSixSevenGreenBBase \crawSixSevenGreenCBase
  \crawSixSevenGreenUBase \crawSixSevenGreenDLobeFive}
\newcommand{\crawRouteSixSevenSixLeft}{%
  \crawSixSevenBlackRear \crawSixSevenGreenGSixLeft
  \crawSixSevenGreenASixLeftRear
  \crawSixSevenBlackFront \crawSixSevenGreenASixLeftFront
  \crawSixSevenGreenBBase \crawSixSevenGreenCBase
  \crawSixSevenGreenULobeSix \crawSixSevenGreenDBase}
\newcommand{\crawRouteSixSevenSixRight}{%
  \crawSixSevenBlackRear \crawSixSevenGreenGSixLeft
  \crawSixSevenGreenBSixRightRear \crawSixSevenGreenCSixRightRear
  \crawSixSevenBlackFront \crawSixSevenGreenABase
  \crawSixSevenGreenBSixRightFront \crawSixSevenGreenCSixRightFront
  \crawSixSevenGreenULobeSix \crawSixSevenGreenDBase}
\newcommand{\crawRouteSixSevenSeven}{%
  \crawSixSevenBlackRear \crawSixSevenGreenGSeven
  \crawSixSevenGreenCSevenRear
  \crawSixSevenBlackFront \crawSixSevenGreenABase
  \crawSixSevenGreenBBase \crawSixSevenGreenCSevenFront
  \crawSixSevenGreenULobeSeven \crawSixSevenGreenDBase}
\newcommand{\crawRouteSixSevenEight}{%
  \crawSixSevenBlackRear \crawSixSevenGreenGEight
  \crawSixSevenGreenAEightRear
  \crawSixSevenBlackFront \crawSixSevenGreenAEightFront
  \crawSixSevenGreenBBase \crawSixSevenGreenCBase
  \crawSixSevenGreenUBase \crawSixSevenGreenDLobeEight}
\newcommand{\crawRouteSixSevenNine}{%
  \crawSixSevenBlackRear \crawSixSevenGreenGNine
  \crawSixSevenGreenCNineRear
  \crawSixSevenBlackFront \crawSixSevenGreenABase
  \crawSixSevenGreenBBase \crawSixSevenGreenCNineFront
  \crawSixSevenGreenUBase \crawSixSevenGreenDLobeNine}

\newcommand{\crawRouteSixBase}{%
  \crawGlobalRear \crawOpenBase \crawUBase \crawDBase}
\newcommand{\crawRouteSixTen}{%
  \crawCandidateTenC \crawOpenAOnly \crawOpenBOnly \crawUBase \crawDBase}
\newcommand{\crawRouteSixRightKink}{%
  \crawGlobalRearRightKink \crawOpenBase \crawUBase \crawDBase}
\newcommand{\crawRouteSixLeftKink}{%
  \crawGlobalRearLeftKink \crawOpenBase \crawUBase \crawDBase}
\newcommand{\crawRouteSixNeck}{%
  \crawGlobalRearRightNeck \crawOpenBase \crawUBase \crawDBase}
\newcommand{\crawRouteSixUTongueR}{%
  \crawGlobalRear \crawOpenBase \crawUOutsideEyeTongue \crawDBase}
\newcommand{\crawRouteSixThreeAUnderEye}{%
  \crawGlobalRear \crawCandidateThreeAUnderEye \crawOpenBOnly \crawOpenCOnly
  \crawUBase \crawDBase}
\newcommand{\crawRouteSixThreeCAboveLowerEye}{%
  \crawGlobalRear \crawOpenAOnly \crawOpenBOnly \crawCandidateThreeCAboveLowerEye
  \crawUBase \crawDBase}
\newcommand{\crawRouteSixDTongueR}{%
  \crawGlobalRear \crawOpenBase \crawUBase \crawDOutsideEyeTongue}
\newcommand{\crawRouteSixUpperEyeOutside}{%
  \crawGlobalRear \crawOpenBase \crawUOutsideEyeBase \crawDBase}
\newcommand{\crawRouteSixLowerEyeOutside}{%
  \crawGlobalRear \crawOpenBase \crawUBase \crawDOutsideEyeBase}
\newcommand{\crawRouteSixUpperEyeInside}{%
  \crawGlobalRearLeftOfUpperEye \crawOpenBase \crawULobeAroundEye \crawDBase}
\newcommand{\crawRouteSixUpperEyeInsideNeck}{%
  \crawGlobalRearLeftOfUpperEyeNeck \crawOpenBase \crawULobeAroundEye \crawDBase}
\newcommand{\crawRouteSixLowerEyeInside}{%
  \crawGlobalRearLeftOfLowerEye \crawOpenBase \crawUBase \crawDLobeAroundEye}
\newcommand{\crawRouteSixUTongueL}{%
  \crawGlobalRear \crawOpenBase \crawUTongueLeft \crawDBase}
\newcommand{\crawRouteSixUTopEar}{%
  \crawGlobalRearRightKink \crawOpenBase \crawUTopEar \crawDBase}
\newcommand{\crawRouteSixEleven}{%
  \crawCandidateElevenA \crawOpenBCOnly \crawUBase \crawDBase}
\newcommand{\crawRouteSixTwelve}{%
  \crawGlobalRear \crawOpenAOnly \crawCandidateTwelveB \crawOpenCOnly \crawUBase}
\newcommand{\crawRouteSixThirteen}{%
  \crawGlobalRear \crawOpenAOnly \crawCandidateThirteenB \crawOpenCOnly \crawDBase}
\newcommand{\crawRouteSixFourteen}{%
  \crawGlobalRear \crawOpenAOnly \crawOpenBOnly \crawCandidateFourteenC \crawUBase}
\newcommand{\crawRouteSixFifteen}{%
  \crawGlobalRear \crawCandidateFifteenA \crawOpenBOnly \crawOpenCOnly \crawDBase}
\newcommand{\crawRouteSixDBottomEar}{%
  \crawGlobalRear \crawOpenBase \crawUBase \crawDBottomEar}
\newcommand{\crawRouteSixUpperMigrated}{%
  \crawGlobalRearRightNeck \crawOpenBase \crawUMigratedRight \crawDBase}
\newcommand{\crawRouteSixLowerMigrated}{%
  \crawGlobalRearRightNeck \crawOpenBase \crawUBase \crawDMigratedRight}
\newcommand{\crawRouteSixBridge}{%
  \crawGlobalRearRightKink \crawOpenBase \crawUBridgeLong \crawDBridgeLong}

%
\newcommand{\crawPillowFixedBlack}{%
  \draw[craw left]
    (-5.03,.29)..controls(-3.86,.32)and(-3.10,1.60)..(-.38,1.60)
    --(crawTopOpen);
  \draw[craw left]
    (-5.03,0)..controls(-3.72,.08)and(-2.96,1.34)..(-.38,1.39)
    --(crawTopGlobalFront);
  \draw[craw left]
    (-5.03,-.29)..controls(-3.85,-.30)and(-3.10,-1.60)..(-.38,-1.60)
    --(crawBottomOpen);
  \draw[craw left]
    (crawTopLocal)--(-.38,1.17)
    ..controls(-1.62,1.18)and(-1.70,.45)..(-.38,.28)
    --(crawMiddleUpper);
  \draw[craw left]
    (crawMiddleOpen)--(-.38,0)
    ..controls(-1.82,-.12)and(-1.92,-1.24)..(-.38,-1.39)
    --(crawBottomGlobalFront);
  \draw[craw left]
    (crawMiddleLowerFront)--(-.38,-.28)
    ..controls(-1.70,-.45)and(-1.62,-1.18)..(-.38,-1.17)
    --(crawBottomLocalFront);}

\newcommand{\crawPillowGreenTOneRone}{%
  \draw[craw right]
    (crawTopOpen)--(.38,1.60)
    ..controls(3.10,1.60)and(3.85,.30)..(5.03,.29);}
\newcommand{\crawPillowGreenTTwoRtwo}{%
  \draw[craw right]
    (crawTopGlobalFront)--(.38,1.39)
    ..controls(2.55,1.42)and(3.65,.10)..(5.03,0);}
\newcommand{\crawPillowGreenDThreeRthree}{%
  \draw[craw right]
    (crawBottomOpen)--(.38,-1.60)
    ..controls(3.10,-1.60)and(3.85,-.30)..(5.03,-.29);}
\newcommand{\crawPillowGreenTThreeMone}{%
  \draw[craw right]
    (crawMiddleUpper)--(.38,.28)
    ..controls(1.28,.44)and(1.28,1.03)..(.38,1.17)
    --(crawTopLocal);}
\newcommand{\crawPillowGreenTThreeMoneMigrated}{%
  \draw[craw right]
    (crawMiddleUpper)..controls(.74,.31)and(1.03,.34)..(1.34,.33)
    ..controls(2.05,.32)and(2.38,.53)..(2.36,.79)
    ..controls(2.33,1.10)and(1.34,1.13)..(crawTopLocal);}
\newcommand{\crawPillowGreenMTwoDtwoFront}{%
  \draw[craw right]
    (crawMiddleOpen)--(.38,0)
    ..controls(.44,-.12)and(.48,-.38)..(.62,-.50)
    ..controls(1.05,-.64)and(1.06,-1.08)..(.62,-1.30)
    ..controls(.50,-1.36)and(.30,-1.39)..(crawBottomGlobalFront);}
\newcommand{\crawPillowGreenMTwoDtwoRear}{%
  \draw[craw right rear]
    (crawMiddleOpenRear)
    ..controls(-.16,-.20)and(-.10,-.36)..(0,-.48)
    ..controls(.78,-.50)and(.78,-.90)..(0,-.92)
    ..controls(-.13,-1.05)and(-.18,-1.24)..(crawBottomGlobalRear);}
\newcommand{\crawPillowGreenMThreeDone}{%
  \draw[craw right]
    (crawBottomLocalFront)--(.38,-1.17)
    ..controls(1.46,-1.03)and(1.46,-.44)..(.38,-.28)
    --(crawMiddleLowerFront);}

\newcommand{\crawPillowGreenPOne}{%
  \crawPillowGreenTOneRone
  \crawPillowGreenTTwoRtwo
  \draw[craw right]
    (crawMiddleOpen)..controls(1.82,.02)and(3.70,-.24)..(5.03,-.29);
  \crawPillowGreenTThreeMone
  \crawPillowGreenMThreeDone
  \draw[craw right]
    (crawBottomGlobalFront)..controls(.72,-1.42)and(.78,-1.59)..(crawBottomOpen);}
\newcommand{\crawRoutePillowPOne}{%
  \crawPillowFixedBlack \crawPillowGreenPOne}

\newcommand{\crawPillowGreenPTwoRear}{%
  \path (0,-.70) ++({.72*cos(52)},{.22*sin(52)})
    coordinate (crawPTwoMThreeTurn);
  \path (0,-.70) ++({.72*cos(-52)},{.22*sin(-52)})
    coordinate (crawPTwoDOneTurn);
  \coordinate (crawPTwoMThreeShoulder) at (3.50327,-1.49538);
  \coordinate (crawPTwoDThreeShoulder) at (1.82,-1.74);
  \draw[craw right rear]
    (crawPTwoMThreeTurn)
    ..controls(1.16,-.63)and(2.55,-1.20)..(crawPTwoMThreeShoulder);
  \draw[craw right rear]
    (crawPTwoDOneTurn)
    ..controls(.82,-1.13)and(1.28,-1.61)..(crawPTwoDThreeShoulder);}
\newcommand{\crawPillowGreenPTwoFront}{%
  \crawPillowGreenTOneRone
  \crawPillowGreenTTwoRtwo
  \crawPillowGreenTThreeMone
  \draw[craw right]
    (crawMiddleLowerFront)..controls(.25,-.31)and(.36,-.43)
    ..(crawPTwoMThreeTurn);
  \draw[craw right]
    (crawPTwoMThreeShoulder)
    ..controls(4.10,-1.34)and(4.18,-.55)..(5.03,-.29);
  \draw[craw right]
    (crawBottomLocalFront)..controls(.23,-1.04)and(.35,-.95)
    ..(crawPTwoDOneTurn);
  \draw[craw right]
    (crawPTwoDThreeShoulder)
    ..controls(1.28,-1.73)and(.70,-1.66)..(.38,-1.60)
    --(crawBottomOpen);
  \draw[craw right]
    (crawMiddleOpen)
    ..controls(.78,-.03)and(1.62,-.18)..(1.78,-.55)
    ..controls(1.98,-.91)and(1.42,-1.28)..(.65,-1.37)
    ..controls(.47,-1.39)and(.30,-1.39)..(crawBottomGlobalFront);}
\newcommand{\crawRoutePillowPTwo}{%
  \crawPillowGreenPTwoRear \crawPillowFixedBlack \crawPillowGreenPTwoFront}

\newcommand{\crawPillowGreenPThreeRear}{%
  \path (0,.70) ++({.72*cos(50)},{.22*sin(50)})
    coordinate (crawPThreeTopOuterTurn);
  \path (0,-.70) ++({.72*cos(-50)},{.22*sin(-50)})
    coordinate (crawPThreeBottomOuterTurn);
  \path (0,.70) ++({.72*cos(-50)},{.22*sin(-50)})
    coordinate (crawPThreeTopInnerTurn);
  \path (0,-.70) ++({.72*cos(50)},{.22*sin(50)})
    coordinate (crawPThreeBottomInnerTurn);
  \draw[craw right rear]
    (crawPThreeTopOuterTurn)
    ..controls(1.34,.65)and(1.34,-.65)..(crawPThreeBottomOuterTurn);
  \draw[craw right rear]
    (crawPThreeTopInnerTurn)
    ..controls(.88,.32)and(.88,-.32)..(crawPThreeBottomInnerTurn);}
\newcommand{\crawPillowGreenPThreeFront}{%
  \crawPillowGreenTOneRone
  \crawPillowGreenTTwoRtwo
  \crawPillowGreenDThreeRthree
  \draw[craw right]
    (crawTopLocal)..controls(.24,1.04)and(.36,.95)
    ..(crawPThreeTopOuterTurn);
  \draw[craw right]
    (crawPThreeBottomOuterTurn)..controls(.36,-.95)and(.24,-1.04)
    ..(crawBottomLocalFront);
  \draw[craw right]
    (crawMiddleUpper)..controls(.25,.31)and(.38,.44)
    ..(crawPThreeTopInnerTurn);
  \draw[craw right]
    (crawPThreeBottomInnerTurn)..controls(.38,-.44)and(.25,-.31)
    ..(crawMiddleLowerFront);
  \draw[craw right,preaction={draw=crawFill,line width=2.15pt}]
    (crawMiddleOpen)--(.38,0)
    ..controls(.56,-.10)and(.72,-.34)..(.78,-.53)
    ..controls(1.15,-.72)and(1.14,-1.08)..(.68,-1.30)
    ..controls(.51,-1.36)and(.30,-1.39)..(crawBottomGlobalFront);}
\newcommand{\crawRoutePillowPThree}{%
  \crawPillowGreenPThreeRear \crawPillowFixedBlack \crawPillowGreenPThreeFront}

\newcommand{\crawPillowGreenPFour}{%
  \crawPillowGreenTOneRone
  \crawPillowGreenTTwoRtwo
  \crawPillowGreenDThreeRthree
  \draw[craw right]
    (crawTopLocal)--(.38,1.17)
    ..controls(2.05,1.05)and(2.34,.46)..(2.34,0)
    ..controls(2.34,-.48)and(2.02,-1.30)..(.38,-1.39)
    --(crawBottomGlobalFront);
  \draw[craw right]
    (crawMiddleUpper)..controls(.72,.24)and(.74,.03)..(crawMiddleOpen);
  \crawPillowGreenMThreeDone}
\newcommand{\crawRoutePillowPFour}{%
  \crawPillowFixedBlack \crawPillowGreenPFour}

\newcommand{\crawPillowGreenPFive}{%
  \crawPillowGreenTOneRone
  \draw[craw right]
    (crawBottomGlobalFront)--(.38,-1.39)
    ..controls(1.72,-1.42)and(2.26,-1.05)..(2.34,-.70)
    ..controls(2.46,-.31)and(3.78,-.08)..(5.03,0);
  \crawPillowGreenDThreeRthree
  \draw[craw right]
    (crawTopGlobalFront)--(.38,1.39)
    ..controls(1.82,1.34)and(1.90,.16)..(.38,0)
    --(crawMiddleOpen);
  \crawPillowGreenTThreeMone
  \crawPillowGreenMThreeDone}
\newcommand{\crawRoutePillowPFive}{%
  \crawPillowFixedBlack \crawPillowGreenPFive}

\newcommand{\crawPillowGreenPSixRear}{%
  \coordinate (crawPillowSixTOneShoulder) at (3.50327,1.49538);
  \coordinate (crawPillowSixDOneShoulder) at (4.44283,.82762);
  \path (0,-.70) ++({.72*cos(62)},{.22*sin(62)})
    coordinate (crawPillowSixMThreeTurn);
  \path (0,-.70) ++({.72*cos(-65)},{.22*sin(-65)})
    coordinate (crawPillowSixDOneTurn);
  \draw[craw right rear]
    (crawPillowSixTOneShoulder)
    ..controls(2.75,1.18)and(1.22,-.10)..(crawPillowSixMThreeTurn);
  \draw[craw right rear]
    (crawPillowSixDOneTurn)
    ..controls(.62,-1.14)and(1.38,-1.14)..(1.82,-.72)
    ..controls(2.62,-.05)and(3.55,.46)..(crawPillowSixDOneShoulder);}
\newcommand{\crawPillowGreenPSixFront}{%
  \draw[craw right]
    (crawTopOpen)--(.38,1.60)
    ..controls(1.30,1.64)and(2.42,1.68)..(crawPillowSixTOneShoulder);
  \draw[craw right]
    (crawPillowSixMThreeTurn)
    ..controls(.23,-.45)and(.18,-.34)..(crawMiddleLowerFront);
  \draw[craw right,preaction={draw=crawFill,line width=2.15pt}]
    (crawMiddleOpen)
    ..controls(.55,-.02)and(1.18,-.22)..(1.38,-.52)
    ..controls(1.62,-.82)and(1.55,-1.16)..(1.18,-1.36)
    ..controls(.86,-1.50)and(.42,-1.45)..(crawBottomGlobalFront);
  \draw[craw right]
    (crawBottomLocalFront)..controls(.13,-1.01)and(.12,-.95)
    ..(crawPillowSixDOneTurn);
  \draw[craw right]
    (crawPillowSixDOneShoulder)..controls(4.66,.64)and(4.88,.39)
    ..(5.03,.29);
  \crawPillowGreenDThreeRthree
  \crawPillowGreenTThreeMone
  \draw[craw right,preaction={draw=crawFill,line width=2.15pt}]
    (crawTopGlobalFront)--(.38,1.39)
    ..controls(2.55,1.42)and(3.65,.10)..(5.03,0);}
\newcommand{\crawRoutePillowPSix}{%
  \crawPillowGreenPSixRear \crawPillowFixedBlack \crawPillowGreenPSixFront}

\newcommand{\crawPillowGreenPSevenL}{%
  \draw[craw right]
    (crawMiddleOpen)
    ..controls(.40,-.03)and(.73,-.16)..(.90,-.38)
    ..controls(1.05,-.62)and(.93,-.98)..(.67,-1.16)
    ..controls(.48,-1.30)and(.30,-1.36)..(crawBottomGlobalFront);}
\newcommand{\crawPillowGreenPSevenBase}{%
  \crawPillowGreenTOneRone \crawPillowGreenTTwoRtwo
  \crawPillowGreenDThreeRthree}
\newcommand{\crawPillowGreenPSevenCoreNoL}{%
  \crawPillowGreenPSevenBase \crawPillowGreenTThreeMone}
\newcommand{\crawPillowGreenPSevenCore}{%
  \crawPillowGreenPSevenCoreNoL \crawPillowGreenPSevenL}
\newcommand{\crawPillowGreenPSeven}{%
  \crawPillowGreenPSevenCore \crawPillowGreenMThreeDone}
\newcommand{\crawRoutePillowPSeven}{%
  \crawPillowFixedBlack \crawPillowGreenPSeven}


\newcommand{\crawPillowRawGTenIFront}{%
  \draw[craw right]
    (crawMiddleLowerFront)--(.38,-.28)
    ..controls(.88,-.13)and(1.66,-.12)..(2.22,-.30)
    ..controls(2.84,-.51)and(2.77,-1.17)..(2.16,-1.37)
    ..controls(1.52,-1.53)and(.73,-1.45)..(.38,-1.39)
    --(crawBottomGlobalFront);}

\newcommand{\crawPillowRawGElevenIFront}{%
  \draw[craw right]
    (crawMiddleLowerFront)--(.38,-.28)
    ..controls(.90,-.16)and(1.38,-.10)..(1.84,-.19)
    ..controls(2.57,-.32)and(3.04,-.70)..(2.82,-1.09)
    ..controls(2.55,-1.51)and(1.30,-1.50)..(.38,-1.39)
    --(crawBottomGlobalFront);}

\newcommand{\crawPillowRawGTenIIRear}{%
  \coordinate (crawPillowTenIIMiddleTurn) at (0,-.48);
  \path (5.20,0) ++({.17*cos(205)},{.52*sin(205)})
    coordinate (crawPillowTenIIOuterTurn);
  \draw[craw right rear]
    (crawPillowTenIIMiddleTurn)
    ..controls(.48,-.44)and(.78,-.18)..(1.24,-.16)
    ..controls(1.88,-.13)and(2.47,-.36)..(2.60,-.67)
    ..controls(3.10,-.88)and(4.25,-.42)..(crawPillowTenIIOuterTurn);}
\newcommand{\crawPillowRawGTenIIFront}{%
  \draw[craw right]
    (crawMiddleLowerFront)
    arc[start angle=-32,end angle=-90,x radius=.17,y radius=.48];
  \draw[craw right]
    (crawPillowTenIIOuterTurn)
    ..controls(4.18,-.43)and(3.35,-1.04)..(2.45,-1.25)
    ..controls(1.67,-1.43)and(.72,-1.40)..(crawBottomGlobalFront);}

\newcommand{\crawPillowRawGElevenIIRear}{%
  \coordinate (crawPillowElevenIIMiddleTurn) at (0,-.48);
  \path (5.20,0) ++({.17*cos(205)},{.52*sin(205)})
    coordinate (crawPillowElevenIIOuterTurn);
  \draw[craw right rear]
    (crawPillowElevenIIMiddleTurn)
    ..controls(.46,-.82)and(1.06,-.89)..(1.62,-.70)
    ..controls(2.10,-.52)and(2.11,-.18)..(2.55,-.14)
    ..controls(3.34,-.08)and(4.28,-.20)..(crawPillowElevenIIOuterTurn);}
\newcommand{\crawPillowRawGElevenIIFront}{%
  \draw[craw right]
    (crawMiddleLowerFront)
    arc[start angle=-32,end angle=-90,x radius=.17,y radius=.48];
  \draw[craw right]
    (crawPillowElevenIIOuterTurn)
    ..controls(4.02,-.51)and(3.20,-1.12)..(2.30,-1.30)
    ..controls(1.51,-1.46)and(.69,-1.40)..(crawBottomGlobalFront);}

\newcommand{\crawPillowRawGTwelveRear}{%
  \coordinate (crawPillowTwelveMiddleTurn) at (0,-.48);
  \path (5.20,0) ++({.17*cos(205)},{.52*sin(205)})
    coordinate (crawPillowTwelveOuterTurn);
  \draw[craw right rear]
    (crawPillowTwelveMiddleTurn)
    ..controls(.50,-.43)and(.83,-.17)..(1.31,-.16)
    ..controls(1.96,-.14)and(2.47,-.38)..(2.57,-.70)
    ..controls(3.08,-.87)and(4.24,-.41)..(crawPillowTwelveOuterTurn);}
\newcommand{\crawPillowRawGTwelveFront}{%
  \draw[craw right]
    (crawMiddleLowerFront)
    arc[start angle=-32,end angle=-90,x radius=.17,y radius=.48];
  \draw[craw right]
    (crawPillowTwelveOuterTurn)
    ..controls(4.15,-.43)and(3.29,-1.06)..(2.39,-1.27)
    ..controls(1.62,-1.44)and(.71,-1.24)..(crawBottomLocalFront);}

\newcommand{\crawPillowRawGThirteenRear}{%
  \path (5.20,0) ++({.17*cos(200)},{.52*sin(200)})
    coordinate (crawPillowThirteenShoulderTurn);
  \path (0,-1.33) ++({.17*cos(-24)},{.41*sin(-24)})
    coordinate (crawPillowThirteenBottomTurn);
  \draw[craw right rear]
    (crawPillowThirteenShoulderTurn)
    ..controls(4.22,.20)and(3.42,1.23)..(2.43,1.50)
    ..controls(1.25,1.70)and(.82,.78)..(1.05,.08)
    ..controls(1.34,-.72)and(.92,-1.28)..(crawPillowThirteenBottomTurn);}
\newcommand{\crawPillowRawGThirteenFront}{%
  \draw[craw right]
    (crawMiddleLowerFront)--(.38,-.28)
    ..controls(1.12,-.08)and(2.54,-.08)..(3.52,-.10)
    ..controls(4.22,-.11)and(4.73,-.14)..(crawPillowThirteenShoulderTurn);
  \draw[craw right]
    (crawPillowThirteenBottomTurn)
    arc[start angle=-24,end angle=38,x radius=.17,y radius=.41];}

\newcommand{\crawPillowPhotoTwelveGRear}{%
  \path (0,-.70) ++({.72*cos(48)},{.22*sin(48)})
    coordinate (crawPillowPhotoTwelveGUpperTurn);
  \path (0,-.70) ++({.72*cos(-48)},{.22*sin(-48)})
    coordinate (crawPillowPhotoTwelveGLowerTurn);
  \draw[craw right rear]
    (crawPillowPhotoTwelveGUpperTurn)
    ..controls(.50,-.25)and(.62,.18)..(.78,.34)
    ..controls(.95,.47)and(1.25,.48)..(1.45,.38)
    ..controls(1.75,.32)and(1.95,.05)..(1.75,-.15)
    ..controls(1.55,-.45)and(1.15,-.72)..(.75,-.80)
    ..controls(.68,-.84)and(.58,-.87)..(crawPillowPhotoTwelveGLowerTurn);}
\newcommand{\crawPillowPhotoTwelveGFront}{%
  \draw[craw right]
    (crawMiddleLowerFront)
    ..controls(.27,-.29)and(.40,-.42)..(crawPillowPhotoTwelveGUpperTurn);
  \draw[craw right]
    (crawPillowPhotoTwelveGLowerTurn)
    ..controls(.40,-.97)and(.28,-1.05)..(crawBottomLocalFront);}
\newcommand{\crawPillowPhotoTwelveUFront}{%
  \draw[craw right,preaction={draw=crawFill,line width=2.15pt}]
    (crawTopLocal)
    ..controls(.42,1.08)and(1.22,1.04)..(1.52,.70)
    ..controls(1.72,.48)and(1.58,.20)..(1.35,.07)
    ..controls(1.05,-.03)and(.55,.04)..(.34,.18)
    ..controls(.25,.22)and(.20,.24)..(crawMiddleUpper);}
\newcommand{\crawPillowPhotoTwelveLFront}{%
  \draw[craw right,preaction={draw=crawFill,line width=2.15pt}]
    (crawMiddleOpen)
    ..controls(.40,-.03)and(.73,-.16)..(.90,-.38)
    ..controls(1.05,-.62)and(.93,-.98)..(.67,-1.16)
    ..controls(.48,-1.30)and(.30,-1.36)..(crawBottomGlobalFront);}
\newcommand{\crawPillowPhotoThirteenURear}{%
  \path (0,.70) ++({.72*cos(48)},{.22*sin(48)})
    coordinate (crawPillowPhotoThirteenUUpperTurn);
  \path (0,.70) ++({.72*cos(-48)},{.22*sin(-48)})
    coordinate (crawPillowPhotoThirteenULowerTurn);
  \draw[craw right rear]
    (crawPillowPhotoThirteenUUpperTurn)
    ..controls(.85,.95)and(1.45,.65)..(1.55,.25)
    ..controls(1.70,.05)and(1.35,-.08)..(1.10,-.02)
    ..controls(.78,-.10)and(.62,.12)..(crawPillowPhotoThirteenULowerTurn);}
\newcommand{\crawPillowPhotoThirteenUFront}{%
  \draw[craw right]
    (crawTopLocal)
    ..controls(.28,1.04)and(.40,.93)..(crawPillowPhotoThirteenUUpperTurn);
  \draw[craw right]
    (crawPillowPhotoThirteenULowerTurn)
    ..controls(.40,.45)and(.28,.31)..(crawMiddleUpper);}
\newcommand{\crawPillowPhotoThirteenGFront}{%
  \draw[craw right]
    (crawMiddleLowerFront)
    ..controls(.40,-.29)and(.72,-.40)..(.84,-.62)
    ..controls(.96,-.83)and(.66,-1.03)..(crawBottomLocalFront);}
\newcommand{\crawPillowPhotoThirteenLFront}{%
  \draw[craw right,preaction={draw=crawFill,line width=2.15pt}]
    (crawMiddleOpen)
    ..controls(.40,.08)and(.65,.32)..(.90,.38)
    ..controls(1.20,.44)and(1.55,.35)..(1.75,.15)
    ..controls(2.10,-.20)and(2.00,-.90)..(1.55,-1.15)
    ..controls(.90,-1.40)and(.40,-1.40)..(crawBottomGlobalFront);}

\newcommand{\crawPillowPhotoFourteenLFront}{%
  \draw[craw right]
    (crawMiddleOpen)
    ..controls(.50,-.02)and(1.20,-.12)..(1.72,-.18)
    ..controls(1.79,-.18)and(1.82,-.34)..(1.82,-.56)
    ..controls(1.82,-.85)and(1.45,-1.13)..(1.02,-1.27)
    ..controls(.68,-1.39)and(.38,-1.41)..(crawBottomGlobalFront);}
\newcommand{\crawPillowPhotoFourteenGFront}{%
  \draw[craw right]
    (crawMiddleLowerFront)
    ..controls(.36,-.29)and(.72,-.37)..(.92,-.54)
    ..controls(1.08,-.72)and(.76,-.98)..(crawBottomLocalFront);}

\newcommand{\crawPillowRawGFifteenRear}{%
  \path (0,-.70) ++({.72*cos(48)},{.22*sin(48)})
    coordinate (crawPillowFifteenUpperTurn);
  \path (0,-.70) ++({.72*cos(-48)},{.22*sin(-48)})
    coordinate (crawPillowFifteenLowerTurn);
  \draw[craw right rear]
    (crawPillowFifteenUpperTurn)
    ..controls(.60,-.43)and(.72,-.22)..(1.02,-.16)
    ..controls(1.25,-.10)and(1.55,-.10)..(1.75,-.17)
    ..controls(2.05,-.27)and(2.08,-.48)..(1.90,-.58)
    ..controls(1.65,-.78)and(.90,-.86)..(crawPillowFifteenLowerTurn);}
\newcommand{\crawPillowRawGFifteenFront}{%
  \draw[craw right]
    (crawMiddleLowerFront)
    ..controls(.27,-.29)and(.40,-.42)..(crawPillowFifteenUpperTurn);
  \draw[craw right]
    (crawPillowFifteenLowerTurn)
    ..controls(.40,-.97)and(.28,-1.05)..(crawBottomLocalFront);}
\newcommand{\crawPillowPhotoFifteenLFront}{%
  \draw[craw right,preaction={draw=crawFill,line width=2.15pt}]
    (crawMiddleOpen)
    ..controls(.40,-.03)and(.73,-.16)..(.90,-.38)
    ..controls(1.05,-.62)and(.93,-.98)..(.67,-1.16)
    ..controls(.48,-1.30)and(.30,-1.36)..(crawBottomGlobalFront);}

\newcommand{\crawPillowPhotoSixteenCRear}{%
  \coordinate (crawPillowSixteenBottomTurn) at (1.25,-1.74);
  \path (5.20,-.52)
    ..controls(4.20,-.62)and(3.92,-1.74)
    ..coordinate[pos=.30] (crawPillowSixteenShoulderTurn) (2.45,-1.74);
  \draw[craw right rear,dash phase=2pt]
    (crawPillowSixteenBottomTurn)
    ..controls(1.05,-1.40)and(.85,-.75)..(1.05,-.28)
    ..controls(1.20,-.02)and(1.65,.02)..(1.92,-.20)
    ..controls(2.35,-.52)and(3.55,-.72)
    ..(crawPillowSixteenShoulderTurn);}
\newcommand{\crawPillowPhotoSixteenCFront}{%
  \draw[craw right]
    (crawBottomOpen)--(.38,-1.60)
    ..controls(.65,-1.62)and(.95,-1.70)
    ..(crawPillowSixteenBottomTurn);
  \draw[craw right]
    (crawPillowSixteenShoulderTurn)
    ..controls(4.70,-.70)and(4.82,-.38)..(5.03,-.29);}
\newcommand{\crawPillowPhotoSixteenLFront}{%
  \draw[craw right,preaction={draw=crawFill,line width=2.15pt}]
    (crawMiddleOpen)
    ..controls(.50,-.02)and(1.20,-.12)..(1.72,-.18)
    ..controls(1.79,-.18)and(1.82,-.34)..(1.82,-.56)
    ..controls(1.82,-.85)and(1.45,-1.13)..(1.02,-1.27)
    ..controls(.68,-1.39)and(.38,-1.41)..(crawBottomGlobalFront);}
\newcommand{\crawPillowPhotoSixteenGFront}{%
  \draw[craw right]
    (crawMiddleLowerFront)
    ..controls(.36,-.29)and(.72,-.37)..(.92,-.54)
    ..controls(1.08,-.72)and(.76,-.98)..(crawBottomLocalFront);}

\newcommand{\crawPillowPhotoSeventeenGRear}{%
  \path (0,-.70) ++({.72*cos(48)},{.22*sin(48)})
    coordinate (crawPillowSeventeenGUpperTurn);
  \path (0,-.70) ++({.72*cos(-48)},{.22*sin(-48)})
    coordinate (crawPillowSeventeenGLowerTurn);
  \draw[craw right rear]
    (crawPillowSeventeenGUpperTurn)
    ..controls(.60,-.43)and(.72,-.22)..(1.02,-.16)
    ..controls(1.25,-.10)and(1.55,-.10)..(1.75,-.17)
    ..controls(2.05,-.27)and(2.08,-.48)..(1.90,-.58)
    ..controls(1.65,-.78)and(.90,-.86)
    ..(crawPillowSeventeenGLowerTurn);}
\newcommand{\crawPillowPhotoSeventeenGFront}{%
  \draw[craw right]
    (crawMiddleLowerFront)
    ..controls(.27,-.29)and(.40,-.42)
    ..(crawPillowSeventeenGUpperTurn);
  \draw[craw right]
    (crawPillowSeventeenGLowerTurn)
    ..controls(.40,-.97)and(.28,-1.05)..(crawBottomLocalFront);}
\newcommand{\crawPillowPhotoSeventeenLFront}{%
  \draw[craw right,preaction={draw=crawFill,line width=2.15pt}]
    (crawMiddleOpen)
    ..controls(.40,-.03)and(.73,-.16)..(.90,-.38)
    ..controls(1.05,-.62)and(.93,-.98)..(.67,-1.16)
    ..controls(.48,-1.30)and(.30,-1.36)..(crawBottomGlobalFront);}
\newcommand{\crawPillowPhotoSeventeenCFront}{%
  \draw[craw right,preaction={draw=crawFill,line width=2.15pt}]
    (crawBottomOpen)--(.38,-1.60)
    ..controls(.75,-1.58)and(1.05,-1.48)..(1.12,-1.15)
    ..controls(1.20,-.85)and(1.15,-.62)..(1.05,-.50)
    ..controls(.98,-.36)and(1.00,-.22)..(1.12,-.14)
    ..controls(1.50,.08)and(2.45,.15)..(3.20,.05)
    ..controls(4.00,-.05)and(4.60,-.22)..(5.03,-.29);}

\newcommand{\crawPillowPhotoEighteenARear}{%
  \coordinate (crawPillowEighteenTopTurn) at (1.25,1.74);
  \path (5.20,.52)
    ..controls(4.20,.62)and(3.92,1.74)
    ..coordinate[pos=.30] (crawPillowEighteenShoulderTurn) (2.45,1.74);
  \draw[craw right rear,dash phase=2pt]
    (crawPillowEighteenTopTurn)
    ..controls(1.50,1.30)and(1.18,.05)..(1.06,-.43)
    ..controls(1.02,-.58)and(1.30,-.72)..(1.72,-.58)
    ..controls(2.10,-.38)and(3.50,.55)
    ..(crawPillowEighteenShoulderTurn);}
\newcommand{\crawPillowPhotoEighteenAFront}{%
  \draw[craw right]
    (crawTopOpen)--(.38,1.60)
    ..controls(.65,1.62)and(.95,1.70)..(crawPillowEighteenTopTurn);
  \draw[craw right]
    (crawPillowEighteenShoulderTurn)
    ..controls(4.70,.70)and(4.82,.38)..(5.03,.29);}
\newcommand{\crawPillowPhotoEighteenTTwoFront}{%
  \draw[craw right,preaction={draw=crawFill,line width=2.15pt}]
    (crawTopGlobalFront)--(.38,1.39)
    ..controls(2.55,1.42)and(3.65,.10)..(5.03,0);}
\newcommand{\crawPillowPhotoEighteenLFront}{%
  \draw[craw right,preaction={draw=crawFill,line width=2.15pt}]
    (crawMiddleOpen)
    ..controls(.50,-.02)and(1.20,-.12)..(1.72,-.18)
    ..controls(1.79,-.18)and(1.82,-.34)..(1.82,-.56)
    ..controls(1.82,-.85)and(1.45,-1.13)..(1.02,-1.27)
    ..controls(.68,-1.39)and(.38,-1.41)..(crawBottomGlobalFront);}
\newcommand{\crawPillowPhotoEighteenGFront}{%
  \draw[craw right]
    (crawMiddleLowerFront)
    ..controls(.36,-.29)and(.72,-.37)..(.92,-.54)
    ..controls(1.08,-.72)and(.76,-.98)..(crawBottomLocalFront);}

\newcommand{\crawPillowPhotoNineteenGRear}{%
  \path (0,-.70) ++({.72*cos(48)},{.22*sin(48)})
    coordinate (crawPillowNineteenGUpperTurn);
  \path (0,-.70) ++({.72*cos(-48)},{.22*sin(-48)})
    coordinate (crawPillowNineteenGLowerTurn);
  \draw[craw right rear]
    (crawPillowNineteenGUpperTurn)
    ..controls(.60,-.43)and(.72,-.22)..(1.02,-.16)
    ..controls(1.25,-.10)and(1.55,-.10)..(1.75,-.17)
    ..controls(2.05,-.27)and(2.08,-.48)..(1.90,-.58)
    ..controls(1.65,-.78)and(.90,-.86)
    ..(crawPillowNineteenGLowerTurn);}
\newcommand{\crawPillowPhotoNineteenGFront}{%
  \draw[craw right]
    (crawMiddleLowerFront)
    ..controls(.27,-.29)and(.40,-.42)
    ..(crawPillowNineteenGUpperTurn);
  \draw[craw right]
    (crawPillowNineteenGLowerTurn)
    ..controls(.40,-.97)and(.28,-1.05)..(crawBottomLocalFront);}
\newcommand{\crawPillowPhotoNineteenLFront}{%
  \draw[craw right,preaction={draw=crawFill,line width=2.15pt}]
    (crawMiddleOpen)
    ..controls(.36,-.03)and(.66,-.16)..(.82,-.38)
    ..controls(.93,-.56)and(.93,-.82)..(.82,-.98)
    ..controls(.65,-1.22)and(.34,-1.36)..(crawBottomGlobalFront);}
\newcommand{\crawPillowPhotoNineteenTTwoFront}{%
  \draw[craw right,preaction={draw=crawFill,line width=2.15pt}]
    (crawTopGlobalFront)--(.38,1.39)
    ..controls(.85,1.38)and(1.35,1.12)..(1.45,.75)
    ..controls(1.55,.38)and(1.05,-.15)..(1.05,-.52)
    ..controls(1.10,-.72)and(1.45,-.76)..(1.78,-.68)
    ..controls(2.70,-.52)and(3.80,-.12)..(5.03,0);}

\newcommand{\crawRoutePillowRawTenI}{%
  \crawPillowFixedBlack \crawPillowGreenPSevenCore \crawPillowRawGTenIFront}
\newcommand{\crawRoutePillowRawElevenI}{%
  \crawPillowFixedBlack \crawPillowGreenPSevenCore \crawPillowRawGElevenIFront}
\newcommand{\crawRoutePillowRawTenII}{%
  \crawPillowRawGTenIIRear \crawPillowFixedBlack
  \crawPillowGreenPSevenCore \crawPillowRawGTenIIFront}
\newcommand{\crawRoutePillowRawElevenII}{%
  \crawPillowRawGElevenIIRear \crawPillowFixedBlack
  \crawPillowGreenPSevenCore \crawPillowRawGElevenIIFront}
\newcommand{\crawRoutePillowRawTwelve}{%
  \crawPillowPhotoTwelveGRear \crawPillowFixedBlack
  \crawPillowGreenPSevenBase \crawPillowPhotoTwelveLFront
  \crawPillowPhotoTwelveUFront
  \crawPillowPhotoTwelveGFront}
\newcommand{\crawRoutePillowRawThirteen}{%
  \crawPillowPhotoThirteenURear \crawPillowFixedBlack
  \crawPillowGreenPSevenBase \crawPillowPhotoThirteenLFront
  \crawPillowPhotoThirteenGFront \crawPillowPhotoThirteenUFront}
\newcommand{\crawRoutePillowRawFourteen}{%
  \crawPillowFixedBlack \crawPillowGreenPSevenCoreNoL
  \crawPillowPhotoFourteenLFront \crawPillowPhotoFourteenGFront}
\newcommand{\crawRoutePillowRawFifteen}{%
  \crawPillowRawGFifteenRear \crawPillowFixedBlack
  \crawPillowGreenPSevenCoreNoL \crawPillowPhotoFifteenLFront
  \crawPillowRawGFifteenFront}
\newcommand{\crawRoutePillowRawSixteen}{%
  \crawPillowPhotoSixteenCRear \crawPillowFixedBlack
  \crawPillowGreenTOneRone \crawPillowGreenTTwoRtwo
  \crawPillowGreenTThreeMone
  \crawPillowPhotoSixteenLFront \crawPillowPhotoSixteenGFront
  \crawPillowPhotoSixteenCFront}
\newcommand{\crawRoutePillowRawSeventeen}{%
  \crawPillowPhotoSeventeenGRear \crawPillowFixedBlack
  \crawPillowGreenTOneRone \crawPillowGreenTTwoRtwo
  \crawPillowGreenTThreeMone \crawPillowPhotoSeventeenGFront
  \crawPillowPhotoSeventeenLFront \crawPillowPhotoSeventeenCFront}
\newcommand{\crawRoutePillowRawEighteen}{%
  \crawPillowPhotoEighteenARear \crawPillowFixedBlack
  \crawPillowPhotoEighteenAFront \crawPillowPhotoEighteenTTwoFront
  \crawPillowGreenDThreeRthree \crawPillowGreenTThreeMone
  \crawPillowPhotoEighteenLFront \crawPillowPhotoEighteenGFront}
\newcommand{\crawRoutePillowRawNineteen}{%
  \crawPillowPhotoNineteenGRear \crawPillowFixedBlack
  \crawPillowGreenTOneRone \crawPillowGreenDThreeRthree
  \crawPillowGreenTThreeMone \crawPillowPhotoNineteenGFront
  \crawPillowPhotoNineteenLFront \crawPillowPhotoNineteenTTwoFront}

\newcommand{\crawPillowGreenPEight}{%
  \crawPillowGreenTOneRone
  \crawPillowGreenTTwoRtwo
  \draw[craw right]
    (crawTopLocal)--(.38,1.17)
    ..controls(.46,1.07)and(.55,.94)..(.58,.84)
    ..controls(.58,.65)and(.60,.48)..(.82,.38)
    ..controls(1.35,.20)and(2.20,.27)..(2.72,.36)
    ..controls(3.55,.43)and(4.25,-.16)..(5.03,-.29);
  \draw[craw right]
    (crawMiddleUpper)..controls(.72,.24)and(.74,.03)..(crawMiddleOpen);
  \crawPillowGreenMThreeDone
  \draw[craw right]
    (crawBottomGlobalFront)..controls(.72,-1.42)and(.78,-1.59)..(crawBottomOpen);}
\newcommand{\crawRoutePillowPEight}{%
  \crawPillowFixedBlack \crawPillowGreenPEight}

\newcommand{\crawPillowGreenPNineRear}{\crawPillowGreenMTwoDtwoRear}
\newcommand{\crawPillowGreenPNineFront}{%
  \crawPillowGreenTOneRone
  \crawPillowGreenTTwoRtwo
  \draw[craw right]
    (crawTopLocal)--(.38,1.17)
    ..controls(.46,1.07)and(.55,.94)..(.58,.84)
    ..controls(.58,.65)and(.60,.48)..(.82,.38)
    ..controls(1.35,.20)and(2.20,.27)..(2.72,.36)
    ..controls(3.55,.43)and(4.25,-.16)..(5.03,-.29);
  \draw[craw right]
    (crawMiddleUpper)--(.38,.28)
    ..controls(1.34,.24)and(1.42,.08)..(1.18,.04)
    ..controls(.78,.00)and(.40,-.06)..(.34,-.12)
    ..controls(.28,-.18)and(.22,-.23)..(crawMiddleLowerFront);
  \draw[craw right]
    (crawBottomLocalFront)..controls(1.58,-1.02)and(1.72,-1.62)..(crawBottomOpen);}
\newcommand{\crawRoutePillowPNine}{%
  \crawPillowGreenPNineRear \crawPillowFixedBlack \crawPillowGreenPNineFront}

\newcommand{\crawPillowGreenPTen}{%
  \draw[craw right]
    (crawBottomLocalFront)..controls(2.10,-1.10)and(3.72,.18)..(5.03,.29);
  \draw[craw right]
    (crawBottomGlobalFront)..controls(2.55,-1.40)and(3.92,-.08)..(5.03,0);
  \crawPillowGreenDThreeRthree
  \draw[craw right]
    (crawTopOpen)..controls(4.02,1.57)and(4.00,-.18)..(crawMiddleLowerFront);
  \draw[craw right]
    (crawTopGlobalFront)..controls(3.16,1.36)and(3.12,.08)..(crawMiddleOpen);
  \crawPillowGreenTThreeMoneMigrated}
\newcommand{\crawRoutePillowPTen}{%
  \crawPillowFixedBlack \crawPillowGreenPTen}

\newcommand{\crawPillowEarlyDisconnectedLeftSurface}{%
  \path[fill=crawFill,draw=none]
    (-5.20,.52)..controls(-4.20,.62)and(-3.92,1.74)..(-2.45,1.74)
    --(0,1.74)
    arc[start angle=90,end angle=270,x radius=.17,y radius=.41]
    ..controls(-.82,.83)and(-.82,.57)..(0,.48)
    arc[start angle=90,end angle=270,x radius=.17,y radius=.48]
    ..controls(-.82,-.57)and(-.82,-.83)..(0,-.92)
    arc[start angle=90,end angle=270,x radius=.17,y radius=.41]
    --(-2.45,-1.74)
    ..controls(-3.92,-1.74)and(-4.20,-.62)..(-5.20,-.52)--cycle;
  \draw[crawEdge,line width=1.05pt,line cap=round,line join=round]
    (-5.20,.52)..controls(-4.20,.62)and(-3.92,1.74)..(-2.45,1.74)--(0,1.74);
  \draw[crawEdge,line width=1.05pt,line cap=round,line join=round]
    (0,.92)..controls(-.82,.83)and(-.82,.57)..(0,.48);
  \draw[crawEdge,line width=1.05pt,line cap=round,line join=round]
    (0,-.48)..controls(-.82,-.57)and(-.82,-.83)..(0,-.92);
  \draw[crawEdge,line width=1.05pt,line cap=round,line join=round]
    (0,-1.74)--(-2.45,-1.74)
    ..controls(-3.92,-1.74)and(-4.20,-.62)..(-5.20,-.52);}

\newcommand{\crawPillowEarlySurfacesSeven}{%
  \path[craw sheet]
    (0,1.74)..controls(1.35,1.70)and(2.25,1.44)..(2.68,1.16)
    ..controls(2.96,.94)and(2.96,.28)..(2.58,.02)
    ..controls(2.12,-.30)and(1.18,-.44)..(0,-.48)
    arc[start angle=-90,end angle=90,x radius=.17,y radius=.48]
    ..controls(.56,.38)and(1.65,.38)..(1.65,.70)
    ..controls(1.65,1.02)and(.56,1.02)..(0,.92)
    arc[start angle=-90,end angle=90,x radius=.17,y radius=.41]--cycle;
  \path[craw sheet]
    (0,-.92)..controls(1.50,-.82)and(3.05,.00)..(4.35,.45)
    ..controls(4.62,.46)and(4.92,.48)..(5.20,.50)
    --(5.20,-.50)..controls(3.30,-.70)and(1.45,-1.58)..(0,-1.74)
    arc[start angle=-90,end angle=90,x radius=.17,y radius=.41]--cycle;}

\newcommand{\crawPillowEarlySurfacesEightNine}{%
  \path[craw sheet]
    (0,1.74)..controls(1.45,1.68)and(3.30,.70)..(5.20,.52)
    --(5.20,-.50)..controls(4.92,-.48)and(4.62,-.46)..(4.35,-.45)
    ..controls(3.10,-.12)and(1.50,.82)..(0,.92)
    arc[start angle=-90,end angle=90,x radius=.17,y radius=.41]--cycle;
  \path[craw sheet]
    (0,.48)..controls(1.42,.44)and(2.32,.12)..(2.72,-.18)
    ..controls(2.96,-.42)and(2.96,-1.10)..(2.58,-1.38)
    ..controls(2.12,-1.68)and(1.18,-1.71)..(0,-1.74)
    arc[start angle=-90,end angle=90,x radius=.17,y radius=.41]
    ..controls(.56,-1.02)and(1.65,-1.02)..(1.65,-.70)
    ..controls(1.65,-.38)and(.56,-.38)..(0,-.48)
    arc[start angle=-90,end angle=90,x radius=.17,y radius=.48]--cycle;}

\newcommand{\crawPillowEarlyGreenSeven}{%
  \draw[craw right]
    (crawBottomLocalFront)..controls(1.48,-.82)and(3.42,.10)..(5.03,.29);
  \draw[craw right]
    (crawBottomGlobalFront)..controls(1.55,-1.16)and(3.48,-.14)..(5.03,0);
  \draw[craw right]
    (crawBottomOpen)..controls(1.55,-1.48)and(3.55,-.42)..(5.03,-.29);
  \draw[craw right]
    (crawTopOpen)..controls(3.45,1.66)and(3.45,-.18)..(crawMiddleLowerFront);
  \draw[craw right]
    (crawTopGlobalFront)..controls(3.15,1.35)and(3.15,.05)..(crawMiddleOpen);
  \draw[craw right]
    (crawTopLocal)..controls(.76,1.08)and(1.56,1.04)..(2.08,.94)
    ..controls(2.44,.86)and(2.44,.54)..(2.08,.46)
    ..controls(1.56,.35)and(.76,.28)..(crawMiddleUpper);}

\newcommand{\crawPillowEarlyThroughTR}{%
  \draw[craw right]
    (crawTopOpen)..controls(1.48,1.57)and(3.55,.43)..(5.03,.29);
  \draw[craw right]
    (crawTopGlobalFront)..controls(1.52,1.35)and(3.58,.16)..(5.03,0);
  \draw[craw right]
    (crawTopLocal)..controls(1.56,1.04)and(3.60,-.12)..(5.03,-.29);}

\newcommand{\crawPillowEarlyGreenEightRear}{%
  \draw[craw right rear]
    (1.45568,.31680)
    ..controls(1.75,.02)and(2.28,-.35)..(1.47094,-.51656);
  \draw[craw right rear]
    (1.47094,-.88344)
    ..controls(2.24,-1.02)and(2.18,-1.58)..(1.28544,-1.68672);}
\newcommand{\crawPillowEarlyGreenEightFront}{%
  \crawPillowEarlyThroughTR
  \draw[craw right]
    (crawMiddleUpper)
    ..controls(.52,.42)and(1.08,.38)..(1.45568,.31680);
  \draw[craw right]
    (1.47094,-.51656)
    ..controls(1.12,-.32)and(.54,-.28)..(crawMiddleLowerFront);
  \draw[craw right]
    (crawBottomLocalFront)
    ..controls(.52,-1.10)and(1.15,-1.10)..(1.47094,-.88344);
  \draw[craw right]
    (1.28544,-1.68672)
    ..controls(.92,-1.70)and(.46,-1.65)..(crawBottomOpen);
  \draw[craw right,preaction={draw=crawFill,line width=2.15pt}]
    (crawMiddleOpen)
    ..controls(.52,.16)and(1.30,.10)..(1.76,-.14)
    ..controls(2.02,-.38)and(2.00,-1.08)..(1.72,-1.30)
    ..controls(1.34,-1.46)and(.58,-1.44)..(crawBottomGlobalFront);}

\newcommand{\crawPillowEarlyGreenNine}{%
  \crawPillowEarlyThroughTR
  \draw[craw right]
    (crawMiddleUpper)..controls(.72,.23)and(.72,.04)..(crawMiddleOpen);
  \draw[craw right]
    (crawMiddleLowerFront)..controls(.76,-.28)and(1.56,-.32)..(2.08,-.43)
    ..controls(2.50,-.52)and(2.50,-.88)..(2.08,-.97)
    ..controls(1.56,-1.05)and(.76,-1.08)..(crawBottomLocalFront);
  \draw[craw right]
    (crawBottomGlobalFront)..controls(.72,-1.40)and(.72,-1.62)..(crawBottomOpen);}

\newcommand{\crawPillowEarlyDisconnectedCard}[3]{%
  \begin{tikzpicture}[x=.555cm,y=.555cm,baseline=(current bounding box.center)]
    \path[use as bounding box] (-5.72,-2.14) rectangle (5.72,2.17);
    \crawPillowEarlyDisconnectedLeftSurface
    \ifnum#2=7\relax
      \crawPillowEarlySurfacesSeven
    \else
      \crawPillowEarlySurfacesEightNine
    \fi
    \crawThreeSeams
    \draw[craw cuff] (-5.20,0) ellipse[x radius=.17,y radius=.52];
    \draw[craw cuff] (5.20,0) ellipse[x radius=.17,y radius=.52];
    \ifnum#2=8\relax \crawPillowEarlyGreenEightRear\fi
    \crawPillowFixedBlack
    \ifnum#2=7\relax
      \crawPillowEarlyGreenSeven
    \else\ifnum#2=8\relax
      \crawPillowEarlyGreenEightFront
    \else
      \crawPillowEarlyGreenNine
    \fi\fi
    \crawCuffLabels
    \crawOriginalNumber{#1}
    \SDSurfaceTitle{#3}
  \end{tikzpicture}%
}

\newcommand{\crawRoutePillowAllFront}{%
  \crawGlobalFront \crawOpenBase \crawUBase \crawDOutsideEyeBase}
\newcommand{\crawRoutePillowDRear}{%
  \crawGlobalFront \crawOpenBase \crawUBase \crawDRearOutsideEye}
\newcommand{\crawRoutePillowUTongue}{%
  \crawGlobalFront \crawOpenBase \crawUTongueRight \crawDBase}
\newcommand{\crawRoutePillowURearBridge}{%
  \crawGlobalFront \crawOpenBase \crawURearBridgeRight \crawDOutsideEyeBase}
\newcommand{\crawRoutePillowDRearBridge}{%
  \crawGlobalFront \crawOpenBase \crawUBase \crawDRearBridgeRight}
\newcommand{\crawRoutePillowGlobalRear}{%
  \crawGlobalRear \crawOpenBase \crawUBase \crawDOutsideEyeBase}
\newcommand{\crawRoutePillowGlobalRightKink}{%
  \crawGlobalRearRightKink \crawOpenBase \crawUBase \crawDOutsideEyeBase}
\newcommand{\crawRoutePillowGlobalLeftKink}{%
  \crawGlobalRearLeftKink \crawOpenBase \crawUBase \crawDOutsideEyeBase}
\newcommand{\crawRoutePillowMigratedFront}{%
  \crawGlobalFront \crawOpenBase \crawUMigratedRight \crawDOutsideEyeBase}
\newcommand{\crawRoutePillowMigratedDRear}{%
  \crawGlobalFront \crawOpenBase \crawUMigratedRight \crawDRearOutsideEye}

\newcommand{\crawCuffLabels}{}

\newcommand{\crawSmallCross}{\textcolor{red!78!black}{$\times$}}

\newcommand{\crawCard}[6]{%
  \begin{tikzpicture}[x=.555cm,y=.555cm,baseline=(current bounding box.center)]
    \path[use as bounding box] (-5.72,-2.14) rectangle (5.72,2.17);
    \csname crawSurface#3\endcsname
    \csname crawThirdHole#6\endcsname
    \csname crawRoute#2\endcsname
    \crawCuffLabels
    \crawOriginalNumber{#1}
    \SDSurfaceTitle{#4}
    #5
  \end{tikzpicture}%
}

%
\newcommand{\crawDisconnectedBlackFixed}{%
  \draw[craw left]
    (-5.03,.29)..controls(-3.85,.30)and(-3.10,1.60)..(crawTopOpen);
  \draw[craw left] (-5.03,0)--(crawMiddleOpen);
  \draw[craw left]
    (-5.03,-.29)..controls(-3.85,-.30)and(-3.10,-1.60)..(crawBottomOpen);
  \draw[craw left]
    (crawTopLocal)--(-.38,1.17)
    ..controls(-1.62,1.18)and(-1.70,.45)..(-.38,.28)
    --(crawMiddleUpper);
  \draw[craw left]
    (crawMiddleLowerFront)--(-.38,-.28)
    ..controls(-1.70,-.45)and(-1.62,-1.18)..(-.38,-1.17)
    --(crawBottomLocalFront);
  \draw[craw left rear]
    (crawTopGlobalRear)..controls(-2.20,1.50)and(-2.20,-1.50)
    ..(crawBottomGlobalRear);
}

\newcommand{\crawDisconnectedSixteenRight}{%
  \draw[craw right]
    (crawTopOpen)..controls(1.55,1.48)and(3.55,.42)..(5.03,.29);
  \draw[craw right]
    (crawTopLocal)..controls(1.55,1.02)and(3.55,.15)..(5.03,0);
  \draw[craw right rear]
    (crawTopGlobalRear)..controls(.75,1.40)and(1.55,1.24)..(2.15,.88)
    ..controls(2.75,.52)and(3.45,-.20)..(crawSixteenTurn);
  \draw[craw right]
    (crawSixteenTurn)..controls(4.70,-.44)and(4.90,-.33)..(5.03,-.29);
  \draw[craw right]
    (crawMiddleOpen)..controls(1.70,.08)and(1.70,.42)..(crawMiddleUpper);
  \draw[craw right]
    (crawMiddleLowerFront)..controls(1.12,-.34)and(1.82,-.62)
    ..(crawBottomLocalFront);
  \draw[craw right]
    (crawBottomOpen)..controls(1.20,-1.68)and(2.03,-1.22)
    ..(crawSixteenInnerTurn);
  \draw[craw right rear]
    (crawSixteenInnerTurn)..controls(1.84,-.75)and(1.08,-1.17)
    ..(crawBottomGlobalRear);
}

\newcommand{\crawDisconnectedSeventeenRight}{%
  \draw[craw right rear]
    (crawBottomGlobalRear)..controls(.75,-1.40)and(1.55,-1.24)..(2.15,-.88)
    ..controls(2.75,-.52)and(3.45,.20)..(crawSeventeenTurn);
  \draw[craw right]
    (crawSeventeenTurn)..controls(4.70,.44)and(4.90,.33)..(5.03,.29);
  \draw[craw right]
    (crawBottomLocalFront)..controls(1.55,-1.02)and(3.55,-.15)..(5.03,0);
  \draw[craw right]
    (crawBottomOpen)..controls(1.55,-1.48)and(3.55,-.42)..(5.03,-.29);
  \draw[craw right]
    (crawTopLocal)..controls(1.12,1.02)and(1.82,.62)..(crawMiddleUpper);
  \draw[craw right]
    (crawMiddleOpen)..controls(1.70,-.08)and(1.70,-.42)
    ..(crawMiddleLowerFront);
  \draw[craw right]
    (crawTopOpen)..controls(1.20,1.68)and(2.03,1.22)
    ..(crawSeventeenInnerTurn);
  \draw[craw right rear]
    (crawSeventeenInnerTurn)..controls(1.84,.75)and(1.08,1.17)
    ..(crawTopGlobalRear);
}

\newcommand{\crawDisconnectedSixCard}[3]{%
  \begin{tikzpicture}[x=.555cm,y=.555cm,baseline=(current bounding box.center)]
    \path[use as bounding box] (-5.72,-2.14) rectangle (5.72,2.17);
    \path[fill=crawFill,draw=none]
      (-5.20,.52)..controls(-4.20,.62)and(-3.92,1.74)..(-2.45,1.74)
      --(0,1.74)
      arc[start angle=90,end angle=270,x radius=.17,y radius=.41]
      ..controls(-.82,.83)and(-.82,.57)..(0,.48)
      arc[start angle=90,end angle=270,x radius=.17,y radius=.48]
      ..controls(-.82,-.57)and(-.82,-.83)..(0,-.92)
      arc[start angle=90,end angle=270,x radius=.17,y radius=.41]
      --(-2.45,-1.74)
      ..controls(-3.92,-1.74)and(-4.20,-.62)..(-5.20,-.52)--cycle;
    \draw[crawEdge,line width=1.05pt,line cap=round,line join=round]
      (-5.20,.52)..controls(-4.20,.62)and(-3.92,1.74)..(-2.45,1.74)--(0,1.74);
    \draw[crawEdge,line width=1.05pt,line cap=round,line join=round]
      (0,.92)..controls(-.82,.83)and(-.82,.57)..(0,.48);
    \draw[crawEdge,line width=1.05pt,line cap=round,line join=round]
      (0,-.48)..controls(-.82,-.57)and(-.82,-.83)..(0,-.92);
    \draw[crawEdge,line width=1.05pt,line cap=round,line join=round]
      (0,-1.74)--(-2.45,-1.74)
      ..controls(-3.92,-1.74)and(-4.20,-.62)..(-5.20,-.52);
    \ifnum#2=16\relax
      \path[craw sheet]
        (0,1.74)..controls(1.45,1.68)and(3.30,.70)..(5.20,.52)
        --(5.20,-.50)..controls(4.92,-.48)and(4.62,-.46)..(4.35,-.45)
        ..controls(3.10,-.12)and(1.50,.82)..(0,.92)
        arc[start angle=-90,end angle=90,x radius=.17,y radius=.41]--cycle;
      \coordinate (crawSixteenTurn) at (4.35,-.45);
      \path[craw sheet]
        (0,.48)..controls(1.55,.46)and(2.35,.16)..(2.35,-.55)
        ..controls(2.35,-1.25)and(1.55,-1.68)..(0,-1.74)
        arc[start angle=-90,end angle=90,x radius=.17,y radius=.41]
        ..controls(.45,-1.00)and(.72,-.91)..(.72,-.70)
        ..controls(.72,-.49)and(.42,-.43)..(0,-.48)
        arc[start angle=-90,end angle=90,x radius=.17,y radius=.48]--cycle;
      \coordinate (crawSixteenInnerTurn) at (2.35,-.55);
    \else\ifnum#2=17\relax
      \path[craw sheet]
        (0,1.74)..controls(1.55,1.68)and(2.35,1.25)..(2.35,.55)
        ..controls(2.35,-.16)and(1.55,-.46)..(0,-.48)
        arc[start angle=-90,end angle=90,x radius=.17,y radius=.48]
        ..controls(.42,.43)and(.72,.49)..(.72,.70)
        ..controls(.72,.91)and(.45,1.00)..(0,.92)
        arc[start angle=-90,end angle=90,x radius=.17,y radius=.41]--cycle;
      \coordinate (crawSeventeenInnerTurn) at (2.35,.55);
      \path[craw sheet]
        (0,-.92)..controls(1.50,-.82)and(3.05,.00)..(4.35,.45)
        ..controls(4.62,.46)and(4.92,.48)..(5.20,.50)
        --(5.20,-.50)..controls(3.30,-.70)and(1.45,-1.58)..(0,-1.74)
        arc[start angle=-90,end angle=90,x radius=.17,y radius=.41]--cycle;
      \coordinate (crawSeventeenTurn) at (4.35,.45);
    \else
      \path[craw sheet]
        (0,1.74)..controls(2.55,1.20)and(2.55,-1.20)..(0,-1.74)
        arc[start angle=-90,end angle=90,x radius=.17,y radius=.41]
        ..controls(1.00,-.72)and(1.00,.72)..(0,.92)
        arc[start angle=-90,end angle=90,x radius=.17,y radius=.41]--cycle;
      \path[craw sheet]
        (0,.48)..controls(1.55,.32)and(3.40,.48)..(5.20,.50)
        --(5.20,-.50)..controls(3.40,-.48)and(1.55,-.32)..(0,-.48)
        arc[start angle=-90,end angle=90,x radius=.17,y radius=.48]--cycle;
    \fi\fi
    \crawThreeSeams
    \draw[craw cuff] (-5.20,0) ellipse[x radius=.17,y radius=.52];
    \draw[craw cuff] (5.20,0) ellipse[x radius=.17,y radius=.52];
    \crawDisconnectedBlackFixed
    \ifnum#2=16\relax
      \crawDisconnectedSixteenRight
    \else\ifnum#2=17\relax
      \crawDisconnectedSeventeenRight
    \else
      \draw[craw right] (crawMiddleUpper)..controls(1.55,.34)and(3.55,.34)..(5.03,.29);
      \draw[craw right] (crawMiddleOpen)--(5.03,0);
      \draw[craw right] (crawMiddleLowerFront)..controls(1.55,-.34)and(3.55,-.34)..(5.03,-.29);
      \draw[craw right]
        (crawTopOpen)..controls(.82,1.52)and(.82,1.18)..(crawTopLocal);
      \draw[craw right]
        (crawBottomOpen)..controls(.82,-1.52)and(.82,-1.18)..(crawBottomLocalFront);
      \draw[craw right rear] (crawTopGlobalRear)..controls(2.35,.92)and(2.35,-.92)..(crawBottomGlobalRear);
    \fi\fi
    \crawOriginalNumber{#1}
    \SDSurfaceTitle{#3}
  \end{tikzpicture}%
}

\newcommand{\crawPillowElevenThroughBundleTen}{%
  \draw[craw right]
    (crawTopOpen)..controls(1.48,1.57)and(3.55,.43)..(5.03,.29);
  \draw[craw right]
    (crawTopGlobalFront)..controls(1.52,1.35)and(3.58,.16)..(5.03,0);
  \draw[craw right]
    (crawTopLocal)..controls(1.56,1.04)and(3.60,-.12)..(5.03,-.29);}
\newcommand{\crawPillowElevenThroughBundleEleven}{%
  \crawPillowElevenThroughBundleTen}

\newcommand{\crawPillowElevenDepthTurns}{%
  \path (0,0) ++({.17*cos(-12)},{.48*sin(-12)})
    coordinate (crawPillowElevenMiddleTurn);
  \path (0,-1.33) ++({.17*cos(-20)},{.41*sin(-20)})
    coordinate (crawPillowElevenBottomTurn);}
\newcommand{\crawPillowElevenMiddleStubs}{%
  \draw[craw right] (crawMiddleOpen)
    arc[start angle=0,end angle=-12,x radius=.17,y radius=.48];
  \draw[craw right] (crawPillowElevenBottomTurn)
    arc[start angle=-20,end angle=-8,x radius=.17,y radius=.41];}

\newcommand{\crawPillowElevenOuterDepthTurns}{%
  \path (0,.48)..controls(1.48,.38)and(2.98,-.04)
    ..coordinate[pos=.05] (crawPillowElevenOuterMiddleTurn) (3.26,-.68);
  \path (0,-1.74)..controls(1.82,-1.68)and(3.31,-1.30)
    ..coordinate[pos=.05] (crawPillowElevenOuterBottomTurn) (3.26,-.68);}
\newcommand{\crawPillowElevenOuterStubs}{%
  \draw[craw right] (crawMiddleUpper)
    ..controls(.15,.33)and(.18,.42)..(crawPillowElevenOuterMiddleTurn);
  \draw[craw right] (crawPillowElevenOuterBottomTurn)
    ..controls(.23,-1.69)and(.17,-1.63)..(crawBottomOpen);}

\newcommand{\crawPillowElevenInnerDepthTurns}{%
  \path (0,0) ++({.17*cos(-46)},{.48*sin(-46)})
    coordinate (crawPillowElevenInnerMiddleTurn);
  \path (0,-1.33) ++({.17*cos(50)},{.41*sin(50)})
    coordinate (crawPillowElevenInnerBottomTurn);}
\newcommand{\crawPillowElevenInnerStubs}{%
  \draw[craw right] (crawMiddleLowerFront)
    arc[start angle=-32,end angle=-46,x radius=.17,y radius=.48];
  \draw[craw right] (crawPillowElevenInnerBottomTurn)
    arc[start angle=50,end angle=38,x radius=.17,y radius=.41];}

\newcommand{\crawPillowElevenViewTwoHoleTurns}{%
  \path (0,-.48)
    ..controls(.34,-.51)and(.52,-.60)
    ..coordinate[pos=.58] (crawPillowElevenViewTwoUpperTurn) (.52,-.75);
  \path (.52,-.75)
    ..controls(.52,-.88)and(.36,-.96)
    ..coordinate[pos=.55] (crawPillowElevenViewTwoLowerTurn) (0,-.92);}
\newcommand{\crawPillowElevenViewThreeHoleTurns}{%
  \path (0,-.48)
    ..controls(.44,-.49)and(.66,-.56)
    ..coordinate[pos=.58] (crawPillowElevenViewThreeUpperTurn) (.66,-.70);
  \path (.66,-.70)
    ..controls(.66,-.88)and(.45,-.96)
    ..coordinate[pos=.55] (crawPillowElevenViewThreeLowerTurn) (0,-.92);}
\newcommand{\crawPillowElevenViewTwoInnerStubs}{%
  \draw[craw right] (crawMiddleLowerFront)
    ..controls(.24,-.29)and(.34,-.42)..(crawPillowElevenViewTwoUpperTurn);
  \draw[craw right] (crawPillowElevenViewTwoLowerTurn)
    ..controls(.33,-.96)and(.23,-1.04)..(crawBottomLocalFront);}
\newcommand{\crawPillowElevenViewThreeInnerStubs}{%
  \draw[craw right] (crawMiddleLowerFront)
    ..controls(.25,-.29)and(.39,-.40)..(crawPillowElevenViewThreeUpperTurn);
  \draw[craw right] (crawPillowElevenViewThreeLowerTurn)
    ..controls(.37,-.97)and(.24,-1.04)..(crawBottomLocalFront);}
\newcommand{\crawPillowElevenViewFourOuterStubs}{%
  \crawPillowElevenOuterStubs}

\newcommand{\crawPillowElevenViewOneOuter}{%
  \draw[craw right]
    (crawMiddleUpper)
    ..controls(.23,.24)and(.33,.21)..(.42,.20)
    ..controls(1.10,.12)and(2.35,-.08)..(3.06,-.30)
    ..controls(3.45,-.44)and(3.45,-1.02)..(3.06,-1.22)
    ..controls(2.10,-1.48)and(.85,-1.54)..(crawBottomOpen);}
\newcommand{\crawPillowElevenViewOneMiddle}{%
  \draw[craw right]
    (crawMiddleOpen)
    ..controls(.23,-.01)and(.33,-.05)..(.42,-.06)
    ..controls(.92,-.10)and(2.10,-.18)..(2.72,-.38)
    ..controls(3.12,-.50)and(3.12,-.94)..(2.72,-1.08)
    ..controls(1.70,-1.30)and(.70,-1.33)..(crawBottomGlobalFront);}
\newcommand{\crawPillowElevenViewOneInner}{%
  \draw[craw right]
    (crawMiddleLowerFront)
    ..controls(.23,-.27)and(.33,-.29)..(.42,-.29)
    ..controls(.78,-.30)and(1.16,-.36)..(1.38,-.49)
    ..controls(1.58,-.58)and(1.58,-.82)..(1.38,-.95)
    ..controls(1.02,-1.03)and(.61,-1.07)..(crawBottomLocalFront);}
\newcommand{\crawPillowElevenViewOneFront}{%
  \crawPillowElevenViewOneOuter
  \crawPillowElevenViewOneMiddle
  \crawPillowElevenViewOneInner}

\newcommand{\crawPillowElevenViewTwoRear}{%
  \draw[craw right rear]
    (crawPillowElevenViewTwoUpperTurn)
    ..controls(.34,-.30)and(.57,-.29)..(.78,-.30)
    ..controls(1.72,-.27)and(2.58,-.28)..(2.96,-.46)
    ..controls(3.18,-.58)and(3.18,-.88)..(2.96,-1.00)
    ..controls(2.25,-1.20)and(1.22,-1.14)..(.76,-1.00)
    ..controls(.48,-.92)and(.37,-.91)
    ..(crawPillowElevenViewTwoLowerTurn);}
\newcommand{\crawPillowElevenViewTwoFront}{%
  \crawPillowElevenViewTwoInnerStubs
  \draw[craw right]
    (crawMiddleUpper)
    ..controls(.23,.24)and(.34,.20)..(.43,.18)
    ..controls(1.08,.08)and(2.40,-.12)..(3.08,-.32)
    ..controls(3.38,-.47)and(3.40,-.98)..(3.08,-1.20)
    ..controls(2.08,-1.50)and(.84,-1.55)..(crawBottomOpen);
  \draw[craw right,preaction={draw=crawFill,line width=2.15pt}]
    (crawMiddleOpen)
    ..controls(.24,-.02)and(.34,-.06)..(.45,-.08)
    ..controls(.82,-.12)and(1.18,-.25)..(1.36,-.45)
    ..controls(1.54,-.58)and(1.54,-.88)..(1.35,-1.02)
    ..controls(1.04,-1.21)and(.58,-1.32)..(crawBottomGlobalFront);}

\newcommand{\crawPillowElevenViewThreeRear}{%
  \draw[craw right rear]
    (crawPillowElevenViewThreeUpperTurn)
    ..controls(.34,-.31)and(.58,-.30)..(.78,-.31)
    ..controls(1.58,-.30)and(2.30,-.34)..(2.58,-.49)
    ..controls(2.78,-.60)and(2.78,-.83)..(2.58,-.96)
    ..controls(1.90,-1.15)and(1.08,-1.09)..(.72,-1.00)
    ..controls(.48,-.94)and(.40,-.92)
    ..(crawPillowElevenViewThreeLowerTurn);}
\newcommand{\crawPillowElevenViewThreeFront}{%
  \crawPillowElevenViewThreeInnerStubs
  \draw[craw right]
    (crawMiddleUpper)
    ..controls(.23,.24)and(.33,.21)..(.42,.20)
    ..controls(.92,.10)and(1.36,-.12)..(1.56,-.34)
    ..controls(1.72,-.49)and(1.72,-.91)..(1.55,-1.10)
    ..controls(1.22,-1.39)and(.76,-1.51)..(crawBottomOpen);
  \draw[craw right,preaction={draw=crawFill,line width=2.15pt}]
    (crawMiddleOpen)
    ..controls(.28,-.08)and(.38,-.10)..(.46,-.10)
    ..controls(.76,-.14)and(1.05,-.28)..(1.22,-.44)
    ..controls(1.38,-.58)and(1.38,-.84)..(1.22,-.99)
    ..controls(.96,-1.21)and(.57,-1.33)..(crawBottomGlobalFront);}

\newcommand{\crawPillowElevenViewFourRear}{%
  \draw[craw right rear,dash phase=2.00pt]
    (crawPillowElevenOuterMiddleTurn)
    ..controls(.24,.30)and(.66,.05)..(1.38,-.12)
    ..controls(1.52,-.32)and(1.52,-1.03)..(1.38,-1.20)
    ..controls(1.04,-1.43)and(.40,-1.53)
    ..(crawPillowElevenOuterBottomTurn);}
\newcommand{\crawPillowElevenViewFourFront}{%
  \crawPillowElevenViewFourOuterStubs
  \draw[craw right]
    (crawMiddleLowerFront)
    ..controls(.28,-.34)and(.56,-.45)..(.78,-.58)
    ..controls(.94,-.68)and(.94,-.82)..(.78,-.91)
    ..controls(.59,-1.00)and(.30,-1.04)..(crawBottomLocalFront);
  \draw[craw right,preaction={draw=crawFill,line width=2.15pt}]
    (crawMiddleOpen)
    ..controls(.28,-.04)and(.52,-.10)..(.78,-.15)
    ..controls(1.45,-.24)and(2.18,-.34)..(2.38,-.51)
    ..controls(2.56,-.64)and(2.56,-.87)..(2.36,-1.02)
    ..controls(1.70,-1.28)and(.78,-1.35)..(crawBottomGlobalFront);}

\newcommand{\crawPillowElevenTopSheetTen}{%
  \path[craw sheet]
    (0,1.74)..controls(1.45,1.68)and(3.30,.70)..(5.20,.52)
    --(5.20,-.50)..controls(4.92,-.48)and(4.62,-.46)..(4.35,-.45)
    ..controls(3.10,-.12)and(1.50,.82)..(0,.92)
    arc[start angle=-90,end angle=90,x radius=.17,y radius=.41]--cycle;}
\newcommand{\crawPillowElevenTopSheetEleven}{%
  \path[craw sheet]
    (0,1.74)..controls(1.30,1.72)and(3.05,.82)..(5.20,.52)
    --(5.20,-.50)..controls(4.86,-.47)and(4.52,-.42)..(4.20,-.39)
    ..controls(2.92,.02)and(1.68,.73)..(0,.92)
    arc[start angle=-90,end angle=90,x radius=.17,y radius=.41]--cycle;}

\newcommand{\crawPillowElevenBentSheetOne}{%
  \path[craw sheet]
    (0,.48)..controls(1.72,.46)and(3.36,.12)..(3.55,-.56)
    ..controls(3.58,-1.30)and(1.95,-1.70)..(0,-1.74)
    arc[start angle=-90,end angle=90,x radius=.17,y radius=.41]
    ..controls(.42,-.94)and(.62,-.86)..(.62,-.72)
    ..controls(.62,-.58)and(.42,-.50)..(0,-.48)
    arc[start angle=-90,end angle=90,x radius=.17,y radius=.48]--cycle;
  \path[craw torus rim] (2.60,-.72) ellipse[x radius=.28,y radius=.18];}
\newcommand{\crawPillowElevenBentSheetTwo}{%
  \path[craw sheet]
    (0,.48)..controls(1.55,.42)and(3.16,.04)..(3.42,-.64)
    ..controls(3.45,-1.35)and(1.75,-1.72)..(0,-1.74)
    arc[start angle=-90,end angle=90,x radius=.17,y radius=.41]
    ..controls(.36,-.96)and(.52,-.88)..(.52,-.75)
    ..controls(.52,-.60)and(.34,-.51)..(0,-.48)
    arc[start angle=-90,end angle=90,x radius=.17,y radius=.48]--cycle;
  \path[craw torus rim,rotate around={-3:(2.46,-.73)}]
    (2.46,-.73) ellipse[x radius=.25,y radius=.18];}
\newcommand{\crawPillowElevenBentSheetThree}{%
  \path[craw sheet]
    (0,.48)..controls(1.62,.45)and(3.18,.08)..(3.60,-.52)
    ..controls(3.80,-.82)and(3.35,-1.37)..(2.62,-1.55)
    ..controls(1.70,-1.73)and(.82,-1.73)..(0,-1.74)
    arc[start angle=-90,end angle=90,x radius=.17,y radius=.41]
    ..controls(.45,-.96)and(.66,-.88)..(.66,-.70)
    ..controls(.66,-.56)and(.44,-.49)..(0,-.48)
    arc[start angle=-90,end angle=90,x radius=.17,y radius=.48]--cycle;
  \path[craw torus rim,rotate around={-8:(2.02,-.70)}]
    (2.02,-.70) ellipse[x radius=.26,y radius=.17];}
\newcommand{\crawPillowElevenBentSheetFour}{%
  \path[craw sheet]
    (0,.48)..controls(1.48,.38)and(2.98,-.04)..(3.26,-.68)
    ..controls(3.31,-1.30)and(1.82,-1.68)..(0,-1.74)
    arc[start angle=-90,end angle=90,x radius=.17,y radius=.41]
    ..controls(.34,-.95)and(.50,-.86)..(.50,-.69)
    ..controls(.50,-.55)and(.32,-.49)..(0,-.48)
    arc[start angle=-90,end angle=90,x radius=.17,y radius=.48]--cycle;
  \path[craw torus rim,rotate around={8:(1.85,-.69)}]
    (1.85,-.69) ellipse[x radius=.24,y radius=.17];}

\newcommand{\crawDisconnectedPillowCard}[3]{%
  \begin{tikzpicture}[x=.555cm,y=.555cm,baseline=(current bounding box.center)]
    \path[use as bounding box] (-5.72,-2.14) rectangle (5.72,2.17);
    \path[fill=crawFill,draw=none]
      (-5.20,.52)..controls(-4.20,.62)and(-3.92,1.74)..(-2.45,1.74)
      --(0,1.74)
      arc[start angle=90,end angle=270,x radius=.17,y radius=.41]
      ..controls(-.82,.83)and(-.82,.57)..(0,.48)
      arc[start angle=90,end angle=270,x radius=.17,y radius=.48]
      ..controls(-.82,-.57)and(-.82,-.83)..(0,-.92)
      arc[start angle=90,end angle=270,x radius=.17,y radius=.41]
      --(-2.45,-1.74)
      ..controls(-3.92,-1.74)and(-4.20,-.62)..(-5.20,-.52)--cycle;
    \draw[crawEdge,line width=1.05pt,line cap=round,line join=round]
      (-5.20,.52)..controls(-4.20,.62)and(-3.92,1.74)..(-2.45,1.74)--(0,1.74);
    \draw[crawEdge,line width=1.05pt,line cap=round,line join=round]
      (0,.92)..controls(-.82,.83)and(-.82,.57)..(0,.48);
    \draw[crawEdge,line width=1.05pt,line cap=round,line join=round]
      (0,-.48)..controls(-.82,-.57)and(-.82,-.83)..(0,-.92);
    \draw[crawEdge,line width=1.05pt,line cap=round,line join=round]
      (0,-1.74)--(-2.45,-1.74)
      ..controls(-3.92,-1.74)and(-4.20,-.62)..(-5.20,-.52);
    \ifnum#2=1\relax
      \crawPillowElevenTopSheetTen \crawPillowElevenBentSheetOne
    \else\ifnum#2=2\relax
      \crawPillowElevenTopSheetEleven \crawPillowElevenBentSheetTwo
    \else\ifnum#2=3\relax
      \crawPillowElevenTopSheetTen \crawPillowElevenBentSheetThree
    \else
      \crawPillowElevenTopSheetEleven \crawPillowElevenBentSheetFour
    \fi\fi\fi
    \crawThreeSeams
    \draw[craw cuff] (-5.20,0) ellipse[x radius=.17,y radius=.52];
    \draw[craw cuff] (5.20,0) ellipse[x radius=.17,y radius=.52];
    \crawPillowElevenDepthTurns
    \crawPillowElevenOuterDepthTurns
    \crawPillowElevenInnerDepthTurns
    \crawPillowElevenViewTwoHoleTurns
    \crawPillowElevenViewThreeHoleTurns
    \ifcase#2\relax
    \or
    \or \crawPillowElevenViewTwoRear
    \or \crawPillowElevenViewThreeRear
    \or \crawPillowElevenViewFourRear
    \fi
    \crawPillowFixedBlack
    \ifnum#2=1\relax
      \crawPillowElevenThroughBundleTen
    \else\ifnum#2=2\relax
      \crawPillowElevenThroughBundleEleven
    \else\ifnum#2=3\relax
      \crawPillowElevenThroughBundleTen
    \else
      \crawPillowElevenThroughBundleEleven
    \fi\fi\fi
    \ifcase#2\relax
    \or \crawPillowElevenViewOneFront
    \or \crawPillowElevenViewTwoFront
    \or \crawPillowElevenViewThreeFront
    \or \crawPillowElevenViewFourFront
    \fi
    \crawCuffLabels
    \crawOriginalNumber{#1}
    \ifblank{#3}{}{\SDSurfaceTitle{#3}}
  \end{tikzpicture}%
}

\newcommand{\crawPlainRelation}{%
  \begin{tikzpicture}[baseline=-.5ex]
    \draw[<->,>=Latex,draw=ForestGreen,line width=.8pt] (-.42,0)--(.42,0);
  \end{tikzpicture}}
\newcommand{\crawZRelation}{%
  \begin{tikzpicture}[baseline=-.5ex]
    \draw[<->,>=Latex,draw=ForestGreen,line width=.8pt] (-.48,0)--(.48,0);
    \node[craw relation,below=1pt] at (0,0) {$\mathbb Z_2$};
  \end{tikzpicture}}
\newcommand{\crawEqualRelation}{\raisebox{1.1cm}{$=$}}
\newcommand{\crawSpacerRelation}{\hspace{1.05cm}}

\newcommand{\crawLedgerHeading}[2]{%
  \begin{center}
    {\large\bfseries #1}\par
    \smallskip
    {\footnotesize #2}\par
  \end{center}
  \smallskip
}

\newcommand{\crawLegend}{%
  \begin{center}
  \begin{tikzpicture}[baseline=-.4ex]
    \filldraw[fill=crawFill,draw=crawEdge] (0,-.10) rectangle (.72,.10);
    \node[craw note,anchor=west] at (.83,0) {boundary surface};
    \draw[craw left] (3.05,0)--(3.75,0);
    \node[craw note,anchor=west] at (3.86,0) {left piece};
    \draw[craw right] (5.70,0)--(6.40,0);
    \node[craw note,anchor=west] at (6.51,0) {right piece};
    \draw[craw left rear] (8.30,0)--(9.00,0);
    \node[craw note,anchor=west] at (9.11,0) {rear sheet};
  \end{tikzpicture}
  \end{center}
}

\newcommand{\crawKFourNativeSurface}{%
  \path[fill=crawFill,draw=none] (-.32,-1.74) rectangle (5.48,1.74);
  \draw[crawEdge,line width=1.02pt] (-.32,-1.74) rectangle (5.48,1.74);
  \draw[->,draw=black!58,line width=.72pt] (1.80,1.74)--(3.20,1.74);
  \draw[->,draw=black!58,line width=.72pt] (1.80,-1.74)--(3.20,-1.74);
  \draw[->,draw=black!58,line width=.72pt] (-.32,-.55)--(-.32,.55);
  \draw[->,draw=black!58,line width=.72pt] (5.48,-.55)--(5.48,.55);}

\newcommand{\crawKFourGGeneric}{%
  \draw[craw right rear]
    (crawBottomGlobalRear)--(.38,-1.39)
    ..controls(2.20,-1.50)and(2.20,1.50)..(.38,1.39)
    --(crawTopGlobalRear);}
\newcommand{\crawKFourUOutsideEye}{%
  \draw[craw right]
    (crawMiddleUpper)--(.38,.27)
    ..controls(.57,.28)and(.76,.37)..(.80,.55)
    ..controls(.86,.82)and(.70,1.10)..(.38,1.17)
    --(crawTopLocal);}
\newcommand{\crawKFourDOutsideEye}{%
  \draw[craw right]
    (crawBottomLocalFront)--(.38,-1.17)
    ..controls(.76,-1.10)and(1.08,-.82)..(1.04,-.55)
    ..controls(1.01,-.40)and(.72,-.29)..(.38,-.27)
    --(crawMiddleLowerFront);}

\newcommand{\crawKFourInColor}[2]{%
  \begingroup\colorlet{crawRight}{#1}#2\endgroup}

\newcommand{\crawRoutePillowColoredTwelve}{%
  \crawKFourInColor{RoyalBlue}{\crawPillowPhotoTwelveGRear}
  \crawPillowFixedBlack
  \crawKFourInColor{ForestGreen}{\crawPillowGreenTOneRone}
  \crawKFourInColor{BrickRed}{\crawPillowGreenTTwoRtwo}
  \crawKFourInColor{Orange}{\crawPillowGreenDThreeRthree}
  \crawKFourInColor{Magenta}{\crawPillowPhotoTwelveLFront}
  \crawKFourInColor{Purple}{\crawPillowPhotoTwelveUFront}
  \crawKFourInColor{RoyalBlue}{\crawPillowPhotoTwelveGFront}}
\newcommand{\crawRoutePillowColoredThirteen}{%
  \crawKFourInColor{Purple}{\crawPillowPhotoThirteenURear}
  \crawPillowFixedBlack
  \crawKFourInColor{ForestGreen}{\crawPillowGreenTOneRone}
  \crawKFourInColor{BrickRed}{\crawPillowGreenTTwoRtwo}
  \crawKFourInColor{Orange}{\crawPillowGreenDThreeRthree}
  \crawKFourInColor{Magenta}{\crawPillowPhotoThirteenLFront}
  \crawKFourInColor{RoyalBlue}{\crawPillowPhotoThirteenGFront}
  \crawKFourInColor{Purple}{\crawPillowPhotoThirteenUFront}}
\newcommand{\crawRoutePillowColoredFourteen}{%
  \crawPillowFixedBlack
  \crawKFourInColor{ForestGreen}{\crawPillowGreenTOneRone}
  \crawKFourInColor{BrickRed}{\crawPillowGreenTTwoRtwo}
  \crawKFourInColor{Orange}{\crawPillowGreenDThreeRthree}
  \crawKFourInColor{Purple}{\crawPillowGreenTThreeMone}
  \crawKFourInColor{Magenta}{\crawPillowPhotoFourteenLFront}
  \crawKFourInColor{RoyalBlue}{\crawPillowPhotoFourteenGFront}}
\newcommand{\crawRoutePillowColoredFifteen}{%
  \crawKFourInColor{RoyalBlue}{\crawPillowRawGFifteenRear}
  \crawPillowFixedBlack
  \crawKFourInColor{ForestGreen}{\crawPillowGreenTOneRone}
  \crawKFourInColor{BrickRed}{\crawPillowGreenTTwoRtwo}
  \crawKFourInColor{Orange}{\crawPillowGreenDThreeRthree}
  \crawKFourInColor{Purple}{\crawPillowGreenTThreeMone}
  \crawKFourInColor{Magenta}{\crawPillowPhotoFifteenLFront}
  \crawKFourInColor{RoyalBlue}{\crawPillowRawGFifteenFront}}
\newcommand{\crawRoutePillowColoredSixteen}{%
  \crawKFourInColor{Orange}{\crawPillowPhotoSixteenCRear}
  \crawPillowFixedBlack
  \crawKFourInColor{ForestGreen}{\crawPillowGreenTOneRone}
  \crawKFourInColor{BrickRed}{\crawPillowGreenTTwoRtwo}
  \crawKFourInColor{Purple}{\crawPillowGreenTThreeMone}
  \crawKFourInColor{Magenta}{\crawPillowPhotoSixteenLFront}
  \crawKFourInColor{RoyalBlue}{\crawPillowPhotoSixteenGFront}
  \crawKFourInColor{Orange}{\crawPillowPhotoSixteenCFront}}
\newcommand{\crawRoutePillowColoredSeventeen}{%
  \crawKFourInColor{RoyalBlue}{\crawPillowPhotoSeventeenGRear}
  \crawPillowFixedBlack
  \crawKFourInColor{ForestGreen}{\crawPillowGreenTOneRone}
  \crawKFourInColor{BrickRed}{\crawPillowGreenTTwoRtwo}
  \crawKFourInColor{Purple}{\crawPillowGreenTThreeMone}
  \crawKFourInColor{RoyalBlue}{\crawPillowPhotoSeventeenGFront}
  \crawKFourInColor{Magenta}{\crawPillowPhotoSeventeenLFront}
  \crawKFourInColor{Orange}{\crawPillowPhotoSeventeenCFront}}
\newcommand{\crawRoutePillowColoredEighteen}{%
  \crawKFourInColor{ForestGreen}{\crawPillowPhotoEighteenARear}
  \crawPillowFixedBlack
  \crawKFourInColor{ForestGreen}{\crawPillowPhotoEighteenAFront}
  \crawKFourInColor{BrickRed}{\crawPillowPhotoEighteenTTwoFront}
  \crawKFourInColor{Orange}{\crawPillowGreenDThreeRthree}
  \crawKFourInColor{Purple}{\crawPillowGreenTThreeMone}
  \crawKFourInColor{Magenta}{\crawPillowPhotoEighteenLFront}
  \crawKFourInColor{RoyalBlue}{\crawPillowPhotoEighteenGFront}}
\newcommand{\crawRoutePillowColoredNineteen}{%
  \crawKFourInColor{RoyalBlue}{\crawPillowPhotoNineteenGRear}
  \crawPillowFixedBlack
  \crawKFourInColor{ForestGreen}{\crawPillowGreenTOneRone}
  \crawKFourInColor{Orange}{\crawPillowGreenDThreeRthree}
  \crawKFourInColor{Purple}{\crawPillowGreenTThreeMone}
  \crawKFourInColor{RoyalBlue}{\crawPillowPhotoNineteenGFront}
  \crawKFourInColor{Magenta}{\crawPillowPhotoNineteenLFront}
  \crawKFourInColor{BrickRed}{\crawPillowPhotoNineteenTTwoFront}}

\newif\ifcrawKFourSolidABCCycle
\crawKFourSolidABCCyclefalse
\newcommand{\crawKFourBlueInColor}[1]{%
  \begingroup
    \colorlet{crawRight}{RoyalBlue}%
    \ifcrawKFourSolidABCCycle
      \tikzset{craw right rear/.style={draw=crawRight,line width=1.08pt,
        line cap=round,line join=round}}%
    \fi
    #1%
  \endgroup}

\newcommand{\crawKFourRouteOne}{%
  \crawSixSevenBlackRear
  \crawKFourBlueInColor{\crawKFourGGeneric}%
  \crawSixSevenBlackFront
  \crawKFourInColor{ForestGreen}{\crawSixSevenGreenABase}%
  \crawKFourInColor{BrickRed}{\crawSixSevenGreenBBase}%
  \crawKFourInColor{Orange}{\crawSixSevenGreenCBase}%
  \crawKFourInColor{Purple}{\crawKFourUOutsideEye}%
  \crawKFourInColor{Magenta}{\crawSixSevenGreenDBase}}
\newcommand{\crawKFourRouteTwo}{%
  \crawSixSevenBlackRear
  \crawKFourBlueInColor{\crawKFourGGeneric}%
  \crawSixSevenBlackFront
  \crawKFourInColor{ForestGreen}{\crawSixSevenGreenABase}%
  \crawKFourInColor{BrickRed}{\crawSixSevenGreenBBase}%
  \crawKFourInColor{Orange}{\crawSixSevenGreenCBase}%
  \crawKFourInColor{Purple}{\crawSixSevenGreenUBase}%
  \crawKFourInColor{Magenta}{\crawKFourDOutsideEye}}
\newcommand{\crawKFourRouteThreeLeft}{%
  \crawSixSevenBlackRear
  \crawKFourBlueInColor{\crawSixSevenGreenGOuter}%
  \crawSixSevenBlackFront
  \crawKFourInColor{ForestGreen}{\crawSixSevenGreenAThree}%
  \crawKFourInColor{BrickRed}{\crawSixSevenGreenBBase}%
  \crawKFourInColor{Orange}{\crawSixSevenGreenCBase}%
  \crawKFourInColor{Purple}{\crawSixSevenGreenUCompact}%
  \crawKFourInColor{Magenta}{\crawSixSevenGreenDBase}}
\newcommand{\crawKFourRouteThreeRight}{%
  \crawSixSevenBlackRear
  \crawKFourBlueInColor{\crawSixSevenGreenGOuter}%
  \crawSixSevenBlackFront
  \crawKFourInColor{ForestGreen}{\crawSixSevenGreenABase}%
  \crawKFourInColor{BrickRed}{\crawSixSevenGreenBBase}%
  \crawKFourInColor{Orange}{\crawSixSevenGreenCThree}%
  \crawKFourInColor{Purple}{\crawSixSevenGreenUBase}%
  \crawKFourInColor{Magenta}{\crawSixSevenGreenDCompact}}
\newcommand{\crawKFourRouteFour}{%
  \crawSixSevenBlackRear
  \crawKFourBlueInColor{\crawSixSevenGreenGUpperLeft}%
  \crawSixSevenBlackFront
  \crawKFourInColor{ForestGreen}{\crawSixSevenGreenABase}%
  \crawKFourInColor{BrickRed}{\crawSixSevenGreenBBase}%
  \crawKFourInColor{Orange}{\crawSixSevenGreenCBase}%
  \crawKFourInColor{Purple}{\crawSixSevenGreenULobeFour}%
  \crawKFourInColor{Magenta}{\crawSixSevenGreenDBase}}
\newcommand{\crawKFourRouteFive}{%
  \crawSixSevenBlackRear
  \crawKFourBlueInColor{\crawSixSevenGreenGLowerLeft}%
  \crawSixSevenBlackFront
  \crawKFourInColor{ForestGreen}{\crawSixSevenGreenABase}%
  \crawKFourInColor{BrickRed}{\crawSixSevenGreenBBase}%
  \crawKFourInColor{Orange}{\crawSixSevenGreenCBase}%
  \crawKFourInColor{Purple}{\crawSixSevenGreenUBase}%
  \crawKFourInColor{Magenta}{\crawSixSevenGreenDLobeFive}}
\newcommand{\crawKFourRouteSixLeft}{%
  \crawSixSevenBlackRear
  \crawKFourBlueInColor{\crawSixSevenGreenGSixLeft}%
  \crawKFourInColor{ForestGreen}{\crawSixSevenGreenASixLeftRear}%
  \crawSixSevenBlackFront
  \crawKFourInColor{ForestGreen}{\crawSixSevenGreenASixLeftFront}%
  \crawKFourInColor{BrickRed}{\crawSixSevenGreenBBase}%
  \crawKFourInColor{Orange}{\crawSixSevenGreenCBase}%
  \crawKFourInColor{Purple}{\crawSixSevenGreenULobeSix}%
  \crawKFourInColor{Magenta}{\crawSixSevenGreenDBase}}
\newcommand{\crawKFourRouteSixRight}{%
  \crawSixSevenBlackRear
  \crawKFourBlueInColor{\crawSixSevenGreenGSixLeft}%
  \crawKFourInColor{BrickRed}{\crawSixSevenGreenBSixRightRear}%
  \crawKFourInColor{Orange}{\crawSixSevenGreenCSixRightRear}%
  \crawSixSevenBlackFront
  \crawKFourInColor{ForestGreen}{\crawSixSevenGreenABase}%
  \crawKFourInColor{BrickRed}{\crawSixSevenGreenBSixRightFront}%
  \crawKFourInColor{Orange}{\crawSixSevenGreenCSixRightFront}%
  \crawKFourInColor{Purple}{\crawSixSevenGreenULobeSix}%
  \crawKFourInColor{Magenta}{\crawSixSevenGreenDBase}}
\newcommand{\crawKFourRouteSeven}{%
  \crawSixSevenBlackRear
  \crawKFourBlueInColor{\crawSixSevenGreenGSeven}%
  \crawKFourInColor{Orange}{\crawSixSevenGreenCSevenRear}%
  \crawSixSevenBlackFront
  \crawKFourInColor{ForestGreen}{\crawSixSevenGreenABase}%
  \crawKFourInColor{BrickRed}{\crawSixSevenGreenBBase}%
  \crawKFourInColor{Orange}{\crawSixSevenGreenCSevenFront}%
  \crawKFourInColor{Purple}{\crawSixSevenGreenULobeSeven}%
  \crawKFourInColor{Magenta}{\crawSixSevenGreenDBase}}
\newcommand{\crawKFourRouteEight}{%
  \crawSixSevenBlackRear
  \crawKFourBlueInColor{\crawSixSevenGreenGEight}%
  \crawKFourInColor{ForestGreen}{\crawSixSevenGreenAEightRear}%
  \crawSixSevenBlackFront
  \crawKFourInColor{ForestGreen}{\crawSixSevenGreenAEightFront}%
  \crawKFourInColor{BrickRed}{\crawSixSevenGreenBBase}%
  \crawKFourInColor{Orange}{\crawSixSevenGreenCBase}%
  \crawKFourInColor{Purple}{\crawSixSevenGreenUBase}%
  \crawKFourInColor{Magenta}{\crawSixSevenGreenDLobeEight}}
\newcommand{\crawKFourRouteNine}{%
  \crawSixSevenBlackRear
  \crawKFourBlueInColor{\crawSixSevenGreenGNine}%
  \crawKFourInColor{Orange}{\crawSixSevenGreenCNineRear}%
  \crawSixSevenBlackFront
  \crawKFourInColor{ForestGreen}{\crawSixSevenGreenABase}%
  \crawKFourInColor{BrickRed}{\crawSixSevenGreenBBase}%
  \crawKFourInColor{Orange}{\crawSixSevenGreenCNineFront}%
  \crawKFourInColor{Purple}{\crawSixSevenGreenUBase}%
  \crawKFourInColor{Magenta}{\crawSixSevenGreenDLobeNine}}

\newcommand{\crawRouteSixColoredOne}{\crawKFourRouteOne}
\newcommand{\crawRouteSixColoredTwo}{\crawKFourRouteTwo}
\newcommand{\crawRouteSixColoredThreeLeft}{\crawKFourRouteThreeLeft}
\newcommand{\crawRouteSixColoredThreeRight}{\crawKFourRouteThreeRight}
\newcommand{\crawRouteSixColoredFour}{\crawKFourRouteFour}
\newcommand{\crawRouteSixColoredFive}{\crawKFourRouteFive}
\newcommand{\crawRouteSixColoredSixLeft}{\crawKFourRouteSixLeft}
\newcommand{\crawRouteSixColoredSixRight}{\crawKFourRouteSixRight}
\newcommand{\crawRouteSixColoredSeven}{\crawKFourRouteSeven}
\newcommand{\crawRouteSixColoredEight}{\crawKFourRouteEight}
\newcommand{\crawRouteSixColoredNine}{\crawKFourRouteNine}

\newcommand{\crawKFourCapVertices}{%
  \coordinate (crawKFourA) at (0,1.33);
  \coordinate (crawKFourB) at (0,0);
  \coordinate (crawKFourC) at (0,-1.33);
  \coordinate (crawKFourT) at (5.22,0);
  \draw[draw=ForestGreen,line width=1.02pt] (crawTopOpen)--(crawKFourA);
  \draw[draw=RoyalBlue,line width=1.02pt] (crawTopGlobalRear)--(crawKFourA);
  \draw[draw=Purple,line width=1.02pt] (crawTopLocal)--(crawKFourA);
  \draw[draw=Purple,line width=1.02pt] (crawMiddleUpper)--(crawKFourB);
  \draw[draw=BrickRed,line width=1.02pt] (crawMiddleOpen)--(crawKFourB);
  \draw[draw=Magenta,line width=1.02pt] (crawMiddleLowerFront)--(crawKFourB);
  \draw[draw=Magenta,line width=1.02pt] (crawBottomLocalFront)--(crawKFourC);
  \draw[draw=RoyalBlue,line width=1.02pt] (crawBottomGlobalRear)--(crawKFourC);
  \draw[draw=Orange,line width=1.02pt] (crawBottomOpen)--(crawKFourC);
  \draw[draw=ForestGreen,line width=1.02pt] (5.03,.29)--(crawKFourT);
  \draw[draw=BrickRed,line width=1.02pt] (5.03,0)--(crawKFourT);
  \draw[draw=Orange,line width=1.02pt] (5.03,-.29)--(crawKFourT);
  \foreach \p/\lab in {crawKFourA/a,crawKFourB/b,crawKFourC/c,crawKFourT/t}{
    \node[circle,draw=black!72,fill=white,line width=.55pt,
      minimum size=3.4mm,inner sep=0pt,font=\tiny] at (\p) {$\lab$};}}

\newcommand{\crawKFourAffineEdge}[5]{%
  \foreach \ii in {-3,...,3}{%
    \foreach \jj in {-3,...,3}{%
      \pgfmathsetmacro{\crawxa}{\crawmxx*(#2+\ii)+\crawmxy*(#3+\jj)+\crawtx}%
      \pgfmathsetmacro{\crawya}{\crawmyx*(#2+\ii)+\crawmyy*(#3+\jj)+\crawty}%
      \pgfmathsetmacro{\crawxb}{\crawmxx*(#4+\ii)+\crawmxy*(#5+\jj)+\crawtx}%
      \pgfmathsetmacro{\crawyb}{\crawmyx*(#4+\ii)+\crawmyy*(#5+\jj)+\crawty}%
      \draw[line width=1.16pt,line cap=round,#1]
        (\crawxa,\crawya)--(\crawxb,\crawyb);}}}

\newcommand{\crawKFourAffineVertex}[3]{%
  \pgfmathsetmacro{\crawvxraw}{\crawmxx*#2+\crawmxy*#3+\crawtx}%
  \pgfmathsetmacro{\crawvyraw}{\crawmyx*#2+\crawmyy*#3+\crawty}%
  \pgfmathsetmacro{\crawvx}{\crawvxraw-floor(\crawvxraw)}%
  \pgfmathsetmacro{\crawvy}{\crawvyraw-floor(\crawvyraw)}%
  \node[circle,draw=black!78,fill=white,line width=.55pt,
    minimum size=3.6mm,inner sep=0pt,font=\tiny]
    at (\crawvx,\crawvy) {$#1$};}

\newcommand{\crawKFourSquareCard}[7]{%
  \begingroup
  \def\crawmxx{#2}\def\crawmxy{#3}%
  \def\crawmyx{#4}\def\crawmyy{#5}%
  \def\crawtx{#6}\def\crawty{#7}%
  \begin{tikzpicture}[x=2.42cm,y=2.42cm,
      baseline=(current bounding box.center)]
    \path[use as bounding box] (-.055,-.055) rectangle (1.055,1.105);
    \path[fill=crawFill] (0,0) rectangle (1,1);
    \begin{scope}
      \clip (0,0) rectangle (1,1);
      \crawKFourAffineEdge{ForestGreen}{.35}{.45}{.65}{.45}
      \crawKFourAffineEdge{RoyalBlue}{.35}{.45}{.50}{.70}
      \crawKFourAffineEdge{Orange}{.50}{.70}{.65}{.45}
      \crawKFourAffineEdge{BrickRed}{.50}{.22}{.65}{.45}
      \crawKFourAffineEdge{Purple}{.35}{.45}{-.50}{.22}
      \crawKFourAffineEdge{Magenta}{.50}{.22}{.50}{-.30}
    \end{scope}
    \crawKFourAffineVertex{a}{.35}{.45}%
    \crawKFourAffineVertex{b}{.50}{.22}%
    \crawKFourAffineVertex{c}{.50}{.70}%
    \crawKFourAffineVertex{t}{.65}{.45}%
    \draw[black!58,line width=.78pt] (0,0) rectangle (1,1);
    \draw[->,draw=black!58,line width=.68pt] (.36,1)--(.64,1);
    \draw[->,draw=black!58,line width=.68pt] (.36,0)--(.64,0);
    \draw[->,draw=black!58,line width=.68pt] (0,.36)--(0,.64);
    \draw[->,draw=black!58,line width=.68pt] (1,.36)--(1,.64);
    \node[font=\scriptsize\bfseries] at (.5,1.055) {#1};
  \end{tikzpicture}%
  \endgroup}

\newcommand{\crawKFourEyeSurface}{%
  \path[craw sheet] (2.20,0)
    ellipse[x radius=3.35,y radius=2.18];}

\newcommand{\crawKFourEyeCaps}{%
  \path[fill=crawFill,draw=none] (0,1.33)
    ellipse[x radius=.29,y radius=.46];
  \path[fill=crawFill,draw=none] (0,0)
    ellipse[x radius=.29,y radius=.53];
  \path[fill=crawFill,draw=none] (0,-1.33)
    ellipse[x radius=.29,y radius=.46];
  \path[fill=crawFill,draw=none] (5.08,0)
    ellipse[x radius=.22,y radius=.56];
  \coordinate (crawEyeA) at (.52,1.33);
  \coordinate (crawEyeB) at (.52,0);
  \coordinate (crawEyeC) at (.52,-1.33);
  \coordinate (crawEyeT) at (5.02,0);
  \draw[ForestGreen,line width=1.05pt] (crawTopOpen)--(crawEyeA);
  \draw[RoyalBlue,line width=1.05pt] (crawTopGlobalRear)--(crawEyeA);
  \draw[Purple,line width=1.05pt] (crawTopLocal)--(crawEyeA);
  \draw[Purple,line width=1.05pt] (crawMiddleUpper)--(crawEyeB);
  \draw[BrickRed,line width=1.05pt] (crawMiddleOpen)--(crawEyeB);
  \draw[Magenta,line width=1.05pt] (crawMiddleLowerFront)--(crawEyeB);
  \draw[Magenta,line width=1.05pt] (crawBottomLocalFront)--(crawEyeC);
  \draw[RoyalBlue,line width=1.05pt] (crawBottomGlobalRear)--(crawEyeC);
  \draw[Orange,line width=1.05pt] (crawBottomOpen)--(crawEyeC);
  \draw[ForestGreen,line width=1.05pt] (5.03,.29)--(crawEyeT);
  \draw[BrickRed,line width=1.05pt] (5.03,0)--(crawEyeT);
  \draw[Orange,line width=1.05pt] (5.03,-.29)--(crawEyeT);
  \foreach \p/\lab in {crawEyeA/a,crawEyeB/b,crawEyeC/c,crawEyeT/t}{%
    \node[circle,draw=black!78,fill=white,line width=.58pt,
      minimum size=4.0mm,inner sep=0pt,font=\tiny\bfseries]
      at (\p) {$\lab$};}}

\newcommand{\crawKFourEyeCard}[3]{%
  \begin{tikzpicture}[x=.43cm,y=.48cm,
      baseline=(current bounding box.center)]
    \path[use as bounding box] (-1.18,-2.35) rectangle (5.62,2.85);
    \begin{scope}
      \clip (-1.16,-2.20) rectangle (5.60,2.20);
      \crawThreeSeams
      \crawKFourEyeSurface
      \csname crawThirdHole#3\endcsname
      \begingroup
        \let\crawSixSevenBlackRear\relax
        \let\crawSixSevenBlackFront\relax
        \crawKFourSolidABCCycletrue
        \csname crawRoute#2\endcsname
      \endgroup
      \crawKFourEyeCaps
    \end{scope}
    \node[font=\scriptsize\bfseries] at (2.20,2.55) {#1};
  \end{tikzpicture}}

\tikzset{
  craw mini edge/.style={line width=1.16pt,line cap=round,line join=round},
  craw mini vertex/.style={circle,draw=black!78,fill=white,line width=.58pt,
    minimum size=3.6mm,inner sep=0pt,font=\tiny\bfseries},
  craw pillow mini vertex/.style={circle,draw=black!78,fill=white,
    line width=.58pt,minimum size=3.6mm,inner sep=0pt,
    font=\tiny\bfseries}
}

\newcommand{\crawKFourMiniEye}[2]{%
  \draw[craw hole,line width=.82pt] (#1,#2)
    ellipse[x radius=.32,y radius=.13];}

\newcommand{\crawKFourMiniNodes}{%
  \foreach \p/\lab in {crawMiniA/a,crawMiniB/b,crawMiniC/c,crawMiniT/t}{%
    \node[craw mini vertex] at (\p) {$\lab$};}}

\newcommand{\crawKFourMiniOne}{%
  \coordinate (crawMiniA) at (.15,1.02);
  \coordinate (crawMiniB) at (-1.55,.02);
  \coordinate (crawMiniC) at (.32,-.98);
  \coordinate (crawMiniT) at (.22,.02);
  \crawKFourMiniEye{-.43}{.03}
  \draw[craw mini edge,draw=Purple] (crawMiniA)
    ..controls(-.55,1.04)and(-1.24,.68)..(crawMiniB);
  \draw[craw mini edge,draw=Magenta] (crawMiniB)
    ..controls(-1.18,-.62)and(-.50,-.98)..(crawMiniC);
  \draw[craw mini edge,draw=RoyalBlue] (crawMiniC)
    ..controls(.78,-.52)and(.78,.58)..(crawMiniA);
  \draw[craw mini edge,draw=ForestGreen] (crawMiniA)
    ..controls(.18,.72)and(.21,.38)..(crawMiniT);
  \draw[craw mini edge,draw=BrickRed] (crawMiniB)
    ..controls(-.92,-.42)and(-.02,-.48)..(crawMiniT);
  \draw[craw mini edge,draw=Orange] (crawMiniC)
    ..controls(.28,-.62)and(.22,-.30)..(crawMiniT);
  \crawKFourMiniNodes}

\newcommand{\crawKFourMiniTwo}{%
  \coordinate (crawMiniA) at (.08,1.02);
  \coordinate (crawMiniB) at (-1.53,.00);
  \coordinate (crawMiniC) at (.34,-.96);
  \coordinate (crawMiniT) at (.22,.08);
  \crawKFourMiniEye{-.43}{-.02}
  \draw[craw mini edge,draw=Purple] (crawMiniA)
    ..controls(-.58,1.10)and(-1.23,.68)..(crawMiniB);
  \draw[craw mini edge,draw=Magenta] (crawMiniB)
    ..controls(-1.17,-.65)and(-.48,-1.00)..(crawMiniC);
  \draw[craw mini edge,draw=RoyalBlue] (crawMiniC)
    ..controls(.80,-.50)and(.76,.58)..(crawMiniA);
  \draw[craw mini edge,draw=ForestGreen] (crawMiniA)
    ..controls(.14,.72)and(.20,.40)..(crawMiniT);
  \draw[craw mini edge,draw=BrickRed] (crawMiniB)
    ..controls(-.90,.48)and(-.04,.52)..(crawMiniT);
  \draw[craw mini edge,draw=Orange] (crawMiniC)
    ..controls(.30,-.56)and(.22,-.25)..(crawMiniT);
  \crawKFourMiniNodes}

\newcommand{\crawKFourMiniThreeI}{%
  \coordinate (crawMiniA) at (.10,1.04);
  \coordinate (crawMiniB) at (-1.55,.00);
  \coordinate (crawMiniC) at (.42,-.94);
  \coordinate (crawMiniT) at (.18,-.10);
  \crawKFourMiniEye{-.45}{.04}
  \draw[craw mini edge,draw=Purple] (crawMiniA)
    ..controls(-.62,1.08)and(-1.28,.70)..(crawMiniB);
  \draw[craw mini edge,draw=Magenta] (crawMiniB)
    ..controls(-1.20,-.68)and(-.46,-1.02)..(crawMiniC);
  \draw[craw mini edge,draw=RoyalBlue] (crawMiniC)
    ..controls(.86,-.46)and(.78,.62)..(crawMiniA);
  \draw[craw mini edge,draw=ForestGreen] (crawMiniA)
    ..controls(-.35,.95)and(-.92,.78)..(-.92,.35)
    ..controls(-.98,.05)and(-.86,-.18)..(-.62,-.22)
    ..controls(-.35,-.24)and(-.08,-.18)..(crawMiniT);
  \draw[craw mini edge,draw=BrickRed] (crawMiniB)
    ..controls(-.86,-.48)and(-.02,-.52)..(crawMiniT);
  \draw[craw mini edge,draw=Orange] (crawMiniC)
    ..controls(.32,-.60)and(.20,-.34)..(crawMiniT);
  \crawKFourMiniNodes}

\newcommand{\crawKFourMiniThreeII}{%
  \coordinate (crawMiniA) at (.32,1.00);
  \coordinate (crawMiniB) at (-1.48,.02);
  \coordinate (crawMiniC) at (.38,-.92);
  \coordinate (crawMiniT) at (.20,.16);
  \crawKFourMiniEye{-.45}{.02}
  \draw[craw mini edge,draw=Purple] (crawMiniA)
    ..controls(-.50,1.12)and(-1.20,.72)..(crawMiniB);
  \draw[craw mini edge,draw=Magenta] (crawMiniB)
    ..controls(-1.16,-.64)and(-.44,-1.00)..(crawMiniC);
  \draw[craw mini edge,draw=RoyalBlue] (crawMiniC)
    ..controls(.84,-.46)and(.84,.64)..(crawMiniA);
  \draw[craw mini edge,draw=ForestGreen] (crawMiniA)
    ..controls(.30,.72)and(.24,.42)..(crawMiniT);
  \draw[craw mini edge,draw=Orange] (crawMiniC)
    ..controls(.05,-.78)and(-.60,-.62)..(-.88,-.34)
    ..controls(-1.00,-.08)and(-.90,.22)..(-.68,.30)
    ..controls(-.40,.36)and(-.08,.28)..(crawMiniT);
  \draw[craw mini edge,draw=BrickRed] (crawMiniB)
    ..controls(-.88,.46)and(-.06,.52)..(crawMiniT);
  \crawKFourMiniNodes}

\newcommand{\crawKFourMiniFour}{%
  \coordinate (crawMiniA) at (-.42,1.02);
  \coordinate (crawMiniB) at (1.10,.12);
  \coordinate (crawMiniC) at (-.28,-1.02);
  \coordinate (crawMiniT) at (2.04,.12);
  \crawKFourMiniEye{-.35}{.02}
  \draw[craw mini edge,draw=Purple] (crawMiniA)
    ..controls(.30,1.06)and(.88,.70)..(crawMiniB);
  \draw[craw mini edge,draw=Magenta] (crawMiniB)
    ..controls(1.06,-.52)and(.44,-.98)..(crawMiniC);
  \draw[craw mini edge,draw=RoyalBlue] (crawMiniC)
    ..controls(-1.02,-.52)and(-1.04,.58)..(crawMiniA);
  \draw[craw mini edge,draw=ForestGreen] (crawMiniA)
    ..controls(.48,1.42)and(1.68,1.20)..(crawMiniT);
  \draw[craw mini edge,draw=BrickRed] (crawMiniB)--(crawMiniT);
  \draw[craw mini edge,draw=Orange] (crawMiniC)
    ..controls(.50,-1.36)and(1.72,-1.10)..(crawMiniT);
  \crawKFourMiniNodes}

\newcommand{\crawKFourMiniSixI}{%
  \coordinate (crawMiniA) at (-.26,.82);
  \coordinate (crawMiniB) at (1.00,.10);
  \coordinate (crawMiniC) at (-.18,-.96);
  \coordinate (crawMiniT) at (2.02,-.22);
  \crawKFourMiniEye{-.28}{.06}
  \draw[craw mini edge,draw=Purple] (crawMiniA)
    ..controls(.36,.90)and(.88,.65)..(crawMiniB);
  \draw[craw mini edge,draw=Magenta] (crawMiniB)
    ..controls(.96,-.48)and(.42,-.92)..(crawMiniC);
  \draw[craw mini edge,draw=RoyalBlue] (crawMiniC)
    ..controls(-.94,-.48)and(-.96,.58)..(crawMiniA);
  \draw[craw mini edge,draw=ForestGreen] (crawMiniA)
    ..controls(-1.35,1.38)and(-1.85,.35)..(-1.70,-.45)
    ..controls(-1.52,-1.25)and(-.65,-1.48)..(.20,-1.48)
    ..controls(1.10,-1.48)and(1.72,-.75)..(crawMiniT);
  \draw[craw mini edge,draw=BrickRed] (crawMiniB)
    ..controls(1.34,.06)and(1.72,-.12)..(crawMiniT);
  \draw[craw mini edge,draw=Orange] (crawMiniC)
    ..controls(.45,-1.18)and(1.45,-.90)..(crawMiniT);
  \crawKFourMiniNodes}

\newcommand{\crawKFourMiniSixII}{%
  \coordinate (crawMiniA) at (-.26,.90);
  \coordinate (crawMiniB) at (1.00,.10);
  \coordinate (crawMiniC) at (-.18,-.90);
  \coordinate (crawMiniT) at (2.05,.06);
  \crawKFourMiniEye{-.28}{.06}
  \draw[craw mini edge,draw=Purple] (crawMiniA)
    ..controls(.38,.98)and(.88,.66)..(crawMiniB);
  \draw[craw mini edge,draw=Magenta] (crawMiniB)
    ..controls(.96,-.46)and(.42,-.86)..(crawMiniC);
  \draw[craw mini edge,draw=RoyalBlue] (crawMiniC)
    ..controls(-.94,-.44)and(-.96,.42)..(crawMiniA);
  \draw[craw mini edge,draw=ForestGreen] (crawMiniA)
    ..controls(.38,1.05)and(1.45,.70)..(crawMiniT);
  \draw[craw mini edge,draw=BrickRed] (crawMiniB)
    ..controls(1.22,-.35)and(1.00,-1.10)..(.18,-1.42)
    ..controls(-1.12,-1.55)and(-1.98,-.88)..(-2.05,.04)
    ..controls(-2.10,1.04)and(-1.25,1.55)..(-.35,1.62)
    ..controls(.72,1.70)and(1.76,1.00)..(crawMiniT);
  \draw[craw mini edge,draw=Orange] (crawMiniC)
    ..controls(-.90,-.98)and(-1.40,-.55)..(-1.45,.08)
    ..controls(-1.45,.90)and(-.75,1.42)..(.10,1.42)
    ..controls(.95,1.44)and(1.72,.72)..(crawMiniT);
  \crawKFourMiniNodes}

\newcommand{\crawKFourMiniSeven}{%
  \coordinate (crawMiniA) at (-.28,1.02);
  \coordinate (crawMiniB) at (1.08,.10);
  \coordinate (crawMiniC) at (-.20,-.98);
  \coordinate (crawMiniT) at (2.02,.10);
  \crawKFourMiniEye{-.30}{.04}
  \draw[craw mini edge,draw=Purple] (crawMiniA)
    ..controls(.36,1.08)and(.92,.72)..(crawMiniB);
  \draw[craw mini edge,draw=Magenta] (crawMiniB)
    ..controls(1.02,-.50)and(.42,-.96)..(crawMiniC);
  \draw[craw mini edge,draw=RoyalBlue] (crawMiniC)
    ..controls(-.96,-.50)and(-.98,.58)..(crawMiniA);
  \draw[craw mini edge,draw=ForestGreen] (crawMiniA)
    ..controls(.35,1.28)and(1.48,1.18)..(crawMiniT);
  \draw[craw mini edge,draw=BrickRed] (crawMiniB)--(crawMiniT);
  \draw[craw mini edge,draw=Orange] (crawMiniC)
    ..controls(-1.70,-1.18)and(-1.78,1.08)..(-.38,1.38)
    ..controls(1.05,1.56)and(2.15,.80)..(crawMiniT);
  \crawKFourMiniNodes}

\newcommand{\crawKFourMiniEight}{%
  \coordinate (crawMiniA) at (-.28,1.00);
  \coordinate (crawMiniB) at (1.08,.10);
  \coordinate (crawMiniC) at (-.18,-.98);
  \coordinate (crawMiniT) at (2.02,.10);
  \crawKFourMiniEye{-.30}{-.02}
  \draw[craw mini edge,draw=Purple] (crawMiniA)
    ..controls(.36,1.06)and(.92,.70)..(crawMiniB);
  \draw[craw mini edge,draw=Magenta] (crawMiniB)
    ..controls(1.02,-.50)and(.42,-.96)..(crawMiniC);
  \draw[craw mini edge,draw=RoyalBlue] (crawMiniC)
    ..controls(-.96,-.50)and(-.98,.56)..(crawMiniA);
  \draw[craw mini edge,draw=ForestGreen] (crawMiniA)
    ..controls(-1.68,1.18)and(-1.78,-1.06)..(-.36,-1.38)
    ..controls(1.08,-1.54)and(2.15,-.64)..(crawMiniT);
  \draw[craw mini edge,draw=BrickRed] (crawMiniB)--(crawMiniT);
  \draw[craw mini edge,draw=Orange] (crawMiniC)
    ..controls(.42,-1.22)and(1.42,-1.00)..(crawMiniT);
  \crawKFourMiniNodes}

\newcommand{\crawKFourMiniNine}{%
  \coordinate (crawMiniA) at (-.38,1.02);
  \coordinate (crawMiniB) at (1.08,.10);
  \coordinate (crawMiniC) at (-.20,-.98);
  \coordinate (crawMiniT) at (2.08,.10);
  \crawKFourMiniEye{-.32}{.00}
  \draw[craw mini edge,draw=Purple] (crawMiniA)
    ..controls(.30,1.10)and(.92,.72)..(crawMiniB);
  \draw[craw mini edge,draw=Magenta] (crawMiniB)
    ..controls(1.02,-.50)and(.42,-.96)..(crawMiniC);
  \draw[craw mini edge,draw=RoyalBlue] (crawMiniC)
    ..controls(-.96,-.50)and(-1.00,.58)..(crawMiniA);
  \draw[craw mini edge,draw=ForestGreen] (crawMiniA)
    ..controls(.38,1.40)and(1.62,1.22)..(crawMiniT);
  \draw[craw mini edge,draw=BrickRed] (crawMiniB)--(crawMiniT);
  \draw[craw mini edge,draw=Orange] (crawMiniC)
    ..controls(-1.72,-1.20)and(-1.88,1.16)..(-.42,1.45)
    ..controls(1.10,1.62)and(2.18,.80)..(crawMiniT);
  \crawKFourMiniNodes}

\newcommand{\crawKFourMiniCard}[2]{%
  \begin{tikzpicture}[x=.63cm,y=.57cm,
      baseline=(current bounding box.center)]
    \path[use as bounding box] (-2.72,-1.88) rectangle (2.72,2.13);
    \path[craw sheet] (0,0) ellipse[x radius=2.55,y radius=1.68];
    \csname crawKFourMini#2\endcsname
    \node[font=\scriptsize\bfseries] at (0,1.98) {#1};
  \end{tikzpicture}}

\global\setbox\crawOpenSixjKFourOutputsBox=\hbox{%
\begin{tabular}{@{}c@{\hspace{7mm}}c@{}}
\crawKFourMiniCard{6j-7-1 (1)}{One} &
\crawKFourMiniCard{6j-7-2 (2)}{Two} \\[2mm]
\crawKFourMiniCard{6j-7-3 (3-I)}{ThreeI} &
\crawKFourMiniCard{6j-7-3 (3-II)}{ThreeII} \\[2mm]
\crawKFourMiniCard{6j-7-3 (4)}{Four} &
\crawKFourMiniCard{6j-7-3 (5)}{Four} \\[2mm]
\crawKFourMiniCard{6j-7-1 (6-I)}{SixI} &
\crawKFourMiniCard{6j-7-1 (6-II)}{SixII} \\[2mm]
\crawKFourMiniCard{6j-7-2 (7)}{Seven} &
\crawKFourMiniCard{6j-7-1 (8)}{Eight} \\[2mm]
\crawKFourMiniCard{6j-7-2 (9)}{Nine} & {}
\end{tabular}}

\newcommand{\crawPillowMiniEye}[2]{%
  \draw[craw hole,line width=.82pt] (#1,#2)
    ellipse[x radius=.32,y radius=.13];}

\newcommand{\crawPillowMiniBCCircle}[4]{%
  \draw[craw mini edge,draw=#1] (crawPillowMiniB)
    ..controls(#3,.95)and(#3,-.95)..(crawPillowMiniC);
  \draw[craw mini edge,draw=#2] (crawPillowMiniB)
    ..controls(#4,.95)and(#4,-.95)..(crawPillowMiniC);}

\newcommand{\crawPillowMiniNodes}{%
  \foreach \p/\lab in {crawPillowMiniA/a,crawPillowMiniB/b,
      crawPillowMiniC/c,crawPillowMiniT/t}{%
    \node[craw pillow mini vertex] at (\p) {$\lab$};}}

\newcommand{\crawPillowMiniTwelve}{%
  \coordinate (crawPillowMiniA) at (-.42,.52);
  \coordinate (crawPillowMiniB) at (.35,1.42);
  \coordinate (crawPillowMiniC) at (.35,-1.42);
  \coordinate (crawPillowMiniT) at (-.42,-.52);
  \crawPillowMiniEye{.52}{0}
  \crawPillowMiniBCCircle{Magenta}{RoyalBlue}{-1.78}{1.55}
  \draw[craw mini edge,draw=Purple] (crawPillowMiniB)
    ..controls(.08,1.18)and(-.22,.85)..(crawPillowMiniA);
  \draw[craw mini edge,draw=Orange] (crawPillowMiniC)
    ..controls(.08,-1.18)and(-.22,-.85)..(crawPillowMiniT);
  \draw[craw mini edge,draw=ForestGreen] (crawPillowMiniA)
    ..controls(-.88,.38)and(-.88,-.38)..(crawPillowMiniT);
  \draw[craw mini edge,draw=BrickRed] (crawPillowMiniA)
    ..controls(-.02,.34)and(-.02,-.34)..(crawPillowMiniT);
  \crawPillowMiniNodes}

\newcommand{\crawPillowMiniThirteen}{%
  \coordinate (crawPillowMiniA) at (-1.65,.52);
  \coordinate (crawPillowMiniB) at (.25,1.35);
  \coordinate (crawPillowMiniC) at (.25,-1.35);
  \coordinate (crawPillowMiniT) at (-1.65,-.52);
  \crawPillowMiniEye{.42}{0}
  \crawPillowMiniBCCircle{RoyalBlue}{Magenta}{-.95}{1.55}
  \draw[craw mini edge,draw=Purple] (crawPillowMiniB)
    ..controls(-.20,1.62)and(-1.30,1.52)..(crawPillowMiniA);
  \draw[craw mini edge,draw=Orange] (crawPillowMiniC)
    ..controls(-.20,-1.62)and(-1.30,-1.52)..(crawPillowMiniT);
  \draw[craw mini edge,draw=ForestGreen] (crawPillowMiniA)
    ..controls(-2.35,.38)and(-2.35,-.38)..(crawPillowMiniT);
  \draw[craw mini edge,draw=BrickRed] (crawPillowMiniA)
    ..controls(-1.18,.34)and(-1.18,-.34)..(crawPillowMiniT);
  \crawPillowMiniNodes}

\newcommand{\crawPillowMiniFourteen}{%
  \coordinate (crawPillowMiniA) at (1.65,.52);
  \coordinate (crawPillowMiniB) at (-.25,1.35);
  \coordinate (crawPillowMiniC) at (-.25,-1.35);
  \coordinate (crawPillowMiniT) at (1.65,-.52);
  \crawPillowMiniEye{-.42}{0}
  \crawPillowMiniBCCircle{RoyalBlue}{Magenta}{-1.55}{.95}
  \draw[craw mini edge,draw=Purple] (crawPillowMiniB)
    ..controls(.20,1.62)and(1.30,1.52)..(crawPillowMiniA);
  \draw[craw mini edge,draw=Orange] (crawPillowMiniC)
    ..controls(.20,-1.62)and(1.30,-1.52)..(crawPillowMiniT);
  \draw[craw mini edge,draw=ForestGreen] (crawPillowMiniA)
    ..controls(2.35,.38)and(2.35,-.38)..(crawPillowMiniT);
  \draw[craw mini edge,draw=BrickRed] (crawPillowMiniA)
    ..controls(1.18,.34)and(1.18,-.34)..(crawPillowMiniT);
  \crawPillowMiniNodes}

\newcommand{\crawPillowMiniFifteen}{%
  \coordinate (crawPillowMiniA) at (.42,.52);
  \coordinate (crawPillowMiniB) at (-.35,1.42);
  \coordinate (crawPillowMiniC) at (-.35,-1.42);
  \coordinate (crawPillowMiniT) at (.42,-.52);
  \crawPillowMiniEye{-.52}{0}
  \crawPillowMiniBCCircle{Magenta}{RoyalBlue}{-1.55}{1.78}
  \draw[craw mini edge,draw=Purple] (crawPillowMiniB)
    ..controls(-.08,1.18)and(.22,.85)..(crawPillowMiniA);
  \draw[craw mini edge,draw=Orange] (crawPillowMiniC)
    ..controls(-.08,-1.18)and(.22,-.85)..(crawPillowMiniT);
  \draw[craw mini edge,draw=ForestGreen] (crawPillowMiniA)
    ..controls(.88,.38)and(.88,-.38)..(crawPillowMiniT);
  \draw[craw mini edge,draw=BrickRed] (crawPillowMiniA)
    ..controls(.02,.34)and(.02,-.34)..(crawPillowMiniT);
  \crawPillowMiniNodes}

\newcommand{\crawPillowMiniSixteen}{%
  \coordinate (crawPillowMiniA) at (-1.65,.52);
  \coordinate (crawPillowMiniB) at (.25,1.35);
  \coordinate (crawPillowMiniC) at (.25,-1.35);
  \coordinate (crawPillowMiniT) at (-1.65,-.52);
  \crawPillowMiniEye{.42}{0}
  \crawPillowMiniBCCircle{RoyalBlue}{Magenta}{-.95}{1.55}
  \draw[craw mini edge,draw=Purple] (crawPillowMiniB)
    ..controls(-.20,1.62)and(-1.30,1.52)..(crawPillowMiniA);
  \draw[craw mini edge,draw=Orange] (crawPillowMiniC)
    ..controls(-.20,-1.62)and(-1.30,-1.52)..(crawPillowMiniT);
  \draw[craw mini edge,draw=ForestGreen] (crawPillowMiniA)
    ..controls(-2.35,.38)and(-2.35,-.38)..(crawPillowMiniT);
  \draw[craw mini edge,draw=BrickRed] (crawPillowMiniA)
    ..controls(-1.18,.34)and(-1.18,-.34)..(crawPillowMiniT);
  \crawPillowMiniNodes}

\newcommand{\crawPillowMiniSeventeen}{%
  \coordinate (crawPillowMiniA) at (-.42,.52);
  \coordinate (crawPillowMiniB) at (.35,1.42);
  \coordinate (crawPillowMiniC) at (.35,-1.42);
  \coordinate (crawPillowMiniT) at (-.42,-.52);
  \crawPillowMiniEye{.52}{0}
  \crawPillowMiniBCCircle{Magenta}{RoyalBlue}{-1.78}{1.55}
  \draw[craw mini edge,draw=Purple] (crawPillowMiniB)
    ..controls(.08,1.18)and(-.22,.85)..(crawPillowMiniA);
  \draw[craw mini edge,draw=Orange] (crawPillowMiniC)
    ..controls(.08,-1.18)and(-.22,-.85)..(crawPillowMiniT);
  \draw[craw mini edge,draw=ForestGreen] (crawPillowMiniA)
    ..controls(-.88,.38)and(-.88,-.38)..(crawPillowMiniT);
  \draw[craw mini edge,draw=BrickRed] (crawPillowMiniA)
    ..controls(-.02,.34)and(-.02,-.34)..(crawPillowMiniT);
  \crawPillowMiniNodes}

\newcommand{\crawPillowMiniEighteen}{%
  \coordinate (crawPillowMiniA) at (0,1.42);
  \coordinate (crawPillowMiniB) at (0,.48);
  \coordinate (crawPillowMiniC) at (0,-.48);
  \coordinate (crawPillowMiniT) at (0,-1.42);
  \crawPillowMiniEye{0}{0}
  \crawPillowMiniBCCircle{RoyalBlue}{Magenta}{-1.35}{1.35}
  \draw[craw mini edge,draw=Purple]
    (crawPillowMiniA)--(crawPillowMiniB);
  \draw[craw mini edge,draw=Orange]
    (crawPillowMiniC)--(crawPillowMiniT);
  \draw[craw mini edge,draw=ForestGreen] (crawPillowMiniA)
    ..controls(-2.12,1.38)and(-2.38,.56)..(-2.28,-.40)
    ..controls(-2.10,-1.16)and(-1.10,-1.42)..(crawPillowMiniT);
  \draw[craw mini edge,draw=BrickRed] (crawPillowMiniA)
    ..controls(2.12,1.38)and(2.38,.56)..(2.28,-.40)
    ..controls(2.10,-1.16)and(1.10,-1.42)..(crawPillowMiniT);
  \crawPillowMiniNodes}

\newcommand{\crawPillowMiniNineteen}{%
  \coordinate (crawPillowMiniA) at (0,1.42);
  \coordinate (crawPillowMiniB) at (0,.48);
  \coordinate (crawPillowMiniC) at (0,-.48);
  \coordinate (crawPillowMiniT) at (0,-1.42);
  \crawPillowMiniEye{0}{0}
  \crawPillowMiniBCCircle{Magenta}{RoyalBlue}{-1.35}{1.35}
  \draw[craw mini edge,draw=Purple]
    (crawPillowMiniA)--(crawPillowMiniB);
  \draw[craw mini edge,draw=Orange]
    (crawPillowMiniC)--(crawPillowMiniT);
  \draw[craw mini edge,draw=BrickRed] (crawPillowMiniA)
    ..controls(-2.12,1.38)and(-2.38,.56)..(-2.28,-.40)
    ..controls(-2.10,-1.16)and(-1.10,-1.42)..(crawPillowMiniT);
  \draw[craw mini edge,draw=ForestGreen] (crawPillowMiniA)
    ..controls(2.12,1.38)and(2.38,.56)..(2.28,-.40)
    ..controls(2.10,-1.16)and(1.10,-1.42)..(crawPillowMiniT);
  \crawPillowMiniNodes}

\newcommand{\crawPillowMiniCard}[2]{%
  \begin{tikzpicture}[x=.63cm,y=.57cm,
      baseline=(current bounding box.center)]
    \path[use as bounding box] (-2.72,-1.88) rectangle (2.72,2.13);
    \path[craw sheet] (0,0) ellipse[x radius=2.55,y radius=1.82];
    \csname crawPillowMini#2\endcsname
    \SDMiniTitle{#1}
  \end{tikzpicture}}

\global\setbox\crawPillowTwelveNineteenOutputsBox=\hbox{%
\begin{tabular}{@{}c@{\hspace{1.8mm}}c@{\hspace{1.8mm}}c@{}}
\crawPillowMiniCard{P-7-3 (12)}{Twelve} & \crawZRelation &
\crawPillowMiniCard{P-7-2 (13)$^{\dagger}$}{Thirteen} \\[2mm]
\crawPillowMiniCard{P-7-3 (14)}{Fourteen} & \crawZRelation &
\crawPillowMiniCard{P-7-2 (15)}{Fifteen} \\[2mm]
\crawPillowMiniCard{P-7-2 (16)\;\textcolor{red}{$\times$}}{Sixteen} &
\crawZRelation &
\crawPillowMiniCard{P-7-3 (17)$^{\dagger}$\;\textcolor{red}{$\times$}}{Seventeen} \\[2mm]
\crawPillowMiniCard{P-7-1 (18)}{Eighteen} & \crawZRelation &
\crawPillowMiniCard{P-7-1 (19)}{Nineteen}
\end{tabular}}

\newcommand{\crawFClassRelation}[2]{%
  \begin{tabular}{@{}c@{}}
    {\large$\xrightarrow{\quad F\quad}$}\\[-.6mm]
    {\tiny local change only}\\[-.5mm]
    {\tiny outside fixed}
  \end{tabular}}

\newcommand{\crawFLocalEye}[2]{%
  \draw[craw hole,line width=.82pt] (#1,#2)
    ellipse[x radius=.25,y radius=.11];}
\newcommand{\crawFLocalEyeX}{\crawFLocalEye{0}{-.055}}
\newcommand{\crawFLocalEyeY}{\crawFLocalEye{-.75}{-.80}}
\newcommand{\crawFLocalEyeZ}{\crawFLocalEye{.75}{-.80}}

\newcommand{\crawFLocalExterior}[1]{%
  \coordinate (crawFBluePort) at (1.28,.56);
  \coordinate (crawFGreenPort) at (.48,.174);
  \coordinate (crawFRedPort) at (-.48,.174);
  \coordinate (crawFMagentaPort) at (-1.28,.56);
  \coordinate (crawFLocalC) at (0,-1.40);
  \coordinate (crawFLocalT) at (0,-.68);
  \csname crawFLocalEye#1\endcsname
  \path[fill=MidnightBlue!4,draw=MidnightBlue!58,densely dashed,
    line width=.62pt] (0,.85) ellipse[x radius=1.40,y radius=.72];
  \draw[craw mini edge,draw=RoyalBlue] (crawFLocalC)
    ..controls(1.90,-1.02)and(1.80,.05)..(crawFBluePort);
  \draw[craw mini edge,draw=Magenta] (crawFLocalC)
    ..controls(-1.90,-1.02)and(-1.80,.05)..(crawFMagentaPort);
  \draw[craw mini edge,draw=ForestGreen] (crawFLocalT)
    ..controls(.48,-.60)and(.50,-.10)..(crawFGreenPort);
  \draw[craw mini edge,draw=BrickRed] (crawFLocalT)
    ..controls(-.48,-.60)and(-.50,-.10)..(crawFRedPort);
  \draw[craw mini edge,draw=Orange] (crawFLocalC)
    --(crawFLocalT);}

\newcommand{\crawFLocalKResolution}{%
  \coordinate (crawFLocalA) at (.70,.85);
  \coordinate (crawFLocalB) at (-.70,.85);
  \draw[craw mini edge,draw=RoyalBlue] (crawFBluePort)--(crawFLocalA);
  \draw[craw mini edge,draw=ForestGreen] (crawFGreenPort)--(crawFLocalA);
  \draw[craw mini edge,draw=Magenta] (crawFMagentaPort)--(crawFLocalB);
  \draw[craw mini edge,draw=BrickRed] (crawFRedPort)--(crawFLocalB);
  \draw[craw mini edge,draw=Purple] (crawFLocalA)--(crawFLocalB);}

\newcommand{\crawFLocalPResolution}{%
  \coordinate (crawFLocalB) at (0,1.21);
  \coordinate (crawFLocalA) at (0,.50);
  \draw[craw mini edge,draw=RoyalBlue] (crawFBluePort)
    ..controls(1.05,.97)and(.70,1.25)..(crawFLocalB);
  \draw[craw mini edge,draw=Magenta] (crawFMagentaPort)
    ..controls(-1.05,.97)and(-.70,1.25)..(crawFLocalB);
  \draw[craw mini edge,draw=ForestGreen] (crawFGreenPort)--(crawFLocalA);
  \draw[craw mini edge,draw=BrickRed] (crawFRedPort)--(crawFLocalA);
  \draw[craw mini edge,draw=Purple] (crawFLocalB)--(crawFLocalA);}

\newcommand{\crawFLocalCard}[3]{%
  \begin{tikzpicture}[x=.63cm,y=.57cm,
      baseline=(current bounding box.center)]
    \path[use as bounding box] (-2.72,-1.88) rectangle (2.72,2.13);
    \path[craw sheet] (0,0) ellipse[x radius=2.55,y radius=1.82];
    \crawFLocalExterior{#2}
    \csname crawFLocal#3Resolution\endcsname
    \foreach \p/\lab in {crawFLocalA/a,crawFLocalB/b,
        crawFLocalC/c,crawFLocalT/t}{%
      \node[craw mini vertex,minimum size=3.2mm] at (\p) {$\lab$};}
    \SDMiniTitle{#1}
  \end{tikzpicture}}

\newcommand{\crawBuildFMoveThreeClassMapBox}{%
\setbox\crawFMoveThreeClassMapBox=\hbox{%
\begin{tabular}{@{}c@{\hspace{2.5mm}}c@{\hspace{2.5mm}}c@{}}
\crawFLocalCard{6j-7-1: $abt$}{X}{K} &
\crawFClassRelation{(1,0,0)}{(-1,1,1)} &
\crawFLocalCard{P-7-1}{X}{P} \\[2mm]
\crawFLocalCard{6j-7-2: $bct$}{Y}{K} &
\crawFClassRelation{(0,1,0)}{(0,1,0)} &
\crawFLocalCard{P-7-2}{Y}{P} \\[2mm]
\crawFLocalCard{6j-7-3: $cat$}{Z}{K} &
\crawFClassRelation{(0,0,1)}{(0,0,1)} &
\crawFLocalCard{P-7-3}{Z}{P}
\end{tabular}}}
\crawBuildFMoveThreeClassMapBox

\global\setbox\crawStepNinePillowRawSurvivorsBox=\hbox{%
\begin{tabular}{@{}c@{\hspace{3mm}}c@{}}
\multicolumn{2}{c}{\small\bfseries $C_2$: raw 1 and raw 2}\\[.6mm]
\crawCard{1}{PillowPFive}{Standard}
  {$P$--5; \texttt{2cycles\_18.png}; idx 726}{}{Hidden} &
\crawCard{2}{PillowPSix}{Standard}
  {$P$--6; \texttt{2cycles\_23.png}; idx 5244}{}{Hidden} \\[1.8mm]
\multicolumn{2}{c}{\small\bfseries $D_1$: raw 7}\\[.6mm]
\multicolumn{2}{c}{%
  \crawPillowEarlyDisconnectedCard{7}{7}
    {$P$--10; \texttt{t\_same\_1cycles\_8.png}; idx 5774}}
\end{tabular}}

\newcommand{\CZeroRawLedgerPages}{%
\crawLedgerHeading{Open-$6j$: $C_2$}{%
  Raw 10--15 are the six distinct connected two-cycle parents.  Each surface
  drawing represents one code graph; its code label and index are printed below the
  drawing.  An external component can absorb a rear or local component, so
  equal endpoint labels do not imply three direct strands.}

\begin{center}
\begin{tabular}{@{}c@{\hspace{1.5mm}}c@{\hspace{1.5mm}}c@{}}
\crawCard{10}{SixTen}{Standard}{$6j$--3; ex.~5/24; idx 331}{}{Hidden} &
\crawSpacerRelation &
\crawCard{11}{SixEleven}{Standard}{$6j$--6; ex.~23/24; idx 5553}{}{Hidden} \\[1mm]
\crawCard{12}{SixTwelve}{Standard}{$6j$--4; ex.~18/24; idx 725}{}{Hidden} &
\crawSpacerRelation &
\crawCard{13}{SixThirteen}{Standard}{$6j$--1; ex.~1/24; idx 99}{}{Hidden} \\[1mm]
\crawCard{14}{SixFourteen}{Standard}{$6j$--2; ex.~7/24; idx 363}{}{Hidden} &
\crawSpacerRelation &
\crawCard{15}{SixFifteen}{Standard}{$6j$--5; ex.~21/24; idx 2926}{}{Hidden}
\end{tabular}
\end{center}

\crawLegend

\clearpage
\crawLedgerHeading{Open-$6j$: $C_3$}{%
  Raw 1--9 are multiple surface presentations of the single code parent
  $6j$--7 (idx 409), not nine different parents.  Raw 3 and raw 6 each have
  two cut-open views.  The labels 6j-7-1, 6j-7-2, and 6j-7-3 identify
  their embedding classes; equality signs
  and red crosses reproduce the marks in the original drawings.  Only the six right-hand
  components have been recolored by endpoint pair.}

\global\setbox\crawOpenSixjOneNineViewsBox=\hbox{%
\begin{tabular}{@{}c@{\hspace{1.5mm}}c@{\hspace{1.5mm}}c@{}}
\crawCard{1}{SixColoredOne}{Standard}{6j-7-1; ex.~1/1; idx 409}{}{UpperInner} &
\crawSpacerRelation &
\crawCard{2}{SixColoredTwo}{Standard}{6j-7-2; ex.~1/1; idx 409}{}{Lower} \\[1mm]
\crawCard{3}{SixColoredThreeLeft}{Standard}{6j-7-3; ex.~1/1; idx 409}{}{ThreeUpper} &
\crawEqualRelation &
\crawCard{3}{SixColoredThreeRight}{Standard}{6j-7-3; ex.~1/1; idx 409}{}{ThreeLower} \\[1mm]
\crawCard{4}{SixColoredFour}{Standard}{6j-7-3; ex.~1/1; idx 409}{}{FourUpper} &
\crawEqualRelation &
\crawCard{5}{SixColoredFive}{Standard}{6j-7-3; ex.~1/1; idx 409\;\crawSmallCross}{}{FiveLower} \\[1mm]
\crawCard{6}{SixColoredSixLeft}{Standard}{6j-7-1; ex.~1/1; idx 409}{}{SixUpper} &
\crawEqualRelation &
\crawCard{6}{SixColoredSixRight}{Standard}{6j-7-1; ex.~1/1; idx 409}{}{SixUpper} \\[1mm]
\crawCard{7}{SixColoredSeven}{Standard}{6j-7-2; ex.~1/1; idx 409}{}{SevenUpper} &
\crawSpacerRelation &
\crawCard{8}{SixColoredEight}{Standard}{6j-7-1; ex.~1/1; idx 409\;\crawSmallCross}{}{EightLower} \\[1mm]
\crawCard{9}{SixColoredNine}{Standard}{6j-7-2; ex.~1/1; idx 409\;\crawSmallCross}{}{NineLower} &
\crawSpacerRelation & {} \\
\end{tabular}}
\begin{center}
\usebox{\crawOpenSixjOneNineViewsBox}
\end{center}

\begin{center}
{\scriptsize
black: fixed left piece; dashed: rear sheet\qquad
colored right-piece edges:\par\smallskip
\textcolor{ForestGreen}{$a_1\!-\!t_1$} (green)\quad
\textcolor{RoyalBlue}{$a_3\!-\!c_1$} (blue)\quad
\textcolor{Purple}{$a_2\!-\!b_1$} (purple)\quad
\textcolor{BrickRed}{$b_2\!-\!t_2$} (red)\quad
\textcolor{Magenta}{$b_3\!-\!c_2$} (magenta)\quad
\textcolor{Orange}{$c_3\!-\!t_3$} (orange)}
\end{center}

\clearpage
\crawLedgerHeading{Open-$6j$: $D_1$}{%
  Raw 16--18 are the three distinct disconnected one-cycle parents.  Each
  has two cylinders: one joins two middle cuffs, while the other joins the
  remaining cuff to the right boundary.}

\begin{center}
\begin{tabular}{@{}c@{\hspace{1.5mm}}c@{\hspace{1.5mm}}c@{}}
\crawDisconnectedSixCard{16}{16}{$6j$--8; disconnected 1/6; idx 1} &
\crawSpacerRelation &
\crawDisconnectedSixCard{17}{17}{$6j$--10; disconnected 5/6; idx 5763} \\[1mm]
\crawDisconnectedSixCard{18}{18}{$6j$--9; disconnected 3/6; idx 2880} &
\crawSpacerRelation & {} \\
\end{tabular}
\end{center}

\medskip
\crawLedgerHeading{Open-$6j$: $D_2$}{%
  There is no retained open-$6j$ parent in this sector.}

\crawLegend

\clearpage
\crawLedgerHeading{Pillow: $C_2$}{%
  Raw 1--6 are the six distinct connected two-cycle parents.  The double
  arrows reproduce the $\mathbb Z_2$ comparisons marked on the hand sheet;
  they do not turn the paired drawings into one code parent.}

\begin{center}
\begin{tabular}{@{}c@{\hspace{1.5mm}}c@{\hspace{1.5mm}}c@{}}
\crawCard{1}{PillowPFive}{Standard}{$P$--5; \texttt{2cycles\_18.png}; idx 726}{}{Hidden} &
\crawZRelation &
\crawCard{2}{PillowPSix}{Standard}{$P$--6; \texttt{2cycles\_23.png}; idx 5244}{}{Hidden} \\[1mm]
\crawCard{3}{PillowPFour}{Standard}{$P$--4; \texttt{2cycles\_12.png}; idx 102}{}{Hidden} &
\crawZRelation &
\crawCard{4}{PillowPThree}{Standard}{$P$--3; \texttt{2cycles\_11.png}; idx 100}{}{Hidden} \\[1mm]
\crawCard{5}{PillowPTwo}{Standard}{$P$--2; \texttt{2cycles\_4.png}; idx 46}{}{Hidden} &
\crawZRelation &
\crawCard{6}{PillowPOne}{Standard}{$P$--1; \texttt{2cycles\_3.png}; idx 30}{}{Hidden}
\end{tabular}
\end{center}

\crawLegend

\clearpage
\crawLedgerHeading{Pillow: $C_3$}{%
  Raw 12--19 are multiple surface presentations of the single code parent
  $P$--7 (idx 92), not eight different parents.  The arrows reproduce the
  recorded $\mathbb Z_2$ comparisons.  Crosses on 16 and 17 are copied from
  the original surface drawings and leave six retained presentations, which Section~\ref{sec:embeddings}
  resolves into P-7-1, P-7-2, and P-7-3.}

\global\setbox\crawPillowTwelveNineteenViewsBox=\hbox{%
\begin{tabular}{@{}c@{\hspace{1.5mm}}c@{\hspace{1.5mm}}c@{}}
\crawCard{12}{PillowColoredTwelve}{Standard}{P-7-3; \texttt{3cycles\_1.png}; idx 92}{}{Central} &
\crawZRelation &
\crawCard{13}{PillowColoredThirteen}{Standard}{P-7-2; \texttt{3cycles\_1.png}; idx 92}{}{Central} \\[1mm]
\crawCard{14}{PillowColoredFourteen}{Standard}{P-7-3; \texttt{3cycles\_1.png}; idx 92}{}{Lower} &
\crawZRelation &
\crawCard{15}{PillowColoredFifteen}{Standard}{P-7-2; \texttt{3cycles\_1.png}; idx 92}{}{Lower} \\[1mm]
\crawCard{16}{PillowColoredSixteen}{Standard}{P-7-2; \texttt{3cycles\_1.png}; idx 92\;\crawSmallCross}{}{Lower} &
\crawZRelation &
\crawCard{17}{PillowColoredSeventeen}{Standard}{P-7-3; \texttt{3cycles\_1.png}; idx 92\;\crawSmallCross}{}{Lower} \\[1mm]
\crawCard{18}{PillowColoredEighteen}{Standard}{P-7-1; \texttt{3cycles\_1.png}; idx 92}{}{Lower} &
\crawZRelation &
\crawCard{19}{PillowColoredNineteen}{Standard}{P-7-1; \texttt{3cycles\_1.png}; idx 92}{}{Lower} \\
\end{tabular}}
\begin{center}
\usebox{\crawPillowTwelveNineteenViewsBox}
\end{center}

\begin{center}
\scriptsize
right-piece edge key:
\textcolor{ForestGreen}{$a_1\!-\!t_1$},
\textcolor{BrickRed}{$a_2\!-\!t_2$},
\textcolor{Purple}{$a_3\!-\!b_1$},
\textcolor{RoyalBlue}{$b_3\!-\!c_1$},
\textcolor{Magenta}{$b_2\!-\!c_2$},
\textcolor{Orange}{$c_3\!-\!t_3$};
dashes retain the raw rear-sheet intervals.
\end{center}

\clearpage
\crawLedgerHeading{Pillow: $D_1$}{%
  Raw 7--9 are the three distinct disconnected one-cycle parents.  Candidate
  7 is the separately boxed two-point-symmetric presentation.}

\begin{center}
\begin{tabular}{@{}c@{\hspace{1.5mm}}c@{\hspace{1.5mm}}c@{}}
\crawPillowEarlyDisconnectedCard{7}{7}{$P$--10; \texttt{t\_same\_1cycles\_8.png}; idx 5774} &
\crawSpacerRelation & {} \\[1mm]
\crawPillowEarlyDisconnectedCard{8}{8}{$P$--9; \texttt{t\_same\_1cycles\_2.png}; idx 4} &
\crawZRelation &
\crawPillowEarlyDisconnectedCard{9}{9}{$P$--8; \texttt{t\_same\_1cycles\_1.png}; idx 0}
\end{tabular}
\end{center}

\medskip
\crawLedgerHeading{Pillow: $D_2$}{%
  The four panels below are projections of the single code parent $P$--11
  (idx 14).  The two panels numbered 10 are duplicate projections of one
  marked theta embedding, and the two numbered 11 are duplicate projections
  of the other.  Thus the repeated drawings represent one parent but reduce
  from four views to two marked embeddings.}

\global\setbox\crawPillowTenElevenViewsBox=\hbox{%
\begin{tabular}{@{}c@{\hspace{1.5mm}}c@{\hspace{1.5mm}}c@{}}
\crawDisconnectedPillowCard{10}{1}{$P$--11; \texttt{t\_same\_2cycles\_7.png}; idx 14} &
\crawZRelation &
\crawDisconnectedPillowCard{11}{2}{$P$--11; \texttt{t\_same\_2cycles\_7.png}; idx 14} \\[2mm]
\crawDisconnectedPillowCard{10}{3}{$P$--11; \texttt{t\_same\_2cycles\_7.png}; idx 14} &
\crawZRelation &
\crawDisconnectedPillowCard{11}{4}{$P$--11; \texttt{t\_same\_2cycles\_7.png}; idx 14} \\
\end{tabular}}
\begin{center}
\usebox{\crawPillowTenElevenViewsBox}
\end{center}

\begin{center}
\begin{minipage}{.92\linewidth}
\footnotesize
\textit{Reading convention.}
The warm region is the boundary surface, black is the fixed left piece, and
dashes mark rear-sheet continuations.  The $C_3$ panels have the additional
small gray handle eye that is absent from $C_2$.  The $D_1$ panels are plain
pairs of disconnected cylinders.  Pillow $D_2$ has the
$(a\!-\!t)\sqcup(b\!-\!c)$ cuff pairing of two $D_1$ drawings but adds one
genuine eye to the bent $b$--$c$ component.  Dashed strokes remain on the
surface and never enter a white region or the three-dimensional bulk.
\end{minipage}
\end{center}

\crawLegend

}

\SDFigureLabels{source numbers=false}
%

\providecolor{sdembFill}{RGB}{244,239,224}
\providecolor{sdembEdge}{RGB}{130,130,130}

\newsavebox{\sdembThetaTwoClassBox}
\newsavebox{\sdembKFourThreeClassBox}

\tikzset{
  sdemb theta torus/.style={draw=sdembEdge,fill=sdembFill,
    line width=.72pt},
  sdemb theta identify/.style={-{Latex[length=1.5mm,width=.9mm]},
    draw=black!58,line width=.52pt},
  sdemb theta edge/.style={line width=1.15pt,line cap=round,
    line join=round},
  sdemb theta blue/.style={sdemb theta edge,draw=RoyalBlue},
  sdemb theta green/.style={sdemb theta edge,draw=ForestGreen},
  sdemb theta purple/.style={sdemb theta edge,draw=Purple},
  sdemb theta vertex/.style={circle,draw=black!85,fill=white,
    line width=.68pt,minimum size=4.2mm,inner sep=0pt,
    font=\scriptsize\bfseries},
  sdemb theta title/.style={font=\footnotesize\bfseries,align=center},
  sdemb kfour sheet/.style={draw=sdembEdge,fill=sdembFill,
    line width=1.05pt,line cap=round,line join=round},
  sdemb kfour eye/.style={draw=sdembEdge,fill=white,
    line width=.82pt,line cap=round,line join=round},
  sdemb kfour edge/.style={line width=1.16pt,line cap=round,
    line join=round},
  sdemb kfour vertex/.style={circle,draw=black!78,fill=white,
    line width=.58pt,minimum size=3.6mm,inner sep=0pt,
    font=\tiny\bfseries},
  sdemb kfour title/.style={font=\scriptsize\bfseries,align=center}
}

\newcommand{\sdembThetaPanel}[2]{%
  \begin{scope}
    \path[sdemb theta torus] (0,0) rectangle (4,2.70);
    \draw[sdemb theta identify] (1.55,2.70)--(2.45,2.70);
    \draw[sdemb theta identify] (1.55,0)--(2.45,0);
    \draw[sdemb theta identify] (0,.92)--(0,1.46);
    \draw[sdemb theta identify] (4,.92)--(4,1.46);
    \coordinate (sdembThetaB) at (2,2.24);
    \coordinate (sdembThetaC) at (2,.46);
    \ifnum#1=10\relax
      \draw[sdemb theta blue]
        (sdembThetaB)..controls(3.08,1.88)and(3.08,.82)..(sdembThetaC);
      \draw[sdemb theta green]
        (sdembThetaB)..controls(2.415,1.88)and(2.415,.82)..(sdembThetaC);
      \draw[sdemb theta purple]
        (sdembThetaB)..controls(1.29,2.38)and(.71,2.54)..(.622,2.70);
      \draw[sdemb theta purple]
        (.622,0)..controls(.71,.16)and(1.29,.32)..(sdembThetaC);
      \fill[Purple] (.622,2.70) circle (.045);
      \fill[Purple] (.622,0) circle (.045);
    \else
      \draw[sdemb theta purple]
        (sdembThetaB)..controls(3.08,1.88)and(3.08,.82)..(sdembThetaC);
      \draw[sdemb theta blue]
        (sdembThetaB)..controls(2.415,1.88)and(2.415,.82)..(sdembThetaC);
      \draw[sdemb theta green]
        (sdembThetaB)..controls(1.29,2.38)and(.71,2.54)..(.622,2.70);
      \draw[sdemb theta green]
        (.622,0)..controls(.71,.16)and(1.29,.32)..(sdembThetaC);
      \fill[ForestGreen] (.622,2.70) circle (.045);
      \fill[ForestGreen] (.622,0) circle (.045);
    \fi
    \node[sdemb theta vertex] at (sdembThetaB) {$b$};
    \node[sdemb theta vertex] at (sdembThetaC) {$c$};
    \SDThetaTitle{#2}
  \end{scope}%
}

\newcommand{\sdembBuildThetaTwoClassBox}{%
\setbox\sdembThetaTwoClassBox=\hbox{%
\begin{tikzpicture}[x=.82cm,y=.82cm]
  \sdembThetaPanel{10}{P-11-1 ($\theta_{10}$)\\blue--green disk}
  \begin{scope}[xshift=5.30cm]
    \sdembThetaPanel{11}{P-11-2 ($\theta_{11}$)\\blue--purple disk}
  \end{scope}
  \node[font=\scriptsize,align=center,anchor=north] at (4.65,-.25) {%
    \textcolor{RoyalBlue}{$b_1\!-\!c_3$}\ (blue)\qquad
    \textcolor{ForestGreen}{$b_2\!-\!c_2$}\ (green)\qquad
    \textcolor{Purple}{$b_3\!-\!c_1$}\ (purple)};
\end{tikzpicture}}}
\sdembBuildThetaTwoClassBox

\newcommand{\sdembKFourEye}[2]{%
  \draw[sdemb kfour eye] (#1,#2)
    ellipse[x radius=.32,y radius=.13];}

\newcommand{\sdembKFourNodes}{%
  \foreach \sdembPoint/\sdembLabel in {
      sdembKFourA/a,sdembKFourB/b,sdembKFourC/c,sdembKFourT/t}{%
    \node[sdemb kfour vertex] at (\sdembPoint) {$\sdembLabel$};}}

\newcommand{\sdembKFourOne}{%
  \coordinate (sdembKFourA) at (.15,1.02);
  \coordinate (sdembKFourB) at (-1.55,.02);
  \coordinate (sdembKFourC) at (.32,-.98);
  \coordinate (sdembKFourT) at (.22,.02);
  \sdembKFourEye{-.43}{.03}
  \draw[sdemb kfour edge,draw=Purple] (sdembKFourA)
    ..controls(-.55,1.04)and(-1.24,.68)..(sdembKFourB);
  \draw[sdemb kfour edge,draw=Magenta] (sdembKFourB)
    ..controls(-1.18,-.62)and(-.50,-.98)..(sdembKFourC);
  \draw[sdemb kfour edge,draw=RoyalBlue] (sdembKFourC)
    ..controls(.78,-.52)and(.78,.58)..(sdembKFourA);
  \draw[sdemb kfour edge,draw=ForestGreen] (sdembKFourA)
    ..controls(.18,.72)and(.21,.38)..(sdembKFourT);
  \draw[sdemb kfour edge,draw=BrickRed] (sdembKFourB)
    ..controls(-.92,-.42)and(-.02,-.48)..(sdembKFourT);
  \draw[sdemb kfour edge,draw=Orange] (sdembKFourC)
    ..controls(.28,-.62)and(.22,-.30)..(sdembKFourT);
  \sdembKFourNodes}

\newcommand{\sdembKFourTwo}{%
  \coordinate (sdembKFourA) at (.08,1.02);
  \coordinate (sdembKFourB) at (-1.53,.00);
  \coordinate (sdembKFourC) at (.34,-.96);
  \coordinate (sdembKFourT) at (.22,.08);
  \sdembKFourEye{-.43}{-.02}
  \draw[sdemb kfour edge,draw=Purple] (sdembKFourA)
    ..controls(-.58,1.10)and(-1.23,.68)..(sdembKFourB);
  \draw[sdemb kfour edge,draw=Magenta] (sdembKFourB)
    ..controls(-1.17,-.65)and(-.48,-1.00)..(sdembKFourC);
  \draw[sdemb kfour edge,draw=RoyalBlue] (sdembKFourC)
    ..controls(.80,-.50)and(.76,.58)..(sdembKFourA);
  \draw[sdemb kfour edge,draw=ForestGreen] (sdembKFourA)
    ..controls(.14,.72)and(.20,.40)..(sdembKFourT);
  \draw[sdemb kfour edge,draw=BrickRed] (sdembKFourB)
    ..controls(-.90,.48)and(-.04,.52)..(sdembKFourT);
  \draw[sdemb kfour edge,draw=Orange] (sdembKFourC)
    ..controls(.30,-.56)and(.22,-.25)..(sdembKFourT);
  \sdembKFourNodes}

\newcommand{\sdembKFourThree}{%
  \coordinate (sdembKFourA) at (-.42,1.02);
  \coordinate (sdembKFourB) at (1.10,.12);
  \coordinate (sdembKFourC) at (-.28,-1.02);
  \coordinate (sdembKFourT) at (2.04,.12);
  \sdembKFourEye{-.35}{.02}
  \draw[sdemb kfour edge,draw=Purple] (sdembKFourA)
    ..controls(.30,1.06)and(.88,.70)..(sdembKFourB);
  \draw[sdemb kfour edge,draw=Magenta] (sdembKFourB)
    ..controls(1.06,-.52)and(.44,-.98)..(sdembKFourC);
  \draw[sdemb kfour edge,draw=RoyalBlue] (sdembKFourC)
    ..controls(-1.02,-.52)and(-1.04,.58)..(sdembKFourA);
  \draw[sdemb kfour edge,draw=ForestGreen] (sdembKFourA)
    ..controls(.48,1.42)and(1.68,1.20)..(sdembKFourT);
  \draw[sdemb kfour edge,draw=BrickRed]
    (sdembKFourB)--(sdembKFourT);
  \draw[sdemb kfour edge,draw=Orange] (sdembKFourC)
    ..controls(.50,-1.36)and(1.72,-1.10)..(sdembKFourT);
  \sdembKFourNodes}

\newcommand{\sdembKFourCard}[2]{%
  \begin{tikzpicture}[x=.63cm,y=.57cm,
      baseline=(current bounding box.center)]
    \path[use as bounding box] (-2.72,-1.88) rectangle (2.72,2.13);
    \path[sdemb kfour sheet] (0,0)
      ellipse[x radius=2.55,y radius=1.68];
    \csname sdembKFour#2\endcsname
    \SDKFourTitle{#1}
  \end{tikzpicture}}

\newcommand{\sdembBuildKFourThreeClassBox}{%
\setbox\sdembKFourThreeClassBox=\hbox{%
\begin{tabular}{@{}c@{\hspace{2mm}}c@{\hspace{2mm}}c@{}}
  \sdembKFourCard{6j-7-1: $abt$}{One} &
  \sdembKFourCard{6j-7-2: $bct$}{Two} &
  \sdembKFourCard{6j-7-3: $cat$}{Three} \\[1mm]
  \multicolumn{3}{c}{\scriptsize
    \textcolor{ForestGreen}{$a_1\!-\!t_1$}\ (green)\quad
    \textcolor{RoyalBlue}{$a_3\!-\!c_1$}\ (blue)\quad
    \textcolor{Purple}{$a_2\!-\!b_1$}\ (purple)}\\[-.2mm]
  \multicolumn{3}{c}{\scriptsize
    \textcolor{BrickRed}{$b_2\!-\!t_2$}\ (red)\quad
    \textcolor{Magenta}{$b_3\!-\!c_2$}\ (magenta)\quad
    \textcolor{Orange}{$c_3\!-\!t_3$}\ (orange)}
\end{tabular}}}
\sdembBuildKFourThreeClassBox

\newcommand{\sdembThetaTwoClassFigure}{%
  \usebox{\sdembThetaTwoClassBox}}
\newcommand{\sdembKFourThreeClassFigure}{%
  \usebox{\sdembKFourThreeClassBox}}

\begingroup
\SDFigureLabels{position=upper-left}
\newcommand{\SDPillowCutOperatorLabels}{%
  \tikzset{sdcut operator/.style={font=\tiny,
    fill=crawFill,inner sep=.35pt,text=black}}
  \node[sdcut operator] at (-.48,1.60) {$I$};
  \node[sdcut operator] at (-.90,1.39) {$J$};
  \node[sdcut operator] at (-.46,1.17) {$N$};
  \node[sdcut operator] at (-.48,.28) {$N$};
  \node[sdcut operator] at (-.90,0) {$K$};
  \node[sdcut operator] at (-.48,-.28) {$L$};
  \node[sdcut operator] at (-.48,-1.17) {$L$};
  \node[sdcut operator] at (-.90,-1.39) {$K$};
  \node[sdcut operator] at (-.48,-1.60) {$M$};
  \node[sdcut operator,fill=white,anchor=west] at (5.47,.35) {$I$};
  \node[sdcut operator,fill=white,anchor=west] at (5.47,0) {$J$};
  \node[sdcut operator,fill=white,anchor=west] at (5.47,-.35) {$M$};
}

\newsavebox{\sdPillowCutOneBox}
\newsavebox{\sdPillowCutTwoBox}
\newsavebox{\sdPillowCutPSevenBox}
\newsavebox{\sdPillowCutSevenBox}
\begingroup
  \let\crawCuffLabels\SDPillowCutOperatorLabels
  \global\setbox\sdPillowCutOneBox=\hbox{%
    \crawCard{1}{PillowPFive}{Standard}
      {P-5}{}{Hidden}}
  \global\setbox\sdPillowCutTwoBox=\hbox{%
    \crawCard{2}{PillowPSix}{Standard}
      {P-6}{}{Hidden}}
  \global\setbox\sdPillowCutPSevenBox=\hbox{%
    \crawCard{18}{PillowColoredEighteen}{Standard}
      {P-7-1}{}{Lower}}
  \global\setbox\sdPillowCutSevenBox=\hbox{%
    \crawPillowEarlyDisconnectedCard{7}{7}
      {P-10}}
\endgroup

\newcommand{\SDPillowCutPanel}[1]{%
  \resizebox{.455\linewidth}{!}{\usebox{#1}}}
\newcommand{\SDPillowCutSurvivorGallery}{%
  \begin{tabular}{@{}c@{\hspace{.02\linewidth}}c@{}}
    \multicolumn{2}{c}{\small\bfseries $C_2$}\\[.6mm]
    \SDPillowCutPanel{\sdPillowCutOneBox} &
    \SDPillowCutPanel{\sdPillowCutTwoBox} \\[1.8mm]
    {\small\bfseries $C_3$} & {\small\bfseries $D_1$}\\[.6mm]
    \SDPillowCutPanel{\sdPillowCutPSevenBox} &
    \SDPillowCutPanel{\sdPillowCutSevenBox}
  \end{tabular}}

\newcommand{\SDOneTwoPointCutsFigure}{%
\begin{figure}[!ht]
  \centering
  \begin{tikzpicture}[x=1cm,y=1cm,
    cutdemo body/.style={draw=CZeroSurfaceEdge,fill=CZeroSurfaceFill,
      line width=.75pt,line join=round},
    cutdemo rim/.style={draw=CZeroSurfaceEdge,line width=.75pt},
    cutdemo rear rim/.style={cutdemo rim,
      dash pattern=on 1.8pt off 1.5pt},
    cutdemo rear cut/.style={draw=BrickRed!65,line width=.95pt,
      dash pattern=on 1.8pt off 1.5pt,line cap=round},
    cutdemo front cut/.style={draw=BrickRed,line width=1.05pt,
      line cap=round},
    cutdemo wire/.style={draw=CZeroRight,line width=1.15pt,
      line cap=round,
      preaction={draw=CZeroSurfaceFill,line width=2.8pt,solid}},
    cutdemo title/.style={font=\small,align=center}]
    \foreach \panelshift/\linecount/\paneltitle in {
      -2.65/1/{(a) One-point cut},
       2.65/2/{(b) Two-point cut}}{
      \begin{scope}[xshift=\panelshift cm]
        \node[cutdemo title] at (0,1.07) {\paneltitle};

        \path[cutdemo body]
          (-1.55,.70)--(1.55,.70)
          arc[start angle=90,end angle=-90,x radius=.23,y radius=.70]
          --(-1.55,-.70)
          arc[start angle=270,end angle=90,x radius=.23,y radius=.70]
          --cycle;
        \draw[cutdemo rear rim] (1.55,.70)
          arc[start angle=90,end angle=270,x radius=.23,y radius=.70];
        \draw[cutdemo rear cut] (0,.70)
          arc[start angle=90,end angle=270,x radius=.23,y radius=.70];

        \ifnum\linecount=1\relax
          \def\wireheights{0}
        \else
          \def\wireheights{-.30,.30}
        \fi
        \foreach \wireheight in \wireheights{
          \pgfmathsetmacro{\frontoffset}{.23*sqrt(1-(\wireheight/.70)^2)}
          \draw[cutdemo wire]
            ({-1.55+\frontoffset},\wireheight)
            --({1.55+\frontoffset},\wireheight);
        }

        \filldraw[cutdemo rim,fill=white]
          (-1.55,0) ellipse[x radius=.23,y radius=.70];
        \draw[cutdemo rim] (1.55,-.70)
          arc[start angle=-90,end angle=90,x radius=.23,y radius=.70];
        \draw[cutdemo front cut] (0,-.70)
          arc[start angle=-90,end angle=90,x radius=.23,y radius=.70];
        \foreach \wireheight in \wireheights{
          \pgfmathsetmacro{\frontoffset}{.23*sqrt(1-(\wireheight/.70)^2)}
          \fill[BrickRed] (\frontoffset,\wireheight) circle[radius=.045];
        }
      \end{scope}
    }
  \end{tikzpicture}
  \caption{Local views of one- and two-point cuts: the red circle meets the
  green surface Wilson lines at the marked points; dashed arcs lie behind
  the surface.}
  \label{fig:one-two-point-cuts}
\end{figure}
}%

\newcommand{\SDSixJOnePointReductionFigure}{%
\begin{figure}[!ht]
  \centering
  \begingroup
  \begin{tikzpicture}[x=.60cm,y=.85cm,
    opr label/.style={font=\scriptsize,fill=crawFill,
      inner sep=.45pt,text=black},
    opr outside/.style={font=\scriptsize,inner sep=1pt,text=black},
    opr red/.style={draw=BrickRed,line width=.95pt,line cap=round},
    opr note/.style={font=\scriptsize,text=BrickRed,inner sep=1pt},
    opr heading/.style={font=\small,align=center}]

    \def\oprCuffLabels{%
      \foreach \ly/\lt in {1.60/I,1.17/N,.28/N,0/J,-.28/K,-1.17/K,-1.60/M}{
        \node[opr label] at (-.42,\ly) {$\lt$};}
      \node[opr label] at (-.70,1.39) {$L$};
      \node[opr label] at (-.70,-1.39) {$L$};}
    \def\oprExternalLabels{%
      \foreach \ly/\lt in {.29/I,0/J,-.29/M}{
        \node[opr outside,anchor=west] at (5.50,\ly) {$\lt$};}}
    \def\oprRightRim{%
      \draw[craw seam rear] (5.20,.52)
        arc[start angle=90,end angle=270,x radius=.17,y radius=.52];
      \draw[craw seam front] (5.20,-.52)
        arc[start angle=-90,end angle=90,x radius=.17,y radius=.52];}
    \def\oprOpenCuffs{%
      \foreach \cy/\ry in {1.33/.41,0/.48,-1.33/.41}{
        \path[fill=white] (0,\cy) ellipse[x radius=.17,y radius=\ry];
        \draw[craw seam rear] (0,{\cy+\ry})
          arc[start angle=90,end angle=270,x radius=.17,y radius=\ry];
        \draw[craw seam front] (0,{\cy-\ry})
          arc[start angle=-90,end angle=90,x radius=.17,y radius=\ry];}}

    \begin{scope}[xshift=3.55cm]
      \crawSurfaceStandard
      \crawRouteSixTen
      \oprCuffLabels
      \foreach \ly/\lt in {.29/I,0/J,-.29/M}{
        \node[opr outside,anchor=east] at (-5.53,\ly) {$\lt$};}
      \draw[opr red,-{Stealth[length=1.6mm]}]
        (2.65,-2.00) .. controls (2.35,-1.29) and (1.75,-.83)
        .. (1.34,-.74);
      \node[opr note,anchor=north] at (2.65,-2.07) {one-point cut};
      \node[opr heading] at (0,2.23) {(a) Full diagram};
      \node[font=\scriptsize\bfseries,text=black!78,
        anchor=west,inner sep=0pt] at (-5.18,1.87) {6j-3};
    \end{scope}

    \node[font=\large] at (12.06,0) {$\longrightarrow$};

    \begin{scope}[xshift=8.00cm]
      \path[craw sheet]
        (0,1.74)--(2.45,1.74)
        .. controls (3.92,1.74) and (4.20,.62) .. (5.20,.52)
        arc[start angle=90,end angle=-90,x radius=.17,y radius=.52]
        .. controls (4.20,-.62) and (3.92,-1.74) .. (2.45,-1.74)
        --(0,-1.74)--(0,-.92)
        .. controls (.96,-.92) and (.96,-.48) .. (0,-.48)
        --(0,.48)
        .. controls (.96,.48) and (.96,.92) .. (0,.92)
        --cycle;
      \crawThreeSeams
      \oprOpenCuffs
      \oprRightRim
      \begin{scope}
        \tikzset{craw left/.append style={draw=none},
          craw left rear/.append style={draw=none}}
        \crawRouteSixTen
      \end{scope}
      \draw[opr red]
        (crawBottomLocalFront)--(.38,-1.17)
        .. controls (1.62,-1.18) and (1.70,-.45) .. (.38,-.28)
        --(crawMiddleLowerFront);
      \oprCuffLabels
      \draw[opr red,-{Stealth[length=1.6mm]}]
        (2.04,-1.95) .. controls (1.95,-1.43) and (1.59,-1.03)
        .. (1.32,-.78);
      \node[opr note,anchor=north] at (2.04,-2.07) {$K=\mathds{1}$};
      \node[opr heading] at (2.35,2.23) {(b) Identity on the cut line};
    \end{scope}

    \node[font=\large] at (19.58,0) {$=$};

    \begin{scope}[xshift=12.65cm]
      \path[craw sheet]
        (0,1.74)--(2.45,1.74)
        .. controls (3.92,1.74) and (4.20,.62) .. (5.20,.52)
        arc[start angle=90,end angle=-90,x radius=.17,y radius=.52]
        .. controls (4.20,-.62) and (3.92,-1.74) .. (2.45,-1.74)
        --(.05,-1.74)
        .. controls (-.92,-1.74) and (-1.02,-.34) .. (0,.48)
        .. controls (.96,.48) and (.96,.92) .. (0,.92)
        --cycle;
      \begin{scope}
        \tikzset{craw seam fill/.append style={fill=none},
          craw seam rear/.append style={draw=none},
          craw seam front/.append style={draw=none}}
        \crawThreeSeams
      \end{scope}
      \path[fill=white] (0,1.33) ellipse[x radius=.17,y radius=.41];
      \draw[craw seam rear] (0,1.74)
        arc[start angle=90,end angle=270,x radius=.17,y radius=.41];
      \draw[craw seam front] (0,.92)
        arc[start angle=-90,end angle=90,x radius=.17,y radius=.41];
      \oprRightRim
      \begin{scope}
        \tikzset{craw left/.append style={draw=none},
          craw left rear/.append style={draw=none}}
        \crawCandidateTenC
        \crawOpenAOnly
        \crawOpenBOnly
        \crawUBase
      \end{scope}
      \draw[craw right] (crawMiddleUpper)
        .. controls (-.45,.19) and (-.45,0) .. (crawMiddleOpen);
      \path (.05,-1.74)
        .. controls (-.92,-1.74) and (-1.02,-.34) .. (0,.48)
        coordinate[pos=.25] (oprCapTurn);
      \draw[craw right rear] (crawBottomGlobalRear)
        .. controls (-.39,-1.39) and (-.49,-1.43) .. (oprCapTurn);
      \draw[craw right] (oprCapTurn)
        .. controls (-.34,-1.61) and (-.02,-1.60) .. (crawBottomOpen);
      \foreach \ly/\lt in {1.60/I,1.17/N}{
        \node[opr label] at (-.42,\ly) {$\lt$};}
      \node[opr label] at (-.70,1.39) {$L$};
      \oprExternalLabels
      \node[opr note] at (.65,-.40) {$N=J$};
      \node[opr note] at (.70,-1.15) {$L=M$};
      \node[opr heading] at (2.15,2.23) {(c) Two-point completion};
    \end{scope}
  \end{tikzpicture}
  \endgroup
  \caption{One-point reduction of 6j-3:
  \(K=\mathds{1}\) leaves the pairings \(N=J\) and \(L=M\).}
  \label{fig:sixj-raw10-one-point}
\end{figure}
}%

\newcommand{\SDDisconnectedPillowCutFigure}{%
\begin{figure}[!ht]
  \centering
  \SDPillowCutPanel{\sdPillowCutSevenBox}
  \caption{Disconnected Pillow completion P-10: the two green
  annuli require \(L=I\) and \(K=J\).}
  \label{fig:disconnected-pillow-two-point}
\end{figure}
}%

\maketitle

\begin{abstract}
We test the proposed matrix--tensor description of three-dimensional gravity
using a Schwinger--Dyson relation in its purely open sector.  We enumerate
and classify the topological contributions to a cylinder amplitude.  Under a
proposed counting prescription motivated by Virasoro TQFT, we find agreement
with the conjectured unit difference between the numbers of contributing manifolds associated to the two interaction terms in the potential.
By contrast, counting all inequivalent surface embeddings separately would instead give a
mismatch. This comparison highlights the importance of specifying the set of bulk topologies to be included in the gravitational path integral.

\end{abstract}

\clearpage
\tableofcontents
\clearpage

\section{Introduction}
\label{sec:introduction}

Three-dimensional anti-de Sitter (AdS) gravity has long provided a setting
for exploring the relation between quantum gravity and conformal field
theory (CFT)~\cite{Maldacena:1997re}.
An important part of this problem is to understand how the gravitational
sum over bulk topologies is encoded in the boundary
data~\cite{Maloney:2007ud,Keller:2014xba}.

Jackiw--Teitelboim (JT)
gravity~\cite{Teitelboim:1983ux,Jackiw:1984je,Almheiri:2014cka} provides
a concrete example in a two-dimensional setting.  Its boundary dynamics shares a low-energy effective
description with the Sachdev--Ye--Kitaev (SYK)
model~\cite{Sachdev:1992fk,Kitaev:2015talks,Maldacena:2016hyu,Kitaev:2017awl,Jensen:2016pah,Maldacena:2016upp}.
The JT amplitudes on surfaces with arbitrary handles and boundaries are
reproduced by the genus expansion of a matrix
integral~\cite{Saad:2019lba,Stanford:2019vob}.
In JT gravity, contributions from nontrivial spacetime topologies provide a
geometric explanation for spectral correlations and the averaged late-time
behavior of matter correlation functions~\cite{Saad:2019lba,Saad:2019pqd}.

A proposed three-dimensional analog of the JT/matrix-integral duality relates
AdS$_3$ gravity to formal ensembles of two-dimensional CFT data.\footnote{These ensembles are specified through statistical properties of CFT data, rather than a microscopic construction of an ensemble of exact CFTs.}
Roughly speaking, offshell wormholes have been associated to random matrix statistics of spectral correlators~\cite{Cotler:2020ugk,Maxfield:2020ale,Cotler:2020hgz,DiUbaldo:2023qli,Boruch:2025ilr,deBoer:2025oge,Jafferis:2025jle}, while onshell wormholes have been largely associated to OPE statistics \cite{Collier:2019weq,Belin:2020hea,Belin:2021ryy,Anous:2021caj,Chandra:2022bqq,deBoer:2024mqg,Wang:2025bcx,Hung:2025vgs,Chandra:2025fef,Hartman:2025cyj,Hartman:2025ula,Wang:2025jgo,Belin:2026pko,Jafferis:2026gzn}. This dichotomy is partly due to an incomplete understanding of how Virasoro topological quantum field theory
(VTQFT)~\cite{Collier:2023fwi,Collier:2024mgv}, which yields finite amplitudes for onshell manifolds, extends to offshell manifolds~\cite{Yan:2023rjh,Yan:2025usw}. One attempt to construct a dual ensemble incorporating both offshell and onshell amplitudes is provided by the matrix--tensor model, with the goal of generating the sum over gravitational topologies through a diagrammatic expansion~\cite{Belin:2023efa,Jafferis:2025vyp,Jafferis:2025yxt,Ambrosini:2026int}.

In this paper, we test the proposed matrix--tensor description of
three-dimensional gravity, beginning with its purely open
version~\cite{Jafferis:2025yxt}, which describes an ensemble of boundary conformal field theories (BCFTs) with only open data, defined in more details in \cite{Jafferis:2026gzn} and briefly reviewed below. We examine a Schwinger--Dyson (SD) identity that predicts a relation
between two classes of topological contributions.  By explicitly
enumerating these contributions, we check the SD equation at the leading order.

In the purely open integral, the only data are the conformal weights
\(h_I\) of boundary operators and their disk three-point functions, or boundary OPE
coefficients
\(B^{(abc)}_{IJK}\).  The uppercase indices label boundary operators, while early-alphabet lower-case indices label the boundary conditions on the three intervals between them.
In the three-dimensional picture, these boundary conditions are associated to
colors of end-of-the-world (EOW)
branes~\cite{Takayanagi:2011zk,Wang:2025bcx,Jafferis:2025yxt}.  In the matrix--tensor model, the correlations between level spacing is described by a matrix, whereas the OPE statistics is described as a rank-three tensor symmetric under cyclic index permutations. The model is defined by
\begin{equation}
  \mathcal{Z}_{\mathrm{open}}
  =\int\mathcal D h\,\mathcal D B\,
  \exp\left[-V_0(h)-\frac1\hbar V_3(h,B)\right],
  \label{eq:intro-open-integral}
\end{equation}
where \(V_0\) provides the leading density of the open spectrum and has no dependence on \(B\), while \(V_3\) measures violations of the boundary four-point
crossing equation on the disk and vanishes when that equation is satisfied.
The parameter \(\hbar\) controls the penalty for violating the bootstrap constraint; taking \(\hbar\to0\)
formally localizes the integral onto data satisfying the constraint.
Other terms in the potential of the full open--closed model, i.e., $V_i$ for $i\in\{1,2,4,5,6\}$ in the notation of \cite{Jafferis:2025yxt}, are absent in this model, as the corresponding bootstrap conditions all involve the closed data~\cite{Jafferis:2025vyp,Jafferis:2025yxt}.

The SD equation expresses a simple fact: an infinitesimal
change of an integration variable cannot change the value of the
integral~\cite{Dyson:1949ha,Schwinger:1951ex}.  Applying this fact to one
\(B\) variable gives a relation among three diagrams: one bare two-point correlator, one
Pillow interaction, and one \(6j\) interaction.  The
Pillow diagram enters with a plus sign and the \(6j\) diagram with a minus
sign. As we derive later in Section~\ref{app:sd}, it states
\begin{equation}
  \OpenBWeightedSDFigure
  \label{eq:sd-picture}
\end{equation}
The left-hand side is the exact two-point correlator.  On the right-hand side, the
first term is the bare two-point function, while the other two terms are the two interaction vertices in $V_3$ attached to exact four-point correlators. The two interaction vertices are known as the Pillow and the open-$6j$ diagrams respectively, and will be reviewed in detail. We have used the gray boxes to represent exact correlators. 

Since the SD equation holds as a fact, our goal is not to check the SD equation as defined by the matrix--tensor model. Instead, we test the conjecture that each gray box can be computed in three dimensions by summing over 3-manifolds with equal weight \cite{Jafferis:2025vyp}. Therefore, with a slight abuse of language, when we say we test the SD equation, what we mean is that we test the duality using the SD equation as a tool.  

The exact two-point function on the left-hand side of equation~\eqref{eq:sd-picture} is a sum over many 3-manifolds. To test the SD equation, we may study it manifold by manifold. In other words, we can pick a particular contribution on the left-hand side, which then requires us to only look at diagrams on the right-hand side that reproduces this contribution. The simplest contribution to the left-hand side is a solid cylinder with three Wilson lines running parallel to the interval direction, as shown below. Due to the VTQFT amplitude associated to it, we will refer to it as the $C_0$ diagram or $C_0$ manifold. The relevant part of the equation is then
\begin{equation}
  \OpenBTopologyProjectionFigure
  \label{eq:sd-sector}
\end{equation}
where the gray blobs now represent the subset of manifolds contributing to the exact four-point correlators that, upon gluing to the pillow or the open-$6j$ vertex, yield the $C_0$ manifold. This equation is the purely open counterpart of equation~(5.5) of
\cite{Jafferis:2025vyp}. Wilson loops form in this process and are treated with surgery, which will be reviewed and demonstrated with examples in Appendix~\ref{app:open-surgery}.

Note that the first term on the right-hand side of \eqref{eq:sd-picture} is now absent in \eqref{eq:sd-sector}. This is because it comes with an extra factor of $\hbar$, and we have taken the limit $\hbar \to 0$ as part of the conjecture. In other words, the conjecture only states that the $\hbar\to0$ limit of the matrix--tensor model describes three-dimensional quantum gravity where each manifold contributes with equal weight. 

In this setup where we have projected the full SD equation onto a single manifold, the check reduces to counting \(\NP\) and \(\Nsix\), the numbers of qualifying manifolds for the two terms respectively. 

Our main result is the following:
\begin{equation}
  \Nsix=0,\qquad \NP=1,
  \qquad \NP-\Nsix=1.
  \label{eq:main-result}
\end{equation}
This passes the check. In arriving at these numbers, we have used a prescription for how to count inequivalent manifolds. The prescription takes its motivation from VTQFT and is explained in Section~\ref{app:no-onepoint}.

Our approach has three preliminary steps:
\begin{enumerate}[label=\textbf{Step \arabic*.},leftmargin=*]
  \item \emph{Enumerate candidates.}  The code~\cite{SDcountcode} generates and selects all
  allowed contributions at the level of abstract graphs formed by the Wilson lines.
  \item \emph{Perform the surgery check.}  The abstract graphs are then embedded in the EOW branes, and we check if gluing and performing the surgery produces the
  \(C_0\) cylinder. Each surviving graph is counted once at this stage.
  \item \emph{Count marked embeddings.}  We count the distinct embeddings
  of each retained graph. In particular, an abstract graph can have more than one embedding.
\end{enumerate}

At this stage, a natural prescription for counting manifolds might be to count the different inequivalent embeddings. In doing so, we would get 
\begin{equation}
  \NP=14, \qquad \Nsix=12,
  \qquad \NP-\Nsix=2.
  \label{eq:intro-marked-count}
\end{equation}
However, once we look at the VTQFT amplitudes defined on these candidate manifolds, it becomes clear that many of them constitute invalid or redundant configurations in the SD equation. Following the VTQFT-motivated prescription, the count changes to $N_{6j}=0$ and $\NP=1$. 

\textbf{Roadmap.}
Section~\ref{app:sd} derives the SD equation and its projection onto the topology we focus on.  Section~\ref{app:no-onepoint} explains the counting prescription. This is presented early given its central role in our analysis, even though it will only be applied after all embeddings are counted. Section~\ref{sec:enumeration} explains the enumeration of all abstract graphs, whereas Section~\ref{sec:embeddings} studies the embeddings of the graphs that turn them into valid candidate contributions. We conclude in Section~\ref{sec:discussion} with a summary of the results and discussion of open
questions.  The appendices include a review of the surgery procedure, a detailed presentation of the full algorithm, details of the code, the list of embeddings, and a technique for recognizing equivalent handlebodies which we call handle migration.

\section{The purely open matrix--tensor model and its
Schwinger--Dyson equation}
\label{app:sd}

In this section, we review the derivation of the counting prediction
\(\NP-\Nsix=1\).  Starting from the purely open model of
\cite{Jafferis:2025yxt}, we identify the Pillow and \(6j\) interactions
and derive the SD equation relating their contributions.  We then select
the cylinder $C_0$ contribution and explain the assumptions needed to turn this
equation into a relation between diagram counts, following the approach of
\cite{Jafferis:2025vyp}.

\subsection{Data retained in the purely open model}

BCFT extends a two-dimensional CFT to
surfaces with conformal boundary conditions~\cite{Cardy:1984bb}.
Besides the bulk data, one specifies boundary conditions,
boundary operator spectra, boundary OPE coefficients, and bulk-to-boundary
couplings~\cite{Cardy:1991tv}.  These data must satisfy crossing symmetry
of four-point disk correlators~\cite{Lewellen:1991tb} and other consistency conditions involving both open and closed data \cite{Cardy:1989ir}.  Recently, analytic BCFT bootstrap has
used these constraints to derive universal high-energy boundary spectra
and asymptotics of averaged boundary and bulk-to-boundary OPE
data~\cite{Kusuki:2021gpt,Numasawa:2022cni}.  The purely open model studied
here retains only the boundary spectra and boundary OPE coefficients, with
disk four-point crossing as the only crossing
constraint~\cite{Jafferis:2025yxt}.

The purely open matrix--tensor model is based on the purely open BCFT. After removing closed data, the remaining variables are just the open spectrum and the boundary-to-boundary OPE coefficients \(B^{(abc)}_{IJK}\), and the model is given by
\cite{Jafferis:2025yxt}
\begin{equation}
\mathcal{Z}_{\mathrm{open}}
=\int\mathcal D h\,\mathcal D B\,
\exp\left[-V_0(h)-\frac1\hbar V_3(h,B)\right].
\label{eq:open-integral}
\end{equation}
The tensor has the following symmetry: 
\begin{equation}
B^{(abc)}_{IJK}=B^{(bca)}_{JKI}=B^{(cab)}_{KIJ},
\qquad
\left(B^{(abc)}_{IJK}\right)^*=B^{(cba)}_{JIK}.
\label{eq:B-cyclic}
\end{equation}

\subsection{Crossing squared potential}

The Pillow and \(6j\) interactions arise from squaring the disk four-point
crossing equation~\cite{Jafferis:2025yxt}.
We write the disk four-point crossing equation using the Virasoro fusion
kernel to relate the two OPE channels~\cite{Ponsot:1999uf,Ponsot:2000mt}.
For reviews of the fusion kernel and its conventions, see \cite{Teschner:2001rv,Eberhardt:2023mrq}.

Let \(\mathcal H^{ab}\) denote the Hilbert space associated to states defined on an interval with boundary conditions \(a\) and \(b\) on its ends, and let \(P_M\) denote the
conformal weight (in the Liouville convention) of the operator \(M\). In terms of the crossing kernel $\mathbb{F}$, the constraint is given by $M^{(abcd)}_{3,IJKL}(P)=0$, where
\begin{align}
\mathcal M^{(abcd)}_{3,IJKL}(P)
={}&\frac{1}{\sqrt{g_ag_c}}
\sum_{M\in\mathcal H^{ac}}
B^{(abc)}_{IJM}B^{(cda)}_{KLM}\,
\delta(P-P_M)\notag\\
&-\frac{1}{\sqrt{g_dg_b}}
\sum_{M\in\mathcal H^{bd}}
B^{(dab)}_{LIM}B^{(bcd)}_{JKM}\,
\mathbb F_{MP}
\begin{bmatrix}J&I\\ K&L\end{bmatrix}.
\label{eq:crossing-mismatch}
\end{align}
The first term is the \(s\)-channel conformal block expansion with an intermediate state of weight $P_M$  projected onto the $s$-channel conformal block with an intermediate state of weight $P$, while the second term is the \(t\)-channel expression with an intermediate state of weight $P_M$ projected to the $s$-channel conformal block with an intermediate state of weight $P$. For each boundary condition $a$, the quantity \(g_a\) is the associated $g$-function, defined to be the disk partition function. 

The potential penalizes violations of the crossing symmetry by taking the square of the constraint~\cite{Jafferis:2025yxt}:
\begin{align}
V_3={}c_3\sum_{a,b,c,d}
 \sum_{I\in\mathcal H^{ab}}
 \sum_{J\in\mathcal H^{bc}}
 \sum_{K\in\mathcal H^{cd}}
 \sum_{L\in\mathcal H^{da}}
 \int_{\mathbb R_+\cup\{\mathds{1}\}}
 \frac{\left|\mathcal M^{(abcd)}_{3,IJKL}(P)\right|^2}
 {\rho_0(P)C_0(IJP)C_0(KLP)}\,\mathrm{d}P .
\label{eq:crossing-square}
\end{align}
The function \(\rho_0\) denotes Cardy density and \(C_0\) is the
DOZZ formula~\cite{Dorn:1994xn,Zamolodchikov:1995aa},
in the normalization of \cite{Collier:2019weq}.  The prefactor $c_3$ is a normalization constant that will be fixed momentarily. 

Expanding the square and performing the integral over $P$, we obtain
\begin{align}
    V_3 = V_{\mathrm{P}}-V_{6j},
\end{align}
where
\begin{align}
V_{\mathrm P}
={}&2c_3\sum_{abcd}\sum_{IJKLMN}
B^{(abc)}_{IJM}B^{(cda)}_{KLM}
B^{(acb)}_{NJI}B^{(cad)}_{NLK}
\frac{\delta(P_M-P_N)}
{g_ag_c\rho_0(M)C_0(IJM)C_0(KLM)},
\label{eq:pillow-kernel}\\[2pt]
V_{6j}
={}&2c_3\sum_{abcd}\sum_{IJKLMN}
B^{(abc)}_{IJM}B^{(cda)}_{KLM}
B^{(dba)}_{NIL}B^{(bdc)}_{NKJ}\notag\\[-2pt]
&\qquad\times
\frac{\begin{Bmatrix}N&L&I\\ M&J&K\end{Bmatrix}}
{\sqrt{g_ag_bg_cg_d}\,
C_0(IJM)C_0(KLM)C_0(ILN)C_0(JKN)}.
\label{eq:sixj-kernel}
\end{align}

These two quartic interaction terms are assigned what we call the boundary Pillow and boundary $6j$ VTQFT diagrams respectively, as shown in
Figure~\ref{fig:source-interactions}. The $6j$ symbol appearing in $V_{6j}$ is the origin of the name we give to this term, while the Pillow takes its name from the shape of the diagram (especially if we shrink the disks to points).

\begin{figure}[!htbp]
\centering
\begin{minipage}[t]{.47\linewidth}
  \centering
  \includegraphics[height=4.2cm,width=.94\linewidth,keepaspectratio]
    {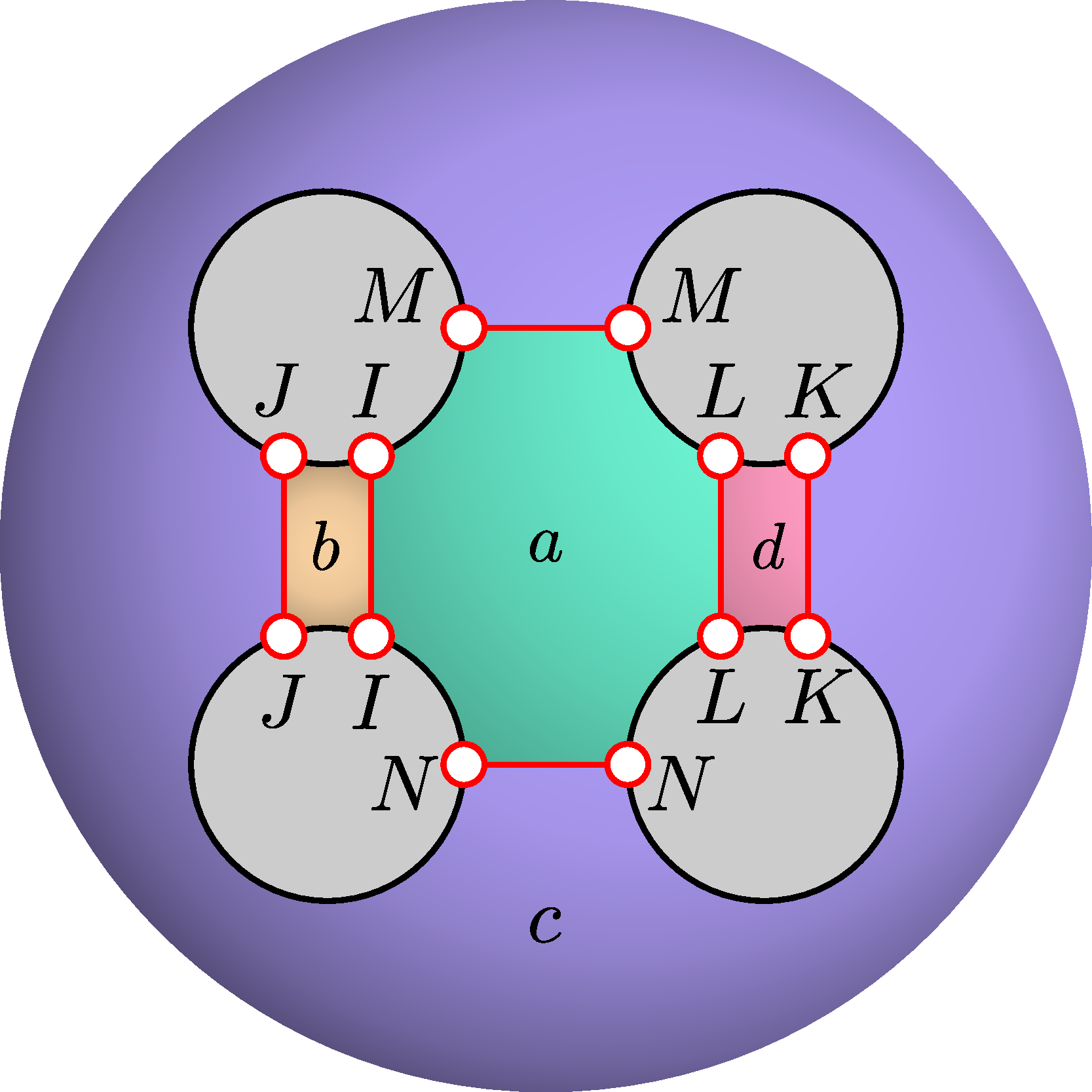}

  \small (a) boundary Pillow
\end{minipage}\hfill
\begin{minipage}[t]{.47\linewidth}
  \centering
  \includegraphics[height=4.2cm,width=.94\linewidth,keepaspectratio]
    {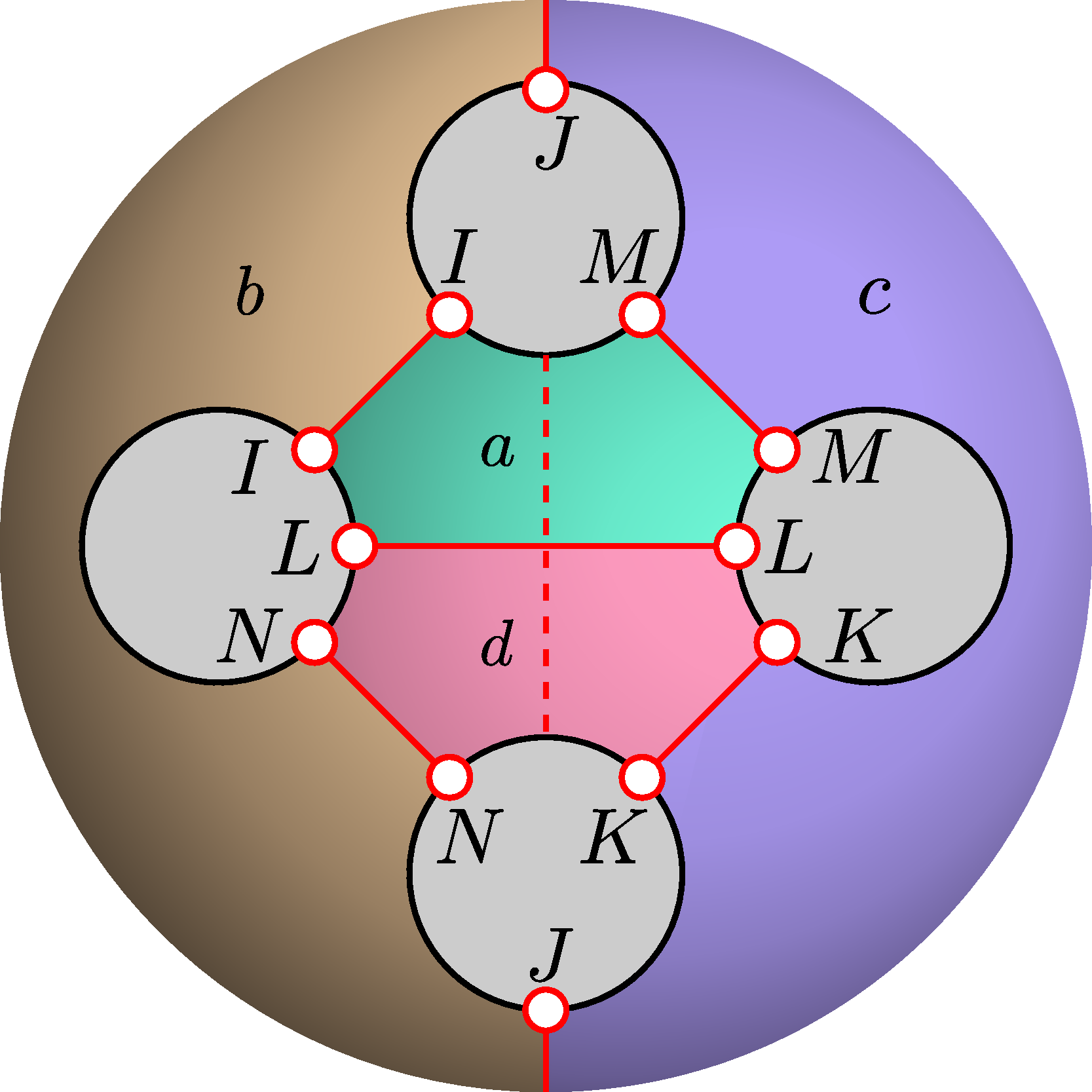}

  \small (b) boundary \(6j\)
\end{minipage}

\caption{The VTQFT diagrams associated to the two quartic interactions in the purely open model, adapted from
\cite{Jafferis:2025yxt}. The red lines are boundary Wilson lines, which lie on the EOW brane and separate it into four segments. Each segment of the brane, here drawn with a distinct color, has an independent tension parameter related to the $g$-function, and all (black) intervals on the boundaries of the disks bordering a brane with a given label have the same boundary condition with that label. Capital letters label both the Wilson lines and the boundary operators connected by them. The gray disks are state faces which can be glued along, and they correspond to the legs of the vertices in the tensor potential.}
\label{fig:source-interactions}
\end{figure}

\subsection{The identity operator and the normalization}
\label{sec:identity-normalization}

So far, we have not discussed the identity operator which appears in the sum, but they have been implicitly included in the sums. Terms containing the identity operator reduce to quadratic and constant terms, and it is convenient to separate them out. 

The Pillow delta function in equation~\eqref{eq:pillow-kernel} imposes \(M=N\).
Consequently, setting one of these two labels to the identity also sets the
other to the identity and gives a constant, which we can throw away without changing the model. In the $6j$ term, setting $N=\mathds{1}$ reduces two of the $B$ terms to 
\begin{equation}
B^{(bba)}_{\mathds{1}\,IL}=\sqrt{g_ag_b}\,\delta_{IL},
\qquad
B^{(bbc)}_{\mathds{1}\,KJ}=\sqrt{g_bg_c}\,\delta_{KJ},
\label{eq:identity-B-data}
\end{equation}
and reduces the Virasoro \(6j\) symbol to
\begin{equation}
\begin{Bmatrix}\mathds{1}&I&I\\ M&J&J\end{Bmatrix}
=C_0(II\mathds{1})C_0(JJ\mathds{1})C_0(IJM).
\label{eq:identity-sixj}
\end{equation}
The Kronecker deltas collapse the corresponding label sums, the factors of
\(g\) cancel, and equation~\eqref{eq:identity-sixj} cancels three of the four
\(C_0\) factors in equation~\eqref{eq:sixj-kernel}.  The sectors
\(M=\mathds{1},N\ne\mathds{1}\) and
\(N=\mathds{1},M\ne\mathds{1}\) are equal after relabeling, and each
contributes \(-2c_3\) times the same \(BB^*\) expression.  Their sum is
\begin{equation}
V_3^{(2)}=-4c_3\sum_{I,J,K}'
\frac{B^{(abc)}_{IJK}B^{(cba)}_{JIK}}{C_0(IJK)}.
\label{eq:quadratic-general}
\end{equation}
The prime means summing over only nonidentity states. When both internal labels are set to the identity, the result is another constant. 

After separating these sectors, the purely open crossing potential is now
\begin{equation}
  V_3=V_3^{(2)}+V_{\mathrm P}-V_{6j}+\text{constant},
\label{eq:potential-signs}
\end{equation}
where \(V_{\mathrm P}\) and \(V_{6j}\) have been redefined to denote the truly quartic pieces in the sense that all external states are now nonidentity states.

As in \cite{Jafferis:2025yxt}, we normalize the quadratic term by choosing
\begin{equation}
  c_3=\frac14,
  \qquad
  V_3^{(2)}=-\sum_{I,J,K}'
  \frac{B^{(abc)}_{IJK}B^{(cba)}_{JIK}}{C_0(IJK)}.
\label{eq:c3-choice}
\end{equation}
It is clear that this overall normalization is not physical but a mere convenience.

\subsection{Deriving the Schwinger--Dyson equation}

We now derive the SD equation. Schwinger--Dyson identities are obtained by integrating a total
derivative~\cite{Dyson:1949ha,Schwinger:1951ex}.  We insert one
factor of \(B^{(abc)}_{IJM}\) with nonidentity index labels and differentiate with respect to this same
variable.  Since \(V_0\) does not depend on
\(B\), only \(V_3\) contributes to the derivative of the
potential.  Assuming a regulator and a contour on
which the boundary term vanishes, and inserting a symmetry factor of $1/3$ for convenience, we obtain
\begin{align}
0
&=
\frac{1}{3\mathcal{Z}_{\mathrm{open}}}\int\mathcal D h\,\mathcal D B\,
\frac{\partial}{\partial B^{(abc)}_{IJM}}
\left(B^{(abc)}_{IJM}e^{-V_0-V_3/\hbar}\right)\notag\\
&=1-\frac1\hbar\left\langle
B^{(abc)}_{IJM}
\frac{\partial V_3}{\partial B^{(abc)}_{IJM}}
\right\rangle .
\label{eq:ibp}
\end{align}
Note that the sum over cyclic permutations
cancels with the symmetry factor.

Using \eqref{eq:potential-signs} to write out $V_3$ and substituting \eqref{eq:quadratic-general}, equation~\eqref{eq:ibp} becomes
\begin{equation}
0
=\frac{\hbar}{3}+\frac{2\left\langle B^{(abc)}_{IJM}B^{(cba)}_{JIM}\right\rangle}
{C_0(IJM)}
-\left\langle B^{(abc)}_{IJM}
\frac{\partial(V_{\mathrm P}-V_{6j})}{\partial B^{(abc)}_{IJM}}
\right\rangle .
\label{eq:sd-before-rearranging}
\end{equation}
Furthermore, using the explicit form of $V_{\mathrm{P}}$ and $V_{6j}$, the SD equation can then be written as
\begin{align} 
\left\langle B^{(abc)}_{IJK}B^{(cba)}_{JIM}\right\rangle
={}&-\frac{\hbar}{6} C_0(IJM)\notag\\
&+\sum_d\sum_{K,L,N}'
\frac{\delta(P_M-P_N)}
{g_ag_c\rho_0(M)C_0(KLM)}\notag\\[-2pt]
&\quad\times\left\langle
B^{(abc)}_{IJM}B^{(cda)}_{KLM}
B^{(acb)}_{NJI}B^{(cad)}_{NLK}\right\rangle\notag\\[2pt]
&-\sum_d\sum_{K,L,N}'
\frac{\begin{Bmatrix}N&L&I\\M&J&K\end{Bmatrix}}
{\sqrt{g_ag_bg_cg_d}\,C_0(KLM)C_0(ILN)C_0(JKN)}\notag\\[-2pt]
&\quad\times\left\langle
B^{(abc)}_{IJM}B^{(cda)}_{KLM}
B^{(dba)}_{NIL}B^{(bdc)}_{NKJ}\right\rangle ,
\label{eq:weighted-sd}
\end{align}
where, more precisely, the second term on the right-hand should be taken to mean the symmetrized expression, i.e., 
\begin{align}
   &
\frac{\delta(P_M-P_N)}
{g_ag_c\rho_0(M)C_0(KLM)}\left\langle
B^{(abc)}_{IJM}B^{(cda)}_{KLM}
B^{(acb)}_{NJI}B^{(cad)}_{NLK}\right\rangle \quad  \longrightarrow \nonumber\\
&
\frac{1}{3}   
\Bigg[ \frac{\delta(P_M-P_N)}
{g_ag_c\rho_0(M)C_0(KLM)}\left\langle
B^{(abc)}_{IJM}B^{(cda)}_{KLM}
B^{(acb)}_{NJI}B^{(cad)}_{NLK}\right\rangle
+
\frac{\delta(P_I-P_N)}
{g_bg_a\rho_0(I)C_0(KLI)}\left\langle
B^{(bca)}_{MIJ}B^{(adb)}_{KLI}
B^{(bac)}_{NMJ}B^{(cad)}_{NLK}\right\rangle\nonumber\\
&\quad +
\frac{\delta(P_J-P_N)}
{g_cg_b\rho_0(J)C_0(KLJ)}\left\langle
B^{(cab)}_{JMI}B^{(bdc)}_{KLJ}
B^{(cba)}_{NIM}B^{(cad)}_{NLK}\right\rangle 
\Bigg].
\end{align}
The left-hand side of \eqref{eq:weighted-sd} is the exact two-point function.  The right-hand side contains the
bare two-point function, the Pillow vectex joined to the exact four-point function, and the
\(6j\) vertex joined to the exact four-point function.

We summarized this diagrammatically in
equation~\eqref{eq:sd-picture}, repeated here:
\begin{equation*}
  \OpenBWeightedSDFigure
  \tag{\ref*{eq:sd-picture}}
\end{equation*}
The diagrams are not necessarily thought of as three-dimensional objects as far as the matrix--tensor model is concerned. Nevertheless, it is useful to identify them with three-dimensional manifolds (with Wilson lines), as the vertex weights match those of VTQFT amplitudes. From now on, these diagrams will be thought of as three-dimensional objects and taken to represent the corresponding VTQFT amplitudes as a function of the weights of the Wilson lines. The benefit of doing so will be come clear momentarily. 

\subsection{Projected SD equation and the cylinder count}
\label{app:topology-projection}

As explained in the introduction, the exact SD equation can be checked manifold by manifold. In other words, pick a particular contribution to the exact two-point function, and project both sides of the SD equation onto it (formally). On the right-hand, this singles out a subset of manifolds contributing to the exact four-point functions. We then check the conjecture that each manifold contributes exactly once in this sector.

Let us label an individual manifold in the exact four-point function by \(p\)
for the Pillow term and by \(q\) for the \(6j\) term. Importantly, they may not (and in fact should not) be the same.
Figure~\ref{fig:operator-port-key} demonstrates how these completions attach.  
\begin{figure}[htbp]
\centering
\begin{tikzpicture}[x=.90cm,y=.92cm]
  \tikzset{
    portdot/.style={circle,fill=RoyalBlue,draw=white,line width=.25pt,
      minimum size=2.9pt,inner sep=0pt},
    portlabel/.style={font=\scriptsize,text=MidnightBlue,
      fill=white,fill opacity=.9,text opacity=1,inner sep=.45pt},
    operatorlabel/.style={font=\small,text=black,
      fill=CZeroSurfaceFill,fill opacity=.95,text opacity=1,inner sep=1pt},
    keysheet/.style={draw=CZeroSurfaceEdge,fill=CZeroSurfaceFill,
      line width=.65pt,line cap=round,line join=round},
    keyseamfront/.style={draw=CZeroSurfaceEdge,line width=.65pt},
    keyseamback/.style={draw=CZeroSurfaceEdge,line width=.7pt,
      dash pattern=on .15pt off 1.7pt,line cap=round},
    keyfront/.style={draw=black!88,line width=1.02pt,
      line cap=round,line join=round},
    keyback/.style={draw=black!88,line width=1.02pt,
      dash pattern=on 3.2pt off 2pt,line cap=round,line join=round},
    keywire/.style={draw=black!78,line width=.82pt,line cap=round},
    keyblob/.style={draw=black!46,fill=black!38,line width=.55pt}
  }

  \def\keycompletionblock#1{%
    \path[keyblob]
      (1.00,-1.66)
      ..controls(.83,-1.29)and(.89,-.88)..(1.00,-.58)
      ..controls(1.11,-.27)and(.93,.06)..(1.06,.36)
      ..controls(1.18,.67)and(.84,1.08)..(1.05,1.46)
      ..controls(1.24,1.78)and(1.68,1.73)..(1.86,1.48)
      ..controls(2.11,1.15)and(1.96,.82)..(2.09,.50)
      ..controls(2.22,.18)and(2.21,-.17)..(2.08,-.50)
      ..controls(1.95,-.83)and(2.11,-1.17)..(1.86,-1.45)
      ..controls(1.64,-1.70)and(1.19,-1.82)..(1.00,-1.66)--cycle;
    \node[font=\small] at (1.52,0) {#1};
    \foreach \yy/\ii in {.28/1,0/2,-.28/3}{
      \draw[keywire] (2.10,\yy)--(2.70,\yy);
      \node[portdot] at (2.70,\yy) {};
      \node[portlabel,anchor=west] at (2.79,\yy) {$t_{\ii}$};
    }
  }

  \begin{scope}[xshift=-3.85cm]
    \node[font=\small\bfseries] at (-.42,2.10)
      {(a) Pillow with completion $p$};
    \path[keysheet]
      (-3.35,.48)..controls(-2.63,.56)and(-2.48,1.63)..(-1.42,1.63)
      --(0,1.63)
      arc[start angle=90,end angle=270,x radius=.16,y radius=.38]
      ..controls(-.73,.78)and(-.73,.55)..(0,.46)
      arc[start angle=90,end angle=270,x radius=.16,y radius=.46]
      ..controls(-.73,-.55)and(-.73,-.78)..(0,-.87)
      arc[start angle=90,end angle=270,x radius=.16,y radius=.38]
      --(-1.42,-1.63)
      ..controls(-2.48,-1.63)and(-2.63,-.56)..(-3.35,-.48)--cycle;
    \draw[keysheet,fill=white] (-3.35,0)
      ellipse[x radius=.15,y radius=.48];
    \foreach \yc/\ry in {1.25/.38,0/.46,-1.25/.38}{
      \path[fill=CZeroSurfaceFill,draw=none]
        (0,\yc) ellipse[x radius=.16,y radius=\ry];
      \draw[keyseamback] (0,{\yc+\ry})
        arc[start angle=90,end angle=270,x radius=.16,y radius=\ry];
      \draw[keyseamfront] (0,{\yc-\ry})
        arc[start angle=-90,end angle=90,x radius=.16,y radius=\ry];
    }
    \coordinate (ps1) at (-3.20,.28);
    \coordinate (ps2) at (-3.20,0);
    \coordinate (ps3) at (-3.20,-.28);
    \coordinate (pa1) at (.13,1.48);
    \coordinate (pa2) at (.16,1.30);
    \coordinate (pa3) at (.13,1.02);
    \coordinate (pb1) at (.135,.22);
    \coordinate (pb2) at (.16,0);
    \coordinate (pb3) at (.135,-.22);
    \coordinate (pc1) at (.13,-1.02);
    \coordinate (pc2) at (.16,-1.30);
    \coordinate (pc3) at (.13,-1.48);
    \draw[keyfront] (ps1)..controls(-2.48,.34)and(-2.25,1.45)
      ..(-.34,1.48)--(pa1);
    \draw[keyfront] (ps2)..controls(-2.42,.08)and(-2.16,1.24)
      ..(-.34,1.30)--(pa2);
    \draw[keyfront] (ps3)..controls(-2.48,-.34)and(-2.25,-1.45)
      ..(-.34,-1.48)--(pc3);
    \draw[keyfront] (pa3)--(-.34,1.08)
      ..controls(-1.42,1.08)and(-1.48,.36)..(-.34,.25)--(pb1);
    \draw[keyfront] (pb2)--(-.34,0)
      ..controls(-1.53,-.12)and(-1.62,-1.15)..(-.34,-1.30)--(pc2);
    \draw[keyfront] (pb3)--(-.34,-.25)
      ..controls(-1.38,-.39)and(-1.34,-1.06)..(-.34,-1.08)--(pc1);
    \foreach \q/\yy in
      {pa1/1.48,pa2/1.30,pa3/1.02,pb1/.22,pb2/0,pb3/-.22,
       pc1/-1.02,pc2/-1.30,pc3/-1.48}
      \draw[keywire] (\q)--(1.03,\yy);
    \keycompletionblock{$p$}
    \foreach \q in
      {ps1,ps2,ps3,pa1,pa2,pa3,pb1,pb2,pb3,pc1,pc2,pc3}
      \node[portdot] at (\q) {};
    \node[portlabel,anchor=east] at (-3.46,.28) {$s_1$};
    \node[portlabel,anchor=east] at (-3.46,0) {$s_2$};
    \node[portlabel,anchor=east] at (-3.46,-.28) {$s_3$};
    \node[portlabel,anchor=west] at (.25,1.48) {$a_1$};
    \node[portlabel,anchor=west] at (.25,1.30) {$a_2$};
    \node[portlabel,anchor=west] at (.25,1.02) {$a_3$};
    \node[portlabel,anchor=west] at (.25,.22) {$b_1$};
    \node[portlabel,anchor=west] at (.25,0) {$b_2$};
    \node[portlabel,anchor=west] at (.25,-.22) {$b_3$};
    \node[portlabel,anchor=west] at (.25,-1.02) {$c_1$};
    \node[portlabel,anchor=west] at (.25,-1.30) {$c_2$};
    \node[portlabel,anchor=west] at (.25,-1.48) {$c_3$};
    \node[operatorlabel] at (-1.66,1.22) {$I$};
    \node[operatorlabel] at (-.95,1.30) {$J$};
    \node[operatorlabel] at (-.95,-1.48) {$M$};
    \node[operatorlabel] at (-.90,.99) {$N$};
    \node[operatorlabel] at (-1.27,-.62) {$K$};
    \node[operatorlabel] at (-.67,-.32) {$L$};
  \end{scope}

  \begin{scope}[xshift=3.85cm]
    \node[font=\small\bfseries] at (-.42,2.10)
      {(b) $6j$ with completion $q$};
    \path[keysheet]
      (-3.35,.48)..controls(-2.63,.56)and(-2.48,1.63)..(-1.42,1.63)
      --(0,1.63)
      arc[start angle=90,end angle=270,x radius=.16,y radius=.38]
      ..controls(-.73,.78)and(-.73,.55)..(0,.46)
      arc[start angle=90,end angle=270,x radius=.16,y radius=.46]
      ..controls(-.73,-.55)and(-.73,-.78)..(0,-.87)
      arc[start angle=90,end angle=270,x radius=.16,y radius=.38]
      --(-1.42,-1.63)
      ..controls(-2.48,-1.63)and(-2.63,-.56)..(-3.35,-.48)--cycle;
    \draw[keysheet,fill=white] (-3.35,0)
      ellipse[x radius=.15,y radius=.48];
    \foreach \yc/\ry in {1.25/.38,0/.46,-1.25/.38}{
      \path[fill=CZeroSurfaceFill,draw=none]
        (0,\yc) ellipse[x radius=.16,y radius=\ry];
      \draw[keyseamback] (0,{\yc+\ry})
        arc[start angle=90,end angle=270,x radius=.16,y radius=\ry];
      \draw[keyseamfront] (0,{\yc-\ry})
        arc[start angle=-90,end angle=90,x radius=.16,y radius=\ry];
    }
    \coordinate (js1) at (-3.20,.28);
    \coordinate (js2) at (-3.20,0);
    \coordinate (js3) at (-3.20,-.28);
    \coordinate (ja1) at (.13,1.48);
    \coordinate (ja3) at (-.16,1.30);
    \coordinate (ja2) at (.13,1.02);
    \coordinate (jb1) at (.135,.22);
    \coordinate (jb2) at (.16,0);
    \coordinate (jb3) at (.135,-.22);
    \coordinate (jc2) at (.13,-1.02);
    \coordinate (jc1) at (-.16,-1.30);
    \coordinate (jc3) at (.13,-1.48);
    \draw[keyback] (ja3)--(-.34,1.30)
      ..controls(-1.92,1.39)and(-1.92,-1.39)..(-.34,-1.30)--(jc1);
    \draw[keyfront] (js1)..controls(-2.48,.34)and(-2.25,1.45)
      ..(-.34,1.48)--(ja1);
    \draw[keyfront] (js2)--(jb2);
    \draw[keyfront] (js3)..controls(-2.48,-.34)and(-2.25,-1.45)
      ..(-.34,-1.48)--(jc3);
    \draw[keyfront] (ja2)--(-.34,1.08)
      ..controls(-1.42,1.08)and(-1.48,.36)..(-.34,.25)--(jb1);
    \draw[keyfront] (jb3)--(-.34,-.25)
      ..controls(-1.48,-.36)and(-1.42,-1.08)..(-.34,-1.08)--(jc2);
    \foreach \q/\yy in
      {ja1/1.48,ja3/1.30,ja2/1.02,jb1/.22,jb2/0,jb3/-.22,
       jc2/-1.02,jc1/-1.30,jc3/-1.48}
      \draw[keywire] (\q)--(1.03,\yy);
    \keycompletionblock{$q$}
    \foreach \q in
      {js1,js2,js3,ja1,ja2,ja3,jb1,jb2,jb3,jc1,jc2,jc3}
      \node[portdot] at (\q) {};
    \node[portlabel,anchor=east] at (-3.46,.28) {$s_1$};
    \node[portlabel,anchor=east] at (-3.46,0) {$s_2$};
    \node[portlabel,anchor=east] at (-3.46,-.28) {$s_3$};
    \node[portlabel,anchor=west] at (.25,1.48) {$a_1$};
    \node[portlabel,anchor=west] at (.25,1.30) {$a_3$};
    \node[portlabel,anchor=west] at (.25,1.02) {$a_2$};
    \node[portlabel,anchor=west] at (.25,.22) {$b_1$};
    \node[portlabel,anchor=west] at (.25,0) {$b_2$};
    \node[portlabel,anchor=west] at (.25,-.22) {$b_3$};
    \node[portlabel,anchor=west] at (.25,-1.02) {$c_2$};
    \node[portlabel,anchor=west] at (.25,-1.30) {$c_1$};
    \node[portlabel,anchor=west] at (.25,-1.48) {$c_3$};
    \node[operatorlabel] at (-1.66,1.22) {$I$};
    \node[operatorlabel] at (-2.35,0) {$J$};
    \node[operatorlabel] at (-.95,-1.48) {$M$};
    \node[operatorlabel] at (-.90,.99) {$N$};
    \node[operatorlabel] at (-.90,-.99) {$K$};
    \node[operatorlabel] at (-1.42,.62) {$L$};
  \end{scope}
\end{tikzpicture}
\caption{Figure~\ref{fig:code-to-surface-key} but with labels matching equation~\eqref{eq:weighted-sd}. The capital letters are Wilson line labels, which connect operators with the corresponding labels. The port labels ($s_i,a_i,b_i$...) are used for coding purposes and specify where each line attaches.}
\label{fig:operator-port-key}
\end{figure}

To emphasize again, the prefactors multiplying the two \(\langle BBBB\rangle\) correlators in
equation~\eqref{eq:weighted-sd} are precisely the analytic weights of the
fixed Pillow and \(6j\) pieces in Figure~\ref{fig:operator-port-key}.
Each correlator supplies the gray completion. The prefactors in the SD equation~\eqref{eq:sd-sector} are therefore already included in the diagrammatic representations, except the \(+\) and \(-\) signs.  The product of the prefactor
and the correlator, summed over the internal labels, gives the resulting function on the right-hand side of the SD equation. 

The conjecture of \cite{Jafferis:2025vyp} states that in the \(\hbar\to0\) limit, each manifold should contribute with weight one. In this limit, the bare term \(-\hbar C_0(IJM)\) vanishes, so the SD equation simplifies further. 

The simplest contribution to the left-hand side of the SD equation is the cylinder manifold \(C_0\). On the right-hand side, we look for candidate contributions to the exact four-point function such that gluing it to the quartic vertex produces the $C_0$ manifold, after performing the annular surgery for each Wilson loop that forms. The reason behind the surgery is provided by the potential $V_0$, which provides the leading spectral density of the dual matrix model \cite{Jafferis:2025jle}. This is reviewed in Appendix~\ref{app:open-surgery}. If the cylinder diagram is produced, we retain the candidate manifold. Define \(\NP\) and \(\Nsix\) to be the valid Pillow and \(6j\)
completions we retain, respectively.

For this particular choice, the projected SD equation was presented in 
equation~\eqref{eq:sd-sector}, repeated here:
\begin{equation*}
  \OpenBTopologyProjectionFigure
  \tag{\ref*{eq:sd-sector}}
\end{equation*}
Here \(1\) counts the selected cylinder, not the bare contact term which has been discarded due to the $\hbar\to0$ limit we have taken.

\section{Counting prescription}
\label{app:no-onepoint}

A main point of this work is to understand when two contributions to the exact four-point correlator should be considered equivalent and when they should not be. Before listing all possible candidate contributions in a later section, we will first address this question by providing a prescription motivated by VTQFT.

As we will see, this prescription selects only one
Pillow contribution and no \(6j\) contribution, yielding
\(\NP-\Nsix=1-0=1\), in agreement with the SD prediction.

The prescription is the following. We propose that we include only manifolds with no \emph{one-point} or \emph{two-point cuts}: VTQFT diagrams that do not have any cross-section that contains a disk with fewer than three boundary operators inserted.\footnote{In the SD equation projected onto a hyperbolic manifold, manifolds admitting zero-point disk cuts do not appear, so we remain agnostic about their treatment in this work.} Figure~\ref{fig:one-two-point-cuts} show the cases with one and two points.

\SDOneTwoPointCutsFigure{}

Before providing evidence for why they should be excluded in the SD equation, let us comment that an intuitive reason for the prescription is related to the Virasoro conformal block that underlies
VTQFT.  In the \(C_0\) normalization used here, taking one entry of the
three-point function to the identity gives~\cite{Collier:2023fwi,Jafferis:2025yxt}
\begin{equation}
  C_0(MN\mathds{1})
  =\frac{\delta(P_M-P_N)}{\rho_0(P_M)},
  \qquad P_M,P_N>0.
  \label{eq:vtqft-two-point-selection}
\end{equation}
The evaluation of $C_0$ with an identity entry gives a distributional limit, not an ordinary value of the function.\footnote{This is also related to the fact that such cross-sections make the manifold pseudo-hyperbolic, rather than hyperbolic, which nevertheless play an important role in other proposed dual descriptions of three-dimensional gravity \cite{Wang:2025jgo,Belin:2026pko}.}

This two-point function pairs only equal representations.
Replacing one of the two insertions by the identity leaves a one-point
function, which vanishes for a nonidentity primary by conformal
invariance.\footnote{Setting \(M\) to the identity means replacing its
insertion by the identity operator.  The resulting one-point function
vanishes for a nonidentity primary; this is not obtained by substituting
into the real-momentum delta function.}

The full list before removing the disallowed contributions will be presented in Sections~\ref{sec:enumeration}
and~\ref{sec:embeddings}, but we will preview some of them here to demonstrate the effect of the prescription.

\subsection{One-point cuts}

A one-point cut forces the label on its
single Wilson line to be the identity.  If this is an internal label,
rather than one of the fixed external labels \(I,J,M\), setting it to the
identity removes two factors of \(B\), since the index occurs
twice.\footnote{For this discussion we set all brane labels equal and omit
them; the argument is the same with those labels restored.}
The completion then has only two boundaries and contributes to the
two-point moment.
It is therefore either already accounted for on the left-hand side of
equation~\eqref{eq:weighted-sd}, or does not belong to the chosen
\(C_0\) contribution.  In either case, the prescription does not count it
as a new four-point completion.

For example, consider 6j-3 (we are using labels for the complete list that appears in later sections), shown with operator labels in
Figure~\ref{fig:sixj-raw10-one-point}.
\SDSixJOnePointReductionFigure{}
Its one-point cut sets \(K\) to the identity, reducing the right half
to a two-point contribution. This does not glue into \(\langle B_{IJM}B_{JIM}\rangle\) to give the
chosen \(C_0\) diagram, so it does not contribute.  The other one-point-cut
diagrams are omitted by the same proposed reasoning.

In this proposed one-point classification, only P-5, P-6, P-7-1, and P-10 survive;
no \(6j\) diagrams remain:
\begin{equation}
\begin{array}{c|rrrr|r}
\toprule
&C_2&C_3&D_1&D_2&\text{total}\\
\midrule
\sixj&0&0&0&0&0\\
\text{Pillow}&2&1&1&0&4\\
\bottomrule
\end{array}
\label{eq:no-onepoint-count}
\end{equation}
where \(C_2,C_3\) mean connected completions with two or three
closed Wilson loops, respectively (for now, we just treat these groups as a convenient way to organize the counting; we will explain the meaning of these classes in later sections); \(D_1,D_2\) mean completions split into
two pieces, with one or two loops, respectively.  Connected or disconnected
refers to the EOW carrying the Wilson lines; loops are counted after gluing it
to interaction vertex.

Figure~\ref{fig:no-onepoint-survivors} displays all four remaining candidates.

\begin{figure}[!ht]
  \centering
  \SDPillowCutSurvivorGallery
  \caption{P-5, P-6, P-7-1, and P-10, the four remaining diagrams after the
  proposed one-point-cut exclusion, with operator labels.}
  \label{fig:no-onepoint-survivors}
\end{figure}

\subsection{Two-point cuts}

We now compare the four surviving Pillow candidates. For a two-point cut, 
equation~\eqref{eq:vtqft-two-point-selection} enforces that the momenta of the Wilson lines passing though the cut be equal.  We take the external
momenta \(P_I,P_J,P_M\) to be pairwise distinct, so a cut requiring two
of them to be equal gives zero.

Let us start with the disconnected completion P-10, shown again in
Figure~\ref{fig:disconnected-pillow-two-point}.  Its two green annuli
already require \(L=I\) and \(K=J\), without an extra cut condition. This contribution is picked out of the sum over possible completions in the SD by a Kronecker-delta function on the weights, since we are implicitly treating the sum over weights as continuous. 

\SDDisconnectedPillowCutFigure{}

Connected completion P-5 already has \(K=J\) from its green connections.  Its
two-point cut crosses the \(I\) and \(L\) lines, requiring \(P_L=P_I\).
Likewise, P-6 already has \(L=I\), and its cut through the \(J\) and
\(K\) lines requires \(P_K=P_J\). Note that these equalities are enforced via a Dirac-delta function, since the delta function part of the definition of a diagram with a two-point cut.

P-7-1 has no new identification from its green connections.  It has two
two-point cuts: one crosses \(I,L\) and requires \(P_L=P_I\), while the
other crosses \(J,K\) and requires \(P_K=P_J\).  At the level of the code
graph, cutting \(I,L\) first gives P-6, while cutting \(J,K\) first gives
P-5.  The second cut then gives P-10 in either order.

Table~\ref{tab:pillow-cut-identifications} collects these conditions.
\begin{table}[H]
  \centering
  \begin{tabular}{@{}ccc@{}}
    \toprule
    Pillow & From the green connections & From the two-point cut \\
    \midrule
    P-5 (connected) & \(K=J\) & \(P_L=P_I\) \\
    P-6 (connected) & \(L=I\) & \(P_K=P_J\) \\
    P-7-1 (connected) & none & \(P_L=P_I,\ P_K=P_J\) \\
    P-10 (disconnected) & \(L=I,\ K=J\) & none needed \\
    \bottomrule
  \end{tabular}
  \caption{Label identifications for the four surviving Pillow completions.}
  \label{tab:pillow-cut-identifications}
\end{table}
All four therefore obey the same momentum constraints:
\(P_L=P_I\) and \(P_K=P_J\). 

Although we see from the table above that the index structure of the four candidates is the same, there is a discrepancy between having continuous and discrete delta functions in the connected and disconnected candidates respectively. Other than this subtlety, the functions corresponding to each diagram are equivalent.

However, the delta functions are in fact supposed to be viewed on the same footing for the following reason. In the sum over manifolds, we are doing a discrete sum and implicitly assuming that the weights are also being discretely summed over, but we should really treat them continuously as we do in the gravity computations. Really, the sum we see in SD is a regulated integral, and so is the integral in the definition of the crossing constraint. Hence, both delta functions are really the same regulated delta functions, and the four diagrams are the same contribution. 
The same argument can be applied to any connected diagram with a 2-point cut: there is always a corresponding disconnected contribution which already accounts for it. Hence, we propose to omit the diagrams with 2-point cuts.

For this reason, P-5, P-6, and P-7-1 are already
represented by P-10, so only the latter needs to be counted, giving
\begin{equation}
  \Nsix=0,\qquad \NP=1,
  \qquad \NP-\Nsix=1.
  \label{eq:cut-reduced-count}
\end{equation}
The following sections construct the unreduced list of candidate contributions.

\section{From code candidates to surface realizations}
\label{sec:enumeration}

We now explain how we find the diagrams to which the cut rules of
Section~\ref{app:no-onepoint} are applied.  The aim is to complete the fixed
Pillow and \(6j\) pieces so that gluing and surgery give the target
\(C_0\) cylinder.  We build the full list in this section and
Section~\ref{sec:embeddings}, before applying for prescription.

There are two steps here.  Section~\ref{sec:graph-screen} uses a python code to
list possible connections and select candidates.  Section~\ref{sec:surface-audit}
then checks by hand which candidates can be drawn on the required surface.
Each pattern of connections counts once at this stage, even if it has
several surface embeddings.  Section~\ref{sec:embeddings} will determine how
many of those embeddings count separately.

\subsection{Enumerating candidate connections}
\label{sec:graph-screen}

Figure~\ref{fig:code-to-surface-key}
shows the Pillow and $6j$ completions with labeled end points we call ports. The three \(s_i\) are the fixed external ports on the left, the three \(t_i\)
are the external ports on the right, and \(a_i,b_i,c_i\) are the attachment
points on the three middle cuffs. The full port list and fixed
edge sets are recorded in Appendix~\ref{app:algorithm}.

\begin{figure}[htbp]
\centering
\begin{tikzpicture}[x=.90cm,y=.92cm]
  \tikzset{
    portdot/.style={circle,fill=RoyalBlue,draw=white,line width=.25pt,
      minimum size=2.9pt,inner sep=0pt},
    portlabel/.style={font=\scriptsize,text=MidnightBlue,
      fill=white,fill opacity=.9,text opacity=1,inner sep=.45pt},
    keysheet/.style={draw=CZeroSurfaceEdge,fill=CZeroSurfaceFill,
      line width=.65pt,line cap=round,line join=round},
    keyseamfront/.style={draw=CZeroSurfaceEdge,line width=.65pt},
    keyseamback/.style={draw=CZeroSurfaceEdge,line width=.7pt,
      dash pattern=on .15pt off 1.7pt,line cap=round},
    keyfront/.style={draw=black!88,line width=1.02pt,
      line cap=round,line join=round},
    keyback/.style={draw=black!88,line width=1.02pt,
      dash pattern=on 3.2pt off 2pt,line cap=round,line join=round},
    keywire/.style={draw=black!78,line width=.82pt,line cap=round},
    keyblob/.style={draw=black!46,fill=black!38,line width=.55pt}
  }

  \def\keycompletionblock#1{%
    \path[keyblob]
      (1.00,-1.66)
      ..controls(.83,-1.29)and(.89,-.88)..(1.00,-.58)
      ..controls(1.11,-.27)and(.93,.06)..(1.06,.36)
      ..controls(1.18,.67)and(.84,1.08)..(1.05,1.46)
      ..controls(1.24,1.78)and(1.68,1.73)..(1.86,1.48)
      ..controls(2.11,1.15)and(1.96,.82)..(2.09,.50)
      ..controls(2.22,.18)and(2.21,-.17)..(2.08,-.50)
      ..controls(1.95,-.83)and(2.11,-1.17)..(1.86,-1.45)
      ..controls(1.64,-1.70)and(1.19,-1.82)..(1.00,-1.66)--cycle;
    \node[font=\small] at (1.52,0) {#1};
    \foreach \yy/\ii in {.28/1,0/2,-.28/3}{
      \draw[keywire] (2.10,\yy)--(2.70,\yy);
      \node[portdot] at (2.70,\yy) {};
      \node[portlabel,anchor=west] at (2.79,\yy) {$t_{\ii}$};
    }
  }

  \begin{scope}[xshift=-3.85cm]
    \node[font=\small\bfseries] at (-.42,2.10)
      {(a) Pillow with completion $p$};
    \path[keysheet]
      (-3.35,.48)..controls(-2.63,.56)and(-2.48,1.63)..(-1.42,1.63)
      --(0,1.63)
      arc[start angle=90,end angle=270,x radius=.16,y radius=.38]
      ..controls(-.73,.78)and(-.73,.55)..(0,.46)
      arc[start angle=90,end angle=270,x radius=.16,y radius=.46]
      ..controls(-.73,-.55)and(-.73,-.78)..(0,-.87)
      arc[start angle=90,end angle=270,x radius=.16,y radius=.38]
      --(-1.42,-1.63)
      ..controls(-2.48,-1.63)and(-2.63,-.56)..(-3.35,-.48)--cycle;
    \draw[keysheet,fill=white] (-3.35,0)
      ellipse[x radius=.15,y radius=.48];
    \foreach \yc/\ry in {1.25/.38,0/.46,-1.25/.38}{
      \path[fill=CZeroSurfaceFill,draw=none]
        (0,\yc) ellipse[x radius=.16,y radius=\ry];
      \draw[keyseamback] (0,{\yc+\ry})
        arc[start angle=90,end angle=270,x radius=.16,y radius=\ry];
      \draw[keyseamfront] (0,{\yc-\ry})
        arc[start angle=-90,end angle=90,x radius=.16,y radius=\ry];
    }
    \coordinate (ps1) at (-3.20,.28);
    \coordinate (ps2) at (-3.20,0);
    \coordinate (ps3) at (-3.20,-.28);
    \coordinate (pa1) at (.13,1.48);
    \coordinate (pa2) at (.16,1.30);
    \coordinate (pa3) at (.13,1.02);
    \coordinate (pb1) at (.135,.22);
    \coordinate (pb2) at (.16,0);
    \coordinate (pb3) at (.135,-.22);
    \coordinate (pc1) at (.13,-1.02);
    \coordinate (pc2) at (.16,-1.30);
    \coordinate (pc3) at (.13,-1.48);
    \draw[keyfront] (ps1)..controls(-2.48,.34)and(-2.25,1.45)
      ..(-.34,1.48)--(pa1);
    \draw[keyfront] (ps2)..controls(-2.42,.08)and(-2.16,1.24)
      ..(-.34,1.30)--(pa2);
    \draw[keyfront] (ps3)..controls(-2.48,-.34)and(-2.25,-1.45)
      ..(-.34,-1.48)--(pc3);
    \draw[keyfront] (pa3)--(-.34,1.08)
      ..controls(-1.42,1.08)and(-1.48,.36)..(-.34,.25)--(pb1);
    \draw[keyfront] (pb2)--(-.34,0)
      ..controls(-1.53,-.12)and(-1.62,-1.15)..(-.34,-1.30)--(pc2);
    \draw[keyfront] (pb3)--(-.34,-.25)
      ..controls(-1.38,-.39)and(-1.34,-1.06)..(-.34,-1.08)--(pc1);
    \foreach \q/\yy in
      {pa1/1.48,pa2/1.30,pa3/1.02,pb1/.22,pb2/0,pb3/-.22,
       pc1/-1.02,pc2/-1.30,pc3/-1.48}
      \draw[keywire] (\q)--(1.03,\yy);
    \keycompletionblock{$p$}
    \foreach \q in
      {ps1,ps2,ps3,pa1,pa2,pa3,pb1,pb2,pb3,pc1,pc2,pc3}
      \node[portdot] at (\q) {};
    \node[portlabel,anchor=east] at (-3.46,.28) {$s_1$};
    \node[portlabel,anchor=east] at (-3.46,0) {$s_2$};
    \node[portlabel,anchor=east] at (-3.46,-.28) {$s_3$};
    \node[portlabel,anchor=west] at (.25,1.48) {$a_1$};
    \node[portlabel,anchor=west] at (.25,1.30) {$a_2$};
    \node[portlabel,anchor=west] at (.25,1.02) {$a_3$};
    \node[portlabel,anchor=west] at (.25,.22) {$b_1$};
    \node[portlabel,anchor=west] at (.25,0) {$b_2$};
    \node[portlabel,anchor=west] at (.25,-.22) {$b_3$};
    \node[portlabel,anchor=west] at (.25,-1.02) {$c_1$};
    \node[portlabel,anchor=west] at (.25,-1.30) {$c_2$};
    \node[portlabel,anchor=west] at (.25,-1.48) {$c_3$};
  \end{scope}

  \begin{scope}[xshift=3.85cm]
    \node[font=\small\bfseries] at (-.42,2.10)
      {(b) $6j$ with completion $q$};
    \path[keysheet]
      (-3.35,.48)..controls(-2.63,.56)and(-2.48,1.63)..(-1.42,1.63)
      --(0,1.63)
      arc[start angle=90,end angle=270,x radius=.16,y radius=.38]
      ..controls(-.73,.78)and(-.73,.55)..(0,.46)
      arc[start angle=90,end angle=270,x radius=.16,y radius=.46]
      ..controls(-.73,-.55)and(-.73,-.78)..(0,-.87)
      arc[start angle=90,end angle=270,x radius=.16,y radius=.38]
      --(-1.42,-1.63)
      ..controls(-2.48,-1.63)and(-2.63,-.56)..(-3.35,-.48)--cycle;
    \draw[keysheet,fill=white] (-3.35,0)
      ellipse[x radius=.15,y radius=.48];
    \foreach \yc/\ry in {1.25/.38,0/.46,-1.25/.38}{
      \path[fill=CZeroSurfaceFill,draw=none]
        (0,\yc) ellipse[x radius=.16,y radius=\ry];
      \draw[keyseamback] (0,{\yc+\ry})
        arc[start angle=90,end angle=270,x radius=.16,y radius=\ry];
      \draw[keyseamfront] (0,{\yc-\ry})
        arc[start angle=-90,end angle=90,x radius=.16,y radius=\ry];
    }
    \coordinate (js1) at (-3.20,.28);
    \coordinate (js2) at (-3.20,0);
    \coordinate (js3) at (-3.20,-.28);
    \coordinate (ja1) at (.13,1.48);
    \coordinate (ja3) at (-.16,1.30);
    \coordinate (ja2) at (.13,1.02);
    \coordinate (jb1) at (.135,.22);
    \coordinate (jb2) at (.16,0);
    \coordinate (jb3) at (.135,-.22);
    \coordinate (jc2) at (.13,-1.02);
    \coordinate (jc1) at (-.16,-1.30);
    \coordinate (jc3) at (.13,-1.48);
    \draw[keyback] (ja3)--(-.34,1.30)
      ..controls(-1.92,1.39)and(-1.92,-1.39)..(-.34,-1.30)--(jc1);
    \draw[keyfront] (js1)..controls(-2.48,.34)and(-2.25,1.45)
      ..(-.34,1.48)--(ja1);
    \draw[keyfront] (js2)--(jb2);
    \draw[keyfront] (js3)..controls(-2.48,-.34)and(-2.25,-1.45)
      ..(-.34,-1.48)--(jc3);
    \draw[keyfront] (ja2)--(-.34,1.08)
      ..controls(-1.42,1.08)and(-1.48,.36)..(-.34,.25)--(jb1);
    \draw[keyfront] (jb3)--(-.34,-.25)
      ..controls(-1.48,-.36)and(-1.42,-1.08)..(-.34,-1.08)--(jc2);
    \foreach \q/\yy in
      {ja1/1.48,ja3/1.30,ja2/1.02,jb1/.22,jb2/0,jb3/-.22,
       jc2/-1.02,jc1/-1.30,jc3/-1.48}
      \draw[keywire] (\q)--(1.03,\yy);
    \keycompletionblock{$q$}
    \foreach \q in
      {js1,js2,js3,ja1,ja2,ja3,jb1,jb2,jb3,jc1,jc2,jc3}
      \node[portdot] at (\q) {};
    \node[portlabel,anchor=east] at (-3.46,.28) {$s_1$};
    \node[portlabel,anchor=east] at (-3.46,0) {$s_2$};
    \node[portlabel,anchor=east] at (-3.46,-.28) {$s_3$};
    \node[portlabel,anchor=west] at (.25,1.48) {$a_1$};
    \node[portlabel,anchor=west] at (.25,1.30) {$a_3$};
    \node[portlabel,anchor=west] at (.25,1.02) {$a_2$};
    \node[portlabel,anchor=west] at (.25,.22) {$b_1$};
    \node[portlabel,anchor=west] at (.25,0) {$b_2$};
    \node[portlabel,anchor=west] at (.25,-.22) {$b_3$};
    \node[portlabel,anchor=west] at (.25,-1.02) {$c_2$};
    \node[portlabel,anchor=west] at (.25,-1.30) {$c_1$};
    \node[portlabel,anchor=west] at (.25,-1.48) {$c_3$};
  \end{scope}
\end{tikzpicture}
\caption{The two fixed inputs to the enumeration.  Black curves belong to
the fixed Pillow and \(6j\) interactions.  The gray regions \(p\) and
\(q\) denote arbitrary completions: they connect the twelve ports
\(t_i,a_i,b_i,c_i\), while the three \(s_i\) are fixed external ports.  Blue
dots mark the physical attachment points, and every attachment is made by its
label.  The dashed black segment is on the rear side of the surface.}
\label{fig:code-to-surface-key}
\end{figure}
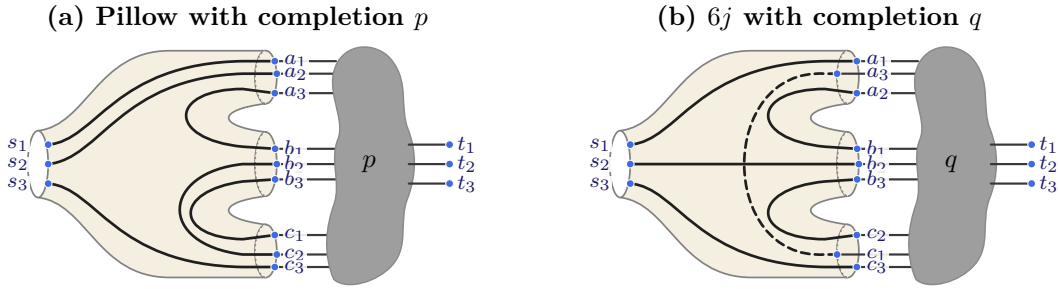

The code generates all 7560 allowed matching options, excluding direct
\(t_i\)--\(t_j\) pairs.  After adding a matching to a fixed half, it ``peels
off'' the closed graph components and checks that the remaining paths connect
\(s_i\) to \(t_i\) for each \(i\).  Exactly 693 matchings pass this test for Pillow and $6j$ each.

The remaining candidates are split into connected and disconnected completion
graphs, treating the three ports on each cuff as belonging to one vertex.
A matching is \textit{disconnected} exactly when all three \(t_i\) attach
directly to one letter family (only one of \(a,b,c\)); otherwise it is
\textit{connected}.  These names refer to the completion ($p$ or $q$), not the final
manifold after gluing.  The connected branch requires forming at least two loops,
whereas the disconnected branch requires at least one.  These conditions
are imposed because we need to remove the genus holes (two in the
connected case, one in the disconnected case) resulting from gluing.
Figure~\ref{fig:candidate-sectors} shows the separate discarding steps.
This leaves 31 \(6j\) and 33 Pillow candidates.  Closed components are
only possible surgery loops: the code does not decide whether they fit on
the required surface or close the correct handles. This latter condition is imposed at the next step.

\begin{figure}[htbp]
\centering
\begin{adjustbox}{max width=\linewidth}
\begin{tikzpicture}[node distance=6mm and 12mm]
  \node[sdstepbox,text width=3.0cm] (generate)
    {generate all 7560\\port matchings};
  \node[sdstepbox,text width=3.25cm,below=6mm of generate] (fixed)
    {add the Pillow or $6j$\\fixed edges};
  \node[sdstepbox,text width=3.45cm,below=6mm of fixed] (cycles)
    {peel and count all\\closed cycle components};
  \node[sdtest,below=7mm of cycles] (paths) {three identity\\paths remain?};
  \node[sdreject,text width=2.35cm,right=18mm of paths] (pathreject)
    {discard};
  \node[sdtest,below=8mm of paths] (specialq) {completion\\disconnected?};
  \node[sdtest,below left=11mm and 23mm of specialq] (ordinarytest)
    {at least two\\closed cycles?};
  \node[sdtest,below right=11mm and 23mm of specialq] (specialtest)
    {at least one\\closed cycle?};
  \node[sdreject,text width=1.75cm,left=11mm of ordinarytest] (ordinaryreject)
    {discard};
  \node[sdreject,text width=1.75cm,right=11mm of specialtest] (specialreject)
    {discard};
  \node[sdaccept,text width=3.5cm,below=7mm of ordinarytest] (ordinarykeep)
    {connected\\25 $6j$, 24 Pillow};
  \node[sdaccept,text width=3.5cm,below=7mm of specialtest] (specialkeep)
    {disconnected\\6 $6j$, 9 Pillow};
  \node[sdaccept,text width=3.25cm]
    (accept) at ($(ordinarykeep.south)!0.5!(specialkeep.south)+(0,-11mm)$)
    {31 $6j$, 33 Pillow\\graph candidates};
  \draw[sdarrow] (generate)--(fixed);
  \draw[sdarrow] (fixed)--(cycles);
  \draw[sdarrow] (cycles)--(paths);
  \draw[sdarrow] (paths)--node[sdbranch,above] {no} (pathreject);
  \draw[sdarrow] (paths)--node[sdbranch,left] {yes: 693 each} (specialq);
  \draw[sdarrow] (specialq)--node[sdbranch,left] {no: connected} (ordinarytest);
  \draw[sdarrow] (specialq)--node[sdbranch,right] {yes} (specialtest);
  \draw[sdarrow] (ordinarytest)--node[sdbranch,above] {no} (ordinaryreject);
  \draw[sdarrow] (specialtest)--node[sdbranch,above] {no} (specialreject);
  \draw[sdarrow] (ordinarytest)--node[sdbranch,left] {yes} (ordinarykeep);
  \draw[sdarrow] (specialtest)--node[sdbranch,right] {yes} (specialkeep);
  \draw[sdarrow] (ordinarykeep.south)--([xshift=-8mm]accept.north);
  \draw[sdarrow] (specialkeep.south)--([xshift=8mm]accept.north);
\end{tikzpicture}
\end{adjustbox}
\caption{The graph search for the cylinder calculation.  After the
identity-path test, connected and disconnected completions are tested
separately with their respective cycle thresholds.  Passing these tests
does not establish a smooth surface realization.}
\label{fig:candidate-sectors}
\end{figure}
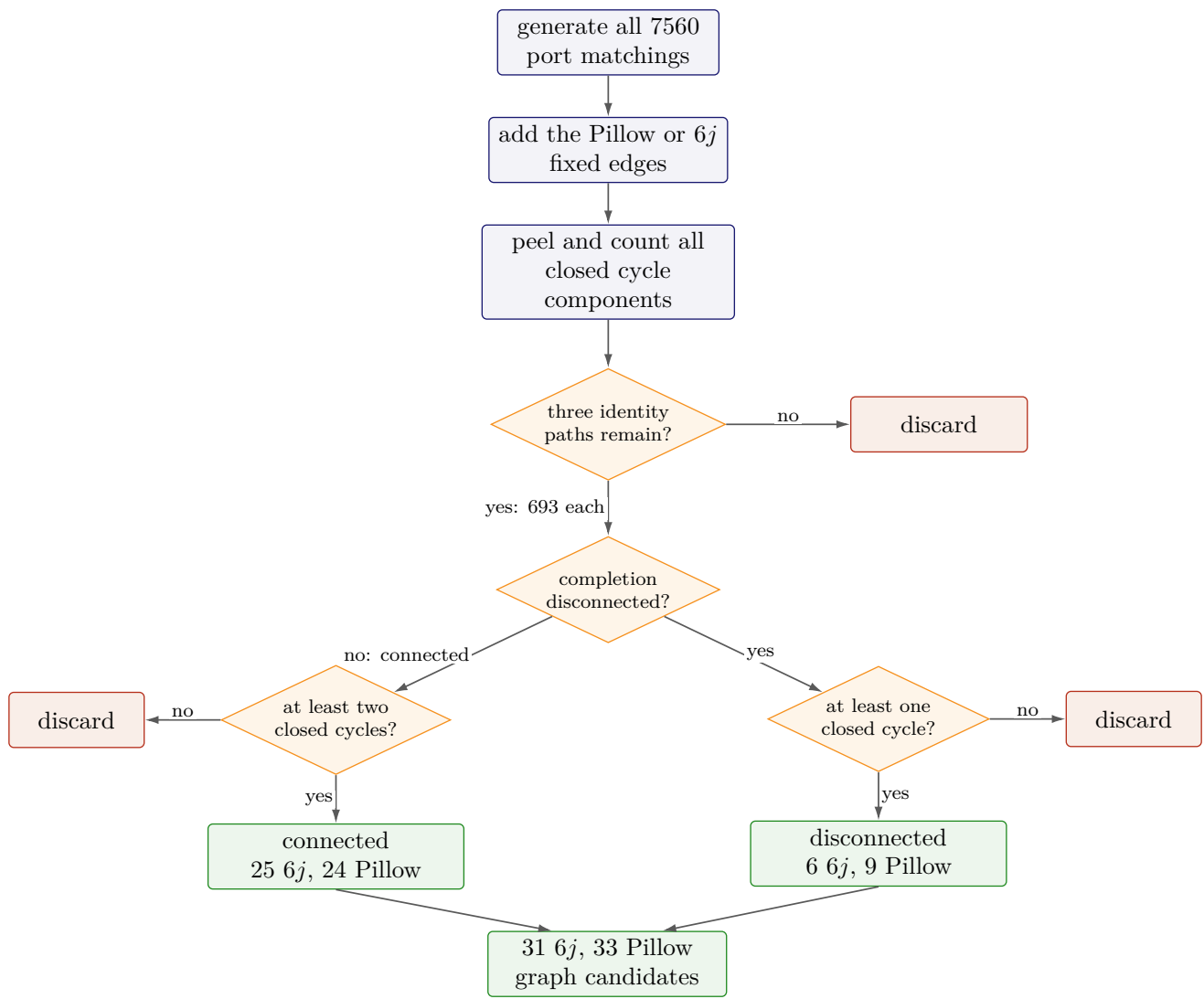

\subsection{Surface drawings of the retained code graphs}
\label{sec:surface-audit}
\label{sec:surface-parents}

We take the 31 \(6j\) and 33 Pillow code outputs and try to place their connections on surfaces.  In Figure~\ref{fig:code-to-surface-key}, the gray
regions \(p,q\) are the unknown completions and the light brown pieces with black lines are fixed.
We glue a completion along the three middle cuffs, joining equally labeled
endpoints, and then perform the annular surgery described in
Appendix~\ref{app:open-surgery}.  Gluing joins the two halves; surgery is the
subsequent operation that reduces the genus of the resulting EOW brane.  The code checks paths and loops,
but does not perform this surface construction.

We group the retained completions into four cases.  \(C\) means that the
surface carrying the green lines is connected; \(D\) means it consists of
two separate pieces.  The subscript counts closed Wilson loops after gluing
to the black half: two or three for \(C_2,C_3\) respectively, and one or two for
\(D_1,D_2\) respectively.  Importantly, the number subscript does \textit{not} correspond to the genus, and the notation \(D\) does \textit{not} mean that the final
cylinder is disconnected.  Figure~\ref{fig:surface-parent-sector-examples}
shows a code graph and one surface drawing for each case.

\begin{figure}[!ht]
\centering
\setlength{\parskip}{0pt}
\begin{adjustbox}{max width=.94\linewidth}
\begin{tabular}{@{}
  >{\centering\arraybackslash}m{.12\linewidth}
  >{\centering\arraybackslash}m{.32\linewidth}
  @{\hspace{1.5mm}\(\longrightarrow\)\hspace{1.5mm}}
  >{\centering\arraybackslash}m{.45\linewidth}@{}}
{\scriptsize\bfseries sector} &
{\scriptsize\bfseries code graph} &
{\scriptsize\bfseries one surface drawing} \\[1mm]
\shortstack{{\bfseries (a) \(C_2\)}\\{\scriptsize connected}\\{\scriptsize two-cycle}} &
\includegraphics[width=\linewidth,height=3.8cm,keepaspectratio]{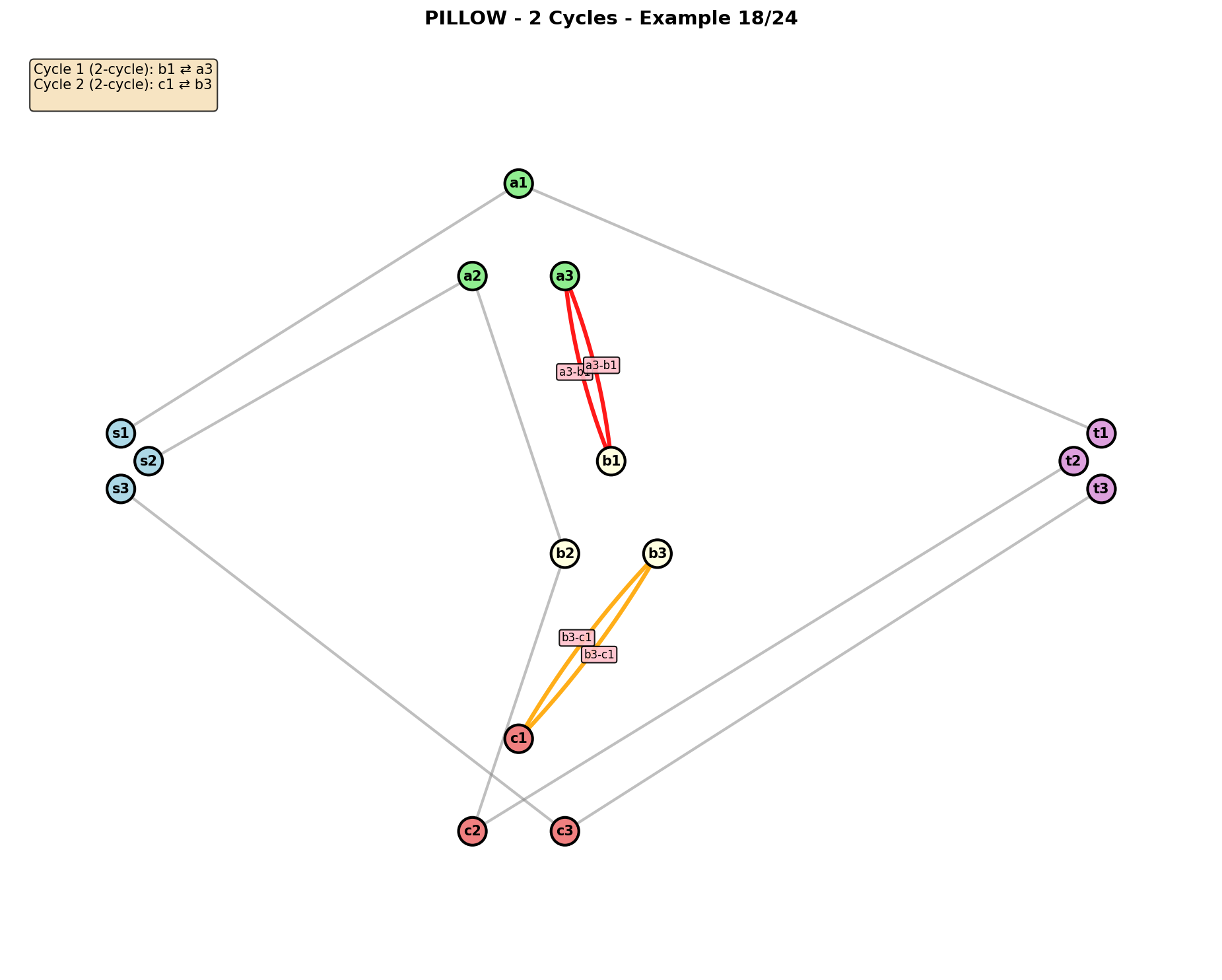}\par
{\scriptsize\textbf{P-5}; idx 726\par}
{\scriptsize\texttt{2cycles\_18.png}\par} &
\adjustbox{max width=\linewidth,max totalheight=3.1cm}{%
  \crawCard{1}{PillowPFive}{Standard}
    {P-5}{}{Hidden}} \\[1mm]
\shortstack{{\bfseries (b) \(C_3\)}\\{\scriptsize connected}\\{\scriptsize three-cycle}} &
\includegraphics[width=\linewidth,height=3.8cm,keepaspectratio]{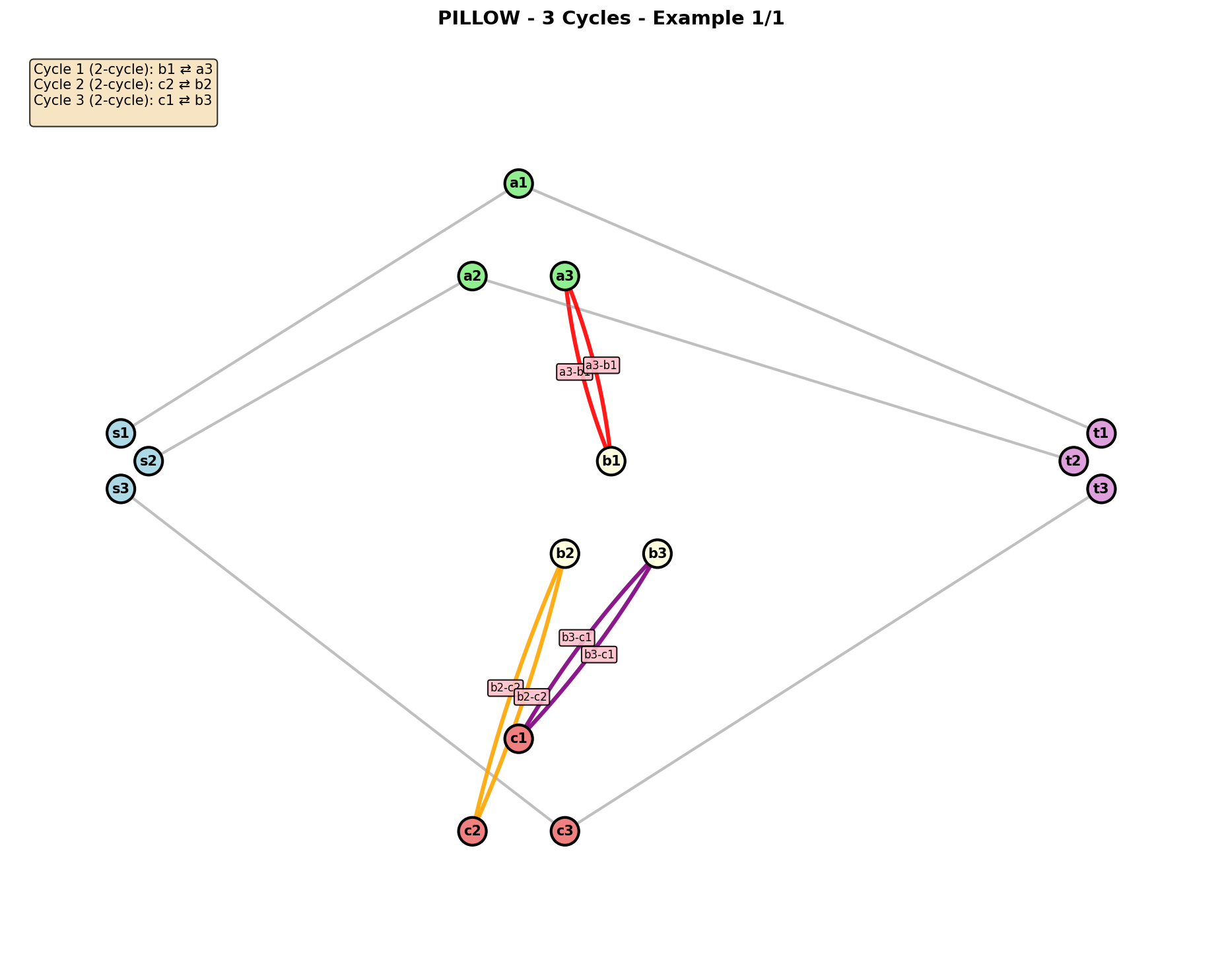}\par
{\scriptsize\textbf{P-7}; idx 92\par}
{\scriptsize\texttt{3cycles\_1.png}\par} &
\adjustbox{max width=\linewidth,max totalheight=3.1cm}{%
  \crawCard{12}{PillowColoredTwelve}{Standard}
    {P-7-3}{}{Central}} \\[1mm]
\shortstack{{\bfseries (c) \(D_1\)}\\{\scriptsize disconnected}\\{\scriptsize one-cycle}} &
\includegraphics[width=\linewidth,height=3.8cm,keepaspectratio]{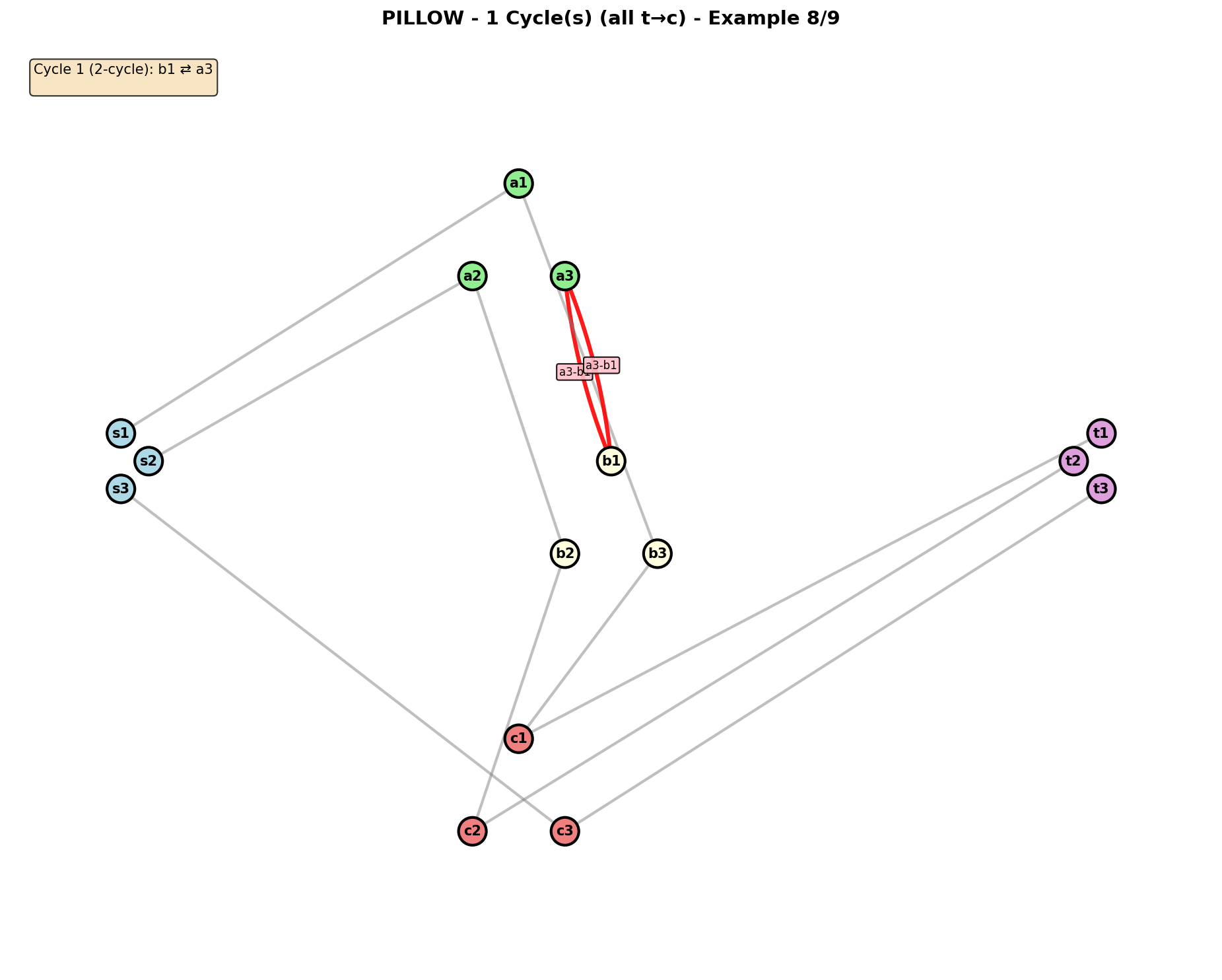}\par
{\scriptsize\textbf{P-10}; idx 5774\par}
{\scriptsize\texttt{t\_same\_1cycles\_8.png}\par} &
\adjustbox{max width=\linewidth,max totalheight=3.1cm}{%
  \crawPillowEarlyDisconnectedCard{7}{7}
    {P-10}} \\[1mm]
\shortstack{{\bfseries (d) \(D_2\)}\\{\scriptsize disconnected}\\{\scriptsize two-cycle}} &
\includegraphics[width=\linewidth,height=3.8cm,keepaspectratio]{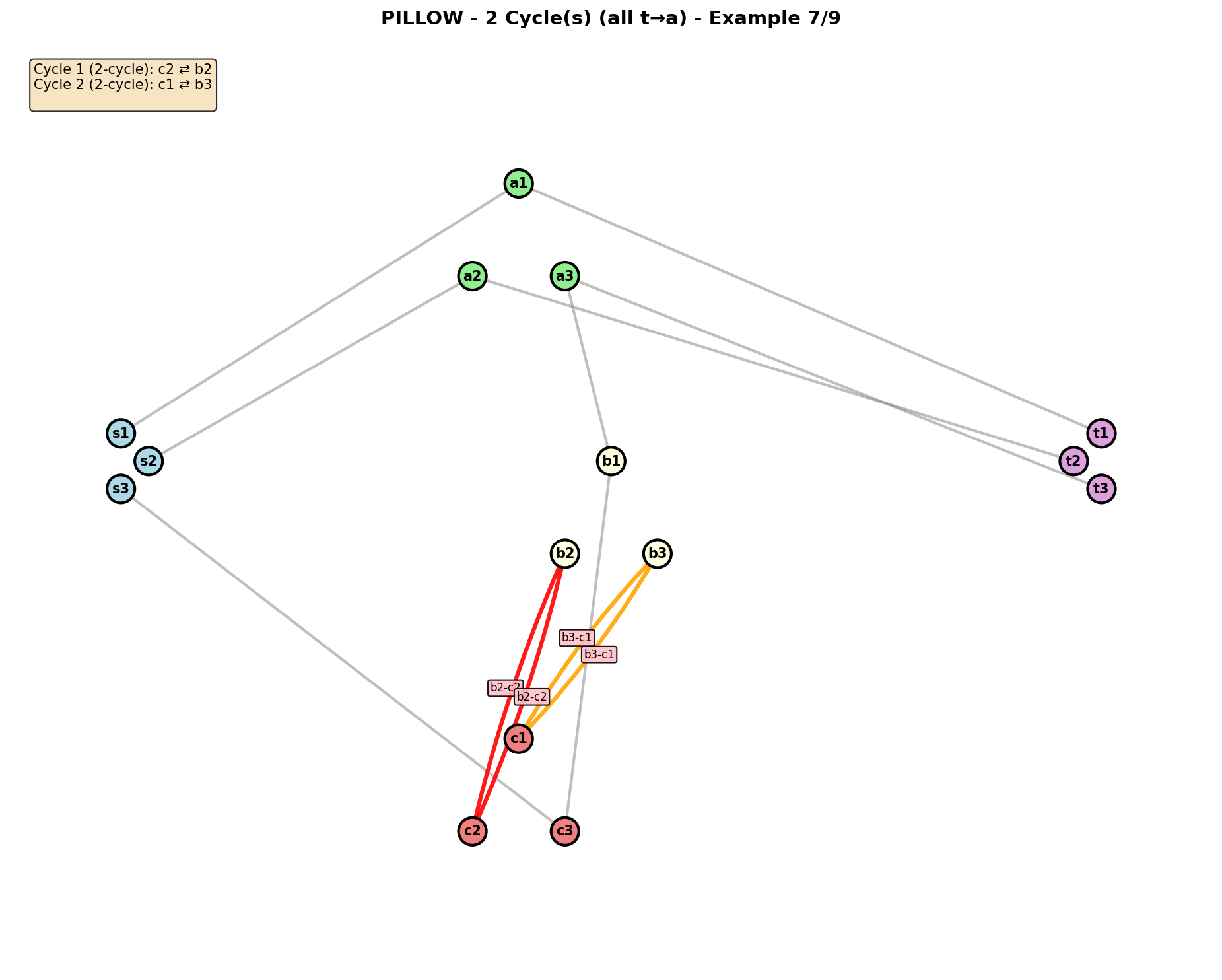}\par
{\scriptsize\textbf{P-11}; idx 14\par}
{\scriptsize\texttt{t\_same\_2cycles\_7.png}\par} &
\adjustbox{max width=\linewidth,max totalheight=3.1cm}{%
  \crawDisconnectedPillowCard{10}{1}
    {P-11-1}}
\end{tabular}
\end{adjustbox}
\caption{One Pillow code graph from each sector and a corresponding
surface drawing.  Section~\ref{sec:surface-audit} counts the code graph, not the number of
surface drawings.  Thus (b) contributes one \(C_3\) parent to
equation~\eqref{eq:surface-parent-count}, although Section~\ref{sec:embeddings} later finds three
marked embeddings of that parent.}
\label{fig:surface-parent-sector-examples}
\end{figure}

The surgery check retains the following code graphs:
\begin{equation}
\begin{array}{c|rrrr|r}
\toprule
&C_2&C_3&D_1&D_2&\text{total}\\
\midrule
\sixj&6&1&3&0&10\\
\text{Pillow}&6&1&3&1&11\\
\bottomrule
\end{array}
\label{eq:surface-parent-count}
\end{equation}
There is one \(C_3\) ``parent'' on each side and one Pillow \(D_2\) ``parent'',
regardless of their different surface drawings.  These are precisely the
parents whose embedding multiplicities must be resolved next.  The complete
catalogue in Appendix~\ref{app:surface-ledger} pairs each retained code
graph with its final surface drawings.

\section{Embeddings and the unreduced count}
\label{sec:embeddings}

The code graphs retained in Section~\ref{sec:surface-audit} specify which lines
connect, but not uniquely how they lie on the surface.  We now count the
distinct surface embeddings of each graph, keeping the labels and
orientation fixed.  Drawings related by a smooth deformation count only once.
For parents with several embeddings, a final hyphenated index distinguishes
the embedding classes: P-7-1, P-7-2, and P-7-3, for example, all belong to
the Pillow parent P-7.  Alternative drawings of the same embedding share
one identifier.

Each of the six \(C_2\) and three \(D_1\) graphs on each side contributes
once.  We therefore focus on the two Pillow \(D_2\) theta embeddings in
Section~\ref{sec:theta-embeddings}, then the three embeddings of each
\(C_3\) graph in Section~\ref{sec:c3-embeddings}.
Section~\ref{sec:marked-count} combines all sectors to give the count
before the cut rules of Section~\ref{app:no-onepoint} are applied.

\subsection{The two theta embeddings of the Pillow
\texorpdfstring{\(D_2\)}{D2} parent}
\label{sec:theta-embeddings}

The Pillow \(D_2\) parent has two marked embeddings, shown in
Figure~\ref{fig:theta-ledger-representatives}.  Capping and shrinking the
\(b,c\) cuffs displays each as a labeled theta graph on a torus (so named due to its resemblance to the letter $\theta$).  The distinction is which pair of edges bounds a disk: blue and
green in \(\theta_{10}\) (P-11-1), but blue and purple in \(\theta_{11}\) (P-11-2).
The fixed edge labels prevent an isotopy from exchanging these two pairs.
These are not a sign pair or two projections of the same marked embedding. 

\begin{figure}[!ht]
\centering
\begin{tabular}{@{}c@{\hspace{5mm}}c@{}}
{\small surface drawing} & {\small theta reduction}\\[1mm]
\resizebox{.42\linewidth}{!}{%
  \crawDisconnectedPillowCard{10}{1}{P-11-1}} &
\resizebox{.23\linewidth}{!}{%
  \begin{tikzpicture}[x=.82cm,y=.82cm,baseline=(thetaMid)]
    \path[use as bounding box] (-.3,-.08) rectangle (4.3,3.45);
    \coordinate (thetaMid) at (2,1.35);
    \sdembThetaPanel{10}{P-11-1}
  \end{tikzpicture}}\\[4mm]
\resizebox{.42\linewidth}{!}{%
  \crawDisconnectedPillowCard{11}{2}{P-11-2}} &
\resizebox{.23\linewidth}{!}{%
  \begin{tikzpicture}[x=.82cm,y=.82cm,baseline=(thetaMid)]
    \path[use as bounding box] (-.3,-.08) rectangle (4.3,3.45);
    \coordinate (thetaMid) at (2,1.35);
    \sdembThetaPanel{11}{P-11-2}
  \end{tikzpicture}}\\[1mm]
\multicolumn{2}{c}{\scriptsize
  \textcolor{RoyalBlue}{$b_1-c_3$} (blue),\quad
  \textcolor{ForestGreen}{$b_2-c_2$} (green),\quad
  \textcolor{Purple}{$b_3-c_1$} (purple)}
\end{tabular}
\caption{The two final Pillow \(D_2\) embeddings, P-11-1 and P-11-2, with their theta
reductions.  The labeled pair bounding the disk distinguishes them.}
\label{fig:theta-ledger-representatives}
\end{figure}
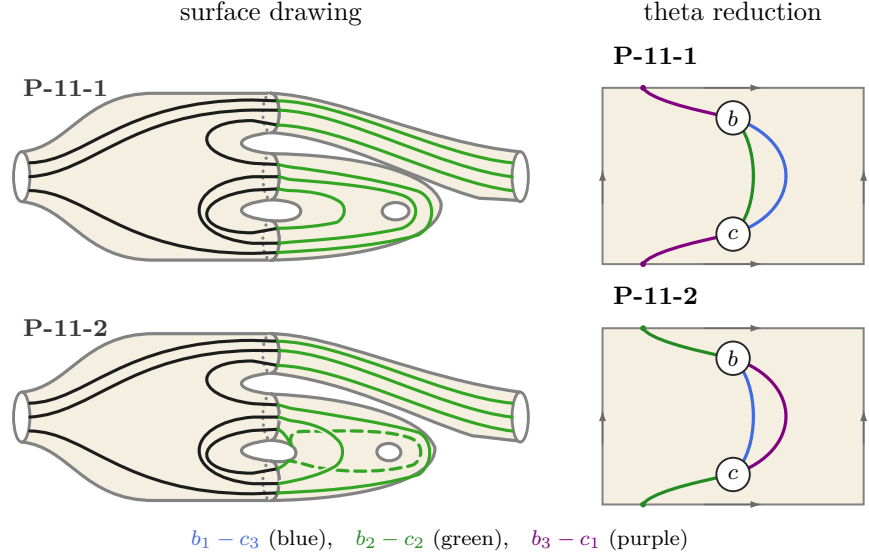

There are naively two possible presentations of each representative, but they only differ by a change in the way they are displayed. The reduction from four to two is discussed in Appendix~\ref{app:embedding-atlas}, with all four views shown. Thus this one $D_2$ parent contributes two,
not four, to the embedding count.

\subsection{The three embeddings of the
\texorpdfstring{\(C_3\)}{C3} parents}
\label{sec:c3-embeddings}

Each \(C_3\) parent has three retained embedding classes.
Figure~\ref{fig:c3-ledger-representatives} lists one representative per
class: 6j-7-1, 6j-7-2, and 6j-7-3,
paired by a local \(F\)-move with P-7-1, P-7-2, and P-7-3, respectively.
The six colors label the same operators on both sides:
green \(I\), red \(J\), orange \(M\), purple \(N\), blue \(L\), and
magenta \(K\); \(N\) is the internal channel changed by the move.

\begin{figure}[!ht]
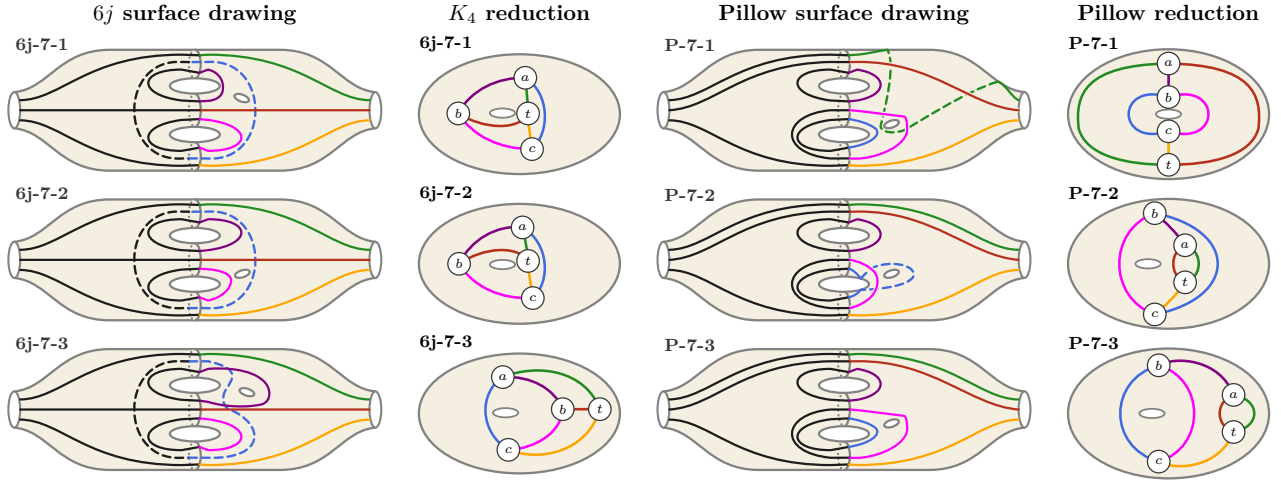

\centering
\resizebox{.98\linewidth}{!}{%
\begin{tabular}{@{}c@{\hspace{3mm}}c@{\hspace{3mm}}c@{\hspace{3mm}}c@{}}
{\small\bfseries \(6j\) surface drawing} & {\small\bfseries \(K_4\) reduction} &
{\small\bfseries Pillow surface drawing} & {\small\bfseries Pillow reduction} \\[1mm]
\crawCard{1}{SixColoredOne}{Standard}
  {6j-7-1}{}{UpperInner} &
\sdembKFourCard{6j-7-1}{One} &
\crawCard{18}{PillowColoredEighteen}{Standard}
  {P-7-1}{}{Lower} &
\crawPillowMiniCard{P-7-1}{Eighteen} \\[1mm]
\crawCard{2}{SixColoredTwo}{Standard}
  {6j-7-2}{}{Lower} &
\sdembKFourCard{6j-7-2}{Two} &
\crawCard{15}{PillowColoredFifteen}{Standard}
  {P-7-2}{}{Lower} &
\crawPillowMiniCard{P-7-2}{Fifteen} \\[1mm]
\crawCard{4}{SixColoredFour}{Standard}
  {6j-7-3}{}{FourUpper} &
\sdembKFourCard{6j-7-3}{Three} &
\crawCard{14}{PillowColoredFourteen}{Standard}
  {P-7-3}{}{Lower} &
\crawPillowMiniCard{P-7-3}{Fourteen}
\end{tabular}}
\caption{The three final \(C_3\) classes: \(6j\) and Pillow surface
drawings, each paired with its reduced graph.  The \(6j\) reduction is
\(K_4\); the Pillow reduction has parallel edges.  Each row is a corresponding
pair under the local \(F\)-move shown in Figure~\ref{fig:three-embeddings}.}
\label{fig:c3-ledger-representatives}
\end{figure}

To distinguish the \(6j\) classes, cap the four cuffs \(a,b,c,t\)
and shrink each cap to a vertex.  The six colored edges form \(K_4\), as
shown in the second column of Figure~\ref{fig:c3-ledger-representatives}.
The labels and the oriented \(abc\) cycle remain fixed, so no permutation
of the four vertices is being taken.  The three cases differ by which
triangle through \(t\), namely \(abt,bct,\) or \(cat\), winds around the
marked handle.

We express this topological character using bracket notation:
\([\gamma]\) records the winding of the oriented cycle \(\gamma\), with
contractible pieces suppressed.\footnote{A contractible piece is one which involves a closed cycle bounding a disk.}  The cycles obey
\begin{equation}
 [abc]=[abt]+[bct]+[cat].
 \label{eq:k4-relative-character}
\end{equation}
For the three retained classes,
\begin{equation}
 \bigl([abt],[bct],[cat]\bigr)
 =\bigl([abc],0,0\bigr),\quad
  \bigl(0,[abc],0\bigr),\quad
  \text{or }\bigl(0,0,[abc]\bigr),
 \label{eq:k4-three-characters}
\end{equation}
where the fixed \([abc]\) is nonzero.  Having different characters distinguishes
the three embeddings.  Reversing orientation would also reverse the
fixed \(abc\) cycle, so mirror drawings do not supply another count of three.

One might try to obtain further embeddings by placing the handle differently
relative to the Wilson lines.  Figure~\ref{fig:handle-migration-example}
shows an apparent extra possibility, but the two drawings represent the same
embedding, 6j-7-3.  Both have character \((0,0,[abc])\); the explicit
handle migration in Appendix~\ref{app:handle-migration}, not the matching
character alone, establishes their equivalence.  This smooth deformation
changes the apparent position of the handle without changing the endpoint
labels or their cyclic orders, so it supplies no additional embedding.

\begin{figure}[!ht]
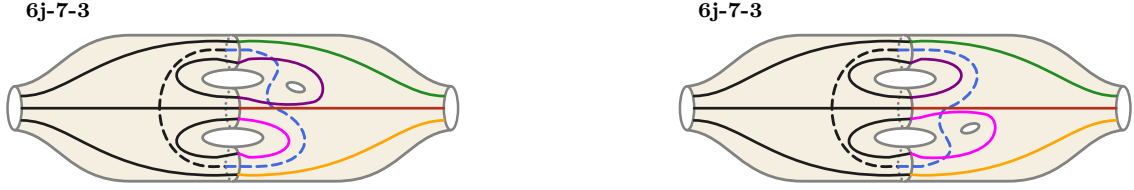

  \centering
  \czheEndpointCell{4}{6j-7-3}\hfill
  \czheEndpointCell{5}{6j-7-3}
  \caption{An apparent extra embedding is another view of 6j-7-3:
  handle migration relates the two drawings, as shown in
  Appendix~\ref{app:handle-migration}.}
  \label{fig:handle-migration-example}
\end{figure}

After this analysis, we find that there is only one representative for each character class in $C_3$ for both Pillow and $6j$ categories.

A local \(F\)-move gives a one-to-one correspondence between the three
\(C_3\) embeddings: 6j-7-1 maps to P-7-1, 6j-7-2 to P-7-2, and
6j-7-3 to P-7-3.  Only the connections inside a small disk change; the
exterior Wilson lines and the handle remain fixed.  The move and its
inverse are explained in Appendix~\ref{app:local-f-move}.

\subsection{The unreduced count}
\label{sec:marked-count}

After the reductions in Sections~\ref{sec:theta-embeddings} and~\ref{sec:c3-embeddings}, the final representatives therefore contribute
\begin{equation}
\begin{aligned}
  \Nsix&=6\ (C_2)+3\ (C_3)+3\ (D_1)=12,\\
  \NP&=6\ (C_2)+3\ (C_3)+3\ (D_1)+2\ (D_2)=14.
\end{aligned}
\label{eq:marked-sector-count}
\end{equation}
Resolving the single \(C_3\) parent adds two on each side, so these additions
cancel in the difference.  The two theta embeddings occur only on the
Pillow side, giving the unreduced result
\begin{equation}
  \NP-\Nsix=14-12=2.
  \label{eq:final-count}
\end{equation}
The stages of the calculation are collected below:
\begin{equation}
\begin{array}{lccc}
\toprule
\text{stage}&\sixj&\text{Pillow}&\text{difference}\\
\midrule
\text{code candidates}&31&33&2\\
\text{embeddable surface parents}&10&11&1\\
\text{marked embeddings}&12&14&2\\
\text{after one-point exclusion}&0&4&4\\
\text{after proposed two-point exclusion}&0&1&1\\
\bottomrule
\end{array}
\label{eq:count-pipeline}
\end{equation}
The first three rows are the enumeration in Sections~\ref{sec:enumeration}
and~\ref{sec:embeddings}. It is only afterwards that we implement the steps of
Section~\ref{app:no-onepoint} on this list, giving the reduced
result \(1-0=1\) as predicted by the conjecture.  The unreduced difference two and the reduced difference
one thus refer to different counting prescriptions.

\section{Summary and outlook}
\label{sec:discussion}

We tested the proposed matrix--tensor description of three-dimensional gravity through a Schwinger--Dyson relation in its purely open sector, projected onto a particular contributing manifold, the $C_0$ or cylinder manifold. The conjecture that the $\hbar\to0$ limit of the matrix--tensor model describes three-dimensional quantum gravity where each 3-manifold is counted with equal weight translates to a relation for the SD equation, which states  $\NP-\Nsix=1$. We enumerated all abstract graphs using code, checked their surface realizations, and
resolved their marked embeddings, obtaining a preliminary counting of \((\Nsix,\NP)=(12,14)\).  Applying the VTQFT-inspired counting prescription, which involves not counting manifolds with one- and two-point cuts, removes most of the contributions, leaving \((\Nsix,\NP)=(0,1)\). 

The same approach we used in this paper applies in the more general setting where other topologies contribute to the exact two-point function, and it would be interesting to test the conjecture further, even in the purely open sector. Moreover, we have picked a hyperbolic contribution to the left-hand side of the SD equation, which is the exact two-point function, and it is natural to pick other hyperbolic manifolds as a next step.  

One- and two-point cuts play an important role in establishing crossing symmetry of OPE statistics \cite{Belin:2026pko,Wang:2025jgo}. It would be interesting to explore the connection between SD equations and crossing symmetry. Relatedly, there is a possibility that at any given order, more than one prescriptions for which manifolds to include can satisfy the SD equation. For example, (the open analog of) the universal sum of topologies generated by the gravitational machine \cite{Belin:2026pko} would be an interesting case to test. We expect that testing to high enough orders can     discriminate between any two prescriptions. 

A natural extension of the story in this paper is to analyze the count for the closed matrix-tensor model. This is where the original conjecture was made. The main difficulty lies in the fact that in the closed case, the Wilson network lives in the bulk of 3-manifolds. In the purely open case, where all Wilson lines lie on the boundary (EOW branes), it is tremendously simpler to classify and distinguish different contributions, and our code has relied heavily on properties of graphs on surfaces. Finding an algorithm in the closed case is therefore not straightforward. More physically, one would need to deal with the infinities that arise due to windings of the bulk Wilson lines and other possible sources of infinity. 

More broadly, one may wonder if the SD equation holds even with offshell manifolds. The distinction between the
VTQFT amplitude and a gravitational amplitude, emphasized for
offshell geometries in \cite{Collier:2023fwi,Jafferis:2025vyp,Yan:2025usw},
is a reason for us to have explicitly presented the counts before and after applying the prescription. It is still an open question how VTQFT can be used to compute offshell amplitudes, potentially by a treatment of the mapping class group.

Finally, recalling that the potential of the matrix--tensor model is built by squaring the bootstrap constraints, we can hope that the approach to the $\hbar=0$ limit should single out a particular solution or a set of solutions to the CFT bootstrap \cite{Belin:2023efa}, and we hope to study it in the future.

\section*{Acknowledgments}
We want to thank Jan de Boer, Alexander Frenkel, Eric Perlmutter, Stephen Shenker, and Douglas Stanford for discussions. LR, DLJ, and CY were supported in part by the Simons Investigator in Physics Award MP-SIP-0001737 and U.S. Department of Energy grants DE-SC0007870 and DE-SC0021013. DW acknowledges support by NSF grant PHY-2207659 and Simons Foundation grant MPS-CH-00001550. This work was performed in part at the Aspen Center for Physics, which is supported by National Science Foundation grant PHY-2210452. We would also like to thank Claude and Codex for assistance with coding and writing of the paper. 

\endgroup 
\appendix

\section{Gluing and Wilson-line surgery}
\label{app:open-surgery}

In this appendix, we review the surgery construction of
Jafferis--Rozenberg--Wang~\cite{Jafferis:2025yxt} in the open matrix--tensor
model.  Their annular surgery on boundary Wilson loops is the open
counterpart of the toroidal surgery on bulk Wilson loops in the closed
model of Jafferis--Rozenberg--Wong~\cite{Jafferis:2025vyp}. 

The annular surgery is a procedure performed to a boundary Wilson loop, with the understanding that when integrated with the measure given by Cardy density,
the resulting manifold has the same VTQFT partition function. It is performed by removing the neighborhood of the loop, leaving an annular cut, which is then glued to a thickened disk. From the dual matrix model perspective, which is the Virasoro minimal string \cite{Collier:2023cyw}, the thickened disk is the leading contribution to the spectral density, precisely given by Cardy formula \cite{Jafferis:2025jle}.

The annular surgery is used in determining if a shape joins a pillow or a 6$j$ and gives the cylinder. In the SD equation \eqref{eq:sd-picture}, loops can form on the right-hand side of the equation when we glue pieces along equally labeled boundary circles and connect the
Wilson-line endpoints in their prescribed cyclic order. This is where the annular surgery is then performed.

We will also show how a doubling procedure relates the annular surgery to the toroidal surgery. See \cite{Burnell:2010mx} 
for an explanation of the logic behind the equivalence of the partition function before and after the surgery in the rational case.

\subsection{Annular surgery}

We now illustrate the annular surgery with two examples: a boundary Wilson loop on
a solid torus, and one on a genus-2 handlebody.  

\begin{figure}[htbp]
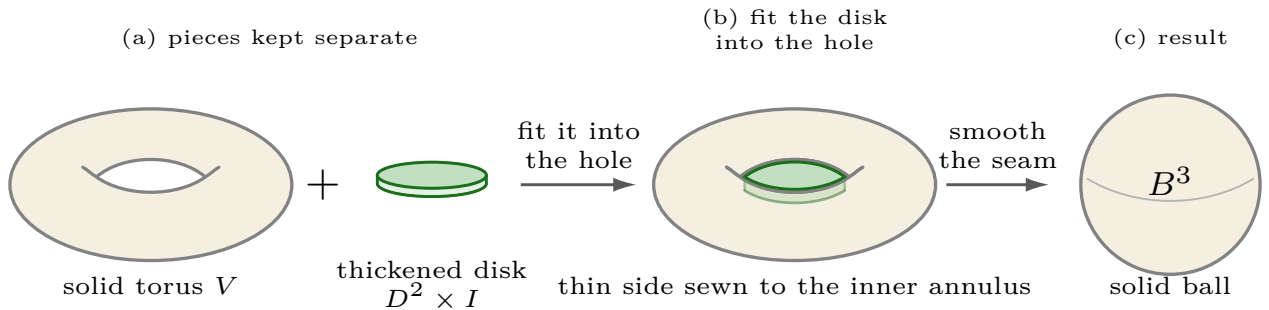

  \centering
  \resizebox{.96\linewidth}{!}{\SDOpenHandleFillingFigure}
  \caption{The geometric attachment used below: a solid torus and a
  thickened disk are joined along the disk's curved side to form a solid ball.}
  \label{fig:open-handle-filling}
\end{figure}

First let us recall the following simple topological fact. Consider a thickened disk \(D^2\times I\). Fitting it into the central hole of a solid
torus \(V\) and sewing its curved side to the inner wall closes the ``tunnel'',
giving a solid ball \(B^3\).  Figure~\ref{fig:open-handle-filling}
illustrates this geometric fact; it does not yet specify a Wilson loop on
which to perform surgery.

\begin{figure}[!htb]
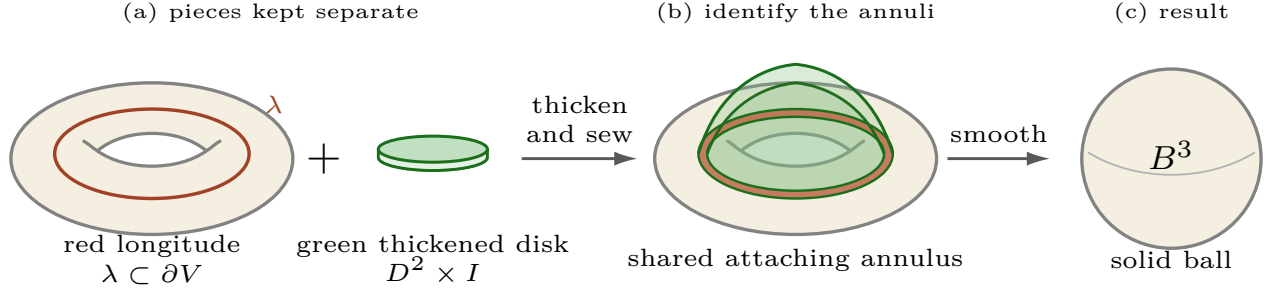

  \centering
  \resizebox{.96\linewidth}{!}{\SDOpenAttachingAnnulusFigure}
  \caption{Annular surgery on the boundary Wilson loop shown in red.  The
  curved side of the green thickened disk is sewn to this strip, while its
  two faces remain on the boundary.  The result is again a solid ball.}
  \label{fig:open-handle-attaching-annulus}
\end{figure}

To turn this attachment into an annular-surgery example, choose a boundary
Wilson loop \(\lambda\) going once around the torus's (genus) hole, as shown in
red in Figure~\ref{fig:open-handle-attaching-annulus}.  Surgery replaces
the strip around this loop by the two faces of a thickened disk, whose
curved side \(\partial D^2\times I\) is sewn to the strip (orange).  The two green
domes are these two faces.  The chosen loop now
bounds a disk, so the handle is removed and the result is a ball, as in Figure~\ref{fig:open-handle-filling}.  The embedding of the loop is important: an arbitrary
annular surgery need not remove a handle.

\begin{figure}[H]
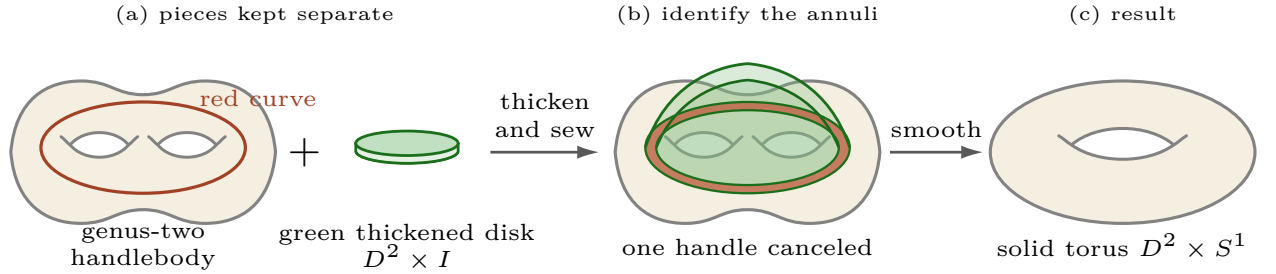

  \centering
  \resizebox{.96\linewidth}{!}{\SDGenusTwoHandleFillingFigure}
  \caption{Annular surgery on the illustrated genus-two handlebody.  Attaching
  the thickened disk along the orange strip removes one handle and leaves a
  solid torus.}
  \label{fig:genus-two-handle-filling}
\end{figure}

\begin{samepage}
The second surgery example starts with two handles, so removing one no
longer leaves a ball.  For the chosen red Wilson loop in
Figure~\ref{fig:genus-two-handle-filling}, the same annular surgery removes
one handle.
The result is therefore a solid torus. \par
\end{samepage}

\subsection{Recovering the closed surgery by doubling}

Gluing two copies of the open construction (it and its mirror image) along the EOW branes gives the closed construction.  This is called
\emph{doubling}.\footnote{It is called the quadrupling trick in \cite{Wang:2025bcx}, as doubling sometimes refers to moving the anti-chiral sector to the mirror copy.} 

As an example, consider the doubling of Figure~\ref{fig:open-handle-attaching-annulus}. First consider the green filling.  Glue two thickened disks along their
two pairs of flat faces.  Their interval directions join into a circle,
so together they form a solid torus:
\begin{equation}
  (D^2\times I)\cup_{D^2\times\partial I}(D^2\times I)
  \cong D^2\times S^1.
  \label{eq:double-thickened-disk}
\end{equation}
This solid torus is then the closed analog of the thickened disk. In this example, the initial solid torus doubles to
\(S^2\times S^1\), while the resulting ball doubles to \(S^3\).
Thus doubling this open surgery gives the closed move
of Jafferis--Rozenberg--Wong~\cite{Burnell:2010mx,Jafferis:2025vyp}:
\begin{equation}
  (S^2\times S^1 -D^2\times S^1) \cup D^2\times S^1
  \longrightarrow{} S^3.
  \label{eq:doubled-open-surgery}
\end{equation}
Here, removing a tubular neighborhood of the bulk Wilson loop leaves a torus boundary, which is then glued to a solid torus with the two cycles exchanged. This exchange of the two torus directions
is also called the \(S\)-gluing and is explained around equation~(3.34) of
\cite{Jafferis:2025vyp}.

\section{Complete graph enumeration and selection rules}
\label{app:algorithm}

This appendix gives the full algorithm summarized in
Section~\ref{sec:graph-screen}, including the fixed port data, every
selection step, the intermediate count tables, and the multigraph example.
The branching flow chart appears in Figure~\ref{fig:candidate-sectors} in
the main text.

\subsection{The completions \texorpdfstring{\(p\)}{p} and
\texorpdfstring{\(q\)}{q} as port matchings}

The two searches use the same completion procedure and differ only in the index structure for the interaction vertices.  The three \(s_i\) are fixed external ports on the left,
the three \(t_i\) are external ports on the right, and
\(a_i,b_i,c_i\) label the nine attachment points along the cut.  Thus the
labeled ports are
\begin{equation}
\begin{aligned}
 &\{s_1,s_2,s_3\},\qquad \{t_1,t_2,t_3\},\\
 &\{a_1,a_2,a_3\},\qquad \{b_1,b_2,b_3\},\qquad
 \{c_1,c_2,c_3\}.
\end{aligned}
\label{eq:ports}
\end{equation}
For this finite search, specifying \(p\) or \(q\) means joining the twelve
loose ports \(t_i,a_i,b_i,c_i\) in six pairs.
Figure~\ref{fig:code-to-surface-key} shows the physical position of every
port and the fixed black line connections.  The six fixed edges that distinguish
the two cases are listed in equation~\eqref{eq:fixed-edges} below.

\begin{equation}
\begin{array}{c|l}
\toprule
\text{fixed half}&\text{fixed edges}\\
\midrule
\sixj&
s_1a_1,\ s_2b_2,\ s_3c_3,\ a_2b_1,\ b_3c_2,\ c_1a_3\\
\text{Pillow}&
s_1a_1,\ s_2a_2,\ s_3c_3,\ a_3b_1,\ b_2c_2,\ b_3c_1\\
\bottomrule
\end{array}
\label{eq:fixed-edges}
\end{equation}
Every subsequent rule is applied identically to the two fixed edge sets.

\subsection{Generate every allowed completion}

A completion is a perfect matching of the twelve ports
\(t_i,a_i,b_i,c_i\).  Direct \(t_i\)--\(t_j\) pairs are excluded because
they would join two external endpoints to one another and could not belong to
any required \(s\)-to-\(t\) path.  The partner of \(t_1\) can therefore be
chosen from the nine \(a\)-, \(b\)-, and \(c\)-ports.  After that choice,
the partners of \(t_2\) and \(t_3\) can be chosen in eight and seven ways.
The six remaining ports can be paired in \(5!!\) ways.  Hence the program
generates
\begin{equation}
  9\cdot8\cdot7\cdot5!!=7560
\end{equation}
labeled completions, each exactly once.

Each matching is added separately to the Pillow and \(6j\) fixed edges
in equation~\eqref{eq:fixed-edges}.  Every
\(a_i,b_i,c_i\) then has degree two, while every \(s_i,t_i\) has degree
one.  Consequently, a completed port graph is a disjoint union of open paths
joining \(s\)-ports to \(t\)-ports and, possibly, closed cycle components.
The implementation treats this as a multigraph: two parallel edges form a
two-edge closed cycle and must not be collapsed into one adjacency.  The
implementation and machine-selected outputs are archived in the
\emph{SDcountcode} GitHub repository~\cite{SDcountcode}.

\subsection{Select the graph-level cylinder candidates}

For each of the 7560 completions and each fixed half, the cylinder program
performs the following tests.

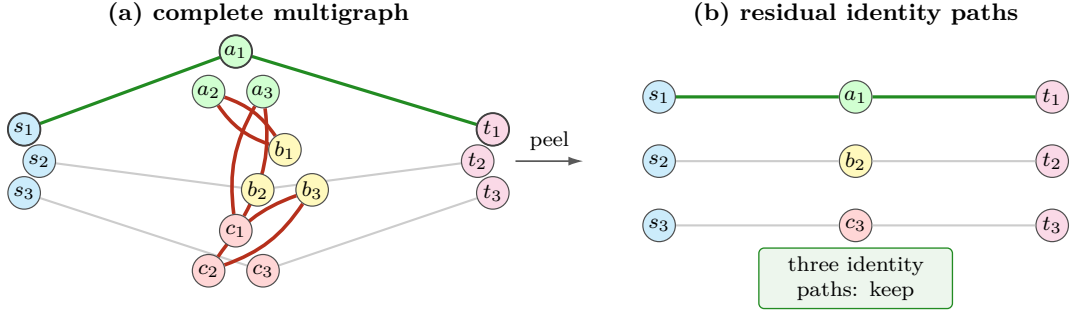
\begin{figure}[htbp]
\centering
\begin{tikzpicture}
  \begin{scope}[xshift=3.3cm]
    \node[sdpaneltitle] at (0,1.95) {(a) complete multigraph};
    \sdcompactports{u}
    \begin{scope}[on background layer]
      \sdedgelist{u}{sdcontext}{s2/b2,b2/t2,s3/c3,c3/t3}
      \sdedgelist{u}{sdpathone}{s1/a1,a1/t1}
      \sdparallel{u}{sdcycle}{a2}{b1}
      \sdparallel{u}{sdcycle}{b3}{c2}
      \sdparallel{u}{sdcycle}{a3}{c1}
    \end{scope}
    \sdoutlineports{u}{s1,a1,t1}
  \end{scope}

  \draw[sdarrow] (6.7,0)--node[above,font=\scriptsize]{peel}(7.6,0);

  \begin{scope}[xshift=11.2cm]
    \node[sdpaneltitle] at (0,1.95) {(b) residual identity paths};
    \node[sds,sdcompact] (q1s) at (-2.6,.85) {$s_1$};
    \node[sda,sdcompact] (q1a) at (0,.85) {$a_1$};
    \node[sdt,sdcompact] (q1t) at (2.6,.85) {$t_1$};
    \draw[sdpathone] (q1s)--(q1a)--(q1t);
    \node[sds,sdcompact] (q2s) at (-2.6,0) {$s_2$};
    \node[sdb,sdcompact] (q2b) at (0,0) {$b_2$};
    \node[sdt,sdcompact] (q2t) at (2.6,0) {$t_2$};
    \draw[sdcontext] (q2s)--(q2b)--(q2t);
    \node[sds,sdcompact] (q3s) at (-2.6,-.85) {$s_3$};
    \node[sdc,sdcompact] (q3c) at (0,-.85) {$c_3$};
    \node[sdt,sdcompact] (q3t) at (2.6,-.85) {$t_3$};
    \draw[sdcontext] (q3s)--(q3c)--(q3t);
    \node[sdaccept,text width=2.35cm,font=\scriptsize]
      at (0,-1.55) {three identity paths: keep};
  \end{scope}
\end{tikzpicture}
\caption{The unique three-cycle \(6j\) matching, with zero-based index
409.  Each doubled red pair is a separate two-edge closed component.  After
the three components are peeled, the residual graph consists of the three
identity-paired paths.  The picture illustrates the multigraph test; it does
not assign a genus to the eventual surface.}
\label{fig:prenlo-three-cycle}
\end{figure}

\begin{enumerate}[leftmargin=*]
  \item Add the six completion edges to the six fixed edges and retain edge
  multiplicity when checking degrees and finding connected components.

  \item Find every closed cycle component, remove it from a copy of the
  graph, and record the number removed.  Here ``two cycles'' means two
  separate closed graph components; it does not mean that the carrier
  surface has genus two.

  \item Keep the matching at this stage only when the remaining open graph
  consists of exactly the three identity-paired paths
  \begin{equation}
    s_1\longrightarrow t_1,\qquad
    s_2\longrightarrow t_2,\qquad
    s_3\longrightarrow t_3.
    \label{eq:identity-paths}
  \end{equation}
  Figure~\ref{fig:prenlo-three-cycle} illustrates this test: removing the
  three closed components leaves the required paths \(s_i\to t_i\).
  Exactly 693 matchings pass this identity-path test for each fixed half.

  \item Classify the completion as \emph{disconnected} when the three \(t_i\) are paired
  directly with the three ports of one letter family: \(a_1,a_2,a_3\),
  \(b_1,b_2,b_3\), or \(c_1,c_2,c_3\).  Otherwise it is \emph{connected}.

  \item Split the 693 survivors into these two branches.  In the connected
  branch, retain only cases with at least two closed cycle components.
  In the disconnected branch, retain
  cases with at least one closed cycle component.  Discard a matching when
  it fails the cycle threshold in its own branch.
\end{enumerate}

To decide whether the completion is connected, use only its six edges, with the three
ports of each cuff grouped into one vertex.  Each of the four vertices
\(a,b,c,t\) has degree three, so each connected component must contain an
even number of vertices.  A disconnected completion therefore splits into
two pairs of cuffs.  Since \(t_i\)--\(t_j\) pairs are forbidden, all three
\(t_i\) must attach to the other cuff in their component.  This proves the
criterion above.  In the code, these are the branches called ``ordinary'' and
``special,'' respectively.

The cycle requirement is a deliberately loose graph screen.  A closed graph
component is a possible Wilson loop for the later surgery, but the abstract
graph does not say whether that loop can be embedded on the required smooth
surface or whether it surrounds the correct hole.  This requirement defines
the broad graph list used as the reference pool for the handwritten
classification; passing it is not a proof of surface realizability.

Exactly 693 completions pass the identity-path test for either fixed half.
Their closed-cycle distribution is
\begin{equation}
\begin{array}{c|rrrr|r}
\toprule
 &0\text{ cycles}&1\text{ cycle}&2\text{ cycles}
 &3\text{ cycles}&\text{total}\\
\midrule
\text{open-}\sixj&480&188&24&1&693\\
\text{Pillow}&480&188&24&1&693\\
\bottomrule
\end{array}
\label{eq:identity-path-count}
\end{equation}
Thus the unrestricted identity-path count is identical on the two sides.
Splitting these 693 survivors into the two disjoint branches gives
\begin{equation}
\begin{array}{ll|rrrr|r|r}
\toprule
\text{branch}&\text{fixed half}
 &0\text{ cycles}&1\text{ cycle}&2\text{ cycles}&3\text{ cycles}
 &\text{total}&\text{retained}\\
\midrule
\text{connected}
 &\text{open-}\sixj&456&182&24&1&663&25\\
 &\text{Pillow}&458&180&23&1&662&24\\
\midrule
\text{disconnected}
 &\text{open-}\sixj&24&6&0&0&30&6\\
 &\text{Pillow}&22&8&1&0&31&9\\
\bottomrule
\end{array}
\label{eq:branch-cycle-count}
\end{equation}

In the connected branch, the retained column sums the two- and three-cycle
columns.  In the disconnected branch, it sums every column with at least one
cycle.  Thus the two retained branch counts are (25,24) and (6,9),
respectively.  Adding the two branches leaves 31 \(6j\) and 33 Pillow
graph candidates for the surface classification.  The zero-cycle entries remain in
the table only to show where the discarded disconnected matchings went.

The program therefore gives 31 \(6j\) and 33 Pillow
labeled graph candidates.  These are the output of the first stage and the
reference pool for the surface classification that we deal with by hand; they are not yet
\(\Nsix\) and \(\NP\), nor are they counts of inequivalent contributions.

\section{Final diagram catalogue}
\label{app:surface-ledger}

This catalogue pairs each retained code graph with its final surface
drawings.  There are ten \(6j\) and eleven Pillow code parents,
but twelve and fourteen marked embeddings: the \(C_3\) code graph has three embeddings, and the Pillow \(D_2\) graph has two.  The \(C_2\) and \(D_1\) parents each have one embedding.
An extra suffix distinguishes embeddings of the same code graph: for example,
P-7-1, P-7-2, and P-7-3 are the three embeddings of P-7.
\texttt{idx} is the
zero-based matching index among the 7560 code outputs.  The classification
arguments are in Appendix~\ref{app:embedding-atlas}.

\begingroup
\SDFigureLabels{position=upper-left}

\newcommand{\SDCatalogueCode}[3]{%
  \includegraphics[height=3.2cm,width=.96\linewidth,keepaspectratio]{#1}\par
  {\scriptsize #2\par}
  {\scriptsize\texttt{#3}\par}%
}
\newcommand{\SDCatalogueSurface}[1]{%
  \resizebox{\linewidth}{!}{#1}\par
}
\newcommand{\SDCatalogueRow}[4]{%
  \par\noindent\hspace*{.035\linewidth}%
  \begin{minipage}[c]{.38\linewidth}
    \centering\setlength{\parskip}{0pt}%
    \SDCatalogueCode{#1}{#2}{#3}%
  \end{minipage}\hfill
  \begin{minipage}[c]{.445\linewidth}
    \centering\setlength{\parskip}{0pt}%
    #4%
  \end{minipage}\hspace*{.035\linewidth}\par
}

\subsection{\texorpdfstring{$6j$}{6j} representatives}
\label{app:final-sixj-catalogue}

Each code graph is shown once, next to its corresponding surface
drawings.  Suffixes on the surface labels distinguish embeddings of the
same parent.  Code labels give the ID, zero-based matching index, and
original filename (local copies add \texttt{6j\_} or \texttt{pillow\_}).

\paragraph{\texorpdfstring{$C_2$: six parents, six embeddings.}{C2: six parents, six embeddings.}}
\mbox{}\par\nopagebreak

\SDCatalogueRow{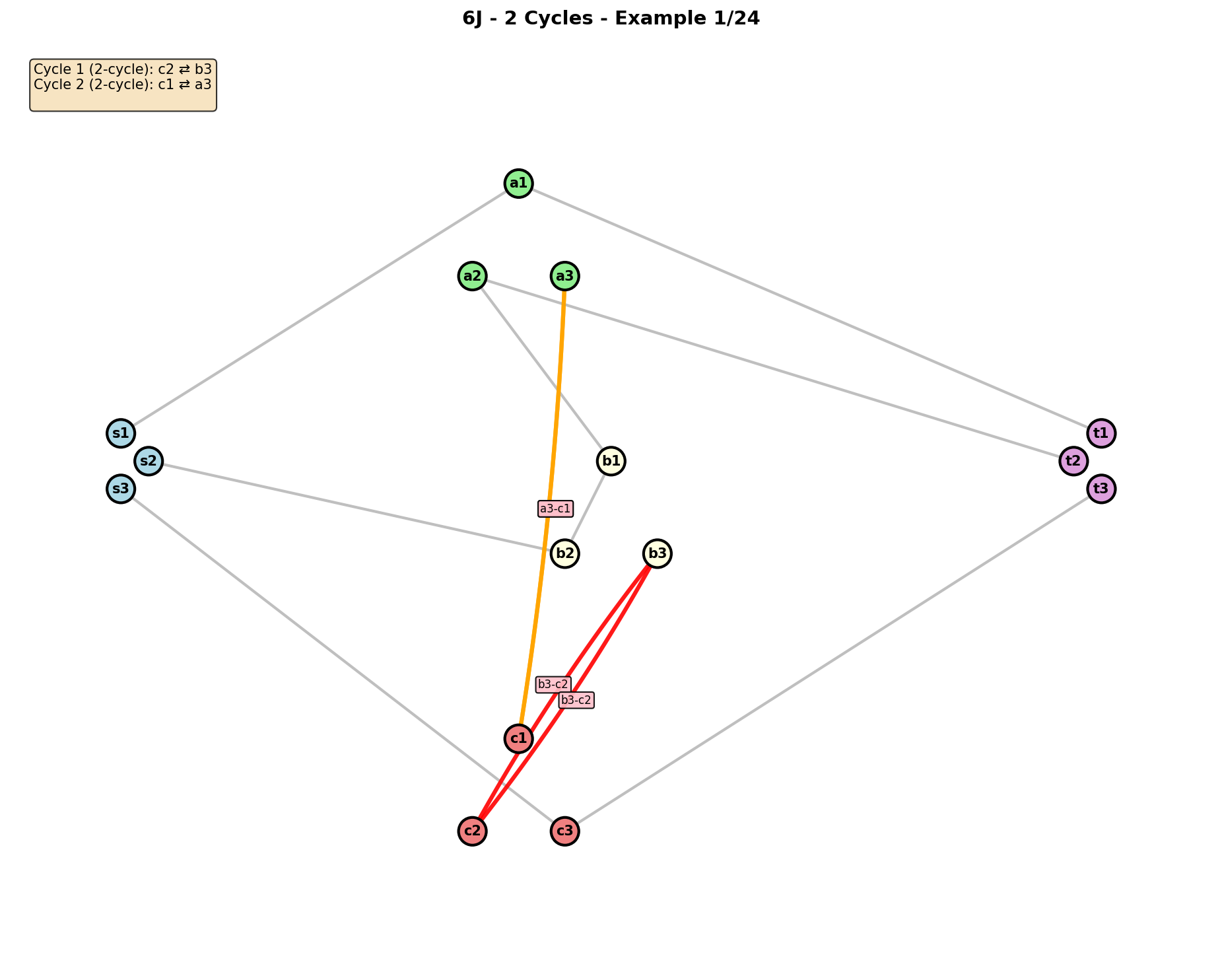}{$6j$--1; idx 99}{2cycles\_1.png}{%
  \SDCatalogueSurface{\crawCard{13}{SixThirteen}{Standard}{6j-1}{}{Hidden}}}
\SDCatalogueRow{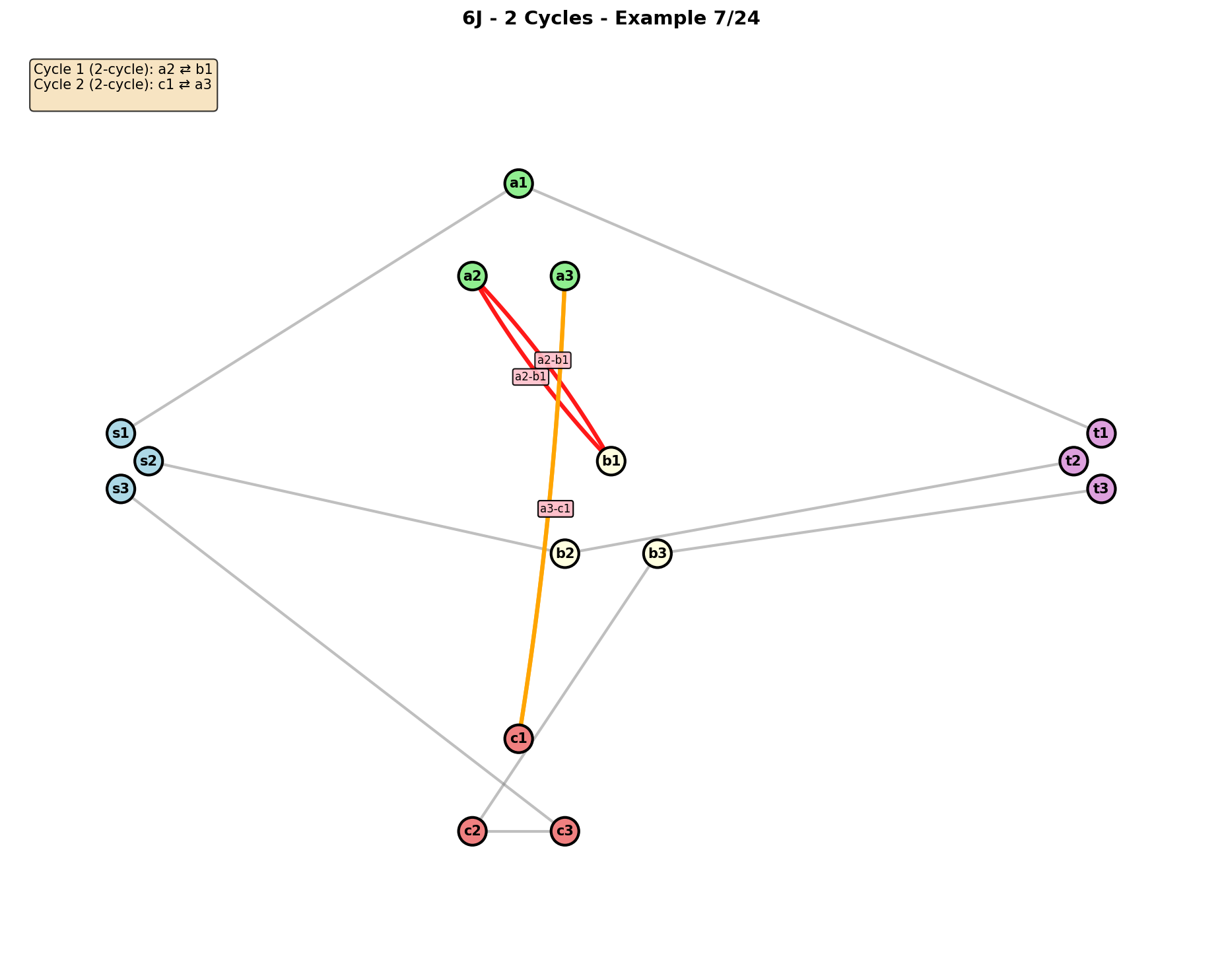}{$6j$--2; idx 363}{2cycles\_7.png}{%
  \SDCatalogueSurface{\crawCard{14}{SixFourteen}{Standard}{6j-2}{}{Hidden}}}
\SDCatalogueRow{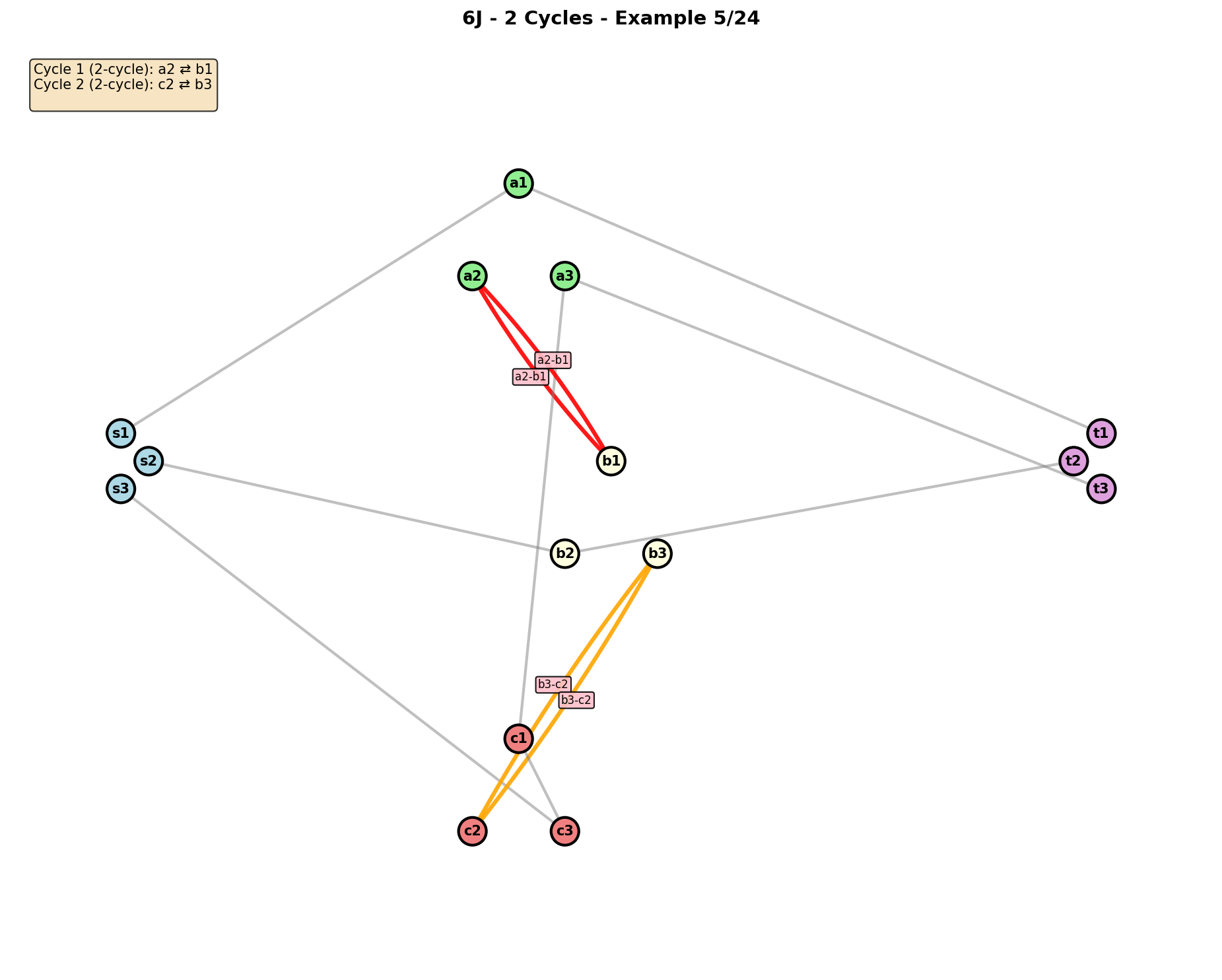}{$6j$--3; idx 331}{2cycles\_5.png}{%
  \SDCatalogueSurface{\crawCard{10}{SixTen}{Standard}{6j-3}{}{Hidden}}}
\SDCatalogueRow{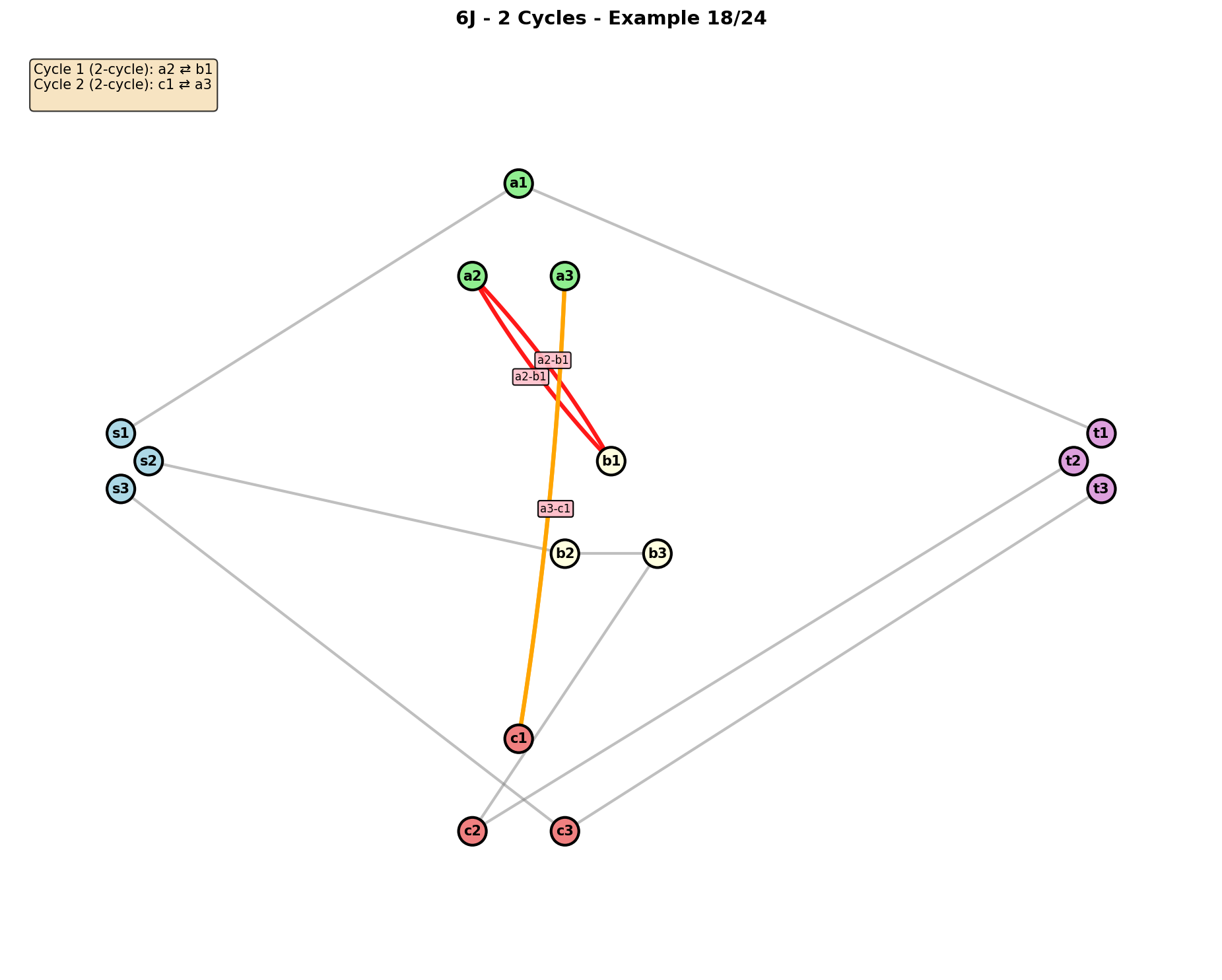}{$6j$--4; idx 725}{2cycles\_18.png}{%
  \SDCatalogueSurface{\crawCard{12}{SixTwelve}{Standard}{6j-4}{}{Hidden}}}
\SDCatalogueRow{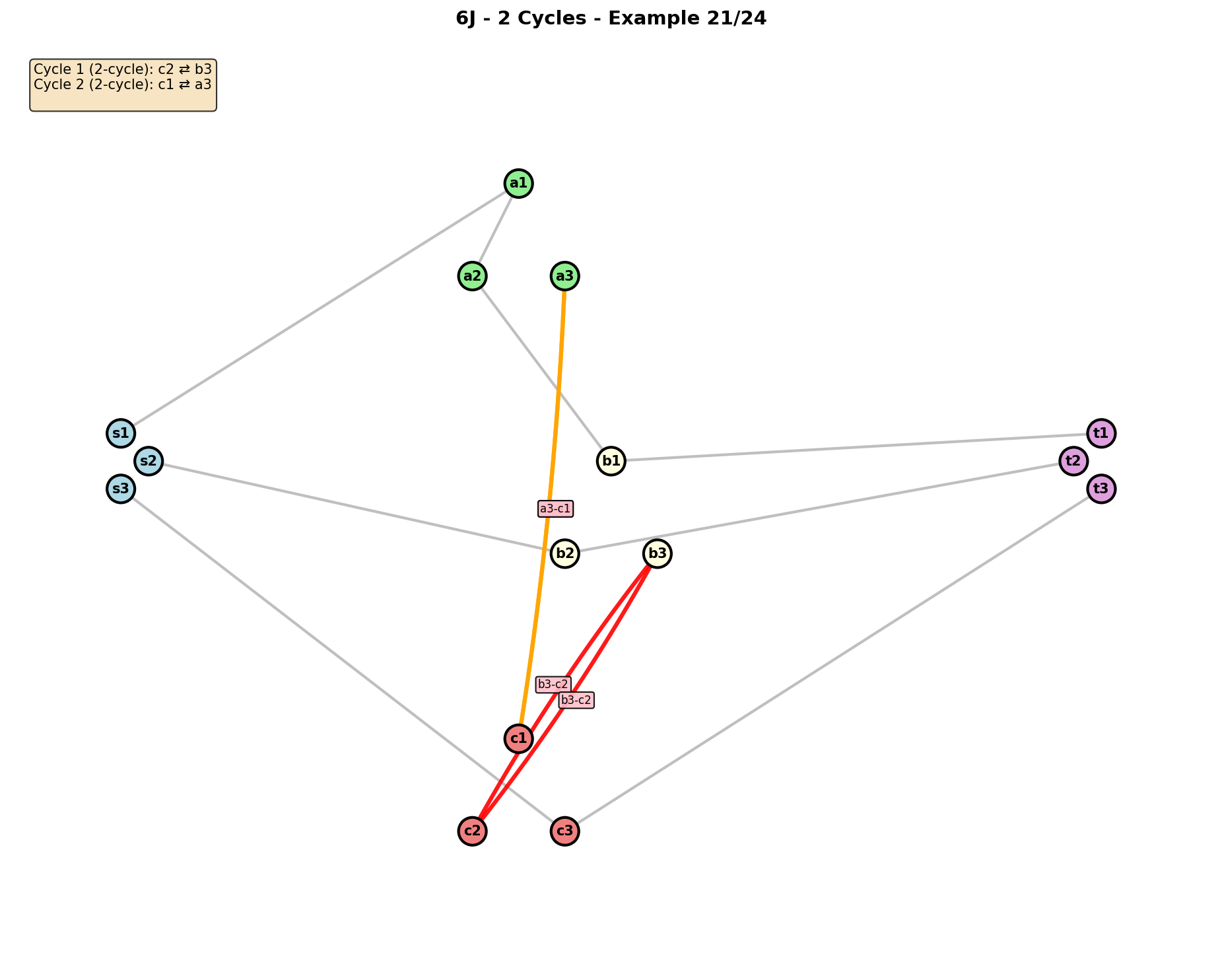}{$6j$--5; idx 2926}{2cycles\_21.png}{%
  \SDCatalogueSurface{\crawCard{15}{SixFifteen}{Standard}{6j-5}{}{Hidden}}}
\SDCatalogueRow{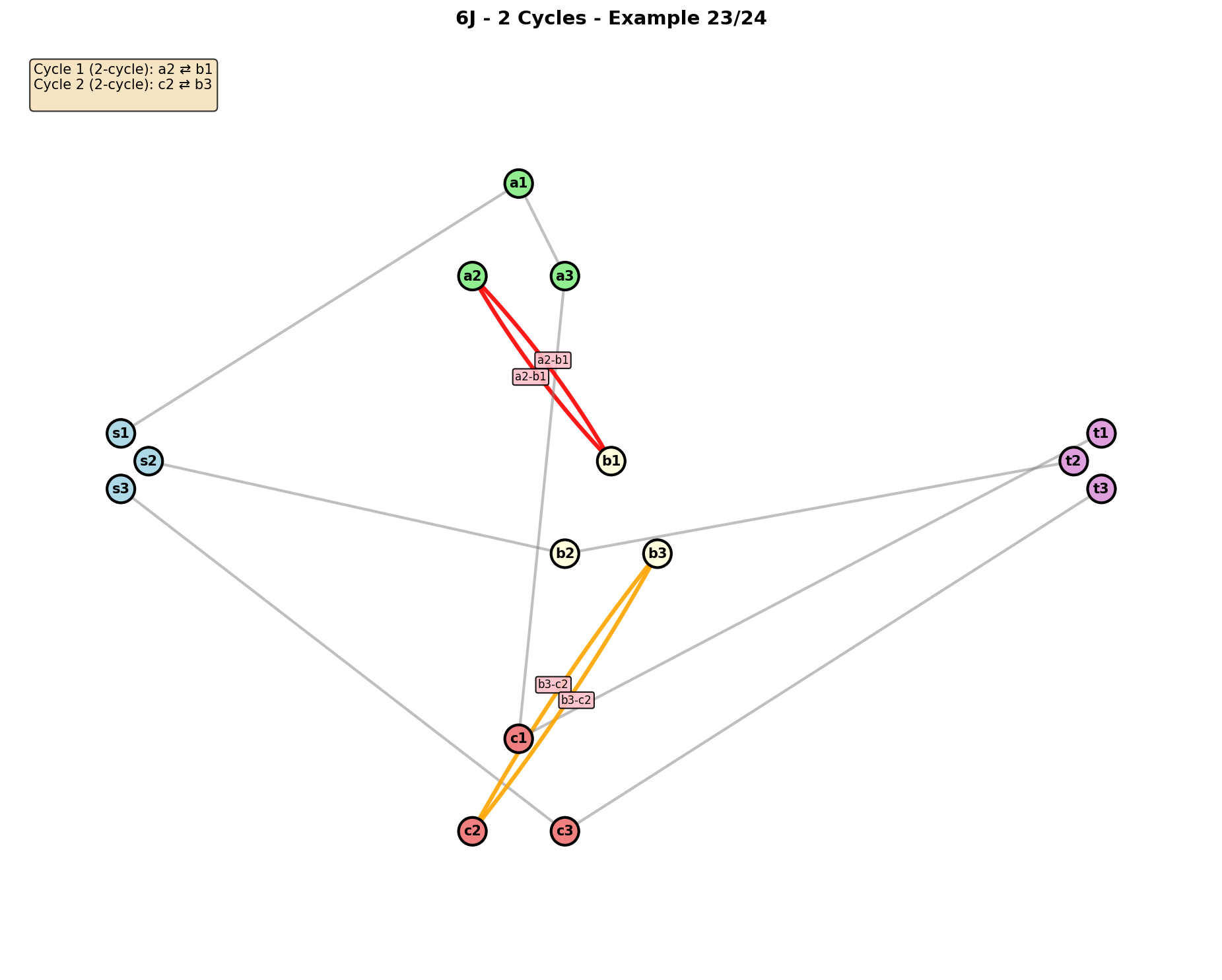}{$6j$--6; idx 5553}{2cycles\_23.png}{%
  \SDCatalogueSurface{\crawCard{11}{SixEleven}{Standard}{6j-6}{}{Hidden}}}

\pagebreak

\paragraph{\texorpdfstring{$C_3$: one parent, three embeddings.}{C3: one parent, three embeddings.}}
The single matching has three embeddings.

\SDCatalogueRow{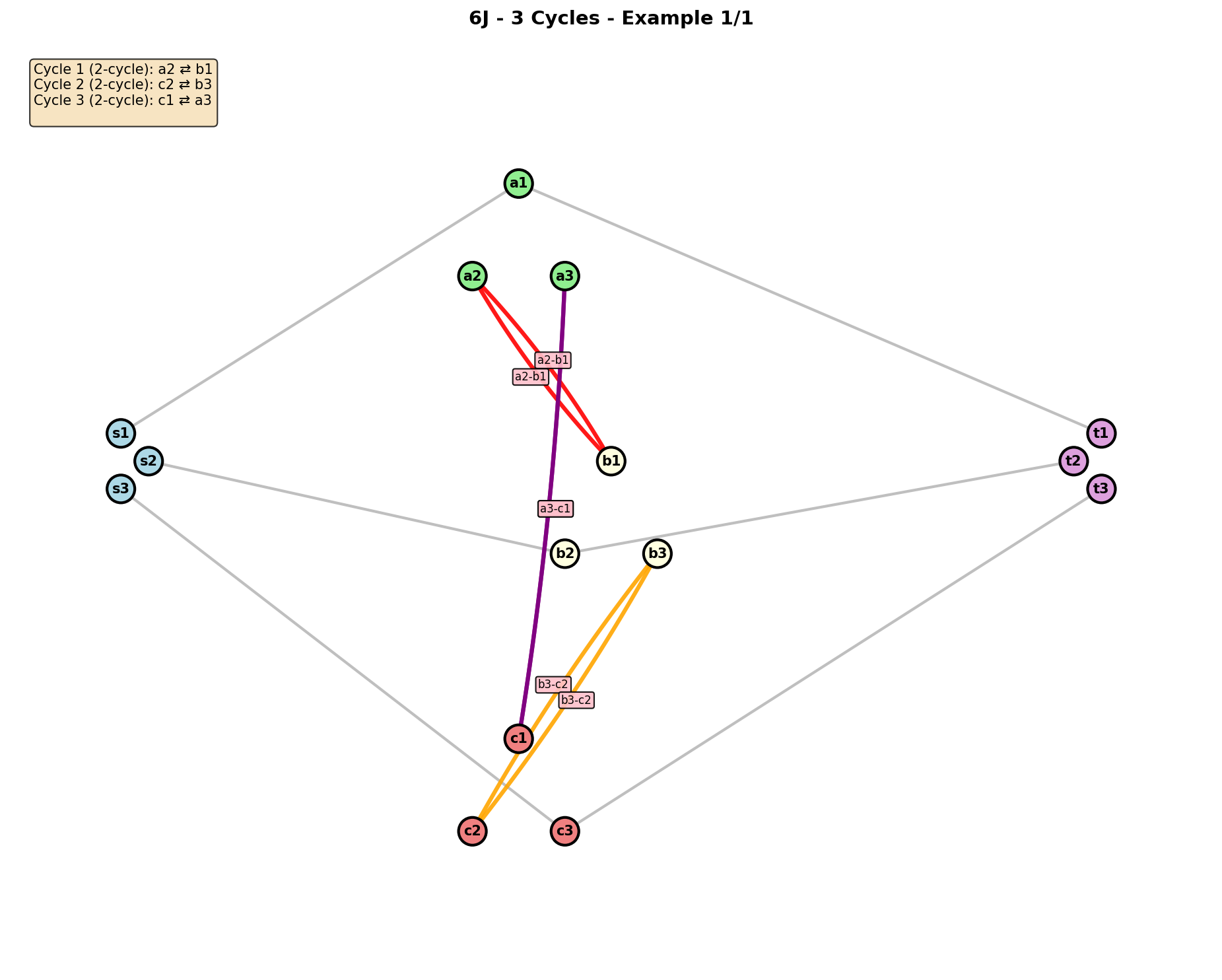}{$6j$--7; idx 409}{3cycles\_1.png}{%
  \SDCatalogueSurface{\crawCard{1}{SixColoredOne}{Standard}{6j-7-1}{}{UpperInner}}
  \SDCatalogueSurface{\crawCard{2}{SixColoredTwo}{Standard}{6j-7-2}{}{Lower}}
  \SDCatalogueSurface{\crawCard{4}{SixColoredFour}{Standard}{6j-7-3}{}{FourUpper}}}

\clearpage
\paragraph{\texorpdfstring{$D_1$: three parents, three embeddings.}{D1: three parents, three embeddings.}}
\mbox{}\par\nopagebreak

\SDCatalogueRow{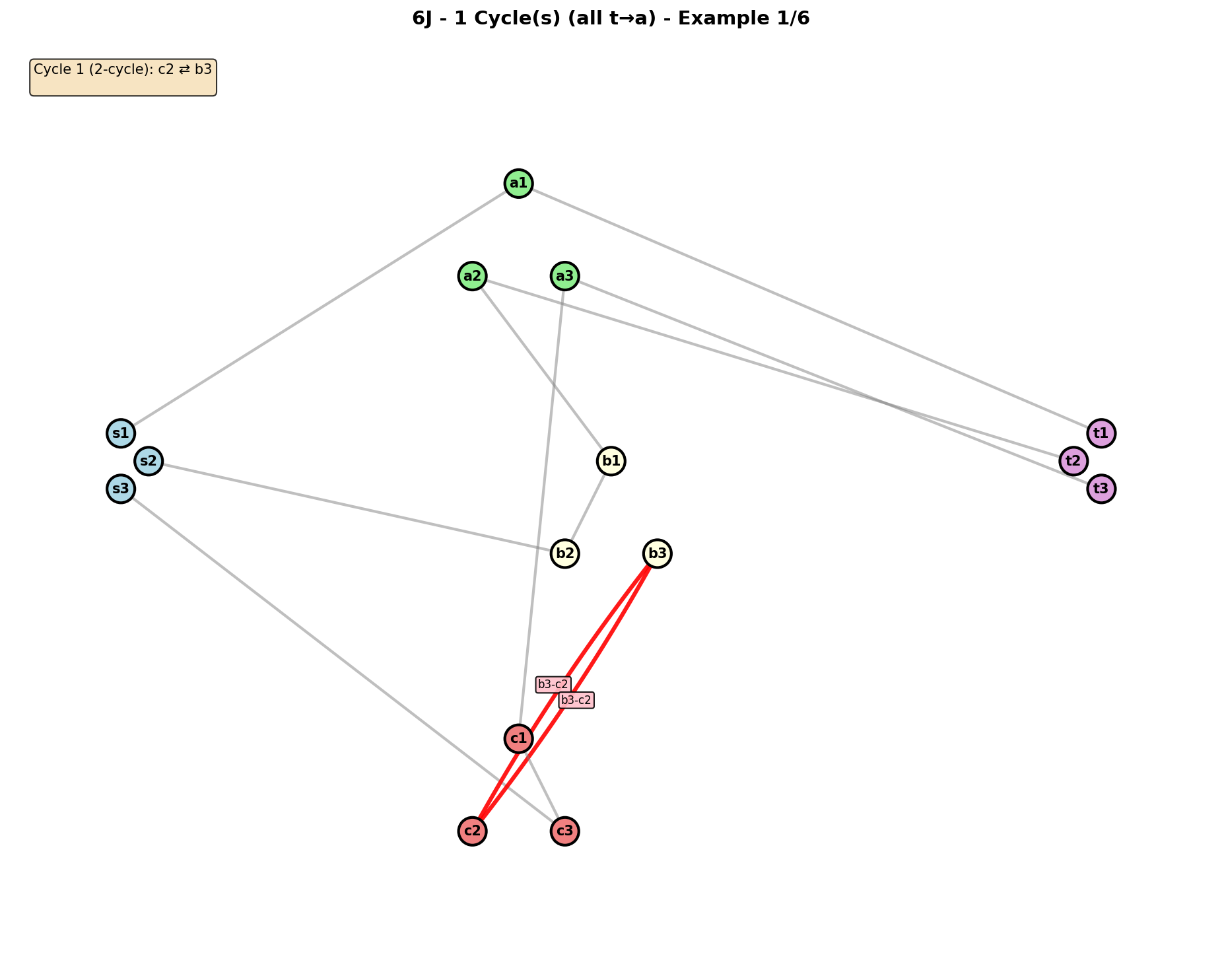}{$6j$--8; idx 1}{t\_same\_1cycles\_1.png}{%
  \SDCatalogueSurface{\crawDisconnectedSixCard{16}{16}{6j-8}}}
\SDCatalogueRow{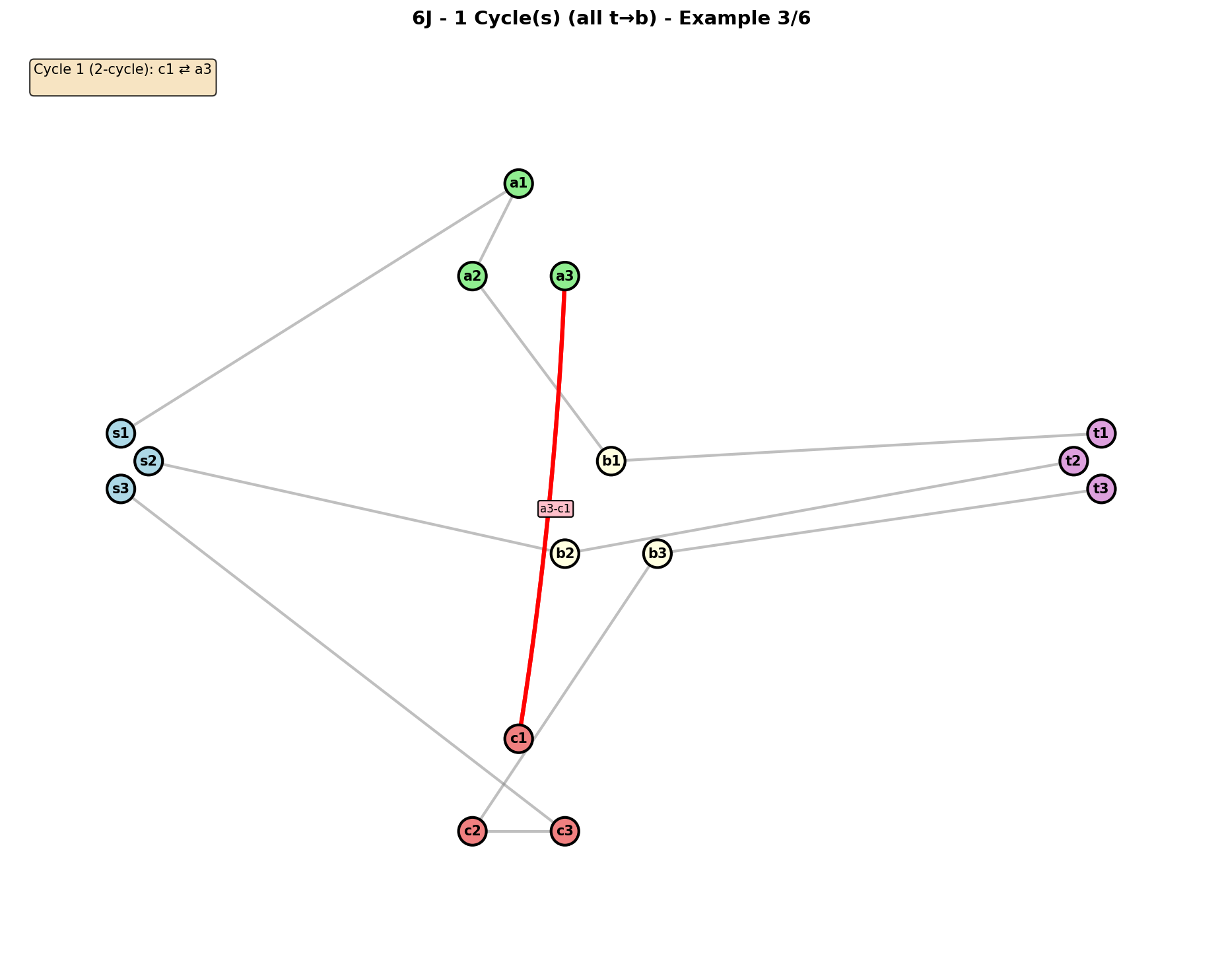}{$6j$--9; idx 2880}{t\_same\_1cycles\_3.png}{%
  \SDCatalogueSurface{\crawDisconnectedSixCard{18}{18}{6j-9}}}
\SDCatalogueRow{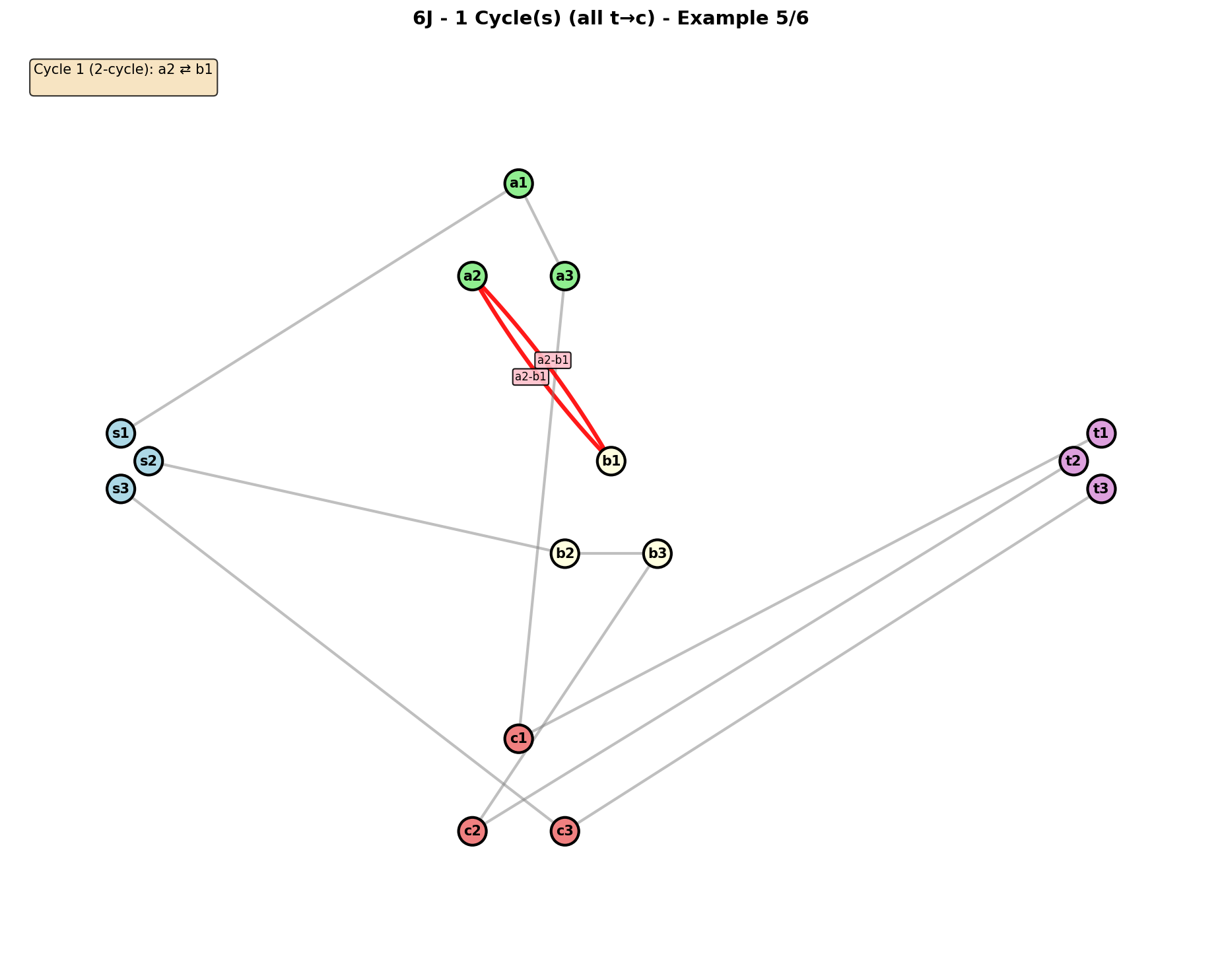}{$6j$--10; idx 5763}{t\_same\_1cycles\_5.png}{%
  \SDCatalogueSurface{\crawDisconnectedSixCard{17}{17}{6j-10}}}

\paragraph{\texorpdfstring{$D_2$: no retained parent.}{D2: no retained parent.}}
Thus the open-$6j$ catalogue contains ten parents and twelve embeddings.

\clearpage
\subsection{Pillow representatives}
\label{app:final-pillow-catalogue}

\paragraph{\texorpdfstring{$C_2$: six parents, six embeddings.}{C2: six parents, six embeddings.}}
\mbox{}\par\nopagebreak

\SDCatalogueRow{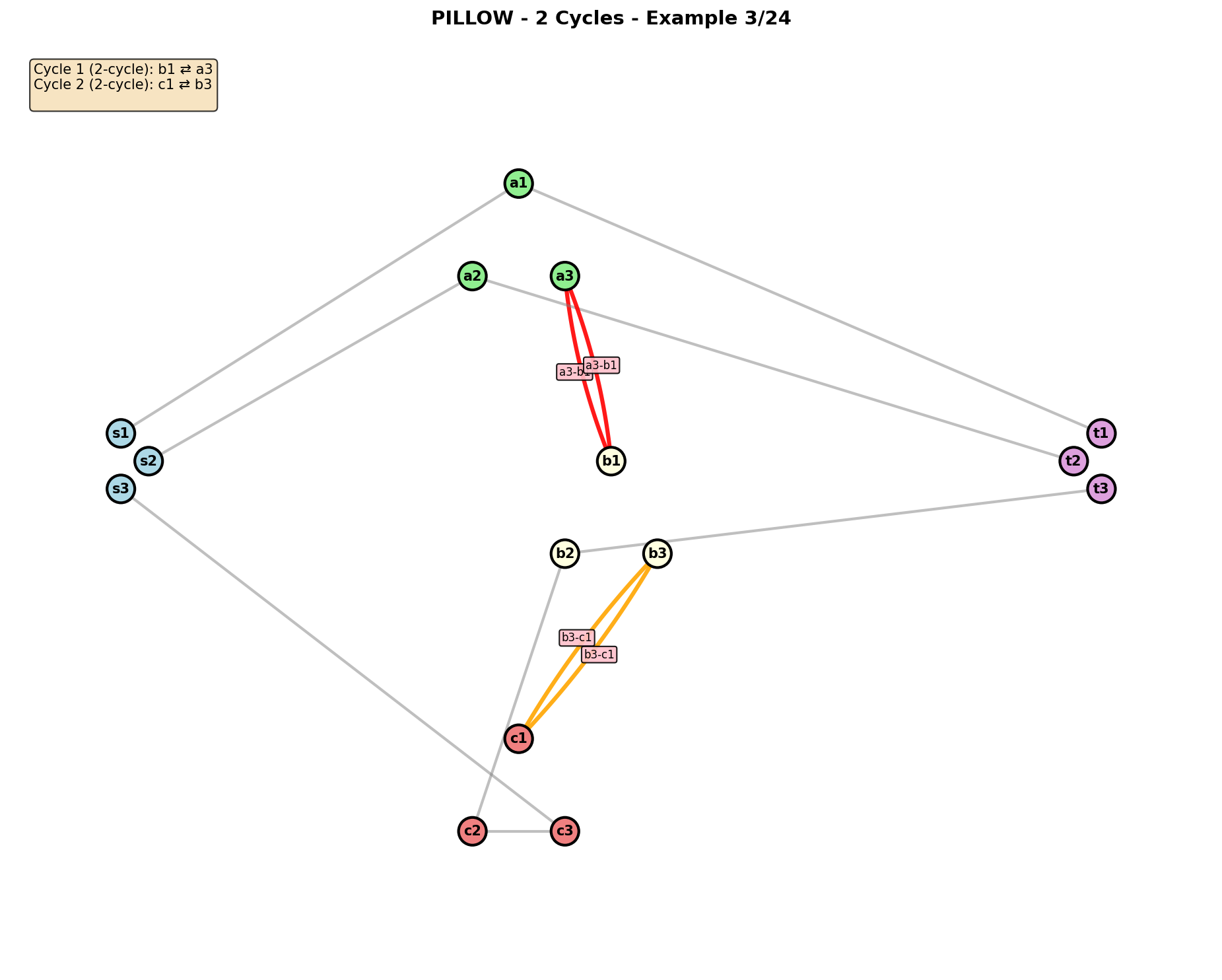}{$P$--1; idx 30}{2cycles\_3.png}{%
  \SDCatalogueSurface{\crawCard{6}{PillowPOne}{Standard}{P-1}{}{Hidden}}}
\SDCatalogueRow{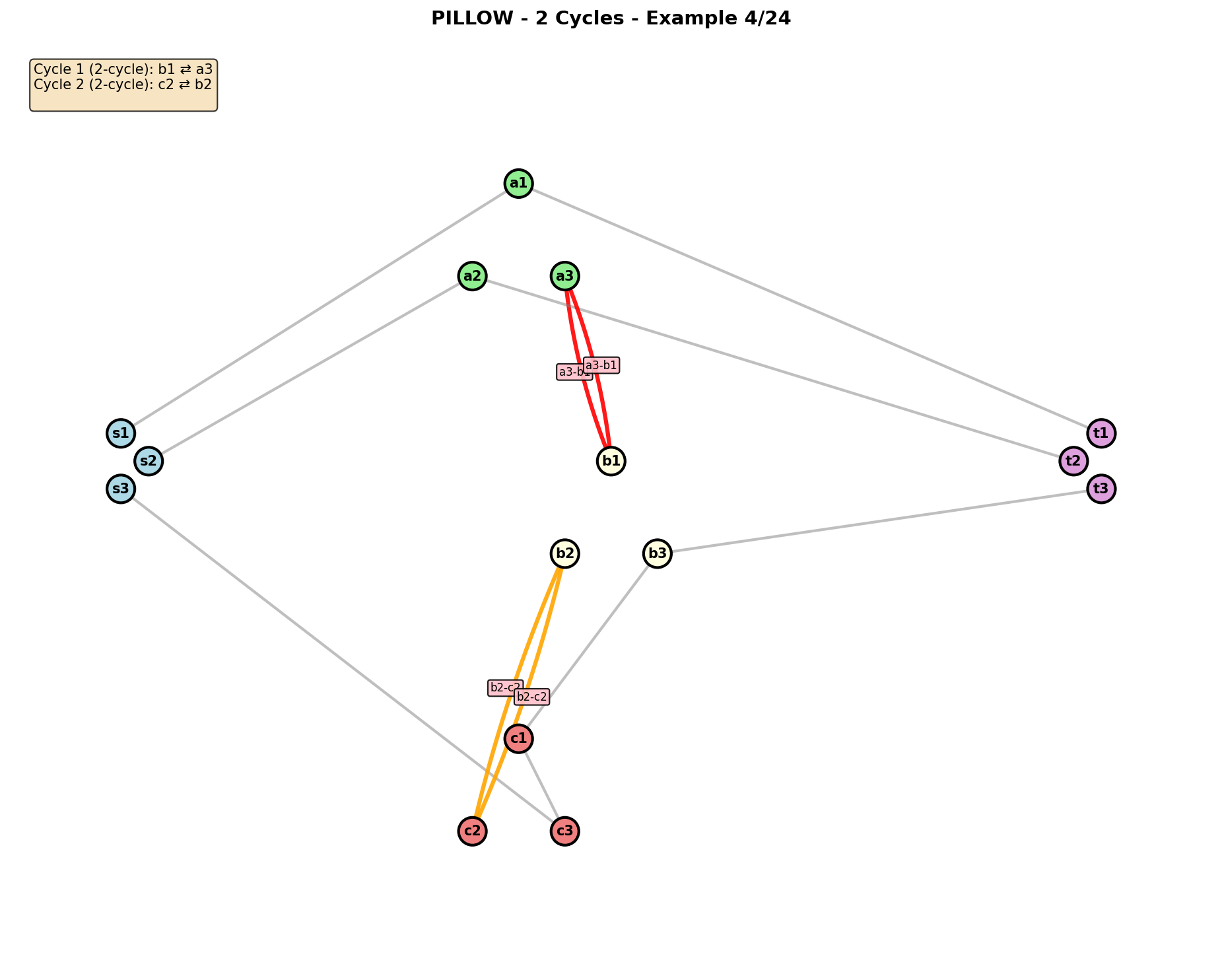}{$P$--2; idx 46}{2cycles\_4.png}{%
  \SDCatalogueSurface{\crawCard{5}{PillowPTwo}{Standard}{P-2}{}{Hidden}}}
\SDCatalogueRow{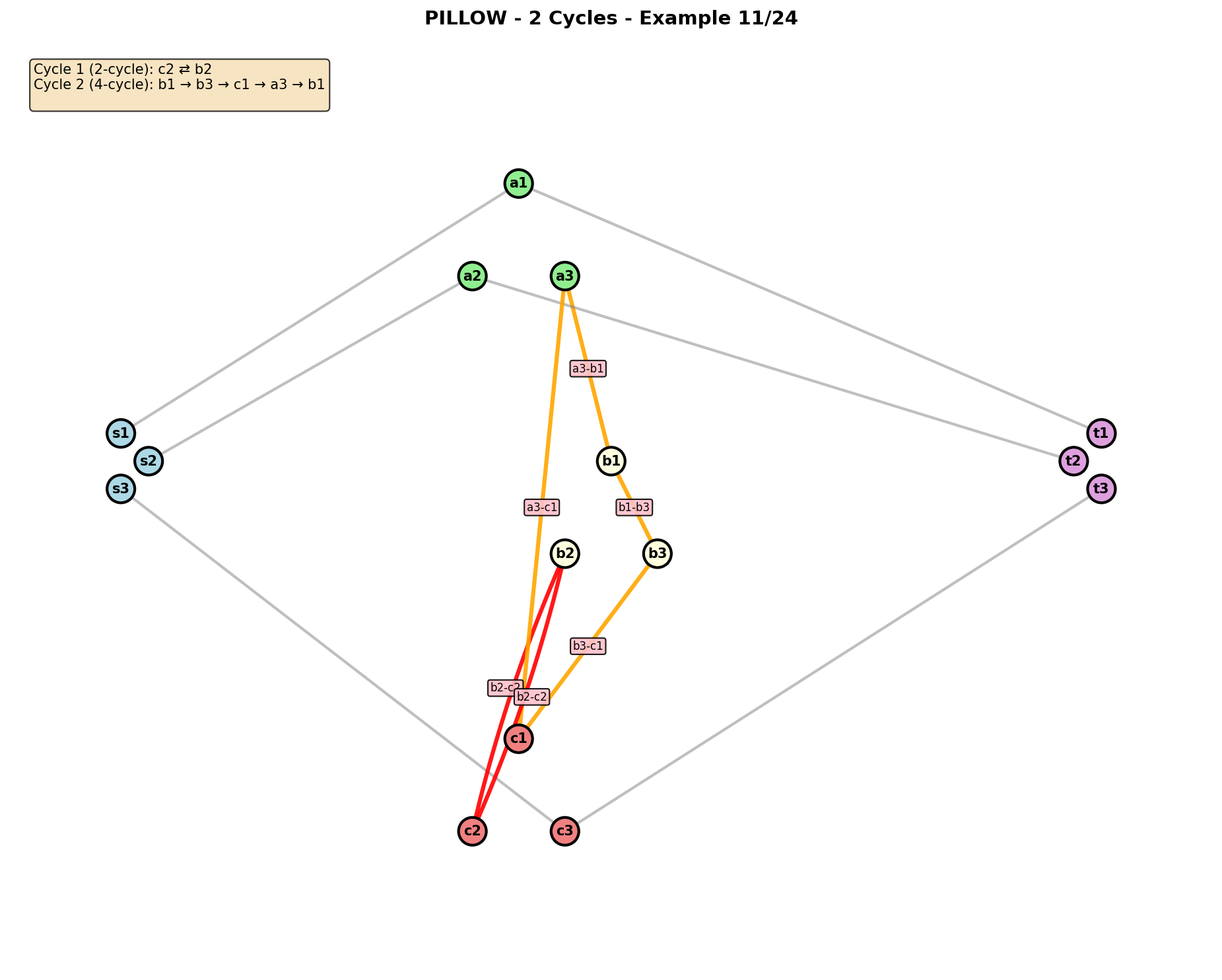}{$P$--3; idx 100}{2cycles\_11.png}{%
  \SDCatalogueSurface{\crawCard{4}{PillowPThree}{Standard}{P-3}{}{Hidden}}}
\SDCatalogueRow{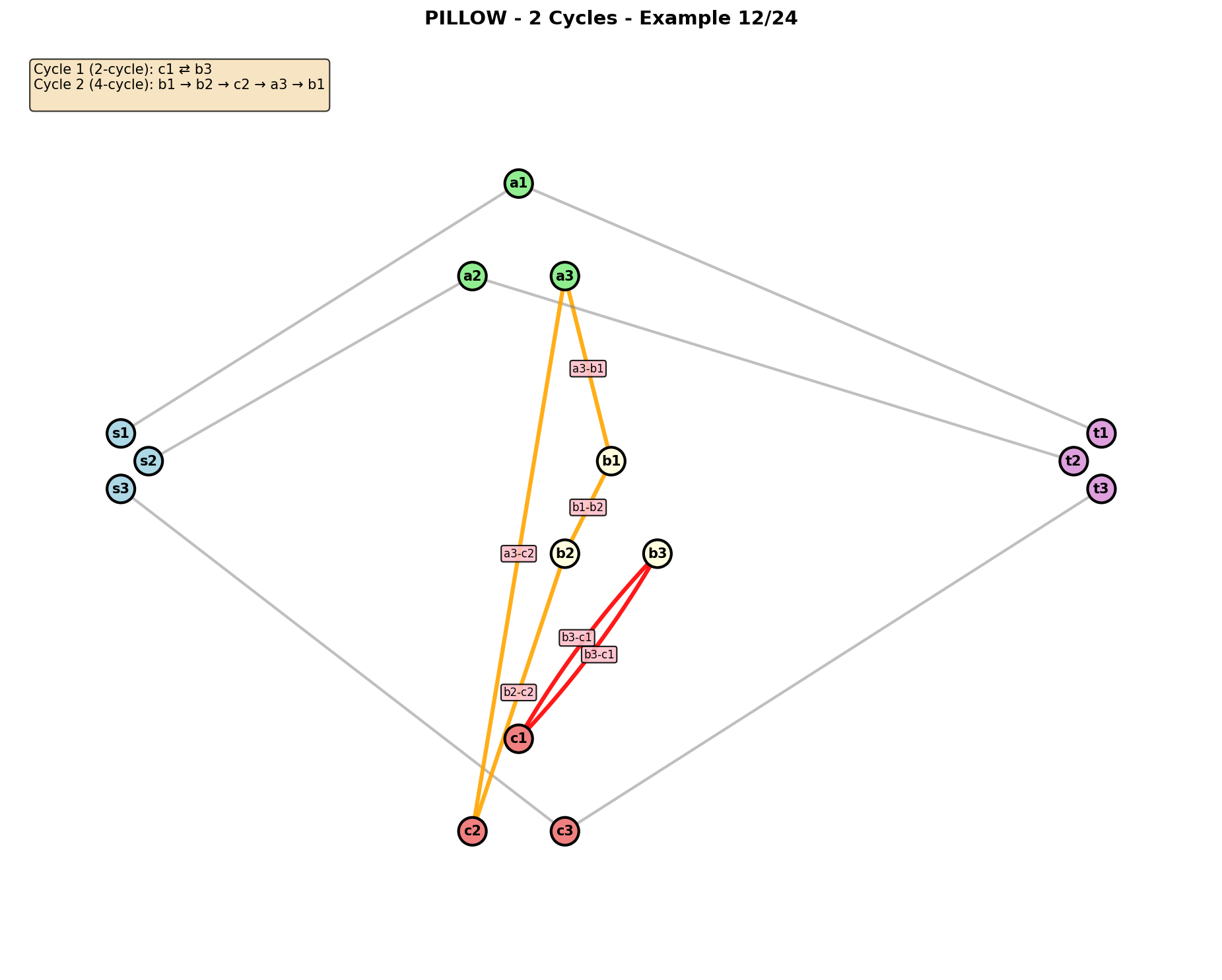}{$P$--4; idx 102}{2cycles\_12.png}{%
  \SDCatalogueSurface{\crawCard{3}{PillowPFour}{Standard}{P-4}{}{Hidden}}}
\SDCatalogueRow{figures/pillow_2cycles_18.png}{$P$--5; idx 726}{2cycles\_18.png}{%
  \SDCatalogueSurface{\crawCard{1}{PillowPFive}{Standard}{P-5}{}{Hidden}}}
\SDCatalogueRow{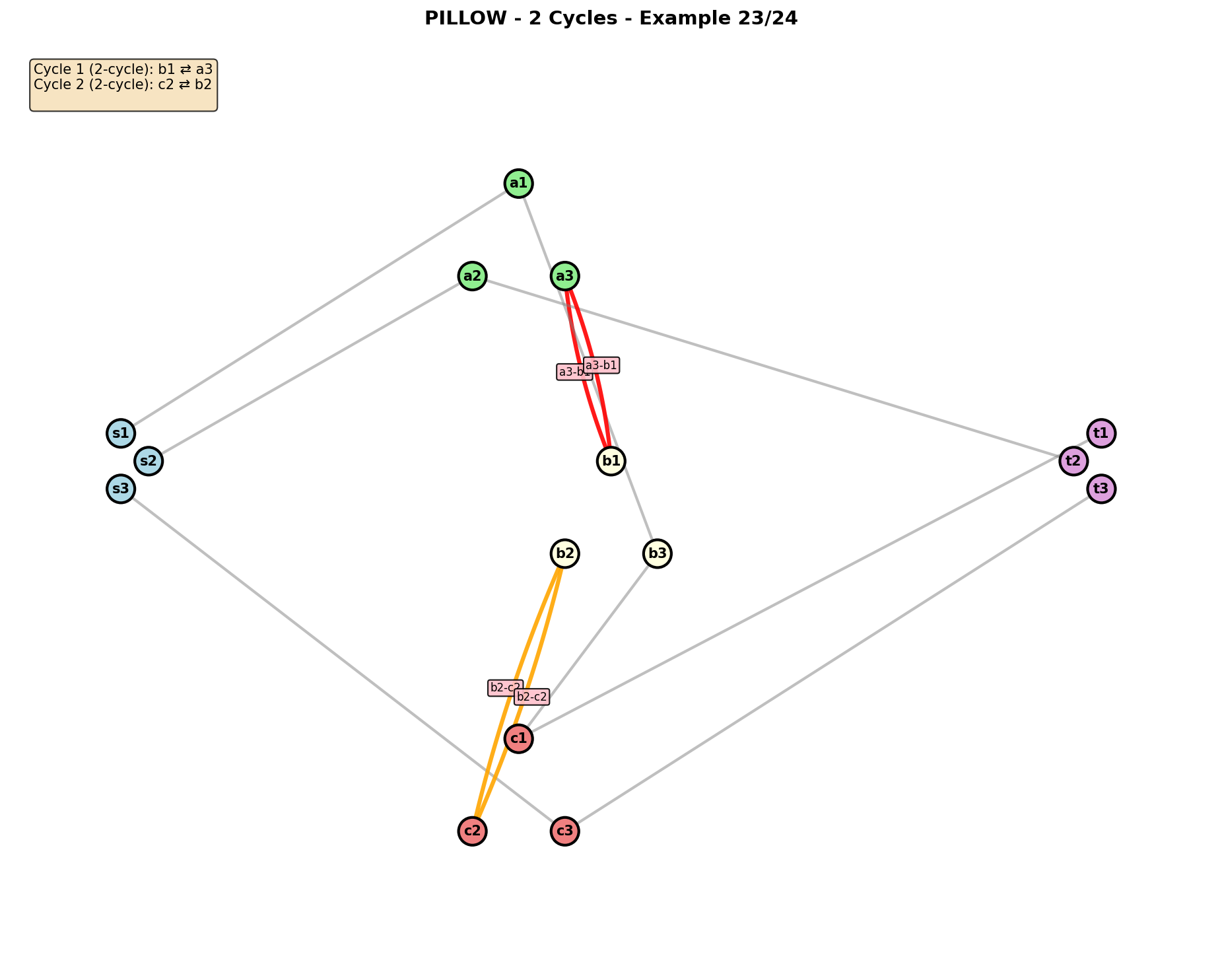}{$P$--6; idx 5244}{2cycles\_23.png}{%
  \SDCatalogueSurface{\crawCard{2}{PillowPSix}{Standard}{P-6}{}{Hidden}}}

\paragraph{\texorpdfstring{$C_3$: one parent, three embeddings.}{C3: one parent, three embeddings.}}
\mbox{}\par\nopagebreak

\SDCatalogueRow{figures/pillow_3cycles_1.png}{$P$--7; idx 92}{3cycles\_1.png}{%
  \SDCatalogueSurface{\crawCard{18}{PillowColoredEighteen}{Standard}{P-7-1}{}{Lower}}
  \SDCatalogueSurface{\crawCard{15}{PillowColoredFifteen}{Standard}{P-7-2}{}{Lower}}
  \SDCatalogueSurface{\crawCard{14}{PillowColoredFourteen}{Standard}{P-7-3}{}{Lower}}}

\clearpage
\paragraph{\texorpdfstring{$D_1$: three parents, three embeddings.}{D1: three parents, three embeddings.}}
\mbox{}\par\nopagebreak

\SDCatalogueRow{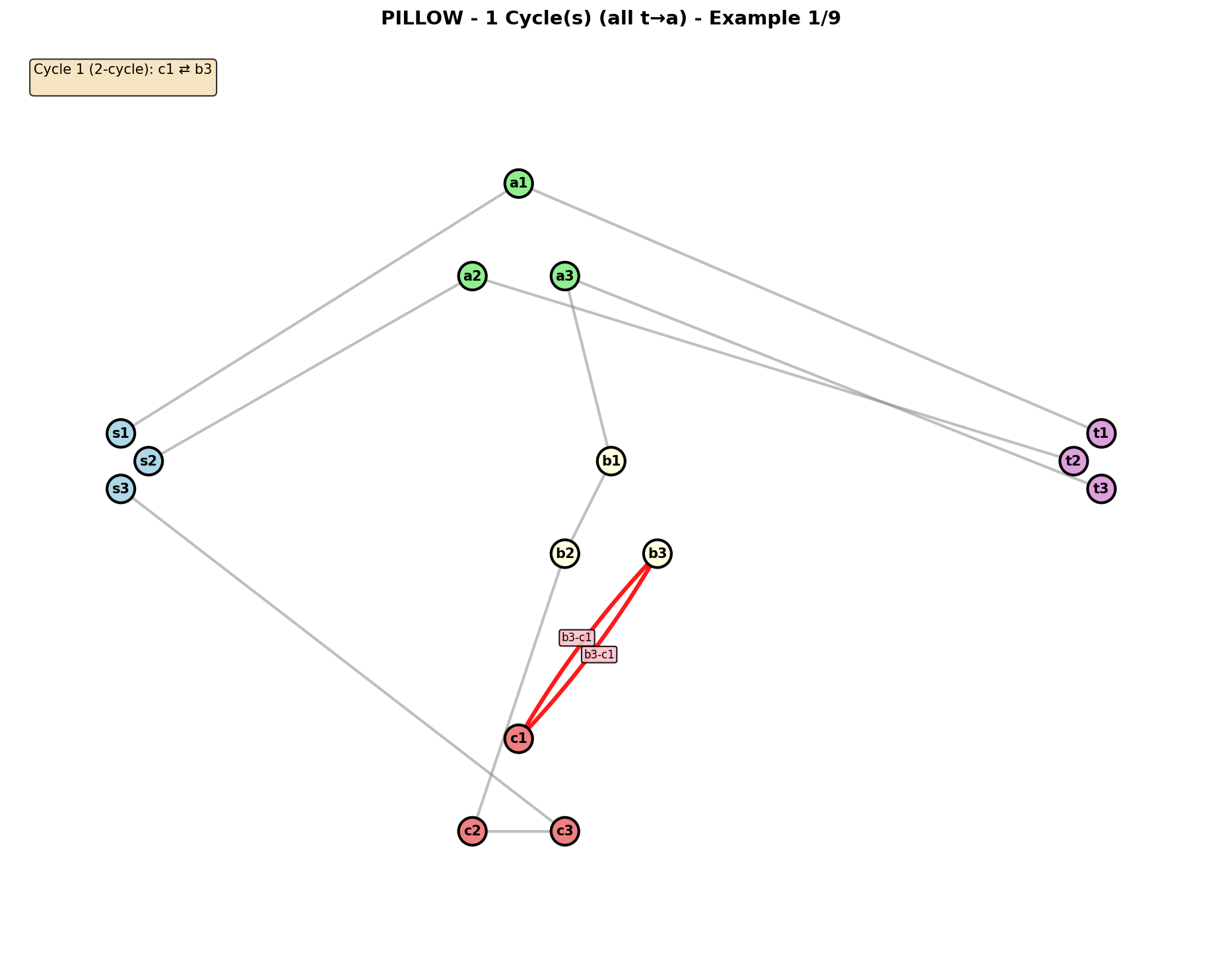}{$P$--8; idx 0}{t\_same\_1cycles\_1.png}{%
  \SDCatalogueSurface{\crawPillowEarlyDisconnectedCard{9}{9}{P-8}}}
\SDCatalogueRow{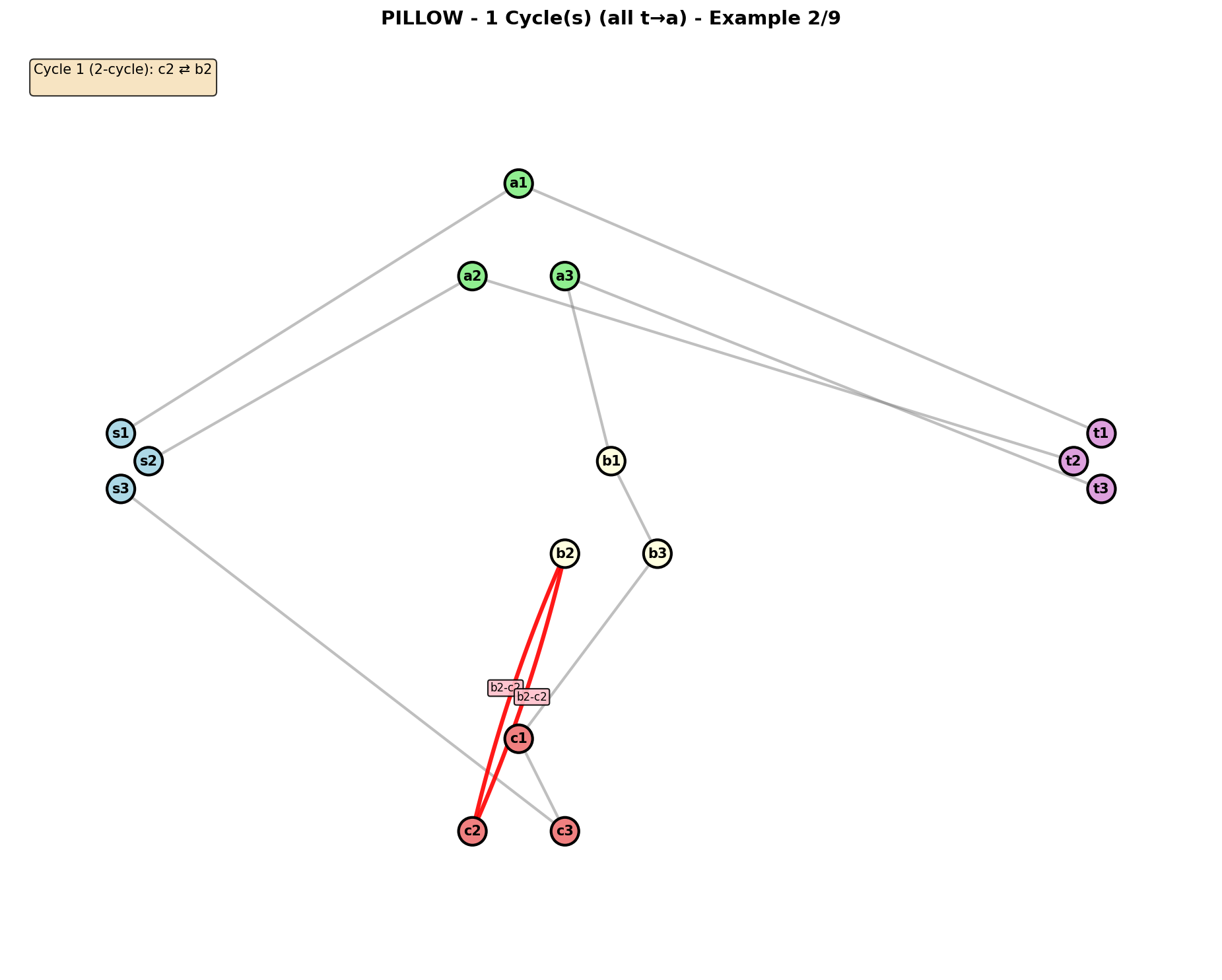}{$P$--9; idx 4}{t\_same\_1cycles\_2.png}{%
  \SDCatalogueSurface{\crawPillowEarlyDisconnectedCard{8}{8}{P-9}}}
\SDCatalogueRow{figures/pillow_t_same_1cycles_8.png}{$P$--10; idx 5774}{t\_same\_1cycles\_8.png}{%
  \SDCatalogueSurface{\crawPillowEarlyDisconnectedCard{7}{7}{P-10}}}

\paragraph{\texorpdfstring{$D_2$: one parent, two embeddings.}{D2: one parent, two embeddings.}}
\mbox{}\par\nopagebreak

\SDCatalogueRow{figures/pillow_t_same_2cycles_7.png}{$P$--11; idx 14}{t\_same\_2cycles\_7.png}{%
  \SDCatalogueSurface{\crawDisconnectedPillowCard{10}{1}{P-11-1}}
  \SDCatalogueSurface{\crawDisconnectedPillowCard{11}{2}{P-11-2}}}

The Pillow catalogue therefore contains eleven parents and fourteen
embeddings.  

\endgroup

\clearpage
\section{Embedding identifications}
\label{app:embedding-atlas}

This appendix explains which surface drawings represent the same embedding
and which do not.  We start with the code graphs that are
retained after the surgery check. Throughout,
the boundary labels, their cyclic orders, and the surface orientation are
held fixed.

We present three results. First, the four views of the Pillow \(D_2\) graph
reduce to two embeddings.  Next, the connected \(6j\) \(C_3\) graph has
three embeddings; handle migration shows explicitly why two different
drawings can belong to the same one.  Finally, we display a local \(F\)-move
on each \(6j\) embedding, establishing its correspondence with the
same-numbered Pillow representative.  The final representatives are listed in
Appendix~\ref{app:surface-ledger}.

\begingroup
\SDFigureLabels{position=upper-left}
\tikzset{sdemb theta title/.append style={align=left,inner xsep=0pt}}
\sdembBuildThetaTwoClassBox
\sdembBuildKFourThreeClassBox
\crawBuildFMoveThreeClassMapBox
\subsection{Pillow \texorpdfstring{\(D_2\)}{D2}: four views, two embeddings}
\label{app:theta-recutting}

The Pillow \(D_2\) code graph, P-11, has two embeddings, P-11-1 and
P-11-2, each shown in two views below.  To see both the equivalence within each pair and the
difference between the pairs, cap the \(b,c\) cuffs and shrink each cap to
a vertex.  This gives the theta graphs embedded on the torus. See the last column below.
\begin{equation}
  \vcenter{\hbox{%
    \begin{tabular}{@{}c@{\hspace{2mm}}c@{\hspace{2mm}}c@{\hspace{2mm}}c@{\hspace{2mm}}c@{}}
      {\small View I} & & {\small View II} & & {\small Theta embedding}\\[1mm]
      \resizebox{.31\linewidth}{!}{\crawDisconnectedPillowCard{10}{1}{P-11-1}} &
      \(\simeq\) &
      \resizebox{.31\linewidth}{!}{\crawDisconnectedPillowCard{10}{3}{P-11-1}} &
      \(\longrightarrow\) &
      \resizebox{.18\linewidth}{!}{%
        \begin{tikzpicture}[x=.82cm,y=.82cm,baseline=(thetaMid)]
          \coordinate (thetaMid) at (2,1.35);
          \sdembThetaPanel{10}{P-11-1}
        \end{tikzpicture}} \\[3mm]
      \resizebox{.31\linewidth}{!}{\crawDisconnectedPillowCard{11}{2}{P-11-2}} &
      \(\simeq\) &
      \resizebox{.31\linewidth}{!}{\crawDisconnectedPillowCard{11}{4}{P-11-2}} &
      \(\longrightarrow\) &
      \resizebox{.18\linewidth}{!}{%
        \begin{tikzpicture}[x=.82cm,y=.82cm,baseline=(thetaMid)]
          \coordinate (thetaMid) at (2,1.35);
          \sdembThetaPanel{11}{P-11-2}
        \end{tikzpicture}}
    \end{tabular}%
  }}
  \label{eq:theta-reduction}
\end{equation}

Within each row, views I and II differ only in where the surface is cut
open to make a flat picture.  After capping the cuffs, moving this cut
changes which parts of the theta graph appear at the front or rear, but
does not move its labeled edges relative to the handle.  Each row therefore
counts as one embedding, not two.

The two rows are genuinely different.  In P-11-1, blue and green bound a
disk; in P-11-2, blue and purple do.
The colors identify fixed labeled edges, so a deformation preserving those
labels cannot exchange the two pairs.  This is not an extra choice of sign
or orientation.

The result is one code graph with two distinct surface embeddings, not four.
Figure~\ref{fig:theta-ledger-representatives} shows one full surface drawing
for each of P-11-1 and P-11-2.

\FloatBarrier
\subsection{\texorpdfstring{\(6j\) \(C_3\)}{6j C3}: three embeddings, with duplicate views}
\label{app:c3-characters}

The connected code graph 6j-7 also has several surface drawings.  Two
different questions arise: how to distinguish its three retained
embeddings, and how to recognize repeated drawings of one embedding.
Winding around the handle answers the first question; the explicit handle
migration below answers the second in a representative example.

\subsubsection{Distinguishing the three embeddings}

Cap the four cuffs \(a,b,c,t\) and shrink each cap to a vertex.  The six
colored Wilson segments then form \(K_4\).  Keep the vertex labels and the
oriented \(abc\) cycle fixed: permuting the vertices would change the
marked data.

As in Section~\ref{sec:c3-embeddings}, \([\gamma]\) records the winding of
an oriented cycle around the marked handle, ignoring contractible pieces.
The three retained embeddings differ by which triangle through \(t\)
carries the fixed nonzero winding \([abc]\): \(abt\), \(bct\), or \(cat\).
These are the three triples in equation~\eqref{eq:k4-three-characters},
subject to \([abc]=[abt]+[bct]+[cat]\) from
equation~\eqref{eq:k4-relative-character}.
Figure~\ref{fig:k4-three-classes} shows them in the same order.

\begin{figure}[!htb]
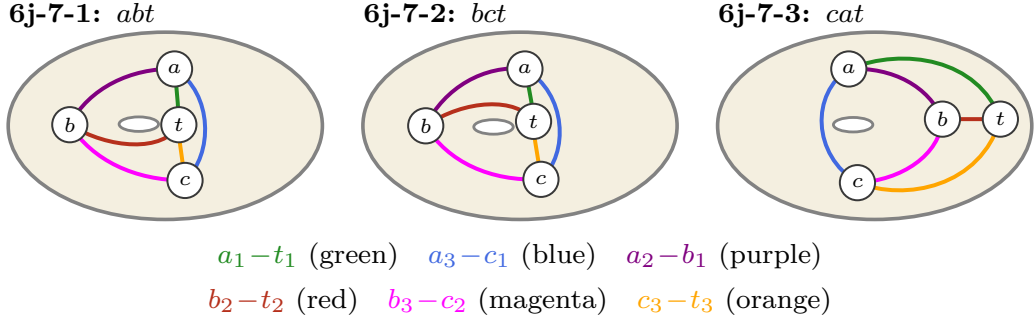

  \centering
  \resizebox{.80\linewidth}{!}{\usebox{\sdembKFourThreeClassBox}}
  \caption{The three \(6j\) embeddings, distinguished by which triangle
  carries the fixed handle winding: \(abt\), \(bct\), or \(cat\).}
  \label{fig:k4-three-classes}
\end{figure}

A deformation preserving the labels cannot change these winding data, so
the three cases are distinct.  Reversing orientation also reverses the
fixed \(abc\) cycle and therefore does not give another three embeddings
in this count.

Different winding data prove that two embeddings differ.  Matching winding
data alone, however, do not prove that two drawings are equivalent; for
that we need a deformation preserving the marked data.

\FloatBarrier
\subsubsection{An apparent extra embedding: handle migration}
\label{app:handle-migration}

Drawing the handle in a different position may seem to produce an embedding
beyond the three already listed.  The sequence below shows why this need
not be new: the first and last pictures look different, but both represent
6j-7-3, with winding data \((0,0,[abc])\).  Lift the handle behind the
visible surface, carry it below the Wilson network, move the lower loop
around it, and lay the handle back into the surface.  This relates the two
views without changing the marked embedding.

\begin{center}
\resizebox{.90\linewidth}{!}{%
\begin{minipage}{\linewidth}
  \czrmTwoCards
    {\CZeroHandleEndpointFour}
    {\CZeroHandleStageTwo}
  \czrmTwoCards
    {\CZeroHandleStageThree}
    {\CZeroHandleStageFour}
  \noindent\hspace*{.2575\linewidth}\CZeroHandleEndpointFive\par
\end{minipage}}
\end{center}

The fixed half remains black; the six completion edges keep the colors of
Figure~\ref{fig:k4-three-classes} throughout the sequence.  In the middle
frames, the paired dotted gray contours show the handle behind the surface,
with the same dashed blue strand between them.  Solid and dashed lines
remain the front and rear portions of the same network: three open lines,
two front loops, and one rear loop.  No line is added, cut, or reconnected,
and endpoint labels and cyclic orders are unchanged.  The frames depict
a smooth deformation of the marked surface (an ambient isotopy), so the
two views count only once.

This equivalence concerns the embedding count, not the code-graph count in
equation~\eqref{eq:surface-parent-count}.  Other apparently new drawings
require their own equivalence check.

\subsection{Local \texorpdfstring{\(F\)}{F}-moves on the three
\texorpdfstring{\(6j\) \(C_3\)}{6j C3} embeddings}
\label{app:local-f-move}

Figure~\ref{fig:three-embeddings} shows the three \(6j\) classes of
Figure~\ref{fig:k4-three-classes} in a layout adapted to the local
\(F\)-move.  The dashed disk contains \(a,b\) and the purple edge;
reconnecting its four colored ends produces the Pillow graph.  The
exterior lines and the handle, marked by the small gray oval, stay fixed.

\begin{figure}[!htb]
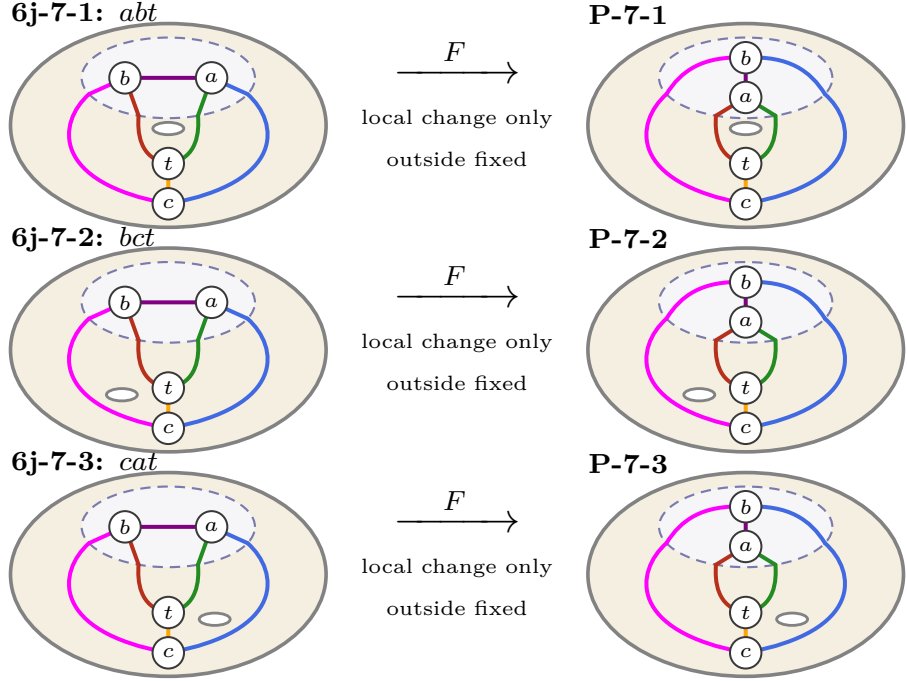

\centering
\resizebox{.70\linewidth}{!}{\usebox{\crawFMoveThreeClassMapBox}}
\caption{The local \(F\)-move pairs each \(6j\) \(C_3\) embedding with the
same-numbered Pillow embedding.}
\label{fig:three-embeddings}
\end{figure}

\FloatBarrier
On the Pillow side, blue--magenta forms the \(b,c\) loop and green--red
forms the \(a,t\) loop.  We use the latter and two larger closed paths to
record the winding.  The arrows and ordered colors below specify each path
unambiguously, including which of the parallel edges to follow:
\begin{equation}
 \begin{array}{ccl}
  \toprule
  \text{Oriented Pillow cycle}&\text{Colors in traversal order}
    &\text{Winding in }6j\text{ cycles}\\
  \midrule
  a\to t\to a&\text{green, red}&-[abt]\\
  a\to b\to c\to t\to a&\text{purple, magenta, orange, green}
    &[abt]+[bct]\\
  b\to a\to t\to c\to b&\text{purple, red, orange, blue}
    &[abt]+[cat]\\
  \bottomrule
 \end{array}
 \label{eq:pillow-cycle-map}
\end{equation}
These relations follow by comparing the unchanged paths outside the disk;
paths inside it add no handle winding.  The first loop traverses the
exterior part of \(abt\) backwards, explaining the minus sign.  The other
two combine the indicated triangle paths.

Reading the \(6j\) windings in the order \((abt,bct,cat)\) and the Pillow
windings in the row order above, the three cases give
\begin{equation}
 \begin{aligned}
 \bigl([abc],0,0\bigr)&\longmapsto
   \bigl(-[abc],[abc],[abc]\bigr),\\
 \bigl(0,[abc],0\bigr)&\longmapsto
   \bigl(0,[abc],0\bigr),\\
 \bigl(0,0,[abc]\bigr)&\longmapsto
   \bigl(0,0,[abc]\bigr).
 \end{aligned}
 \label{eq:k4-pillow-character-map}
\end{equation}
The three local outputs are therefore distinct.\footnote{This comparison
concerns the graphs; the analytic \(F\)-move also includes a fusion kernel
and an internal sum or integral~\cite{Moore:1988qv,Ponsot:2000mt}.}
They are P-7-1, P-7-2, and P-7-3 in
Figure~\ref{fig:c3-ledger-representatives}.  The graph and its complements agree, not just
the winding data.  The inverse move recovers 6j-7-\(i\) from P-7-\(i\),
establishing the one-to-one correspondence for \(i=1,2,3\).
\FloatBarrier

\endgroup

\clearpage
\bibliographystyle{JHEP}
\bibliography{references}

\end{document}